\documentclass[10ptm, twoside, toc=flat]{article}

\usepackage{geometry}

\usepackage{graphicx}
\usepackage{graphics}
\usepackage{amsmath}
\usepackage[obeyspaces,hyphens]{url}
\usepackage[font={footnotesize}]{caption}
\usepackage{subcaption}
\usepackage{parskip}
\usepackage{longtable}
\usepackage{booktabs}
\usepackage{colortbl}
\usepackage{float}
\usepackage{afterpage}
\usepackage{csquotes}
\usepackage{siunitx}
\usepackage{fancyhdr}

\usepackage{abstract}

\usepackage[dvipsnames]{xcolor}
\usepackage[textsize=footnotesize, backgroundcolor=white, bordercolor=red]{todonotes}

\usepackage{endnotes}
\makeatletter
\def\enoteheading{\section*{\notesname
  \@mkboth{\MakeUppercase{\notesname}}{\MakeUppercase{\notesname}}}%
  \mbox{}\par\vskip-2.3\baselineskip\noindent\rule{.5\textwidth}{0.4pt}\par\vskip\baselineskip}
\makeatother

\title{OpenPBR: Novel Features and Implementation Details}

\author{
JAMIE PORTSMOUTH, Autodesk \\[.1cm]
PETER KUTZ,       Adobe    \\[.1cm]
STEPHEN HILL,     Lucasfilm
}

\date{\today}

\usepackage[
  pdfusetitle,
  pdfsubject={Physically Based Shading},
  pdfkeywords={Shading, Rendering},
  breaklinks,
  colorlinks=true,
  linkcolor=NavyBlue,
  citecolor=NavyBlue,
  urlcolor=NavyBlue
]{hyperref}

\usepackage[style=alphabetic,maxbibnames=99,giveninits=true,backend=biber]{biblatex}
\usepackage[tt=false, type1=true]{libertine}
\usepackage[libertine]{newtxmath}
\usepackage{sectsty}

\newcommand*{\titlefont}{\LARGE\bfseries\sffamily}
\newcommand*{\authorfont}{\Large\sffamily}

\sectionfont{\normalfont\large\sffamily\MakeUppercase}
\subsectionfont{\normalfont\large\sffamily}
\subsubsectionfont{\normalfont\sffamily\itshape}

\usepackage{listings}

\definecolor{backcolour}{rgb}{1, 1, 1}
\definecolor{codegreen}{rgb}{0,0.6,0}
\definecolor{codegray}{rgb}{0.5,0.5,0.5}
\definecolor{codepurple}{rgb}{0.58,0,0.82}
\definecolor{codeblue}{rgb}{0,0.3,0.6}
\definecolor{darkblue}{rgb}{0,0.25,0.5}

\lstnewenvironment{code}[1][]
{
  \noindent
  \minipage{\linewidth}
  \vspace{0.5\baselineskip}
  \lstset{frame=single, #1}}
{\endminipage}

\newenvironment{inputcode}[1][]
{
  \noindent
  \minipage{\linewidth}
  \vspace{0.5\baselineskip}
  \lstset{frame=single, #1}}
{\endminipage}

\newcommand{\vecv}{\mathbf{v}}
\newcommand{\vecvv}{\mathbf{v'}}
\newcommand{\vecl}{\mathbf{l}}

\newcommand{\vecn}{\mathbf{n}}

\newcommand{\vech}{\mathbf{h}}
\newcommand{\vecb}{\mathbf{b}}

\newcommand\shorttitle{OpenPBR: Novel Features and Implementation Details}
\newcommand\authors{Portsmouth, Kutz, and Hill}

\fancypagestyle{plain}{
  \fancyhf{} 
}

\fancypagestyle{title}{
  \fancyhf{} 
  \fancyfoot[R]{Part of \textit{Physically Based Shading in Theory and Practice}, SIGGRAPH 2025 Course.}
  \fancyfootalign{-1cm}
}

\begin{document}

\thispagestyle{title} 


\makeatletter
{
  \noindent \titlefont \@title}
\normalsize
\\[3ex]
{\authorfont \@author}
\makeatother
\normalsize
\\[1ex]

\begin{figure}[htb]
  \centering
  \begin{subfigure}{.48\textwidth}
    \centering
    \includegraphics[width=\linewidth]{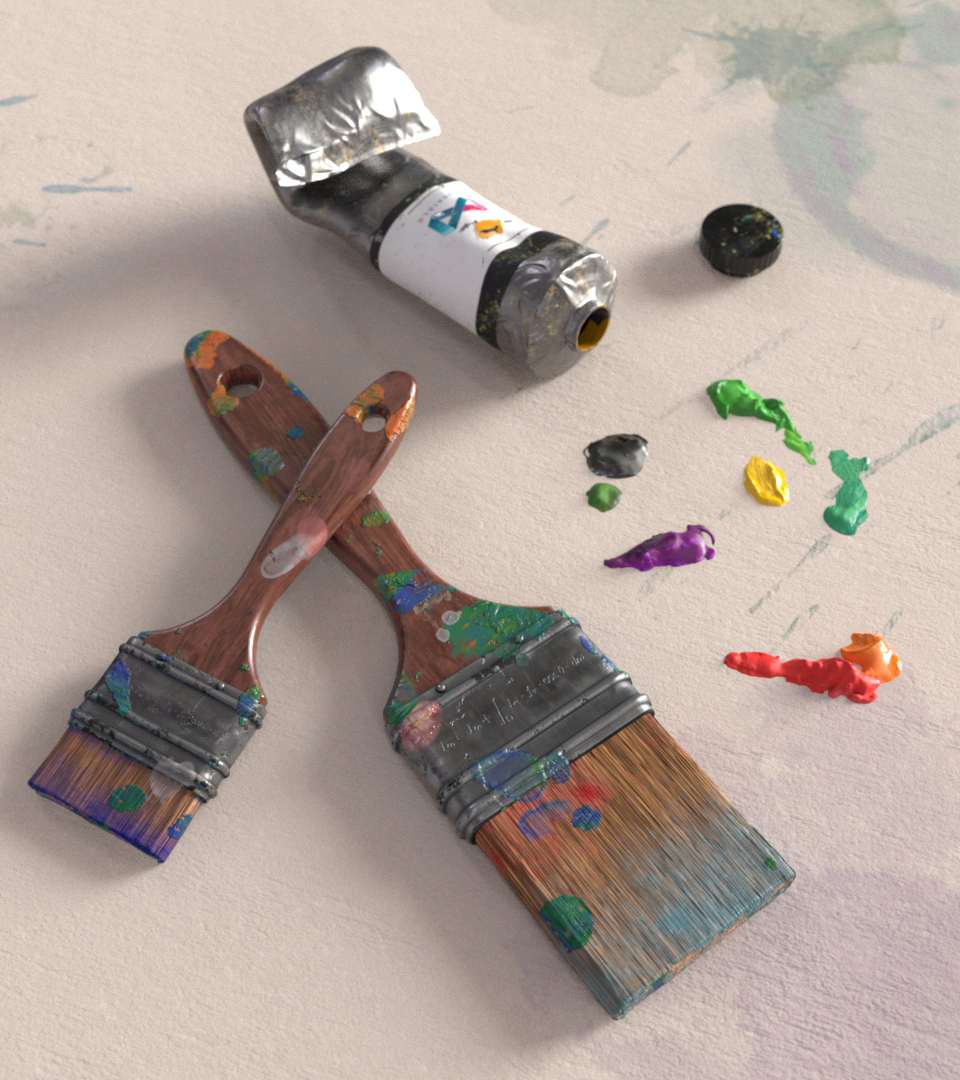}
  \end{subfigure}
  \begin{subfigure}{.48\textwidth}
    \centering
    \includegraphics[width=\linewidth]{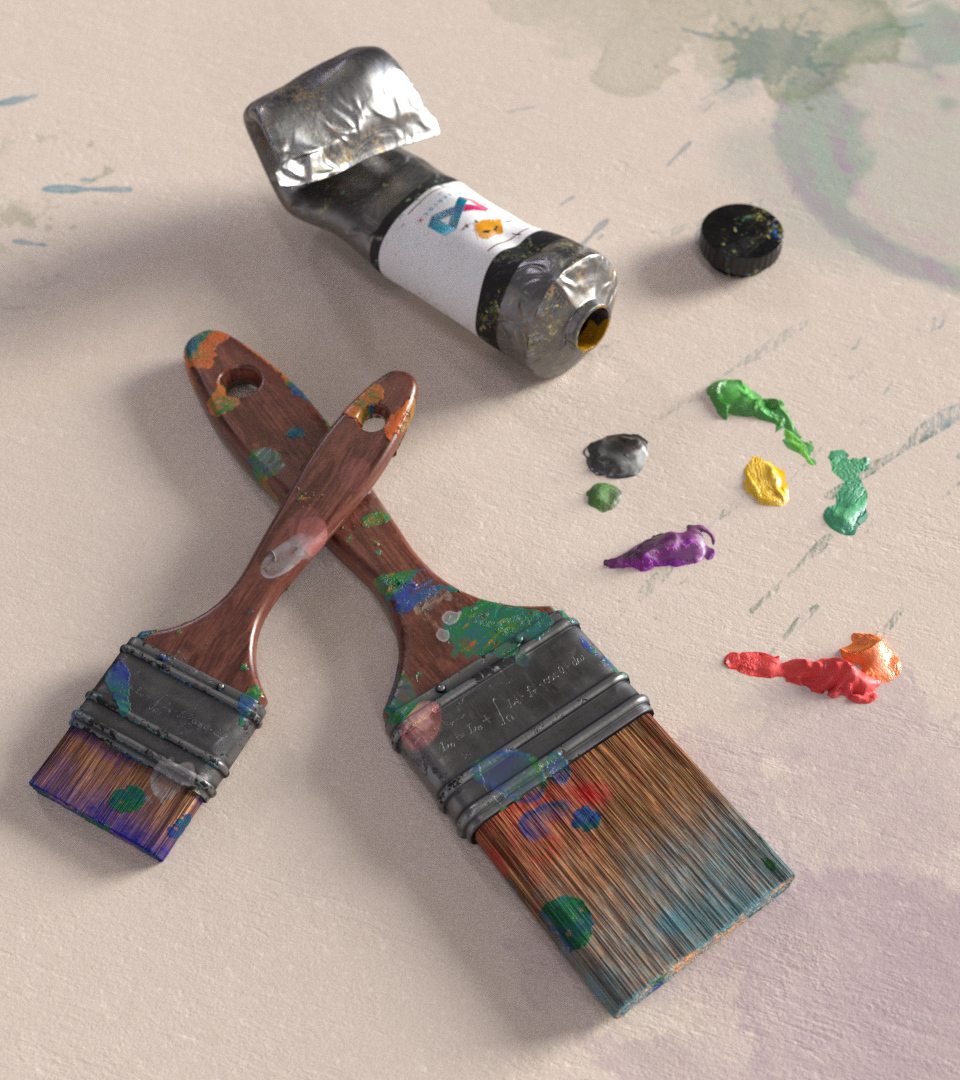}
  \end{subfigure}
  \caption{OpenPBR materials rendered in two separate systems (Arnold left, and Adobe's proprietary renderer right).}
  \label{fig:render_comparison_teaser}
\end{figure}

\begin{abstract}
\noindent
OpenPBR is a physically based, standardized uber-shader developed for interoperable material authoring and rendering across VFX, animation, and design visualization workflows. This document serves as a companion to the official specification, offering deeper insight into the model's development and more detailed implementation guidance, including code examples and mathematical derivations.

We begin with a description of the model's formal structure and theoretical foundations -- covering slab-based layering, statistical mixing, and microfacet theory -- before turning to its physical components. These include metallic, dielectric, subsurface, and glossy-diffuse base substrates, followed by thin-film iridescence, coat, and fuzz layers. A special-case mode for rendering thin-walled objects is also described.

Additional sections explore technical topics in greater depth, such as the decoupling of specular reflectivity from transmission, the choice of parameterization for subsurface scattering, and the detailed physics of coat darkening and thin-film interference. We also discuss planned extensions, including hazy specular reflection and retroreflection.
\end{abstract}

\enlargethispage{4\baselineskip}

\newpage

\vspace*{-6\baselineskip}

\enlargethispage{4\baselineskip}

\thispagestyle{plain} 

\tableofcontents

\medskip

\newpage


\pagestyle{fancy}

\section{Introduction}

The OpenPBR project originated in 2023 as a collaboration between Autodesk and Adobe to consolidate and refine their existing shading models, the Autodesk Standard Surface \cite{Georgiev2019} and Adobe Standard Material \cite{Kutz2021}, respectively.
The goal was to create a single, open, and standardized physically based shading model that would be suitable for the most common use cases in media and entertainment applications, while also being practical for real-time rendering.

OpenPBR is not intended to be a general framework for building material models (which exists in various forms \cite{Gritz2010, Kettner2015, Smythe2016, Hillaire2023}), but a reasonably unambiguous description of a particular form of pre-built ``uber-shader'', which experience has shown is practically very useful.

\begin{figure}[!tb]
  \centering
\includegraphics[width=0.48\linewidth]{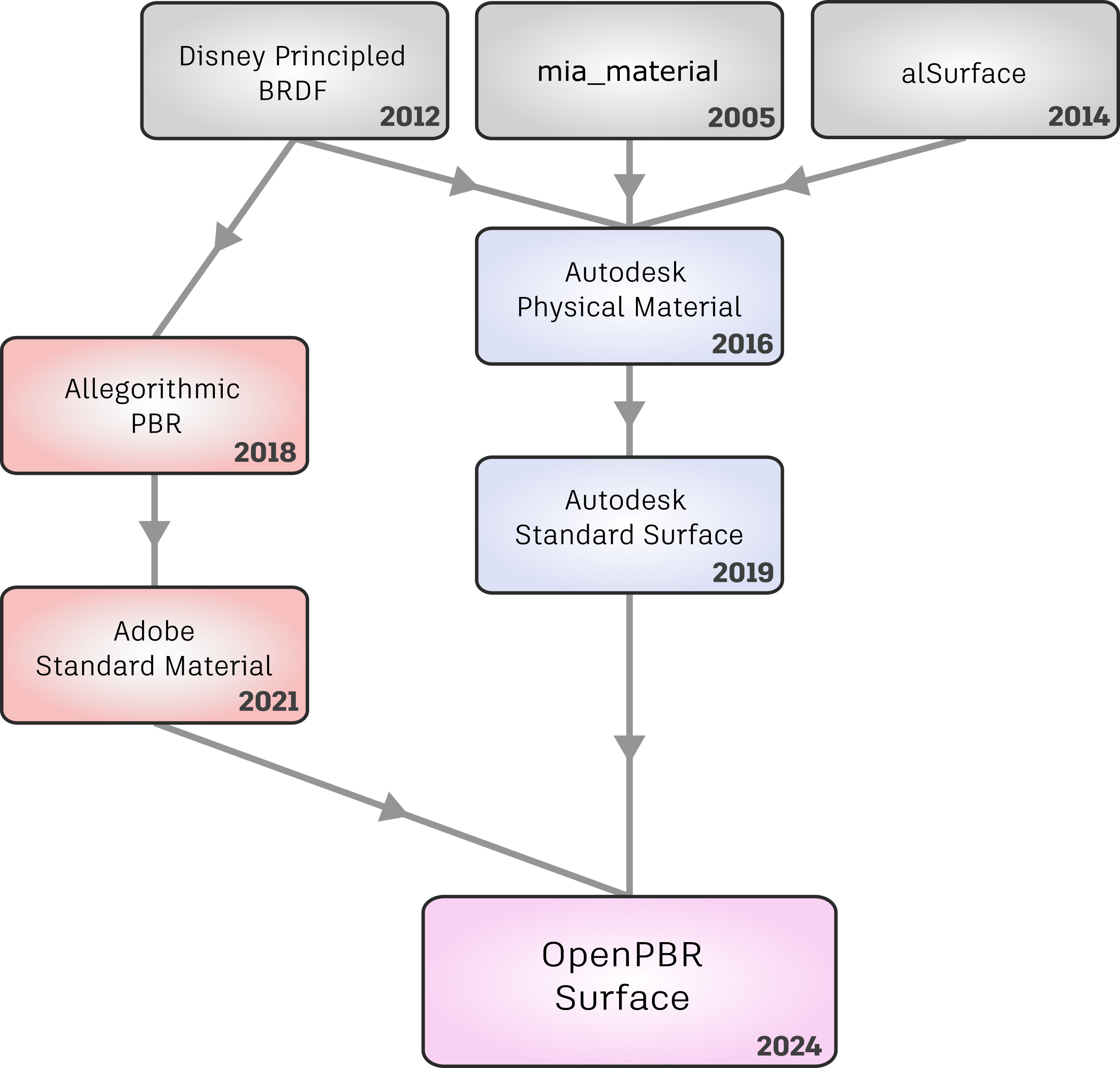}
\caption{The lineage of shading models leading to OpenPBR Surface.}
\label{fig:openpbr_lineage}
\end{figure}

There is a long history of previous proposals for such a physically based uber-shader. Figure~\ref{fig:openpbr_lineage} shows an approximate genealogy, including Disney's ``Principled'' shader \cite{Burley2012}, Allegorithmic's PBR shading model \cite{McDermott2018}, and most recently Autodesk Standard Surface \cite{Georgiev2019} and Adobe Standard Material \cite{Kutz2021}.
These most recent models have been found to be generally useful and practical for media and entertainment, so we based OpenPBR on them.

There are several advantages to defining such a general-purpose uber-shader model, including:

\begin{itemize}

\item \emph{Standardization}: If the model is widely used to define material properties, it facilitates the exchange of assets between different facilities and applications. It is also useful for artists to be presented with a familiar common interface and parameters, which helps with user onboarding and education. Building such a complicated model from scratch can be quite challenging and it is obviously convenient to provide a standardized model that can be used as a starting point.

\item \emph{Simplicity}: The model provides a fairly simple interface (of sliders and color pickers) that is sufficient for most typical use cases. This can be more convenient than a fully general material system. For the more specialized use cases it does not cover (for example very high-end skin, hair, cloth or volume shading), one may need to use a renderer-specific shader, or build a bespoke shading network.

\item \emph{Predictability}: The user-specified values and colors are designed to produce a reasonably intuitive resulting appearance, with linear behavior, while avoiding unexpected color shifts.

\item \emph{Plausibility}: The model is defined in terms of physical light transport, so that its appearance is plausibly close to the reality of a physical material and energy conserving. This physically plausible appearance should be the default, so the user does not have to work hard to achieve it. However, in some well-defined cases the user may want to explicitly ``break'' physics, to achieve some useful visual effects.

\end{itemize}

Another positive outcome of defining a practical general-purpose model is the assurance that the existing standard general material frameworks \cite{Gritz2010, Kettner2015, Smythe2016, Hillaire2023} have the required capabilities to implement it, and if not, arguing for and ensuring their adoption. In several cases, the effort has stimulated research into refinements of existing models, and generally helped to identify areas where further research on the underlying BSDF models is warranted.

A key difference between OpenPBR and most previous models mentioned is that in OpenPBR the material is specified as a particular physical structure consisting of layered slabs of material with specified BSDFs. The ground-truth appearance of the model is then defined as the result of light transport through this structure, apart from some well-defined edge cases where physics is broken, as noted. For implementers, this then makes the target appearance unambiguous so that there is clarity about how it should look, even if it may not always be practical to achieve this except approximately. Each implementation is free to make approximations (e.g., for modeling the layering) as needed based on standard theory, according to the constraints imposed by their use case (e.g., real-time rendering).

This document is a technical companion to the OpenPBR specification. It provides an in-depth explanation of the physical and mathematical foundations, implementation guidance, and rationale behind the OpenPBR layered material model. Unlike the official OpenPBR specification, which is a concise reference for the model's structure and parameters, this document covers the underlying theory, practical approximations, parameterization pitfalls, and implementation tips for production renderers. It also discusses edge cases, future improvements, and gives more detailed derivations and example code.

In Section~\ref{sec:overview}, we give a brief overview of the model structure. We describe the \hyperref[sec:layer_formalism]{layering} and \hyperref[sec:microfacet]{microfacet} formalism used to define the model, the \hyperref[sec:model-structure]{model structure} specified via this formalism, and the user-facing \hyperref[sec:parameterization]{parameterization}. In Section~\ref{sec:base_substrate}, we proceed to describe the model structure in detail, from the bottom up, starting with the base substrate (consisting of four slabs: \hyperref[sec:metallic-base]{metal}, \hyperref[sec:translucent-base]{translucent base}, \hyperref[sec:subsurface]{subsurface}, and \hyperref[sec:glossy-diffuse]{glossy-diffuse}). We also discuss the \hyperref[sec:emission]{emission} properties of the base substrate, and two upcoming improvements still being actively worked on: \hyperref[sec:hazy-specular]{hazy-specular} and \hyperref[sec:retro-reflection]{retro-reflection}. In Section~\ref{sec:thin-film}, we discuss the \hyperref[sec:thin-film]{thin-film iridescence} model, and in Sections ~\ref{sec:coat}, \ref{sec:fuzz} discuss the \hyperref[sec:coat]{coat} and \hyperref[sec:fuzz]{fuzz} layers, respectively. Finally, the special-case \hyperref[sec:thin-walled]{thin-walled} mode is covered in Section~\ref{sec:thin-walled}.

The current specification is available at \url{https://github.com/AcademySoftwareFoundation/OpenPBR}, including the reference MaterialX graph.
These notes pertain specifically to the OpenPBR 1.1 model, which is the current version as of August 2025.
The model is still evolving, and we expect to add new features in future releases, so please check the repository for the latest version. We also encourage those interested in contributing to the development of the model to join the discussion on the OpenPBR GitHub repository and in our ASWF Slack channel.

\clearpage


\section{An overview of the OpenPBR model}

\label{sec:overview}

The specific form of material structure in OpenPBR is a distillation of the aforementioned models. See Figure~\ref{fig:layer_structure} for a schematic of the layered structure:
\begin{figure}[!hb]
  \centering
\includegraphics[width=0.95\linewidth]{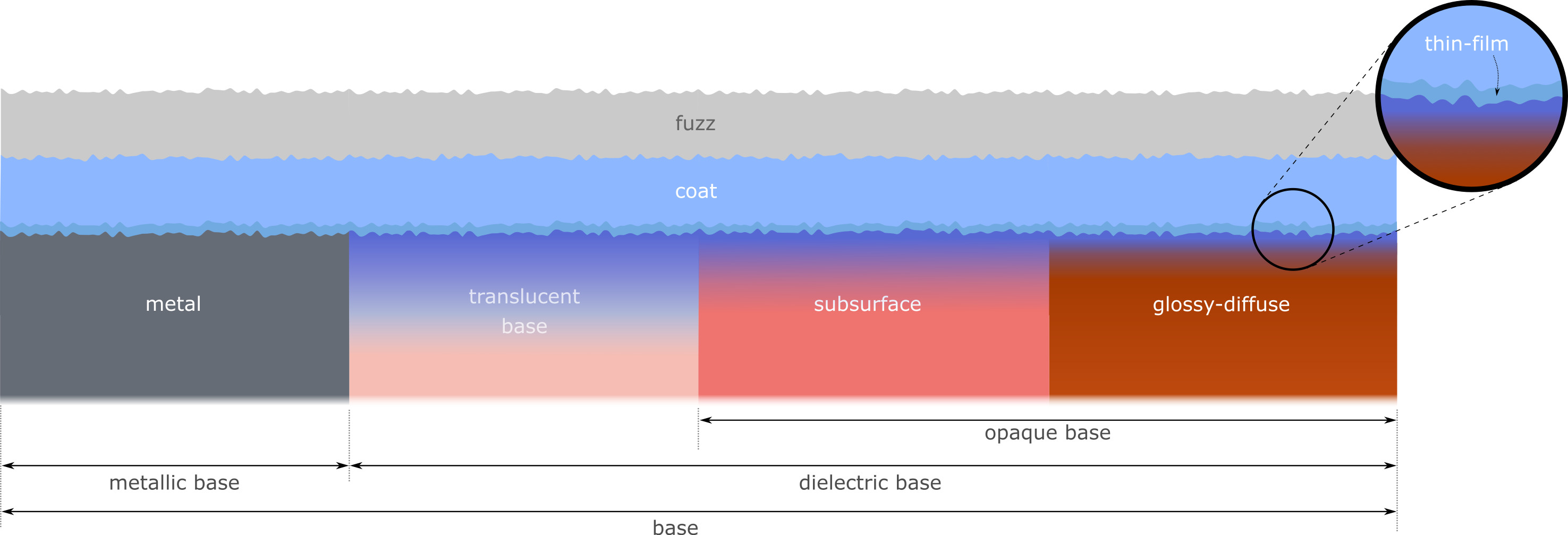}
\caption{Schematic of the layer structure.}
\label{fig:layer_structure}
\end{figure}

In outline the material consists of:
\begin{itemize}
   \item A \hyperref[sec:base_substrate]{base substrate} made of a mixture of \hyperref[sec:metallic-base]{metal} or \hyperref[sec:dielectric-base]{dielectric}. The interface (dielectric or metal) of this base layer produces the primary specular reflection lobe. The dielectric base represents either of three components, which can be statistically mixed:
   \begin{itemize}
        \item \hyperref[sec:glossy-diffuse]{Glossy-diffuse}: Dielectric with opaque internal media (e.g., wood, granite, concrete, cardboard, and wall paint).
        \item \hyperref[sec:subsurface]{Subsurface}: Dielectric with dense highly scattering internal media (e.g., plastic, marble, skin, vegetation, and food).
        \item \hyperref[sec:translucent-base]{Translucent base}: Dielectric with translucent internal media (e.g., glass, crystals, and liquids).
    \end{itemize}
   \item \hyperref[sec:thin-film]{Thin-film}: A layer of optically thin dielectric applied to the base microfacet surface, may be applied optionally to generate iridescence via wave interference (for soap bubbles, oily surfaces, tempered metals, etc.).
   \item \hyperref[sec:coat]{Coat}: On top of the base substrate sits a further optional
    layer of dielectric acting as a coating, which may have an absorbing medium. The dielectric interface of this coat layer provides a secondary specular lobe.
   \item \hyperref[sec:fuzz]{Fuzz}: An optional layer representing the reflection from micro-fibers (such as fine hair, peach fuzz, textile strands, and dust grains) on top of everything else.
\end{itemize}

We now describe in more detail the formalism we used to define the structure technically.



\subsection{Model formalism}

\label{sec:layer_formalism}

To define the aforementioned layered material structure in detail, we introduced a rather simple formalism that describes an abstract physical material built out of slabs, which while simple is powerful enough for our purposes. As noted, OpenPBR is not a general framework for building material models, but we found it helpful to write down the detailed assumptions in the formalism, which we will now describe.

While writing down this formalism is arguably not strictly necessary, it does help to clarify the structure of the model and how it is intended to be implemented in general material frameworks, and it also provides a common language for implementers to discuss the model. It was helpful to resolve potential ambiguities in the intended physical structure and meaning of the parameters, and to ensure that the model is well-defined in terms of light transport through the structure. Our model was designed explicitly to be compatible and consistent with the MaterialX framework \cite{Smythe2016} (indeed, we contributed various changes to MaterialX to ensure compatibility), so the formalism we describe is therefore rather similar to what is provided in MaterialX.

\subparagraph{Slabs}

We conceptualize the material as consisting of slabs of material which are composed by layering.
In describing the model structure, we are working at a ``mesoscopic'' scale where the layers are considered to be thick enough that geometrical optics is valid, but thin enough they are not explicitly visible in the render. This scale is also considered to be small relative to the scale of variations of the material properties, so we can consider the slabs to be locally homogeneous.

Each slab consists of a homogeneous dielectric or conducting medium $V$ with conceptually unbounded extent horizontally and a finite vertical extent, bounded above by a surface with a given interface BSDF $f$. The BSDF may not necessarily correspond exactly to that of a physical dielectric or conductor (for example, to be able to say that the BSDF is simply the Lambert or Oren--Nayar model, for example). In a physical slab, the interface BSDF is just what happens where the slab of dielectric or conductor ends, so should be self-consistent with it (although we don't enforce that in our formalism). We denote a general slab, therefore, as follows:
\begin{equation}
\quad S = \mathrm{Slab}(f, V) \ ,
\end{equation}
where $f$ is the interface BSDF and $V$ is the medium VDF (as defined below).

A slab does not itself specify anything about other slabs in relation to itself (e.g., its substrate slab or overlying slab). The adjacent medium above and below the slab will depend on where it sits in the eventual layer structure. The ambient dielectric medium at the very top of the entire structure (and bottom if thin-walled) is also assumed to be given and unspecified by the model. If the renderer keeps track of the dielectric medium in which the surface is embedded (via a scheme such as ``nested dielectrics'' \cite{Schmidt:2002:SND}) -- which may be the interior dielectric bulk of some transparent object in the scene, such as a piece of glass or body of water -- then the surrounding ambient medium is a dielectric whose IOR we denote $n_\mathrm{ambient}$. Alternatively, if dielectric medium tracking is not performed, then $n_\mathrm{ambient}$ can be assumed to be 1 corresponding to air or vacuum.

\subparagraph{BSDFs}

Each slab has a top interface whose  BSDF is a function of input and output directions $f(\omega_i, \omega_o)$.\footnote{Where, following the usual convention, these both point away from the surface vertex, with $\omega_o$ in the direction of the outgoing light and $\omega_i$ opposite to the direction of the incident light.} Physically, the BSDF should be energy conserving and reciprocal.

Energy conservation of a BRDF $f(\omega_i, \omega_o)$ amounts to the requirement that the \emph{directional reflectance} (or \emph{directional albedo}) $E(\omega_o) \le 1$ for all output ray directions, where $E(\omega_o)$ is defined as the integral of the BRDF over directions in the same hemisphere $\mathcal{H}_+$ as $\omega_o$ with projected solid angle measure
\begin{equation} \label{directional-reflectance-definition}
E(\omega_o) \equiv \int_{\mathcal{H}_+} f(\omega_i, \omega_o) \,\mathrm{d}\omega^\perp_i \ .
\end{equation}
More generally, for transmissive BSDFs there is a similar constraint $E(\omega_o) + T(\omega_o)\le 1$, where the \emph{directional transmittance} $T(\omega_o)$ is defined as the integral of the BSDF over the opposite hemisphere $\mathcal{H}_-$ to $\omega_o$. All BSDFs used are energy conserving in this sense. Additionally, a BSDF that does not \emph{dissipate} any energy, such as a dielectric interface or a diffuse surface with a white albedo, is said to be \emph{energy preserving} and satisfies $E(\omega_o) + T(\omega_o) = 1$. (A technicality to be noted is that in order for the stated energy preservation to hold, the light transport has to be done in terms of the so-called ``basic radiance`` $L / n^2$, where $n$ is the local index of refraction. This accounts for the squeezing of light rays as they enter a medium with a higher index of refraction. This is accounted for in standard derivations of BSDFs and implementation of light transport \cite{Walter2007,Pharr2023}.)

Reciprocity is the requirement that the BSDF is symmetric under interchange of the arguments, i.e., $\omega_i \leftrightarrow \omega_o$.\footnote{Technically, the reciprocity condition also involves the index of refraction \cite{Veach1998}.} In some cases this is relaxed and a non-reciprocal BSDF or formula is used for simplicity or efficiency, as in practice it does not lead to serious issues (particularly if only doing unidirectional path tracing).

If the bounding BSDFs of a slab are completely non-transmissive and opaque (e.g., metallic or diffuse), the internal medium is irrelevant for light transport purposes. For convenience, it can be omitted:
\begin{equation}
S_\mathrm{opaque} = \mathrm{Slab}(f) \ .
\end{equation}

\subparagraph{VDFs}

If the slab is non-opaque (i.e., translucent) then the interior medium of the slab is taken to be a homogeneous dielectric that optionally contains a specified homogeneous volumetric medium (or ``VDF'') $V$. By VDF we mean the set of quantities (which, in general, are spatially varying fields) that define a volumetric optical medium, i.e., the absorption and scattering coefficients (alternatively, extinction coefficient and scattering albedo), phase function, and the index of refraction (IOR) and dispersion of the embedding dielectric medium (grouped with the VDF as ``bulk'' properties of the slab).
A semi-infinite slab at the bottom of the material structure is denoted with a bulk medium $V^\infty$, which has no bottom interface:
\begin{equation}
S_\mathrm{bulk} = \mathrm{Slab}(f, V^\infty) \ .
\end{equation}
We assume that the renderer is aware that the volumetric properties may vary from point to point, inherited from the surface parameters, and how the fields are ``filled in'' in the space surrounding the surfaces is a matter for implementations to deal with.

\begin{figure}[!bt]
  \centering
  \begin{subfigure}{.47\textwidth}
    \centering
    \includegraphics[width=\linewidth]{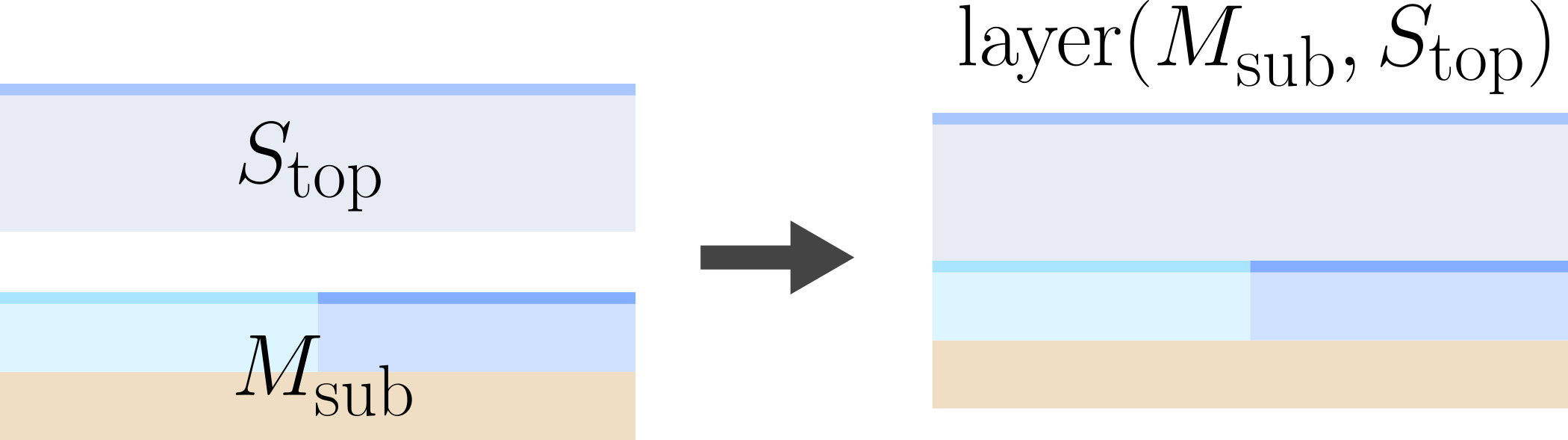}
    \caption{Layer operation.}
    \label{fig:layer_op}
  \end{subfigure}
  \hspace{0.2cm}
  \begin{subfigure}{.47\textwidth}
    \centering
    \includegraphics[width=\linewidth]{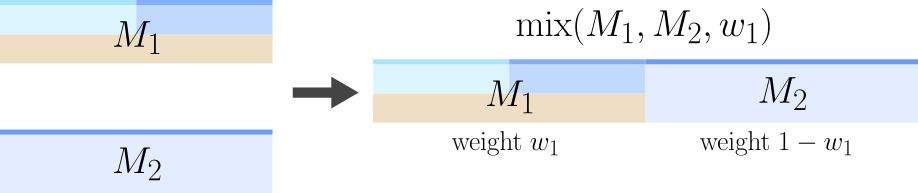}
    \caption{Statistical mix operation.}
    \label{fig:mix_op}
  \end{subfigure}
  \caption{Basic operations on slabs in the OpenPBR slab formalism.}
  \label{fig:slab_formalism}
\end{figure}

Given constituent slabs, we then build a more complex composite material by ``vertically'' layering and ``horizontally'' mixing slabs.
Figure~\ref{fig:slab_formalism} shows a schematic of the slab formalism, with the basic operations on slabs.

\subparagraph{Vertical layering}

\begin{figure}[!b]
  \centering
  \begin{subfigure}{.25\textwidth}
    \includegraphics[width=\linewidth]{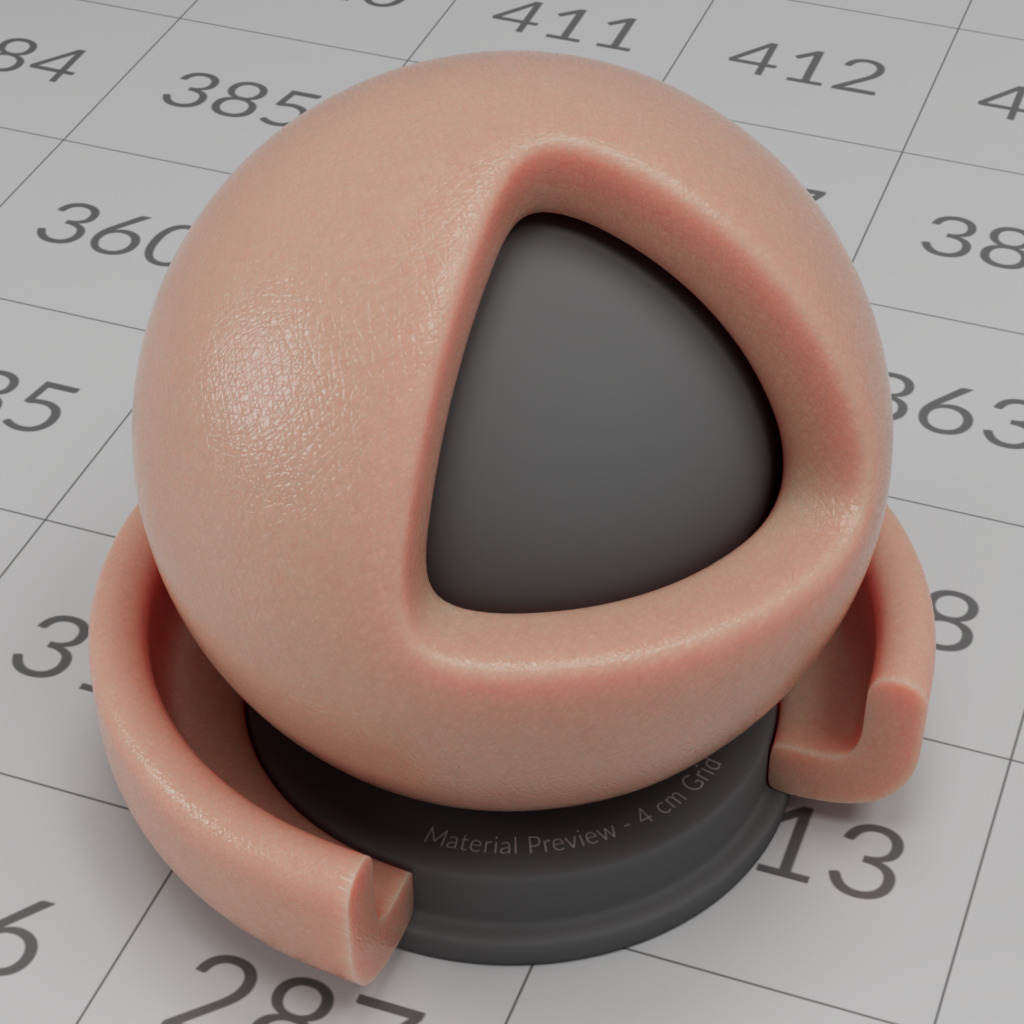}
  \end{subfigure}
  \hspace{0.01\textwidth}
  \begin{subfigure}{.25\textwidth}
    \includegraphics[width=\linewidth]{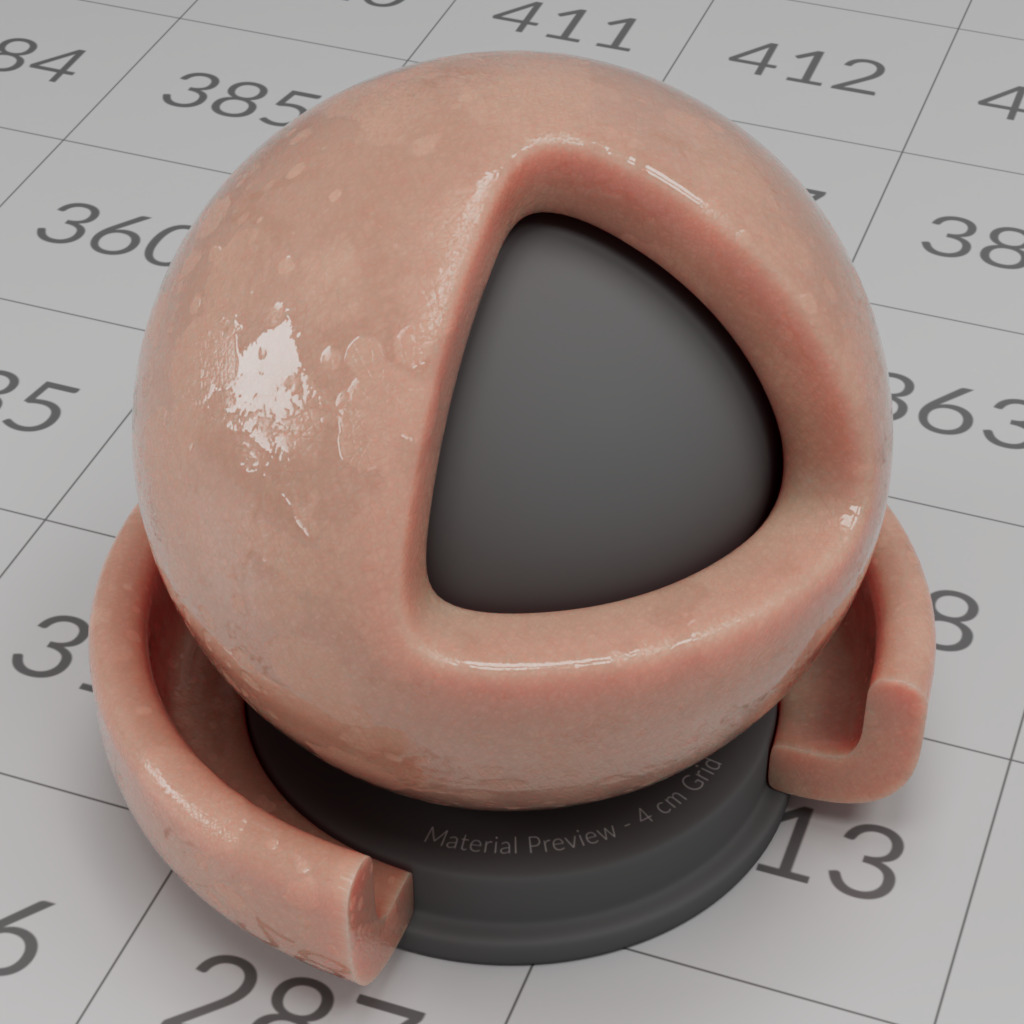}
  \end{subfigure}
  \hspace{0.01\textwidth}
  \begin{subfigure}{.25\textwidth}
    \includegraphics[width=\linewidth]{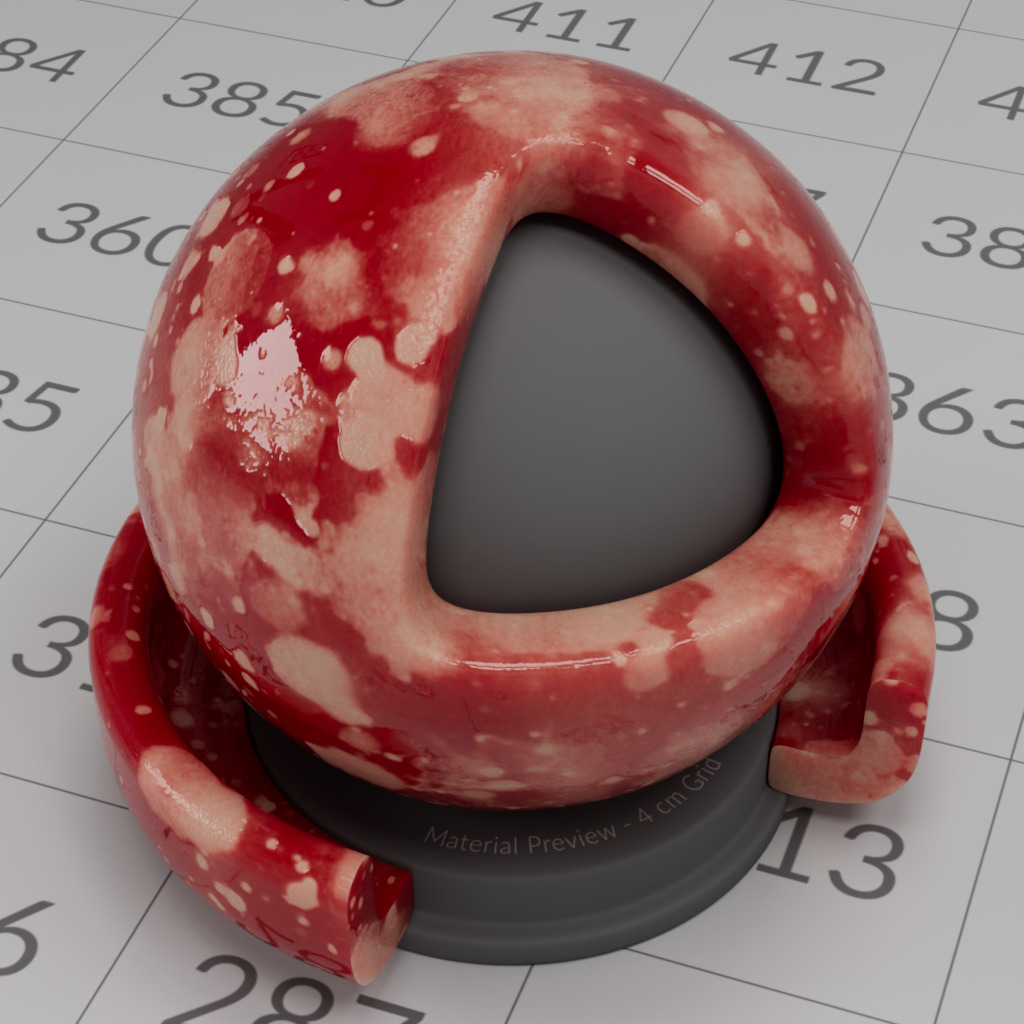}
  \end{subfigure}
  \caption{A simple example of layering: adding a coat layer to skin in order to model wet or bloody skin. \label{fig:skin_coat}}
\end{figure}

The layer operation generates a composite material by depositing a slab \mbox{$S_\mathrm{coat} = \mathrm{Slab}(f_\mathrm{coat}, V_\mathrm{coat})$} on top of another substrate slab $S_\mathrm{sub} = \mathrm{Slab}(f_\mathrm{sub}, V_\mathrm{sub})$ or composite material, conceptually bonding the base of $S_\mathrm{coat}$'s dielectric medium to the surface of $S_\mathrm{sub}$, so that $V_\mathrm{coat}$ becomes the adjacent medium to the top interface of $S_\mathrm{sub}$. The physical act of bringing the two independent slabs $S_\mathrm{sub}$ and $S_\mathrm{coat}$ together and bonding them produces a new composite material $L$ that is ``vertically'' heterogeneous on the mesoscopic scale, denoted by
\begin{equation}
L = \mathrm{\mathbf{layer}}(S_\mathrm{sub}, S_\mathrm{coat}) \ .
\end{equation}
Figure~\ref{fig:skin_coat} shows an example of how a \hyperref[sec:coat]{coat} layer could be added to a skin material to model effects such as water or blood stains. The partially present coat adds strong Fresnel reflection at grazing angles, and modifies the look of the underlying skin due to absorption, darkening, and roughening effects.

In OpenPBR, we are mostly concerned with defining the layer structure rather than requiring any particular method be used to solve for the light transport, the exact solution of which is supposed to correspond to the desired ground truth. However, for practical purposes, it is important to give some suggested approximate schemes, at least as a starting point. In practice, even rather crude approximations can be acceptable if they yield plausible results.
There is a large body of work in the graphics literature on modeling and rendering of layered material structures \cite{Hanrahan1993, Ashikhmin2000, Kelemen2001, Weidlich2007, Jakob2014, Guo2018, Zeltner2018, Belcour2018}. Accurately solving for the light transport through such layered structures is a complex problem, which in general requires solving the radiative transfer equation (RTE) accounting for the scattering at the interfaces as well as inside the internal volumetric media. This is a non-trivial task that in general requires numerical methods such as Monte Carlo integration, or solving the RTE via various approximations such as diffusion theory or discrete ordinates methods \cite{Pharr2023}.

For media and entertainment applications, however, we are often interested in a practical approximation which need not be very accurate, but must be fast to compute while still producing visually plausible results. In this context, the most common approach is to use a simplified model that approximates the light transport through the layered structure by combining the BSDFs of the individual layers in a linear combination, with various heuristics to account for the effects of the layering structure on the weights and parameters of the BSDFs in this combination.

In this spirit, the most simple representation of a layered configuration amounts to simply making a linear combination of the interface BSDFs $f_\mathrm{sub}$ and $f_\mathrm{coat}$. A common approach is the so-called \emph{albedo-scaling} approximation \cite{Smythe2016}, where the total BSDF of this layered configuration is given by summing $f_\mathrm{coat}$ and $f_\mathrm{sub}$, with the substrate lobe weighted by a factor depending on the directional reflectance of the coat, which is designed to ensure that the resulting BSDF is energy conserving:
\begin{equation} \label{non-reciprocal-albedo-scaling}
f_\mathrm{layer}(\omega_i, \omega_o) = f_\mathrm{coat}(\omega_i, \omega_o) + \bigl(1 - E_\mathrm{coat}(\omega_o)\bigr) \,f_\mathrm{sub}(\omega_i, \omega_o) \ ,
\end{equation}
where the directional albedo integral $E_\mathrm{coat}(\omega_o)$ can be either precomputed and tabulated, or computed on the fly via \href{https://www.pbr-book.org/3ed-2018/Light_Transport_I_Surface_Reflection/Sampling_Reflection_Functions#Application:EstimatingReflectance}{Monte Carlo}.
This form ensures that if the directional albedos of the coat and substrate BSDFs are energy conserving (i.e., $E_\mathrm{coat}(\omega_o) \le 1$, $E_\mathrm{sub}(\omega_o) \le 1$), then the combined BSDF is also energy conserving, since
\begin{equation}
E_\mathrm{layer}(\omega_o) \,=\, E_\mathrm{coat}(\omega_o)  +  \bigl(1 - E_\mathrm{coat}(\omega_o)\bigr) \,E_\mathrm{sub}(\omega_o) \;\; \le \, 1  \ .
\end{equation}
It also ensures that if the substrate BSDF perfectly preserves energy (i.e., $E_\mathrm{sub}(\omega_o) = 1$) then the layer BSDF does also, ensuring that a ``white furnace'' test would pass. It ignores the physical requirement of reciprocity but captures the essential view dependence of the coat and substrate BSDFs, and is a good approximation for many practical cases. The Autodesk Standard Surface model \cite{Georgiev2019}, for example, uses this formulation for its layering.

This albedo-scaling approximation is somewhat limited as it does not explicitly take into account the effect of multiple light bounces back and forth between the interfaces, or absorption and scattering in the volumetric medium of the coat $V_\mathrm{coat}$. In general, of course, the resulting BSDF lobe shape will not be a simple linear combination of the interface BSDFs. However, to some extent, these effects can be adequately accounted for within the albedo-scaling model by making various approximations. For example, the effect of the volumetric transmittance through the coat in the incident and output directions could be approximately modeled as
\begin{equation} \label{non-reciprocal-albedo-scaling-with-T}
f_\mathrm{layer}(\omega_i, \omega_o) = f_\mathrm{coat}(\omega_i, \omega_o) + T^2_\mathrm{coat}(\omega_i, \omega_o) \bigl(1 - E_\mathrm{coat}(\omega_o)\bigr) \,f_\mathrm{sub}(\omega_i, \omega_o) \ ,
\end{equation}
where $T^2_\mathrm{coat}(\omega_i, \omega_o)$ accounts for the total volumetric absorption of the coat along the input and output rays. Similarly, if the coat is rough, it will effectively also roughen the substrate BSDF lobe, which can be approximately accounted for via various heuristics.

\begin{figure}[!b]
  \centering
\includegraphics[width=0.75\linewidth]{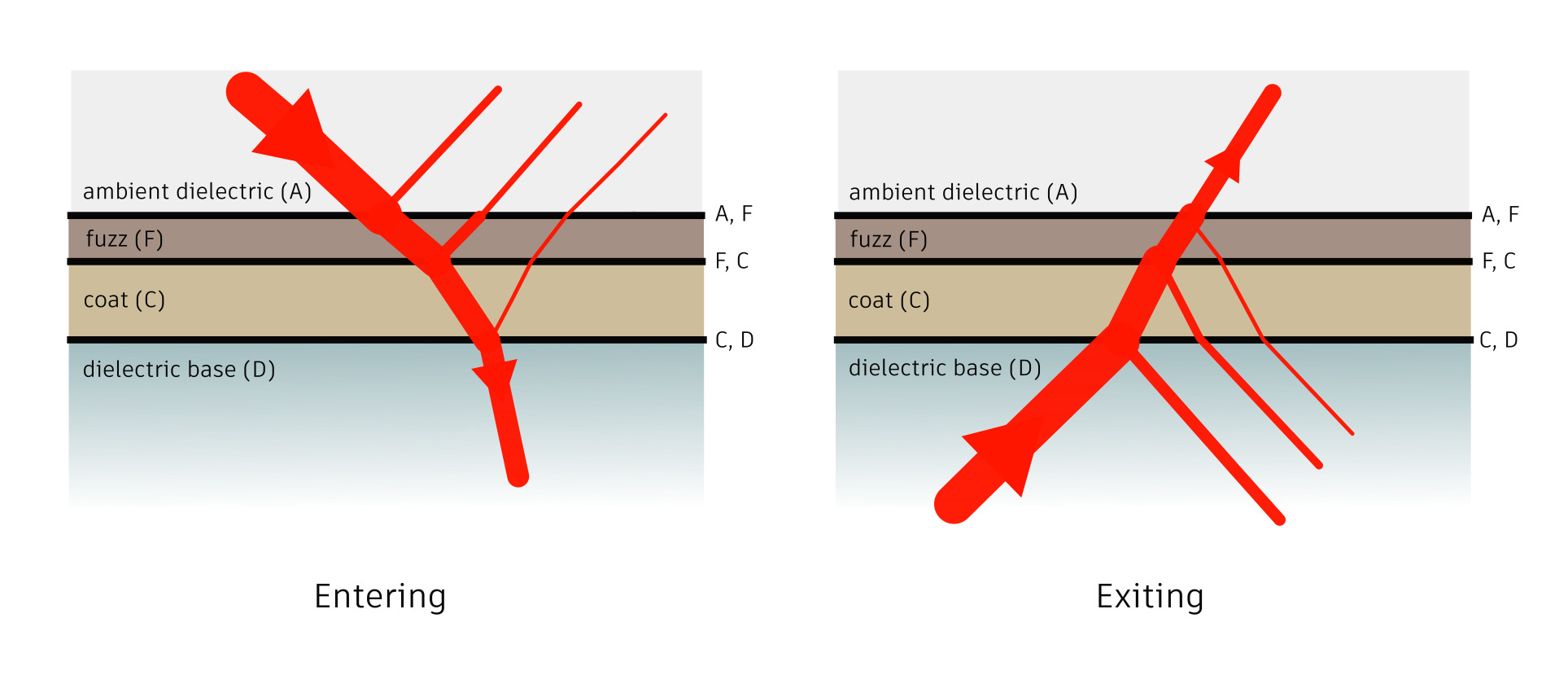}
\caption{Difference between entering and exiting rays.}
\label{fig:entering_exiting}
\end{figure}

The layering system also must handle the different physical effects of the layering for rays exiting and entering the surface. For example, consider the case of a glass object, with a coat and fuzz (see Figure~\ref{fig:entering_exiting}). Rays entering from the exterior (i.e., the ambient dielectric medium) will enter through the fuzz, then the coat, then transmit into the base glass. While rays that hit the surface from the interior of the glass (having refracted into the glass at some earlier point in the path) instead hit the bottom side of the coat, then the fuzz, then transmit into the ambient medium.

For rays entering from the top (ambient medium) and exiting through the dielectric base, the fuzz reflection is un-tinted by the coat absorption, the coat reflection is dimmed and roughened by the fuzz, and the dielectric reflection and transmission are both dimmed and roughened by both the coat and the fuzz. While for rays entering from the internal dielectric and exiting from the top, the fuzz reflection (viewed from inside) will now be tinted by the coat absorption, and the reflection from the coat ``top'' interface (with the fuzz) is not dimmed or roughened by the fuzz. This reflection is also an internal one, so it has a different Fresnel factor. The dielectric transmission is dimmed and roughened by the coat and fuzz, while the reflection is unaffected (except for being an internal reflection as well).


A related issue is how we deal with rays that are exiting from the surface (having entered the base dielectric elsewhere) when the surface is locally opaque (e.g., metallic). Arguably one can make sense of this by assuming that the metal must be a thin sheet of foil covering the interior dielectric, so we would reflect from this foil back into the interior. This would allow, for instance, the rendering of bottles with metallic labels, with the label visible through the glass (without the need for modeling a separate label). In general, however, the interior properties are (currently) ill-defined if different parts of the surface specify different base interiors (i.e., metal, diffuse, subsurface, volume, or a mixture thereof). For the moment it is implementation-dependent as to how this is handled.

\subparagraph{Horizontal mixing}

\begin{figure}[!tb]
  \centering
  \hfill
  \begin{subfigure}{.19\textwidth}
    \includegraphics[width=\linewidth]{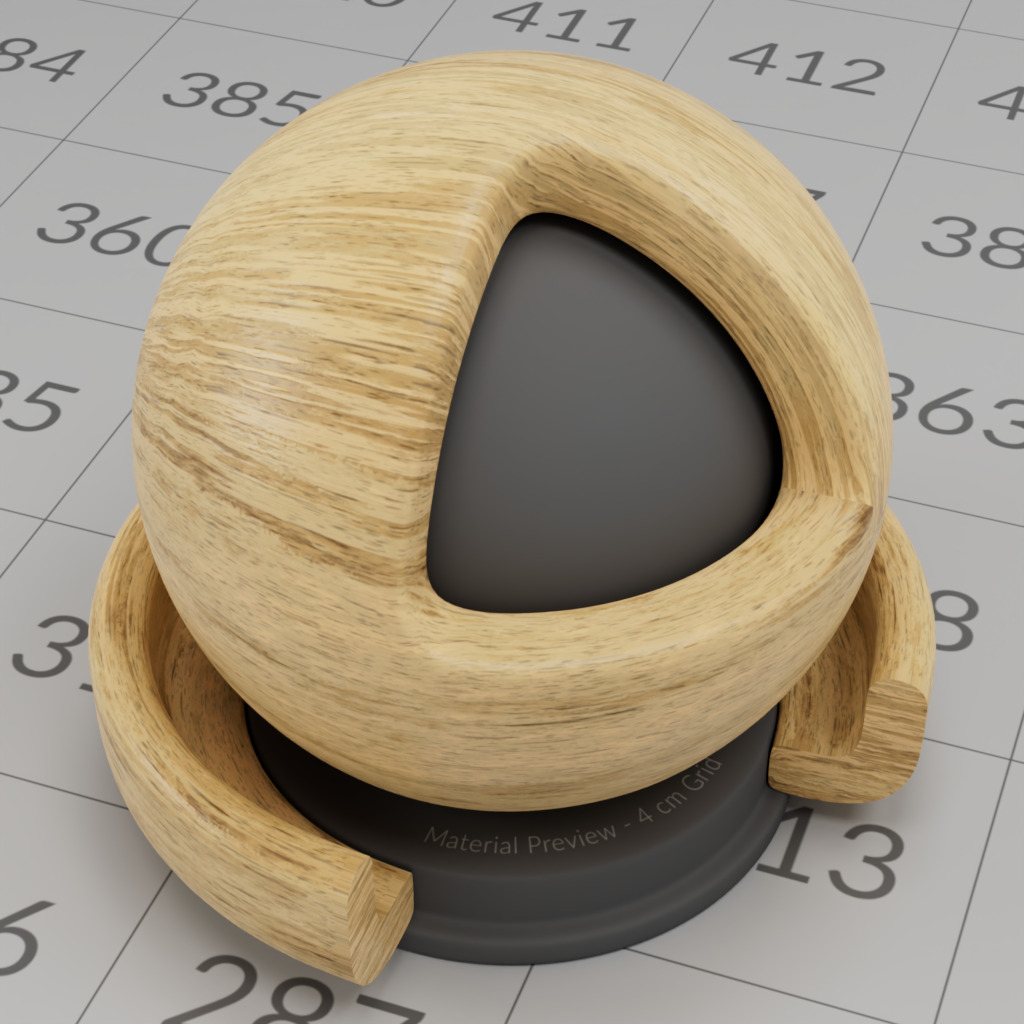}
  \end{subfigure}
  \hfill
  \begin{subfigure}{.19\textwidth}
    \includegraphics[width=\linewidth]{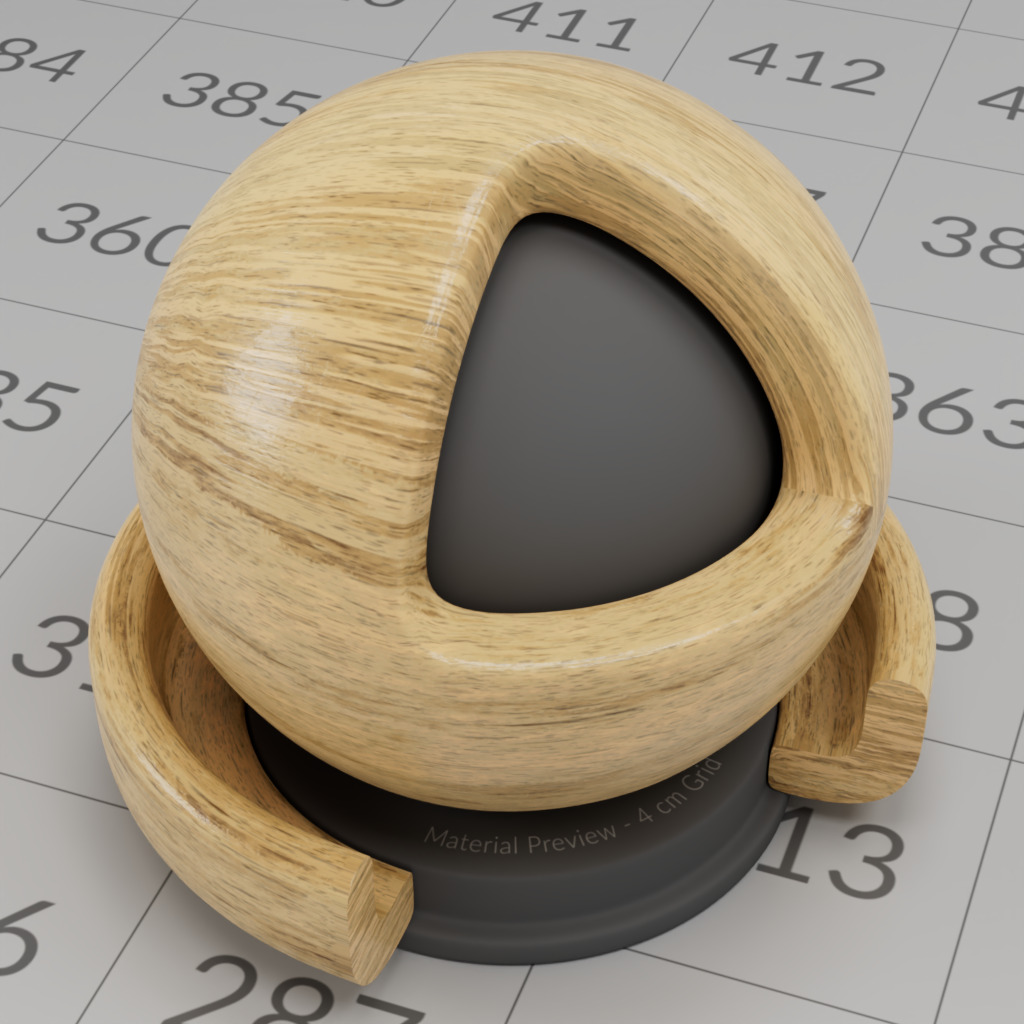}
  \end{subfigure}
  \hfill
  \begin{subfigure}{.19\textwidth}
    \includegraphics[width=\linewidth]{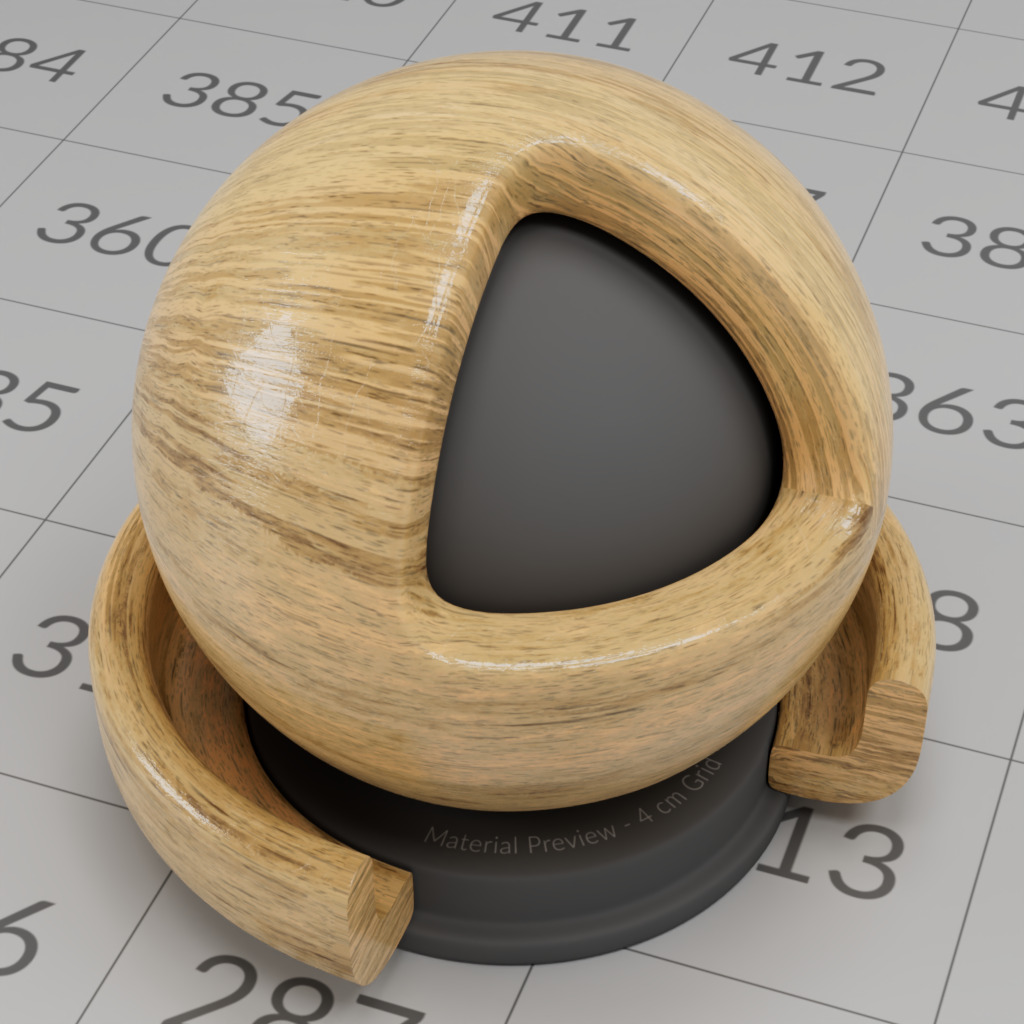}
  \end{subfigure}
  \hfill
  \begin{subfigure}{.19\textwidth}
    \includegraphics[width=\linewidth]{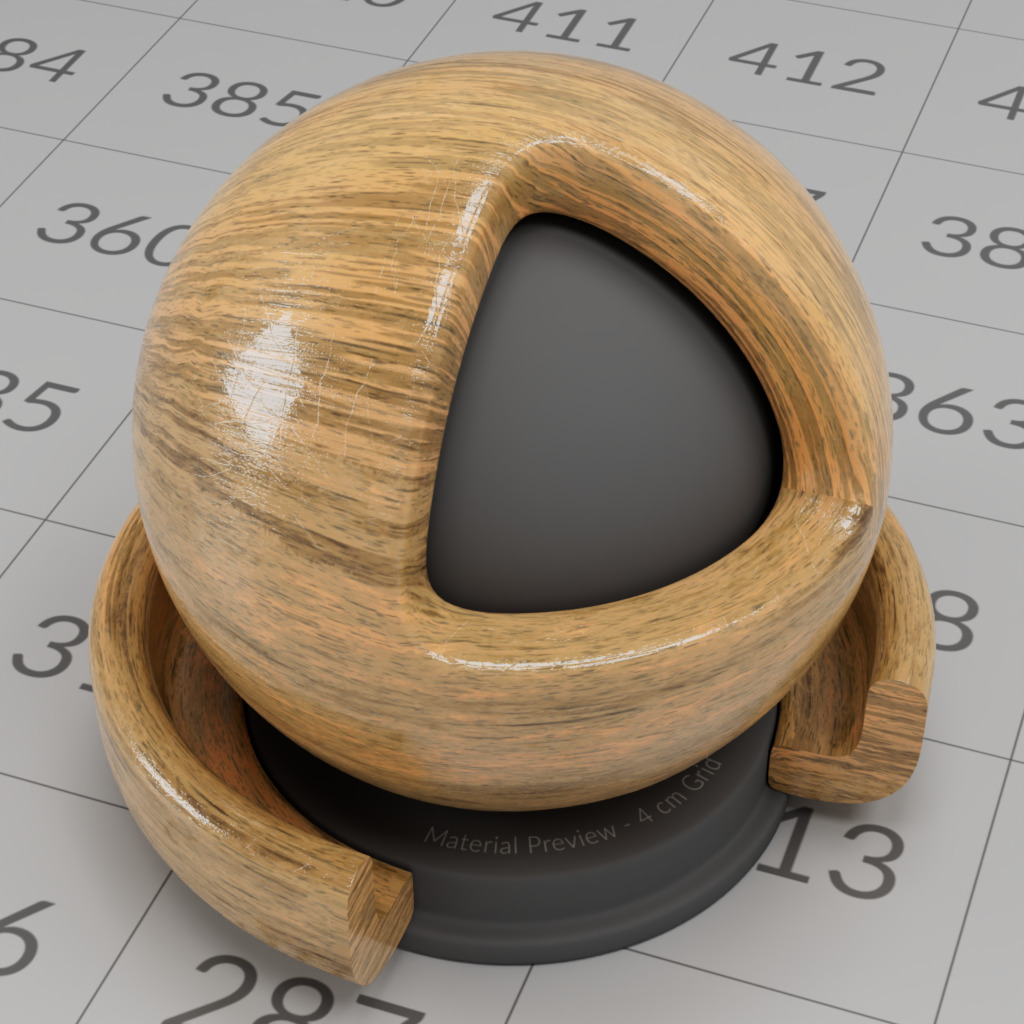}
  \end{subfigure}
  \hfill
    \begin{subfigure}{.19\textwidth}
    \includegraphics[width=\linewidth]{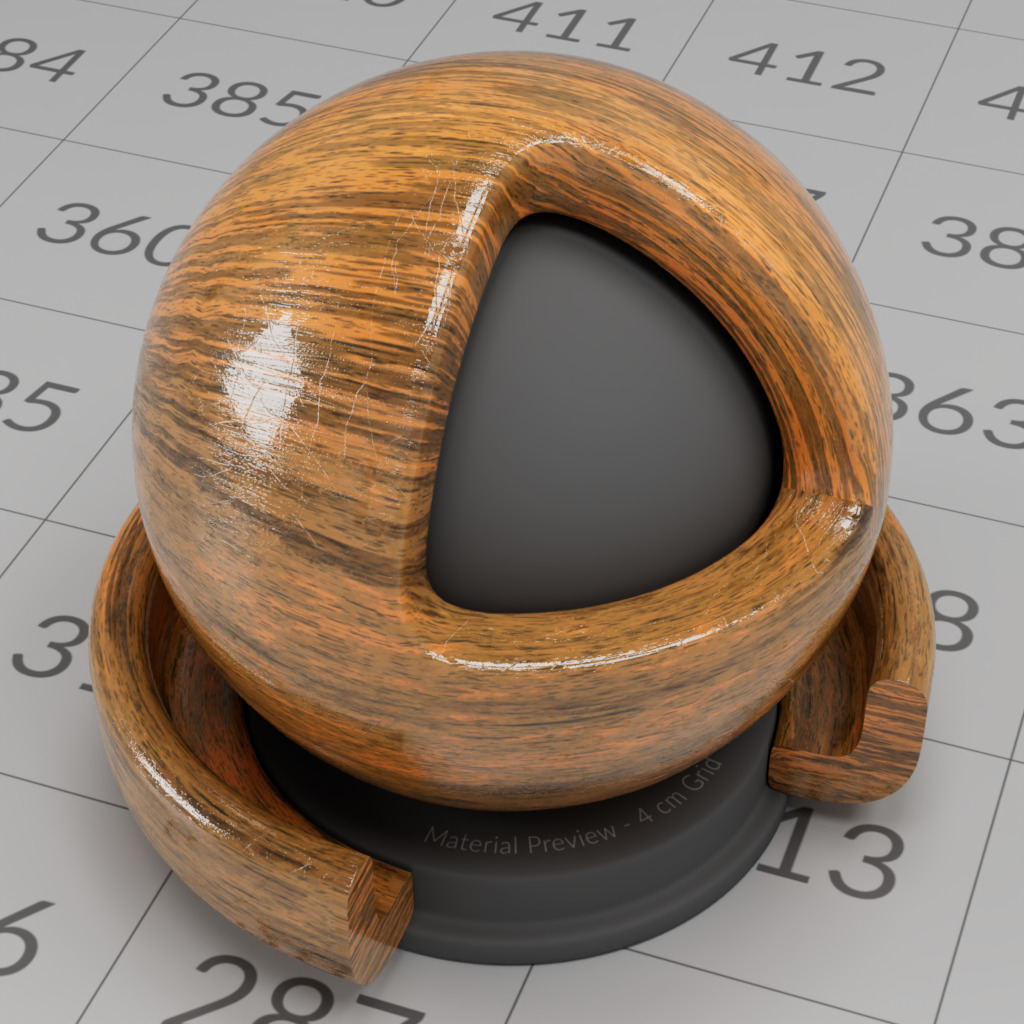}
  \end{subfigure}
  \hfill
  \caption{Varying coat mix (or presence) weight $w_\mathrm{coat}$ from 0 to 1, the surface transitions smoothly from an uncoated state to fully coated, with a partially present coat in between. Note that the \hyperref[sec:coat_darkening]{darkening} effect of the coat becomes more prominent as the presence weight increases. \label{fig:coat_mix_weight}}
\end{figure}

The surface properties and layer structure of a material obviously vary depending on the position, often with large patches of one type of material abruptly changing to another type. In the transition regions, it is often useful to be able to smoothly interpolate between the two materials, so that the transition is not abrupt and does not produce aliasing artifacts.

This is achieved with a \emph{horizontal mixing} operation which models a statistical blend between two materials at the mesoscopic level. The physical picture is that the mesosurface consists of randomly distributed patches of each material in proportion to the blend weight, which thus exhibits heterogeneity in the ``horizontal'' direction. The mix operation of two slabs $S_0$ and $S_1$ with weight $w_1$, generating a new ``horizontally'' heterogeneous composite material $M$, is denoted:
\begin{equation}
M = \mathrm{\mathbf{mix}}(S_0, S_1, w_1)
\end{equation}
where $M \rightarrow S_0$ as $w_1 \rightarrow 0$, and $M \rightarrow S_1$ as $w_1 \rightarrow 1$.

The mix operation is also used to describe the probability that a particular layer structure exists in the material at a particular point. For example, a coating $S_\mathrm{coat}$ may cover the substrate layers $S_\mathrm{sub}$ but be intermittent and applied only to some fraction $w_\mathrm{coat}$ (the coat \emph{presence} or \emph{coverage} weight) of the substrate (Figure~\ref{fig:coat_mix_weight}). This would be specified as
\begin{equation}
\mathrm{\mathbf{mix}}(S_\mathrm{sub}, \mathrm{\mathbf{layer}}(S_\mathrm{sub}, S_\mathrm{coat}), w_\mathrm{coat}) \ .
\end{equation}
For convenience, this may also be written more concisely as a weighted layer operator which covers a given fraction of the substrate with the coat layer:
\begin{equation}
\mathrm{\mathbf{layer}}(S_\mathrm{sub}, S_\mathrm{coat}, w_\mathrm{coat}) \ .
\end{equation}
This physical picture of the mix operation becomes somewhat unrealistic in some cases where the bottom-most bulk materials being blended are not obviously consistent (e.g., blending dielectric and metallic bulks), but in such cases it is understood that the implementation should do the best it can to make sense of the physics. For example, the metal bulk could be considered to actually be surface metallic flakes on top of a single, consistent dielectric.

The implementation of the mix operation in a renderer depends on how the material model is approximated. Typically the model will be reduced to an effective BSDF per slab being mixed, in which case the mix can be implemented simply as a linear blend of those BSDFs. That is, if slabs $S_0$, $S_1$ have BSDFs $f_0$, $f_1$ respectively, then the BSDF of $\mathrm{\mathbf{mix}}(S_0, S_1, w_1)$ can be mapped to\footnote{Where $\mathrm{lerp}(a, b, t) \equiv (1 - t) a + tb$.}
\begin{equation}
f_\mathrm{mix} = (1 - w_1) f_0 + w_1 f_1 = \mathrm{lerp}(f_0, f_1, w_1) \ .
\end{equation}
In the case of the weighted layer operation, if the BSDF of the substrate is $f_\mathrm{sub}$ then $\mathrm{\mathbf{layer}}(S_\mathrm{sub}, S_\mathrm{coat}, w_\mathrm{coat})$ maps to
\begin{equation}
f_\mathrm{weighted-layer} = (1 - w_\mathrm{coat}) f_\mathrm{sub} + w_\mathrm{coat} f_\mathrm{layer} \ ,
\end{equation}
where $f_\mathrm{layer}$ is the BSDF corresponding to $\mathrm{\mathbf{layer}}(S_\mathrm{sub}, S_\mathrm{coat})$.
If we use the non-reciprocal albedo-scaling approach described in the Layering section (Equation~\ref{non-reciprocal-albedo-scaling-with-T}), then $f_\mathrm{layer}$ can be expressed as
\begin{equation}
f_\mathrm{layer} = f_\mathrm{coat} + T^2_\mathrm{coat} \left(1 - E_\mathrm{coat}\right) \,f_\mathrm{sub} \ ,
\end{equation}
where $f_\mathrm{coat}$ is the BSDF of the coat and $T^2_\mathrm{coat}$ is its transmittance (accounting for the entry and exit of the path, assuming normal incidence for simplicity). Thus combining these gives
\begin{equation} \label{coat_layering_formula_with_albedo_scaling}
f_\mathrm{weighted-layer} = w_\mathrm{coat} f_\mathrm{coat} +  \mathrm{lerp}\bigl(1, T^2_\mathrm{coat} (1 - E_\mathrm{coat}),  w_\mathrm{coat}\bigr) f_\mathrm{sub} \ ,
\end{equation}
which, for example, is the formula used in the Autodesk Standard Surface \cite{Georgiev2019} coat layer.

In this fashion, the abstract mix and layer operations can be mapped to a more computationally convenient approximate representation in terms of a weighted sum of BSDF lobes.
In Appendix~\ref{sec:mixture_model}, a brief example derivation of the complete form of the OpenPBR model as such a linear mixture model of BSDFs is presented.

\clearpage


\subsection{Microfacet models}

\label{sec:microfacet}

Here we provide general assumptions about the form and parameterization of the BSDFs which describe the interfaces in the model formalism outlined in the previous sections.

The BSDFs $f_\mathrm{conductor}$, $f_\mathrm{dielectric}$, $f_\mathrm{coat}$ and $f_\mathrm{diffuse}$ of the \hyperref[sec:metallic-base]{metal}, \hyperref[sec:dielectric-base]{dielectric}, \hyperref[sec:coat]{coat} and \hyperref[sec:glossy-diffuse]{glossy-diffuse} slabs, respectively, are each assumed to be described by a standard \emph{microfacet model}. This is a widely used physical picture \cite{Heitz2014, Pharr2023} in which the surface is assumed to be composed of a heightfield consisting of an ensemble of smooth microfacets of either metal, dielectric or Lambertian material, where the statistical distribution of the normal of these facets, termed the \emph{micronormal}, determines the surface roughness characteristics at the macroscopic scale. (The \hyperref[sec:fuzz]{fuzz} model is distinct and based on a volumetric ``microflake'' model \cite{Heitz2015}.) At the present time, microfacet models are effectively the lowest-level physical description in the CG literature of the surface properties of materials. (There is interesting work on more sophisticated proposed formalisms that, for example, account for wave optical effects \cite{Steinberg2022, Steinberg2024}, but these are not yet widely used in practice.)

A microfacet BRDF has the standard single-scattering form \cite{Walter2007,Pharr2023} consisting of a product of the Fresnel factor, the masking-shadowing function, and the Normal Distribution Function (NDF): \footnote{Also a Jacobian factor not shown here.}
\begin{equation}
\label{microfacet_brdf_ss}
f(\omega_i, \omega_o) \propto F(\omega_i, h) \; D(h) \; G(\omega_i, \omega_o) \ ,
\end{equation}
where $h = (\omega_i+\omega_o)/|\omega_i+\omega_o|$ is the half-vector (i.e., the micronormal), which mirror reflects $\omega_i$ into $\omega_o$. For dielectrics there is also a BTDF, i.e., the portion of the BSDF where the input and output directions lie in opposite rather than the same hemispheres, which has a similar form to the BRDF except with a modified half-vector, Fresnel factor, and Jacobian \cite{Walter2007}.

The \emph{Fresnel factor} $F(\omega_i, h)$ is determined by the complex index of refraction (IOR) of the reflecting material of each microfacet (or, more technically, the ratio of this to the exterior IOR), and its form differs depending on whether the material is a dielectric or conductor \cite{Walter2007}.

The \emph{masking-shadowing function} $G(\omega_i, \omega_o)$ accounts for the probability that the input and output directions are occluded by the microsurface. It is usually derived using the Smith model, which determines $G$ given the NDF, and for the GGX NDF Equation~\ref{GGX} the masking-shadowing function then has a well-known form \cite{Heitz2014}.

The \emph{Normal Distribution Function} (NDF) $D(m)$ describes the relative probability of occurrence of micronormal $m$ on the surface, and thus the roughness characteristics. We assume that the NDF is the so-called GGX distribution, which closely approximates the roughness of real materials (the name GGX derives from ``ground glass'', but the formula was originally due to Trowbridge and Reitz \cite{Trowbridge1975, Walter2007,Burley2012,Heitz2014,Pharr2023}), which has the basic form
\begin{equation} \label{GGX}
D_\mathrm{GGX}(m) \propto \left( 1 + \frac{\tan^2\theta_m}{\alpha^2} \right)^{-2} \ ,
\end{equation}
where $\theta_m$ is the angle between $m$ and the (macroscopic) surface normal, and the parameter $\alpha$ controls the apparent roughness of the microsurface. As $\alpha \rightarrow 0$, the distribution of normals becomes highly peaked around $\theta_m=0$ so the microsurface is mostly flat, leading to a smooth appearance, while as $\alpha$ grows the microsurface becomes increasingly jagged, leading to a rough appearance.

In practice, we restrict to the range $\alpha \in [0,1]$, as $\alpha > 1$ does not produce a plausible rough appearance. Following Disney's ``Principled'' shader \cite{Burley2012}, we set (in the isotropic case)
\begin{equation}
\alpha = r^2 \ ,
\end{equation}
where $r \in [0,1]$ is the user-facing \emph{roughness}, as this produces a more perceptually linear resulting change in apparent roughness as $r$ is varied.

The single-scattering microfacet BRDF of Equation~\ref{microfacet_brdf_ss} does not conserve energy, as it neglects to account for multiple scattering between the microfacets. In OpenPBR, we generally consider it important that a material which is supposed to be non-absorbing (e.g., a metal with Fresnel factor close to 1) should conserve energy and pass a white furnace test. Thus an implementation should account for multiple scattering via one of a number of schemes, otherwise the reflection from rough metals and dielectrics will be dimmer and less saturated than it should be. An accurate approach based on the Smith microsurface model is described in \textcite{Heitz2016a}. Simpler approximate models are presented in \textcite{Kulla2017} (which functions by adding compensation lobes to account for the missing energy), and \textcite{Turquin2019} (which scales the albedo of the lobe to maintain energy preservation at the expense of reciprocity).

\subparagraph{Roughness anisotropy}

\begin{figure}[!tb]
  \centering
  \begin{subfigure}{.3\textwidth}
    \includegraphics[width=\linewidth]{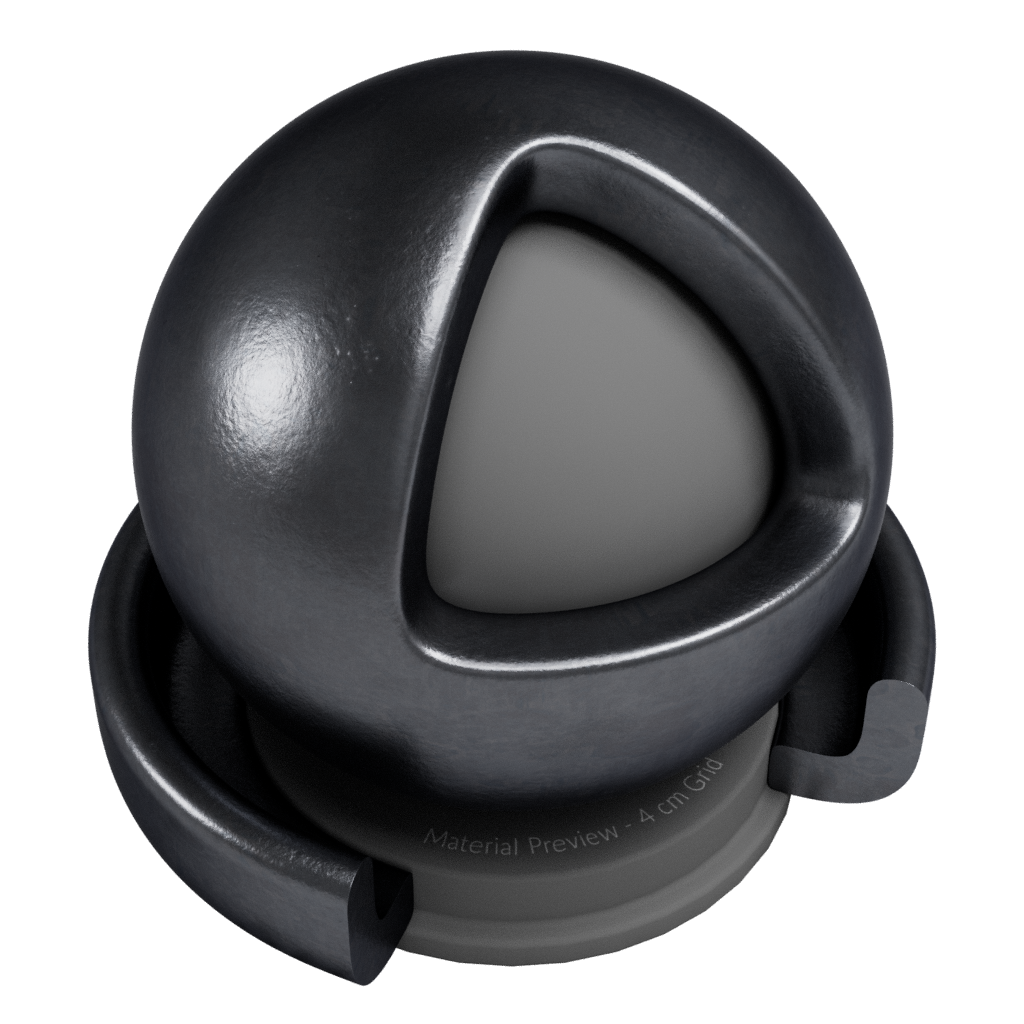}
  \end{subfigure}
  \hspace{0.02\textwidth}
  \begin{subfigure}{.3\textwidth}
    \includegraphics[width=\linewidth]{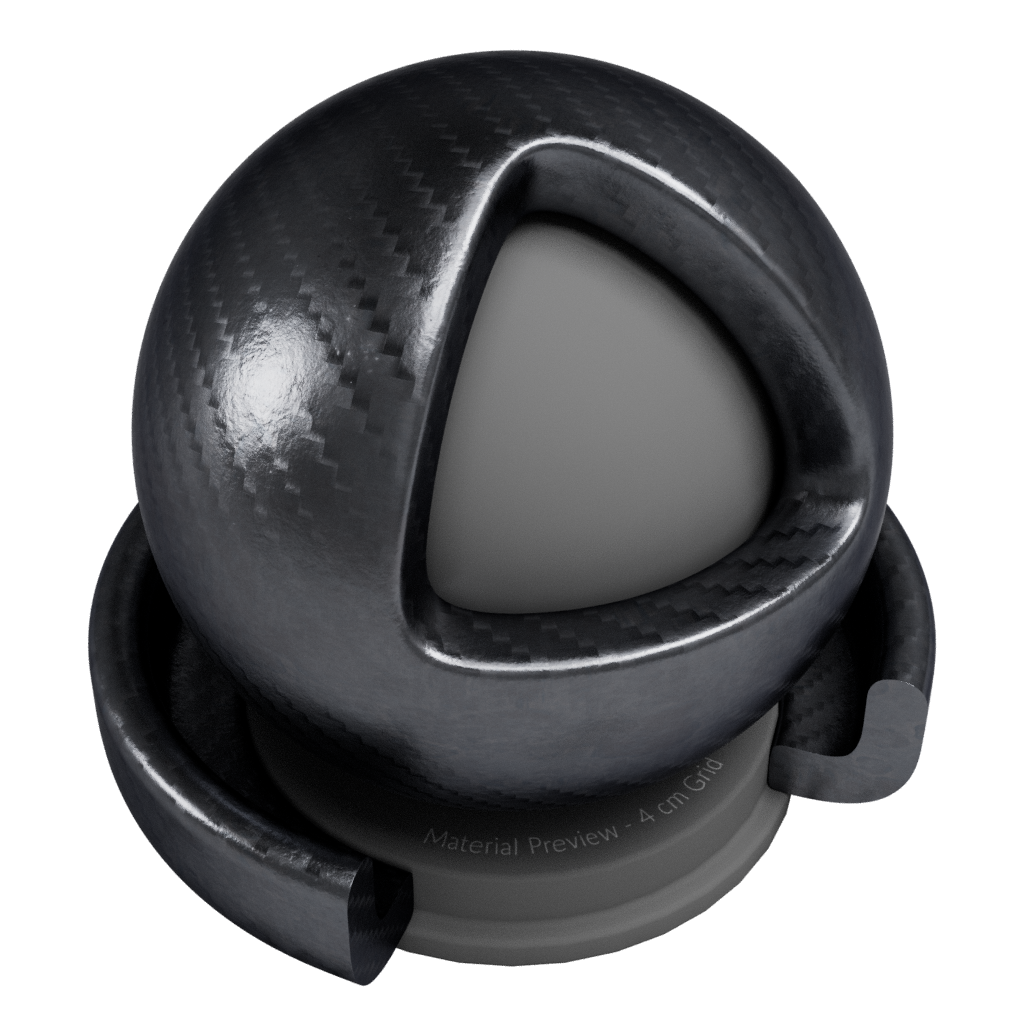}
  \end{subfigure}
  \hspace{0.02\textwidth}
  \begin{subfigure}{.3\textwidth}
    \includegraphics[width=\linewidth]{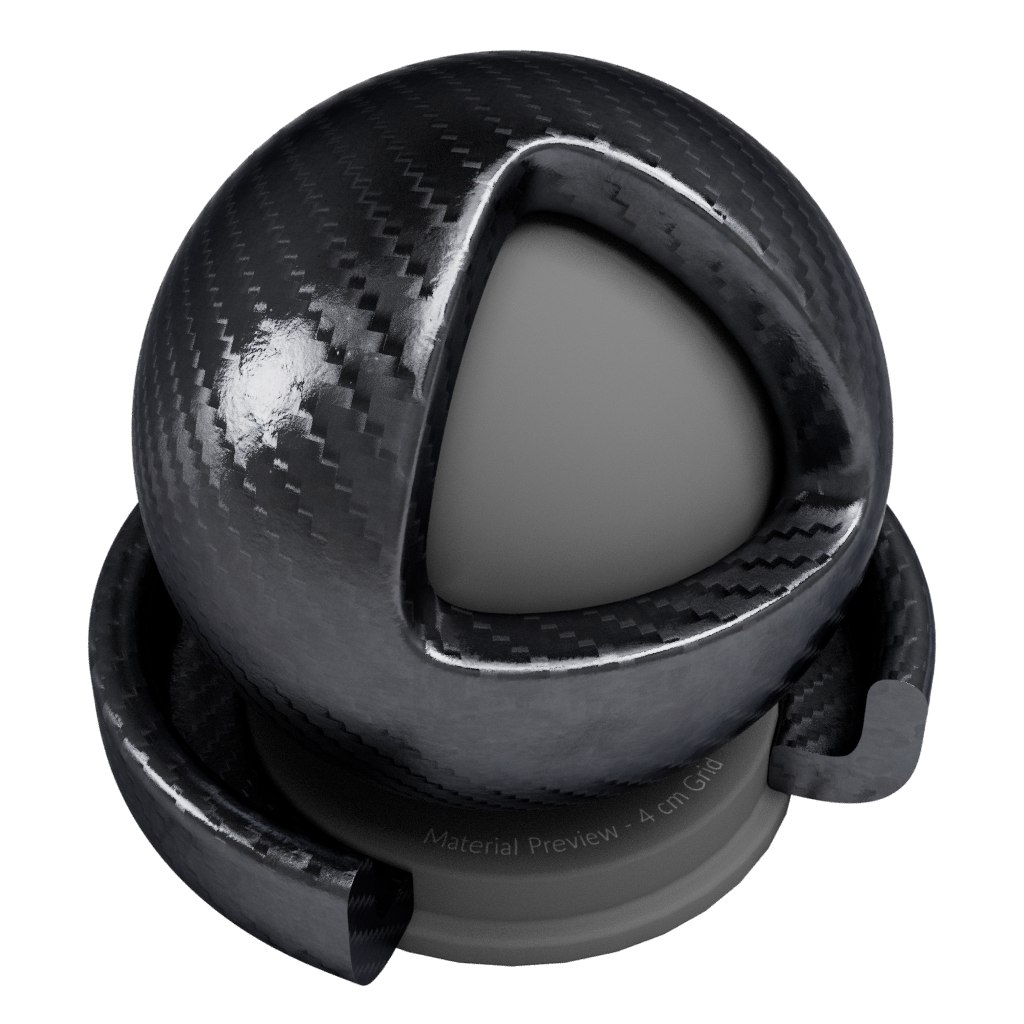}
  \end{subfigure}
  \caption{Comparison of carbon fiber materials with varying $\mathtt{specular\_roughness\_anisotropy}$ $a$. The left image shows the isotropic case ($a=0.0$), the middle image shows moderate anisotropy ($a=0.5$), and the right image shows high anisotropy ($a=1.0$).\label{fig:carbon_fiber}}
\end{figure}

In the general case, the roughness is anisotropic, that is the NDF is not circularly symmetric but stretched along some direction in the surface plane, producing an elongation of the specular highlight along that direction. This simulates coherent microscale groove geometry due to processes such as brushing, scratches, or materials with inherent directionality such as carbon fiber (Figure~\ref{fig:carbon_fiber}).
It is assumed that a reference tangent vector field is defined (via \verb|geometry_tangent| and \verb|geometry_coat_tangent|). The reference tangent vector indicates the direction along which the NDF is stretched, meaning the microscale grooves tend to be aligned with the orthogonal bitangent.
Note that usually the normal map will be generated assuming some fixed per-vertex reference tangents (computed often from the UVs), but those tangents will not necessarily correspond to the desired anisotropy flow.

In practice, the renderer will have the per-vertex reference normals $N$ and tangents $T$ (from UVs), and normal maps are generated relative to these reference $N$, $T$ (and orthonormal bitangent $B$). The anisotropy tangents can then be defined via a ``flow map'' that specifies a 2d vector relative to the reference $B, T$. From this, one can then construct a consistent vertex frame with an orthonormal $(N', T')$ shading normal and anisotropy direction. The GGX distribution in the anisotropic case is then parameterized by two separate $\alpha$-roughnesses $\alpha_t$ and $\alpha_b$ along the tangent and bitangent vectors as follows, given the polar angle $\phi_m$ of the micronormal $m$ relative to the tangent (counterclockwise about the normal):
\begin{equation}
D_\mathrm{GGX}(m) \propto \left( 1 + \tan^2\theta_m \Bigl(\frac{\cos^2\phi_m}{\alpha_t^2} + \frac{\sin^2\phi_m}{\alpha_b^2}\Bigr) \right)^{-2} \ ,
\end{equation}
which reduces to the isotropic form when $\alpha_t = \alpha_b = \alpha$.
Efficient techniques for sampling BSDFs employing the anisotropic GGX microfacet model are presented in \textcite{Heitz2018, Dupuy2023}.

\begin{figure}[!tb]
  \centering
  \begin{subfigure}{.48\textwidth}
    \centering
    \includegraphics[width=\linewidth]{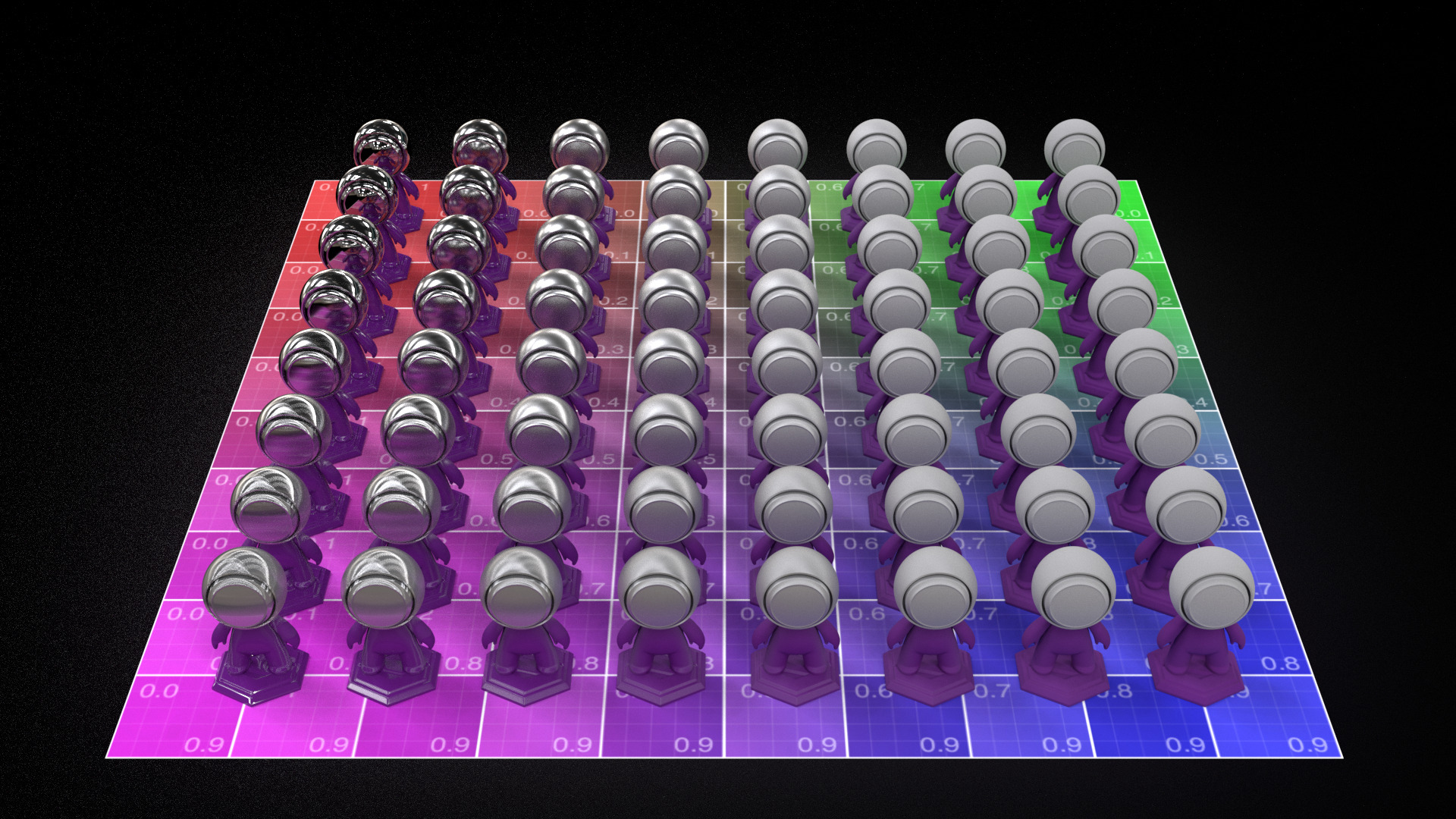}
  \end{subfigure}
  \begin{subfigure}{.48\textwidth}
    \centering
    \includegraphics[width=\linewidth]{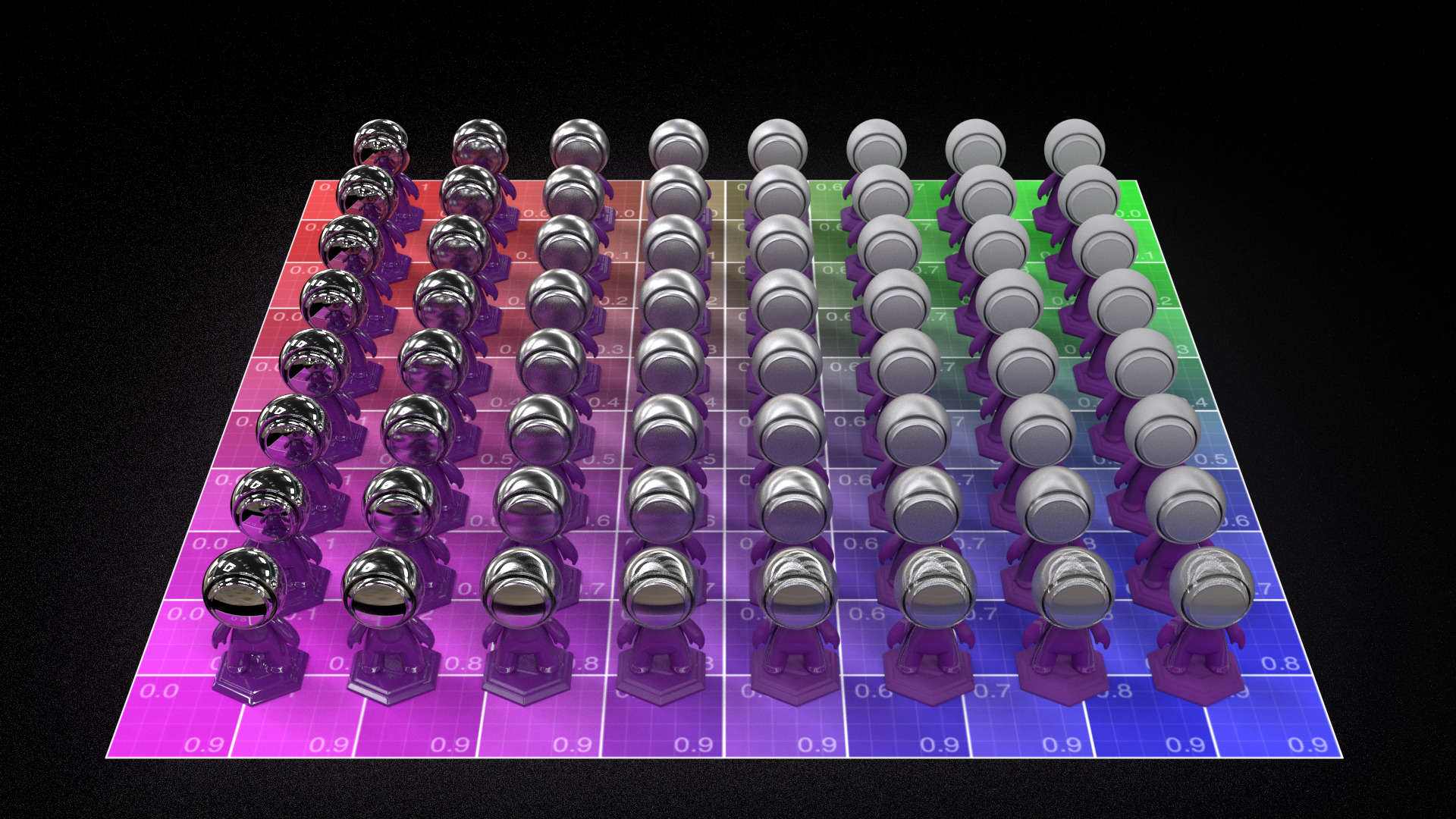}
  \end{subfigure}
  \caption{Wedges of roughness and anisotropy, for the Adobe Standard Material model (left), and OpenPBR (right). The roughness goes from 0 on the left to 1 on the right, and the anisotropy goes from 0 at the top, to 1 at the bottom.}
  \label{fig:ndf_anisotropy_grid}
\end{figure}

The NDF terms $\alpha_t$ and $\alpha_b$ are more conveniently parameterized as the total roughness $r$ (\verb|specular_roughness|) and an anisotropy $a \in [0, 1]$ (\verb|specular_roughness_anisotropy|). In OpenPBR we require the following mapping from $r, a$ to $\alpha_t, \alpha_b$:
\begin{eqnarray}  \label{openpbr-anisotropy-formula}
\alpha_t &=& r^2 \sqrt{\frac{2}{1 + (1 - a)^2}} \;\ \nonumber \\
\alpha_b &=& (1 - a) \, \alpha_t \ .
\end{eqnarray}
This formulation satisfies $\alpha_t^2 + \alpha_b^2 = 2\alpha^2$, to preserve the average roughness regardless of the anisotropy. A rationale is that if a renderer doesn't support anisotropy (or if the feature is turned off for performance considerations, such as level of detail), using only the roughness parameter should result in an isotropic specular highlight perceptually close to the original anisotropic one. Figure~\ref{fig:ndf_anisotropy_grid} shows the resulting behavior for various roughness and anisotropy values according to this formulation (right panel), which was considered to behave more intuitively than the parameterization used in Adobe Standard Material (left panel).

Summarizing the NDF parameterization, the \hyperref[sec:dielectric-base]{dielectric-base} BSDF $f_\mathrm{dielectric}$ and \hyperref[sec:metallic-base]{metallic-base} BRDF $f_\mathrm{conductor}$ share the same roughness parameters (\verb|specular_roughness| and \verb|specular_roughness_anisotropy|), while the \hyperref[sec:coat]{coat} BSDF $f_\mathrm{coat}$ uses an independent set (\verb|coat_roughness| and \verb|coat_roughness_anisotropy|).

\clearpage

\subsection{Model structure}

\label{sec:model-structure}

\begin{figure}[!htb]
  \centering
\includegraphics[width=0.98\linewidth]{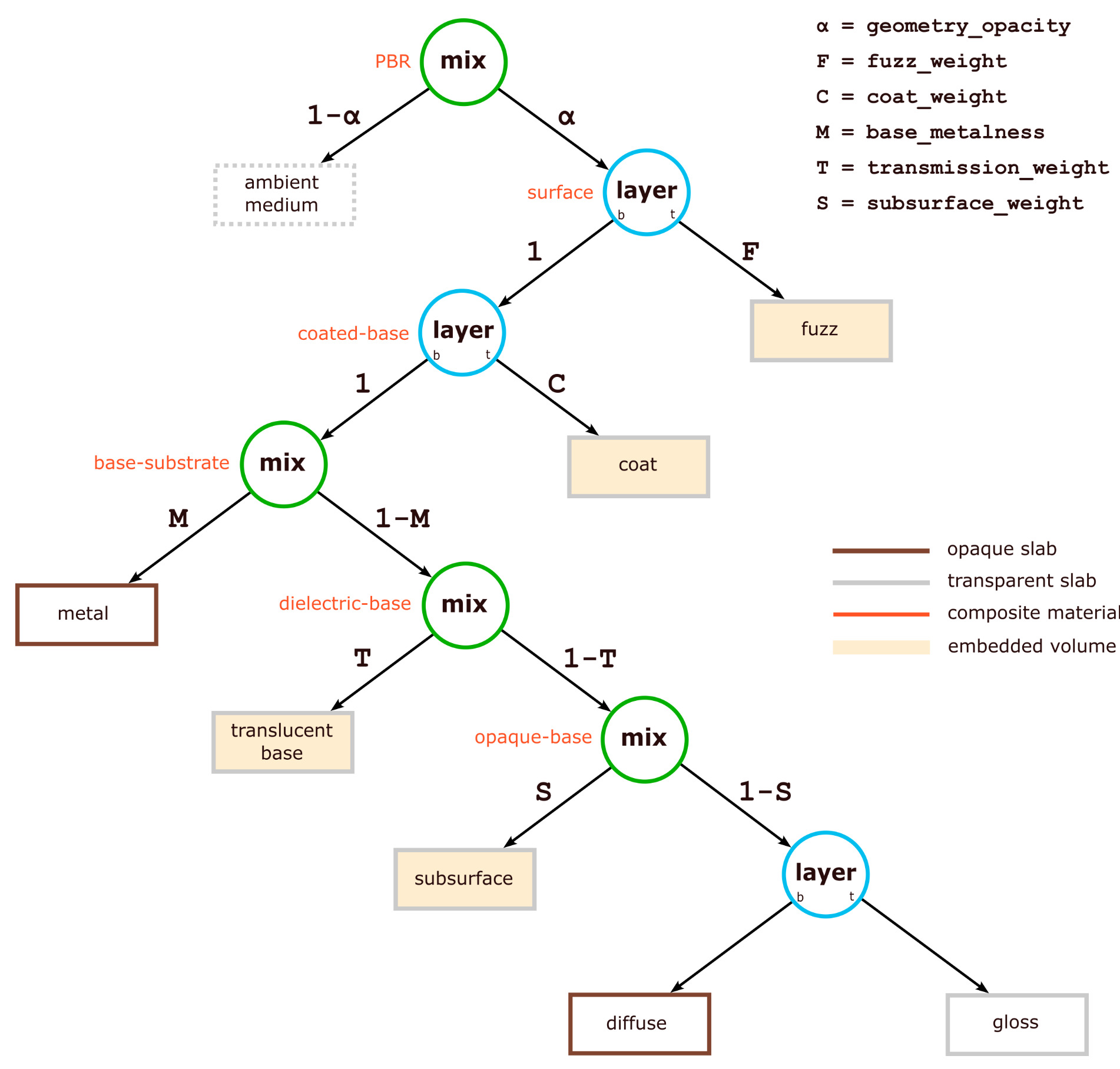}
\caption{Tree structure of the OpenPBR shading model.}
\label{fig:tree_structure}
\end{figure}

In summary, the formal structure consists of the following slabs:
\begin{align}
S_\textrm{ambient-medium}   &= \mathrm{Slab}(\emptyset)                                           &                                                          \nonumber \\
S_\textrm{fuzz}             &= \mathrm{Slab}(f_\mathrm{fuzz}, V_\mathrm{fuzz})                    & \text{\hyperref[sec:fuzz]{fuzz}}                         \nonumber \\
S_\textrm{coat}             &= \mathrm{Slab}(f_\mathrm{coat}, V_\mathrm{coat})                    & \text{\hyperref[sec:coat]{coat}}                         \nonumber \\
S_\textrm{metal}            &= \mathrm{Slab}(f_\mathrm{conductor})                                & \text{\hyperref[sec:metallic-base]{metal}}               \nonumber \\
S_\textrm{translucent-base} &= \mathrm{Slab}(f_\mathrm{dielectric}, V^\infty_\mathrm{dielectric}) & \text{\hyperref[sec:translucent-base]{translucent-base}} \nonumber \\
S_\textrm{subsurface}       &= \mathrm{Slab}(f_\mathrm{dielectric}, V^\infty_\mathrm{subsurface}) & \text{\hyperref[sec:subsurface]{subsurface}}             \nonumber \\
S_\textrm{gloss}            &= \mathrm{Slab}(f_\textrm{dielectric}, V_\mathrm{dielectric})        & \text{\hyperref[sec:glossy-diffuse]{glossy-diffuse}}     \nonumber \\
S_\textrm{diffuse}          &= \mathrm{Slab}(f_\textrm{diffuse})                                  & \text{\hyperref[sec:glossy-diffuse]{glossy-diffuse}}
\end{align}
These are composed to build the material structure, denoted $M_\textrm{PBR}$ below, as follows:
\begin{align}
M_\textrm{PBR}             &= \mathrm{\mathbf{mix}}  (S_\textrm{ambient-medium} , M_\textrm{surface},          \mathtt{\alpha}) \quad\quad &\mathrm{where} \; \mathtt{\alpha} &= \mathtt{geometry \_ opacity}    \nonumber \\
M_\textrm{surface}         &= \mathrm{\mathbf{layer}}(M_\textrm{coated-base}    , S_\textrm{fuzz},             \mathtt{F})      \quad\quad &\mathrm{where} \; \mathtt{F}      &= \mathtt{fuzz \_ weight}         \nonumber \\
M_\textrm{coated-base}     &= \mathrm{\mathbf{layer}}(M_\textrm{base-substrate} , S_\textrm{coat},             \mathtt{C})      \quad\quad &\mathrm{where} \; \mathtt{C}      &= \mathtt{coat \_ weight}         \nonumber \\
M_\textrm{base-substrate}  &= \mathrm{\mathbf{mix}}  (M_\textrm{dielectric-base}, S_\textrm{metal},            \mathtt{M})      \quad\quad &\mathrm{where} \; \mathtt{M}      &= \mathtt{base \_ metalness}      \nonumber \\
M_\textrm{dielectric-base} &= \mathrm{\mathbf{mix}}  (M_\textrm{opaque-base}    , S_\textrm{translucent-base}, \mathtt{T})      \quad\quad &\mathrm{where} \; \mathtt{T}      &= \mathtt{transmission \_ weight} \nonumber \\
M_\textrm{opaque-base}     &= \mathrm{\mathbf{mix}}  (M_\textrm{glossy-diffuse} , S_\textrm{subsurface},       \mathtt{S})      \quad\quad &\mathrm{where} \; \mathtt{S}      &= \mathtt{subsurface \_ weight}   \nonumber \\
M_\textrm{glossy-diffuse}  &= \mathrm{\mathbf{layer}}(S_\textrm{diffuse}        , S_\textrm{gloss})
\end{align}
While this compactly describes things mathematically, it can be helpful to view the composition as a tree formed from the layer and mix operations, as shown in Figure~\ref{fig:tree_structure}, since it visually connects the equations with the physical layer structure (Figure~\ref{fig:layer_structure}).

The task of the implementation is to compute the BSDF of the material $M_\textrm{PBR}$ given this stated structure. Complete conformance to the specification is defined as reproducing all the physical inter-layer light transport effects in the structure thus described. However, enforcing a particular implementation would make the use of the material model impractical for certain classes of renderers and ultimately make it less useful. For this reason, we consider the choice of a specific implementation of the final BSDF to be outside the scope of the specification. In practice, each implementation must decide what level of approximation to use for the light transport within layers, trading off accuracy for efficiency according to its own particular use case.

For convenience and efficiency, at present it is most likely to be mapped to a model consisting of a mixture of BSDF lobes similar to the Autodesk Standard Surface shader \cite{Georgiev2019} and its representation in MaterialX. An example derivation of such a model is provided in Appendix~\ref{sec:mixture_model}. In the official repository we provide a \href{https://github.com/AcademySoftwareFoundation/OpenPBR/blob/main/reference/open_pbr_surface.mtlx}{reference implementation} in MaterialX, which has a structure similar to this derivation.

\clearpage
\subsection{Geometry}

\label{sec:geometry}

We define a number of parameters which are intended to control geometrical aspects of the surface:

\begin{itemize}

\item \verb|geometry_opacity|: the presence weight of the entire surface, thus functioning effectively as a linear transparency ``alpha blend''.

\item \verb|geometry_normal|, \verb|geometry_coat_normal|: these normal inputs define the perturbation of the shading normal for the base and coat BSDF models, respectively. Separately perturbing the coat normal allows for the appearance of a finite thickness coat on top of the base.

\item \verb|geometry_tangent|, \verb|geometry_coat_tangent|: these define the direction of the microfacet anisotropy on the base layer and coat layer, for effects such as brushed metal. Similar to normal mapping, the tangent may be transformed with a 2D vector map to alter the direction of the anisotropy.

\item \verb|geometry_thin_walled|: a boolean parameter which indicates whether the surface is thin-walled, i.e., the thickness of the material is negligible compared to the size of the surface. This is described in more detail in \hyperref[sec:thin-walled]{Section~\ref{sec:thin-walled}}.

\end{itemize}

The perturbed normals and tangents will typically be specified by input textures (or possibly procedurally). The particular parameterization which maps the contents of the texture to the perturbation of the normal or tangent in the shading frame is not defined in the model itself, as this can be done in multiple ways. It is instead assumed that the material is packaged with metadata that makes this mapping unambiguous.

\subparagraph{Opacity}

Opacity is implemented by defining the surface as a mix of the material bulk with the ambient medium:
\begin{equation}
M_\textrm{PBR} = \mathrm{\mathbf{mix}}(S_\textrm{ambient-medium} , M_\textrm{surface}, \mathtt{\alpha}) \quad\quad \mathrm{where} \; \mathtt{\alpha} = \mathtt{geometry\_opacity} \ .
\end{equation}
Note that in the case of a non-thin-walled material, $\mathtt{\alpha} < 1$ doesn't make strict physical sense unless the entire surface is removed, whereas in the \hyperref[sec:thin-walled]{thin-walled} mode the opacity has a clear physical interpretation as the presence weight of the wall (or ``cutout'' areas where $\mathtt{\alpha} = 0$) like in the leaf render below.

We generally leave it as an implementation detail for a renderer to determine how connections to light sources be made through the surface. However a very common approximation used by many renderers is ``transparent shadows'', where a straight-line connection is made to lights and the contribution of the light determined by the total transmittance along the ray, ignoring any refraction events. We give here a suggested form for this shadow ray transmittance.
The computed transmittance should take into account the presence weight of the entire geometry (\verb|geometry_opacity|), and the transmittance through the geometry if present. The latter is the transmittance through the base dielectric only (as the metal is opaque), denoted $\mathbf{T}_\mathrm{dielectric}$, which should take into account the Fresnel factor of the dielectric interface and the extinction in the volumetric media  (in general a statistical mixture of the subsurface medium $V^\infty_\mathrm{subsurface}$ and translucent-base medium $V^\infty_\mathrm{dielectric}$).
The mix weight of the base dielectric is $w_\mathrm{dielectric} = 1 - \mathtt{M}$, where $\mathtt{M} = \mathtt{base\_metalness}$.
The total transmittance can thus be approximated as
\begin{eqnarray}
\mathbf{T}_\mathrm{pbr} &=& \mathrm{lerp}(1,  w_\mathrm{dielectric} \mathbf{T}_\mathrm{dielectric}, \mathtt{\alpha}) \nonumber \\
                        &=& 1 - \mathtt{\alpha} \bigl(1 - (1 - M) \mathbf{T}_\mathrm{dielectric}\bigr) \ .
\end{eqnarray}

\clearpage

\subsection{Parameterization}

\label{sec:parameterization}

There is a suite of parameters exposed to the artist which effectively define the properties of all the slab BSDFs and volumes, as well as the mixture weights of the tree structure defined in Figure~\ref{fig:tree_structure}. The full parameter set is detailed in Appendix~\ref{sec:parameters}.

The parameterization follows several guiding principles to ensure ease of use and consistency.

\begin{itemize}

  \item Each parameter has a unique identifying name. Consistent with Autodesk Standard Surface and Adobe Standard Material, the parameter names are grouped intuitively according to the BSDF lobe whose appearance they most influence, with a corresponding prefix such as \verb|specular| or \verb|subsurface|. It is suggested to use this prefix to group parameters in the user interface. There is an associated, simpler and capitalized label to be used in the user interface.

  \item  Most parameters range preferably between 0 and 1 for ease of texturing and manipulation. In the same spirit, each color is systematically broken down into an albedo and an intensity, with the suffix \verb|color| and \verb|weight|, respectively. This also avoids having to insert a shader between a color texture and its input to modulate its intensity, thus it doesn't require adopting renderers to implement shader networks. Both absolute and suggested useful (i.e., soft) ranges are specified to help ensure a consistent user experience.

  \item The default value is chosen so that the model behaves physically by default, with the possibility for the user to drift away from a realistic look, as is sometimes needed in production. Units are specified for each parameter to facilitate easy conversions between different formats.

 \item  The specification also tries to avoid parameters that require setting another parameter to see an effect. As the parameterization grew more complex, it was considered important to resist the urge to add modal parameters enabled or disabled by dropdowns, which would unnecessarily complexify user experience.

 \item We also provided the logic to determine which parameters can currently be disabled (e.g., presented as ``grayed out'' in the user interface) based on the current state of the model. For example, if the coat weight is set to zero, then the rest of the coat parameters are disabled as they have no effect on the model. This automatic hiding of irrelevant parameters is important to avoid overwhelming the user with too many options, and to ensure that the user can focus on the parameters that are relevant to the current configuration of the material.

\end{itemize}

We now proceed to describe all of the component slabs and their parameterizations in detail, working from the base of the material towards the top.

\clearpage

\section{Base substrate}

\label{sec:base_substrate}

\begin{figure}[!b]
  \centering
  \begin{subfigure}{.3\textwidth}
    \includegraphics[width=\linewidth]{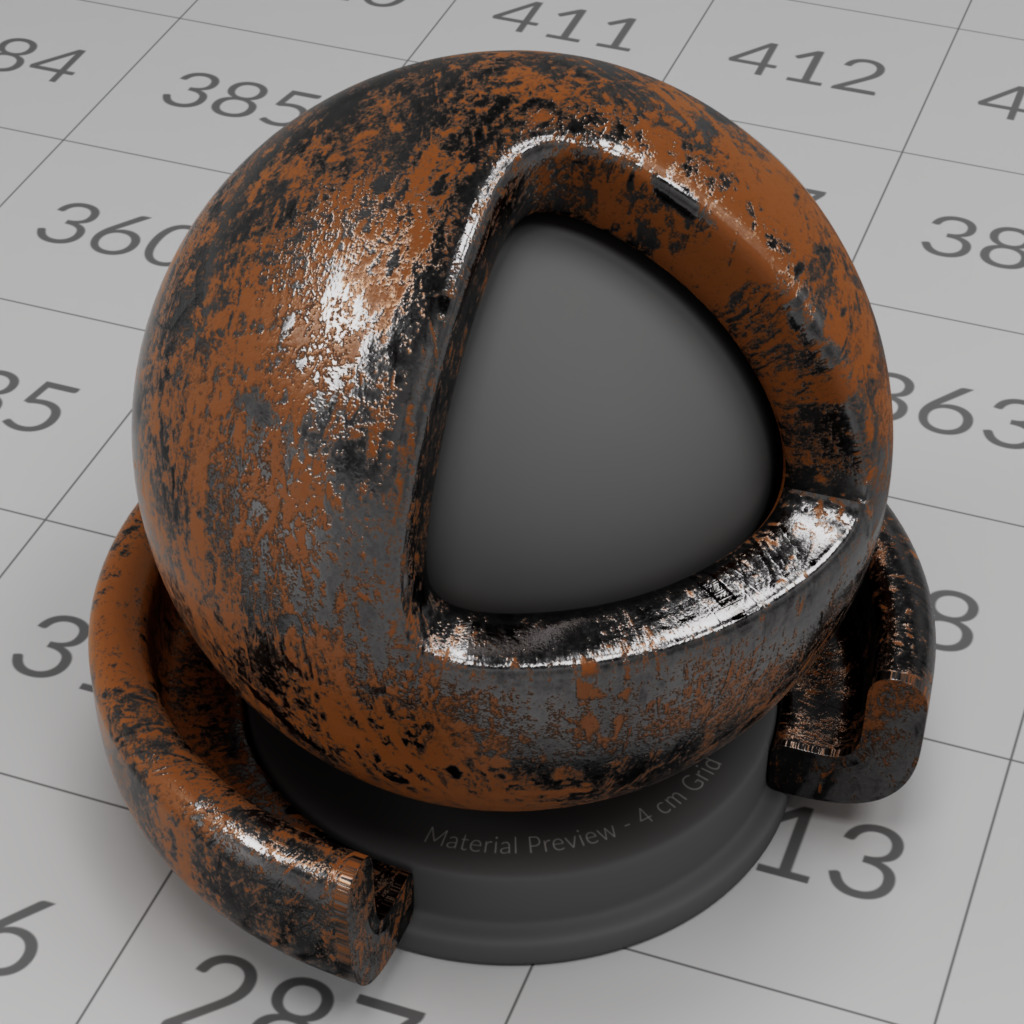}
  \end{subfigure}
  \hspace{0.02\textwidth}
  \begin{subfigure}{.3\textwidth}
    \includegraphics[width=\linewidth]{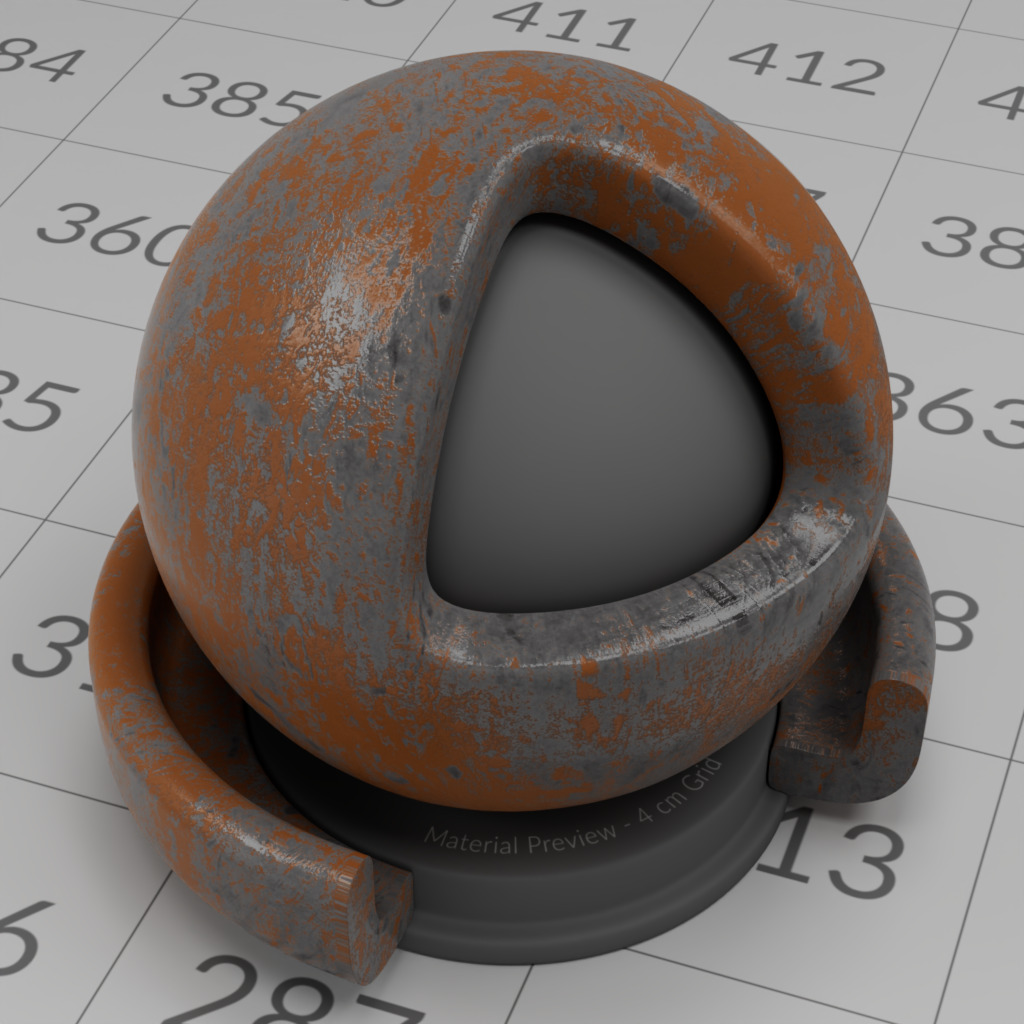}
  \end{subfigure}
  \hspace{0.02\textwidth}
  \begin{subfigure}{.3\textwidth}
    \includegraphics[width=\linewidth]{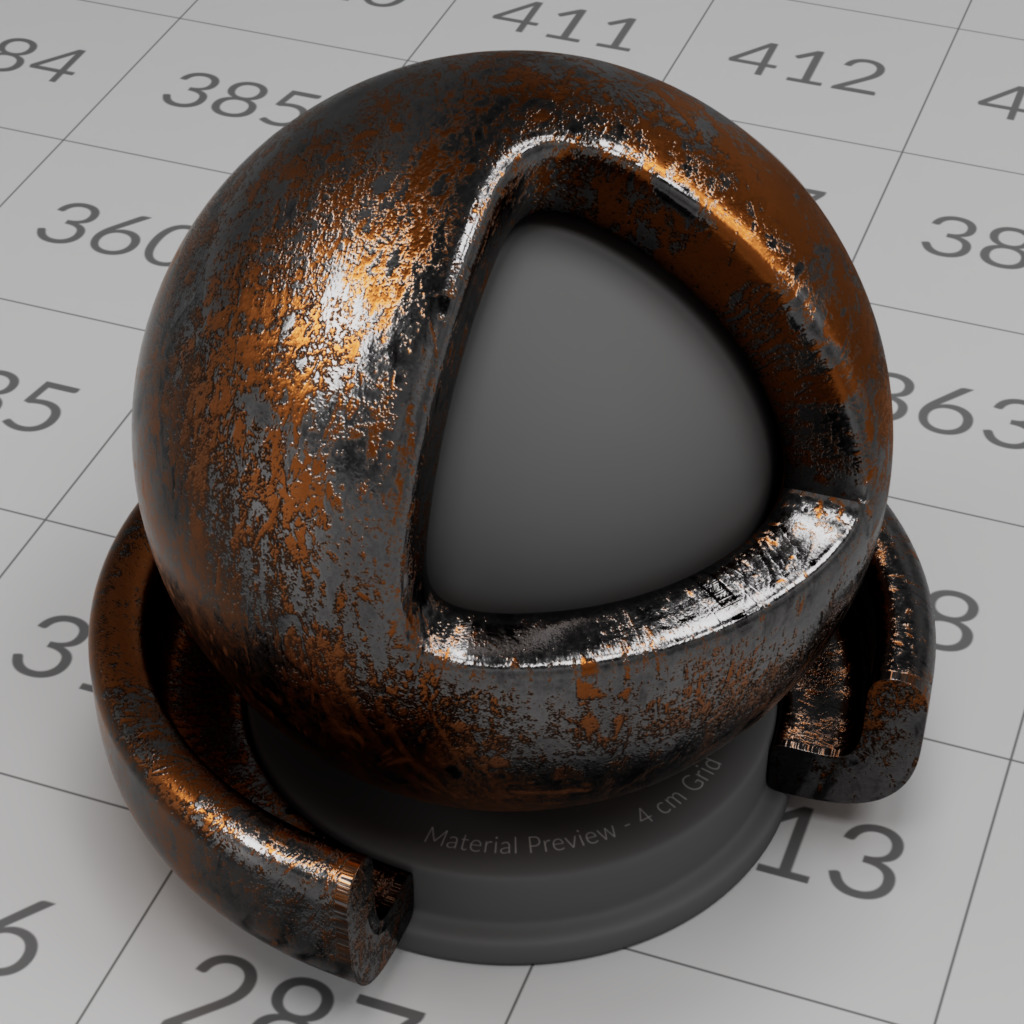}
  \end{subfigure}
  \caption{The use of textured \texttt{base\_metalness} to achieve a rusty metal look (left), and the equivalent with \texttt{base\_metalness} 0 and 1. \label{fig:rust_example}}
\end{figure}

The bulk at the bottom of the material structure, termed the base substrate, consists of a statistical mix of \hyperref[sec:metallic-base]{metal} and \hyperref[sec:dielectric-base]{dielectric} semi-infinite slabs:
\begin{eqnarray}
M_\textrm{base-substrate} &=& \mathrm{\mathbf{mix}}  (M_\textrm{dielectric-base}, S_\textrm{metal}, \mathtt{M}) \ ,
\end{eqnarray}
where $\mathtt{M} = \mathtt{base\_metalness}$, which controls the fraction of microsurface area that is metallic.
As described in the following section, the \hyperref[sec:dielectric-base]{dielectric base} is a mixture of three different components (\hyperref[sec:translucent-base]{translucent base}, \hyperref[sec:subsurface]{subsurface}, and \hyperref[sec:glossy-diffuse]{glossy-diffuse}).
Areas of surface will normally be either fully metallic or fully dielectric, but values of \verb|base_metalness| between 0 and 1 can be used to simulate smoothly blended transitions between areas of bare metal and areas of dielectric (modeling, for example, opaque rust or paint on top of the metal, as in Figure~\ref{fig:rust_example}).

For both the metal and dielectric cases, the primary specular lobe shape is controlled by the roughness properties of the surface, parameterized by \verb|specular_roughness| and \verb|specular_roughness_anisotropy| (as described in Section~\ref{sec:microfacet}).

\begin{figure}[!tb]

  \centering
  \hfill
  \begin{subfigure}{.24\textwidth}
    \captionsetup{font={normalsize}}
    \includegraphics[width=\linewidth]{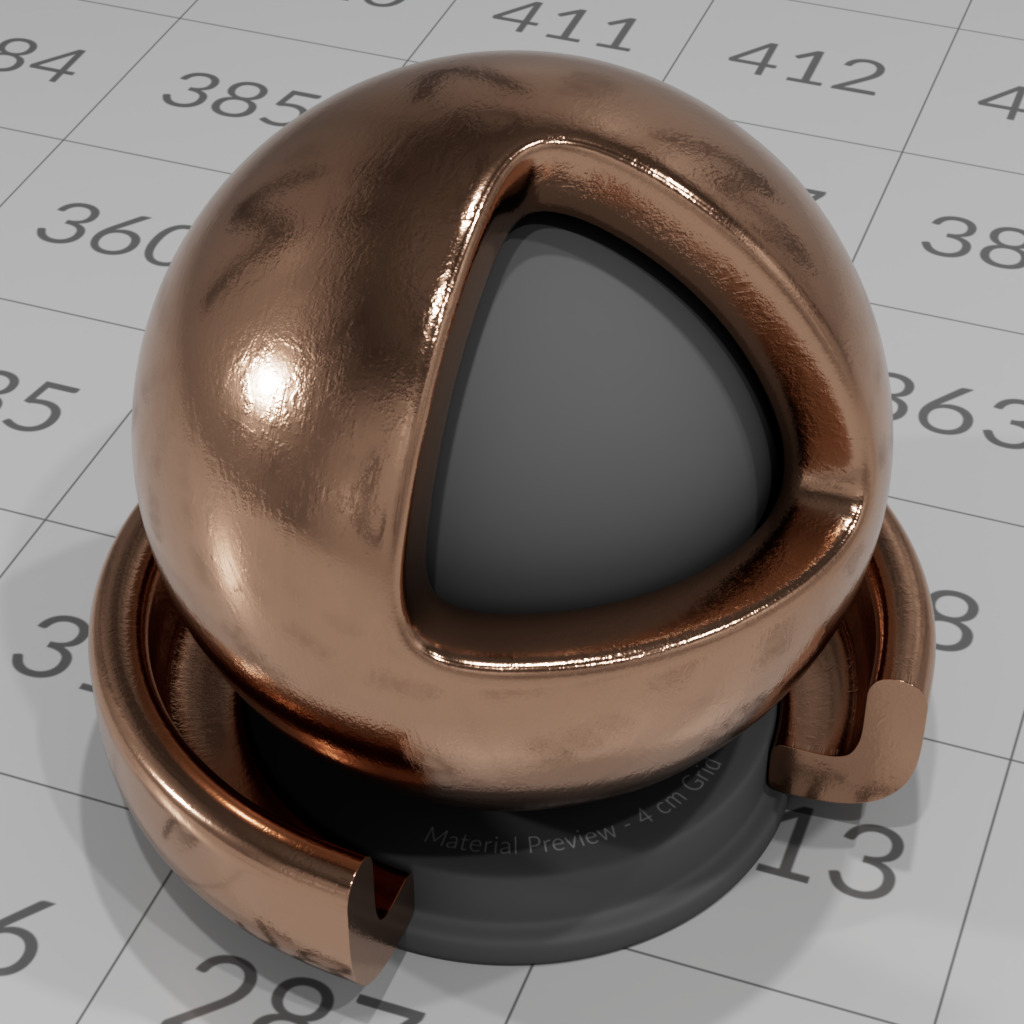}
    \caption{\hyperref[sec:metallic-base]{metal}}
  \end{subfigure}
  \hfill
  \begin{subfigure}{.24\textwidth}
    \captionsetup{font={normalsize}}
    \includegraphics[width=\linewidth]{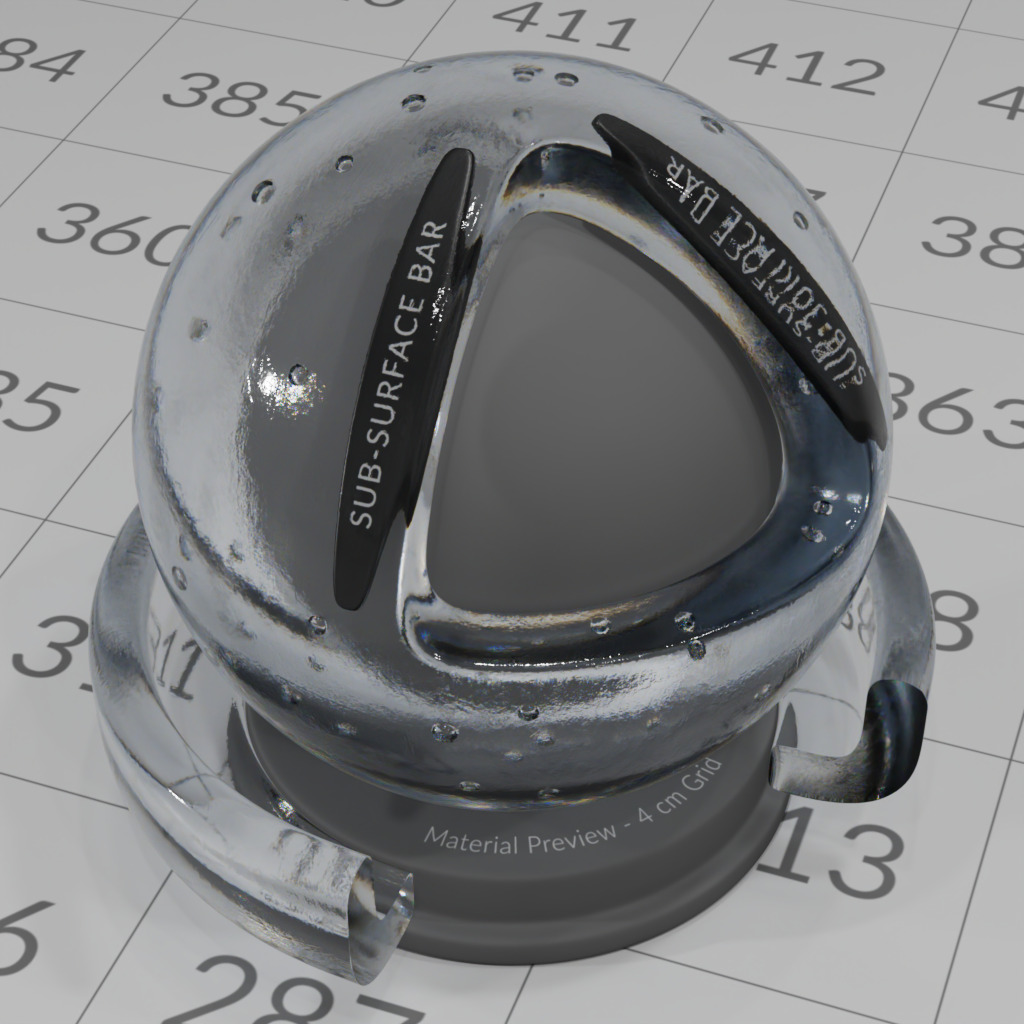}
    \caption{\hyperref[sec:translucent-base]{translucent base}}
  \end{subfigure}
  \hfill
  \begin{subfigure}{.24\textwidth}
    \captionsetup{font={normalsize}}
    \includegraphics[width=\linewidth]{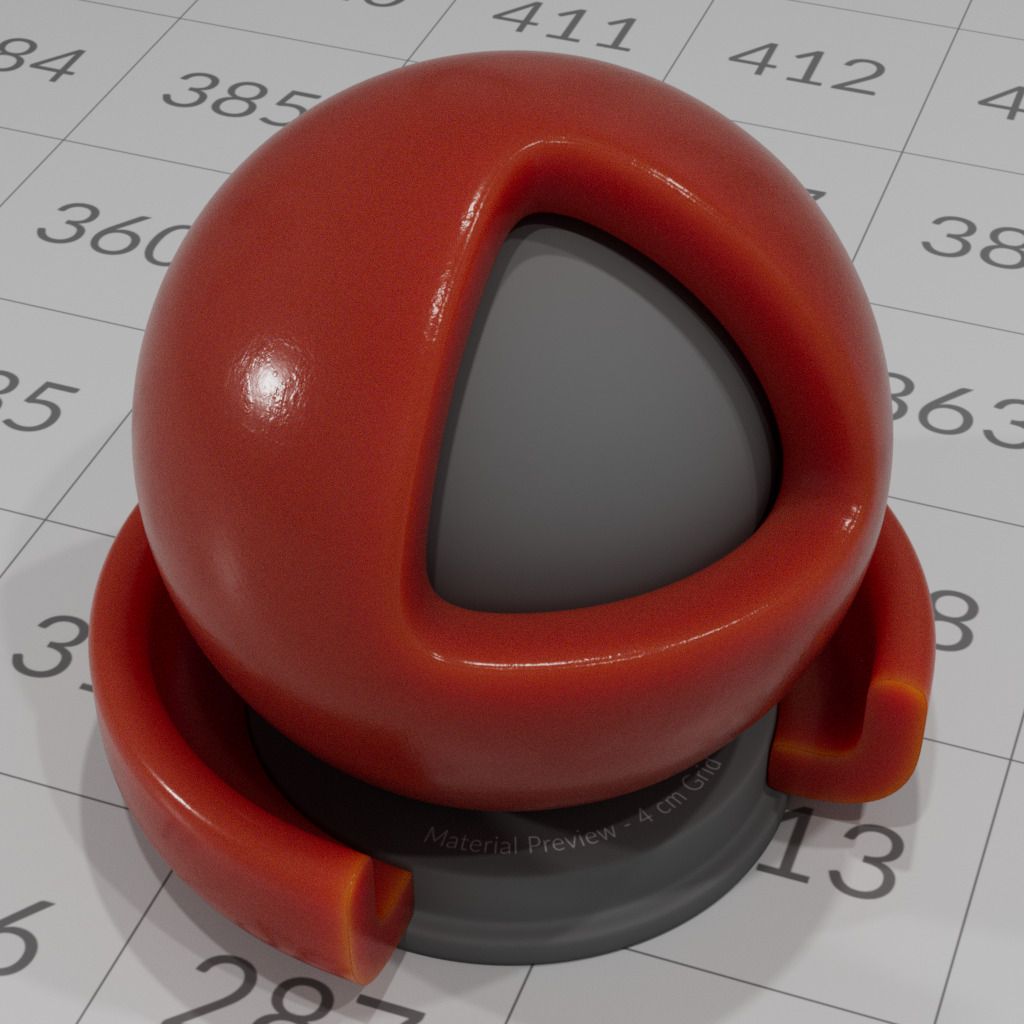}
    \caption{\hyperref[sec:subsurface]{subsurface}}
  \end{subfigure}
  \hfill
  \begin{subfigure}{.24\textwidth}
    \captionsetup{font={normalsize}}
    \includegraphics[width=\linewidth]{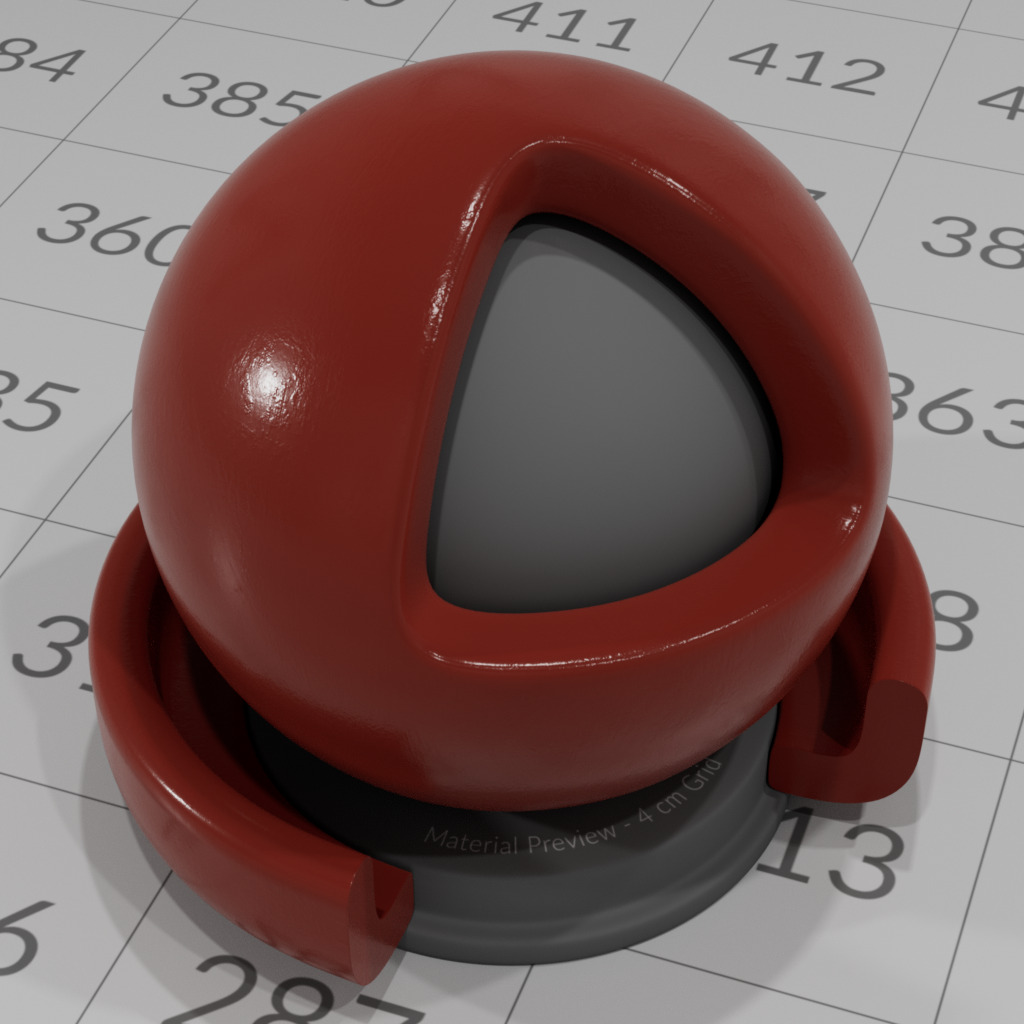}
    \caption{\hyperref[sec:glossy-diffuse]{glossy-diffuse}}
  \end{subfigure}
  \hfill
  \caption{The component slabs of the base substrate (where (b), (c) and (d) collectively constitute the dielectric base.\label{fig:base_components}}
\end{figure}

Figure~\ref{fig:base_components} shows all four component slabs of the base.

\subsection{Dielectric base}

\label{sec:dielectric-base}

The dielectric base is assumed to have a surface BSDF described by a rough GGX dielectric \hyperref[sec:microfacet]{microfacet} model, and a bulk volumetric medium supporting absorption and scattering (whether physically due to the inherent molecular properties of the dielectric as in water or, for instance, a dispersion of embedded particles or flakes as in paint).

However, we distinguish between a separate opaque and \hyperref[sec:translucent-base]{translucent} dielectric bulk, which are placed in a statistical mix:
\begin{equation}
M_\textrm{dielectric-base} = \mathrm{\mathbf{mix}}(M_\textrm{opaque-base}, S_\textrm{translucent-base}, \mathtt{T})
\end{equation}
where $\mathtt{T} = \mathtt{transmission\_weight}$.
This mirrors the usual workflow of artists where they are typically either modeling an opaque surface potentially with some specularity and dense subsurface scattering (such as rock, plastic, skin etc.), or a translucent material with some limited amount of volumetric absorption and scattering (such as glass, liquids, organic matter etc.). These use cases require different parameterizations to effectively control, so it is convenient to split them into separate slabs. Both the opaque and translucent dielectric base share the same dielectric interface BSDF $f_\mathrm{dielectric}$.

The opaque-base substrate is assumed to be a dielectric with dense subsurface volumetric absorption and scattering, which tends to an idealized \hyperref[sec:glossy-diffuse]{glossy-diffuse} BSDF in the limit of infinite density medium. In some cases a blend of subsurface and completely opaque glossy-diffuse scattering is desired, for example in skin rendering where the diffuse component provides the surface details of the skin (freckles, blemishes, makeup, etc.) and the subsurface component provides the color detail of the underlying veins and tissue. To support this, we define the opaque-base substrate as a statistical mix of \hyperref[sec:glossy-diffuse]{glossy-diffuse} and \hyperref[sec:subsurface]{subsurface} models:
\begin{eqnarray}
M_\textrm{opaque-base}  &=& \mathrm{\mathbf{mix}}(M_\textrm{glossy-diffuse}, S_\mathrm{subsurface}, \mathtt{S}) \ ,
\end{eqnarray}
where $\mathtt{S} = \mathtt{subsurface\_weight}$.

\subparagraph{Behavior of \texttt{specular\_weight}}

The \verb|specular_ior| parameter controls the index of refraction (IOR) of the base dielectric (but is ignored for metal). The \verb|specular_weight| parameter provides a convenient, texturable linear $[0, 1]$ multiplier of the base dielectric reflectivity at normal incidence via reduction of this IOR below the reference value set by \verb|specular_ior|. As a convenience, we also allow the \verb|specular_weight| to exceed $1$, thus increasing the reflectivity via an increase of the IOR above the reference value. When \verb|specular_weight| is $0$, the specular reflection disappears entirely as the IOR of the dielectric is then equal to that of the surrounding medium. Equation~\ref{modulated_ior} gives the formula for the applied IOR modulation. We considered multiple different possible interpretations of the \verb|specular_weight|, but found this approach the most appealing as it modulates the Fresnel in a physically correct manner.

\begin{figure}[!tb]
  \centering
  \begin{subfigure}{.3\textwidth}
    \includegraphics[width=\linewidth]{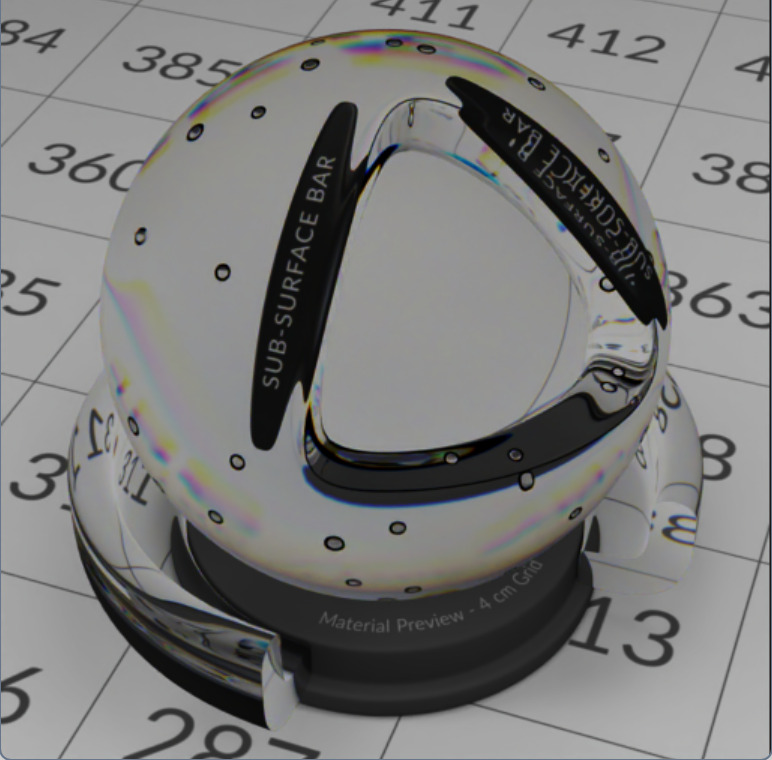}
  \end{subfigure}
  \hspace{0.02\textwidth}
  \begin{subfigure}{.3\textwidth}
    \includegraphics[width=\linewidth]{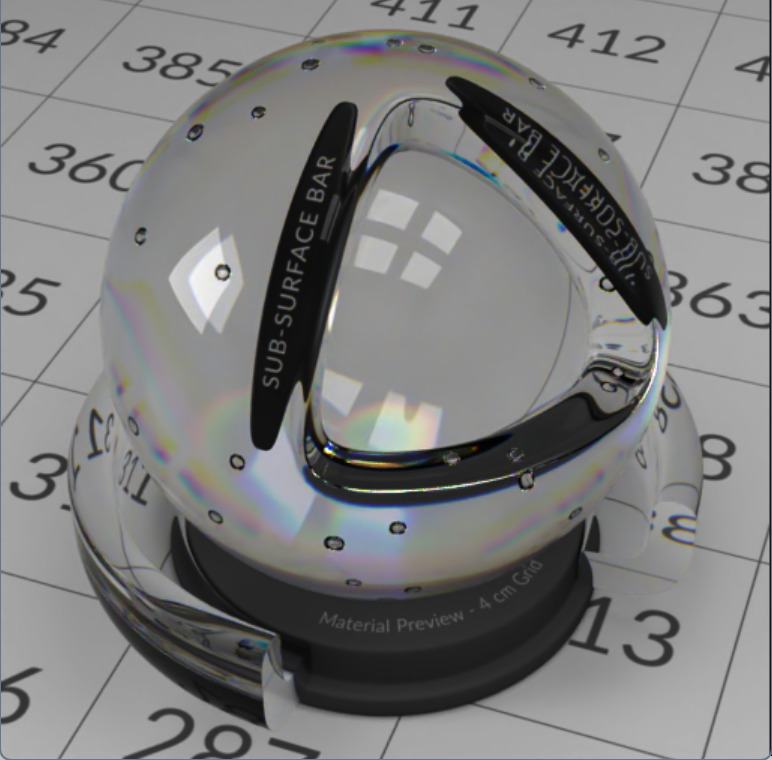 }
  \end{subfigure}
  \hspace{0.02\textwidth}
  \begin{subfigure}{.3\textwidth}
    \includegraphics[width=\linewidth]{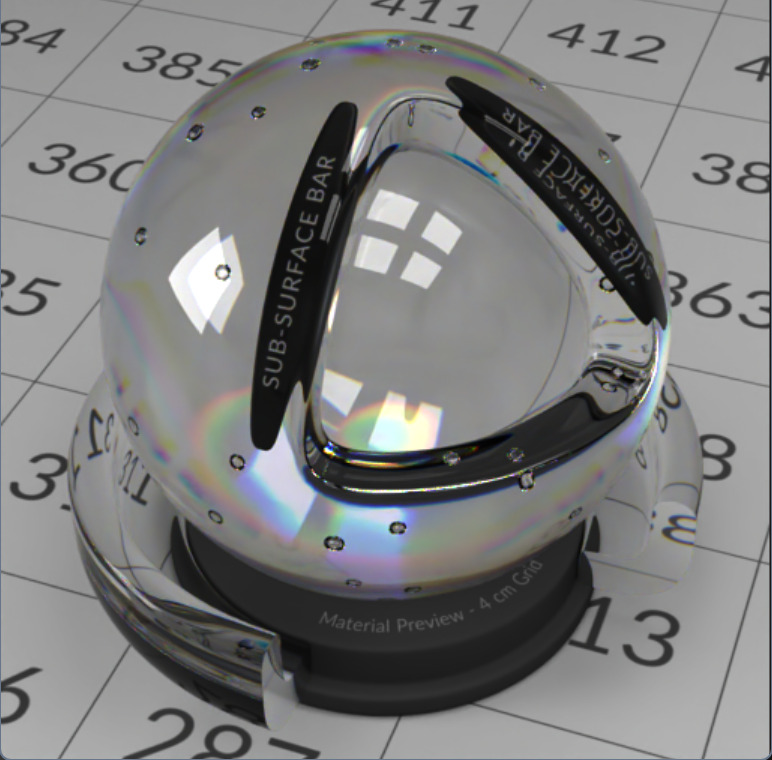 }
  \end{subfigure}
  \caption{Varying \texttt{specular\_weight} with a refractive material (left $0$, center $0.5$, right $1$). \label{fig:specular_weight_refraction}}
\end{figure}

A complication though is that we do \emph{not} want the refraction direction to be modified by \verb|specular_weight|, as this control is designed to vary only the reflectivity without disturbing the refraction appearance, a behavior which was specifically requested by artists. Thus the reflection highlight and the refraction direction are effectively \emph{decoupled}, in an unphysical but artistically convenient way (see Figure~\ref{fig:specular_weight_refraction}).

\begin{figure}[!b]
  \centering
\includegraphics[width=0.9\linewidth]{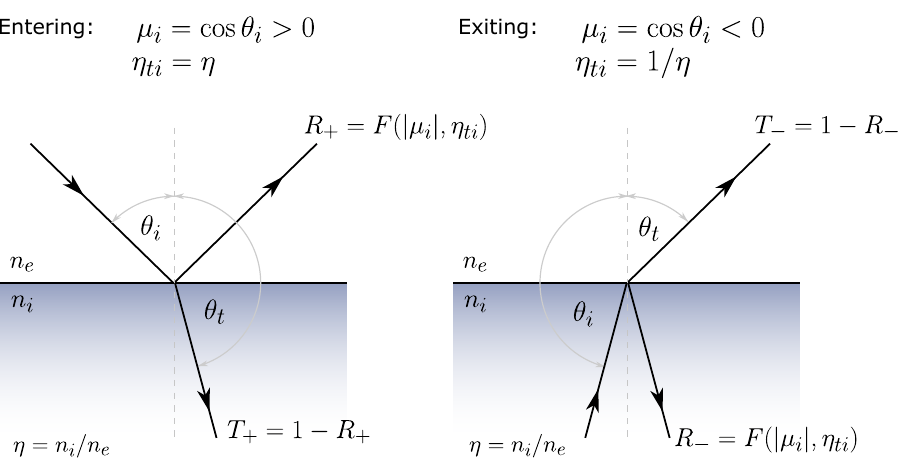}
\caption{Fresnel factors for rays entering (left) and exiting (right) the base dielectric. \label{fig:dielectric_fresnel}}
\end{figure}

The required modification of the Fresnel factors due to \verb|specular_weight| to achieve this decoupling works as follows (see Figure~\ref{fig:dielectric_fresnel} for reference). The physical situation is that the interior of the dielectric has IOR $n_i$ = \verb|specular_ior|, while the exterior has IOR $n_e$, which should take into account the presence of the \hyperref[sec:coat]{coat}. The interior to exterior IOR ratio is $\eta = n_i / n_e$. Incident rays are either entering the base dielectric from the exterior (or coat), or exiting the base dielectric from the interior. For incident rays, the physical reflection Fresnel factors $R_+, R-$ (for entering and exiting incident rays, respectively) are given by $F(\mu_i, \eta_\mathrm{ti})$, where $\eta_\mathrm{ti}$ is the IOR ratio between the transmitted and incident hemispheres (i.e., $\eta_\mathrm{ti} = \eta$ for rays incident from the exterior, and $\eta_\mathrm{ti} = 1/\eta$ for rays incident from the interior). The corresponding transmission Fresnel factors are $T_+ = 1 - R_+$ and $T_- = 1 - R_-$. The dielectric Fresnel reflection factor at normal incidence is given by (for both entering and exiting rays)
\begin{equation} \label{unmodulated_fresnel}
F_0 = \left|\frac{1 - \eta_\mathrm{ti}}{1 + \eta_\mathrm{ti}}\right|^2 \ .
\end{equation}
Given the angle cosine $\mu_i$ of rays incident to the base, the angle cosine $\mu_t$ of the refracted rays follows from the IOR ratio $\eta_\mathrm{ti}$:
\begin{equation} \label{refracted_dir}
  \mu_t^2 = 1 - (1 - \mu_i^2) / \eta^2_\mathrm{ti} \ .
\end{equation}
This refraction direction remains unmodified by \verb|specular_weight| (only the Fresnel factors change).

Now the modification of the Fresnel factors due to \verb|specular_weight| of this standard physical setup is as follows. The transmitted-incident IOR ratio $\eta_\mathrm{ti}$ is modified to $\eta^\prime_\mathrm{ti}$, in such a way as to scale the Fresnel factor at normal incidence by multiplying by $\xi_s = \mathtt{specular\_weight}$ (with $\xi_s \in [0,1]$):
\begin{equation} \label{modulated_ior}
\eta^\prime_\mathrm{ti} = \frac{1 + \epsilon} {1 - \epsilon} \quad \mathrm{with} \quad \epsilon = \mathrm{sgn}(\eta_\mathrm{ti} - 1)\sqrt{\mathrm{min}(\xi_s F_0, 1)}  \ .
\end{equation}
For convenience, we also allow $\xi_s = \mathtt{specular\_weight}$ to exceed 1 so that the reflectivity is increased above the level set by \verb|specular_ior|. The clamp inside the square root ensures that the scaled reflection coefficient cannot exceed $1$.

\begin{figure}[!tb]
\begin{inputcode}
\begin{lstlisting}[frame=!tb, label=listing:ior_decoupling, frame=trBL, caption={IOR decoupling example code, using $\mathtt{specular\_weight}$.}]
// Standard dielectric Fresnel function, depending on:
//   - mu_i = magnitude of angle cosine of incident ray (mu_i >= 0)
//   - eta_ti = (IOR in hemisphere of transmitted light) / (IOR in hemisphere of incident light)
float DielectricFresnel(float mu_i, float eta_ti);

//  specular_weight (>= 0) modulates the Fresnel F0 linearly, without disturbing TIR/refraction
float DielectricFresnel(float mu_i, float eta_ti, float specular_weight=1.f)
{
   if (specular_weight == 1.f)
      return DielectricFresnel(mu_i, eta_ti);

   // Compute modified IOR ratio
   float F0 = sqr((eta_ti - 1.f)/(eta_ti + 1.f));
   float eps = copysignf(min(1.f, sqrtf(specular_weight * F0)), eta_ti - 1.f);
   float eta_ti_prime = (1.f + eps) / max(FLT_EPSILON, 1.f - eps);

   if (eta_ti_prime >= 1.f) // (No TIR possible)
      return DielectricFresnel(mu_i, eta_ti_prime);

   // In the possible-TIR case, check for TIR and if not, use the "un-squeezed" Fresnel curve.
   float mu2_t = 1.f - (1.f - sqr(mu_i))/sqr(eta_ti);
   if (mu2_t <= 0.f)
      return 1.f; // (TIR occurs)
   return DielectricFresnel(sqrtf(mu2_t), 1.f/eta_ti);
}
\end{lstlisting}
\end{inputcode}
\end{figure}

To deal with transmission, there are two cases to consider:

\begin{itemize}

\item If no TIR is possible (i.e., $\eta_\mathrm{ti}, \eta^\prime_\mathrm{ti} \ge 1$) then the modulated $\eta^\prime_\mathrm{ti}$ is used in the angle-dependent dielectric Fresnel formula $F(\mu_i, \eta^\prime_\mathrm{ti})$, producing the desired reflectivity modulation at any incident angle cosine $\mu_i$.

\item However, if TIR \emph{is} possible (i.e., $\eta_\mathrm{ti}, \eta^\prime_\mathrm{ti} < 1$) then a different procedure is used. If the incident ray direction generates TIR then the reflection is unmodified, with Fresnel factor 1 (and no transmission occurs). However, if the incident ray does \emph{not} undergo TIR, then we apply the physical \href{https://en.wikipedia.org/wiki/Stokes_relations}{Stokes relations} (valid for a smooth dielectric interface):
\begin{equation} \label{stokes_reciprocity}
F(\mu_i, \eta_\mathrm{ti}) = F(\mu_t, 1/\eta_\mathrm{ti}) \ ,
\end{equation}
which means the ray that enters along the refraction direction \emph{reversed} reflects the same amount as the original reflection. We maintain this relation after the IOR modulation (ensuring reciprocity), setting the reflection Fresnel factor to $F(\mu_t, 1/\eta^\prime_\mathrm{ti})$, where (since $\eta^\prime_\mathrm{ti} < 1$) this maps to a Fresnel curve without TIR. Thus the modulated reflection curve for entering and exiting rays has the same shape, except with the curve for rays incident from the higher IOR hemisphere squeezed into a cone (with TIR occuring outside this cone).

\end{itemize}

In both cases, the associated transmission Fresnel factor for the refracted ray is given by one minus the reflection Fresnel factor. However, as noted, the refracted ray direction is computed using the \emph{unmodified} IOR ratio $\eta_\mathrm{ti}$. Listing~\ref{listing:ior_decoupling} gives an example implementation.

For physical consistency, we assume that this modification of the Fresnel factor happens at the microfacet level of the base dielectric interface. Thus, in principle, the multiple scattering lobe should accordingly be calculated taking this into account. In practice, it is a good enough approximation to simply apply the usual microfacet multiple-scattering compensation schemes, with the modulated IOR ratio. As noted below, a further tint factor of \verb|specular_color| is applied to the entire macroscopic dielectric BRDF (for both entering and exiting rays), while the BTDF is unaffected by this tint.

\subparagraph{Behavior of \texttt{specular\_color}}

\begin{figure}[!tb]
  \centering
  \begin{subfigure}{.22\textwidth}
    \includegraphics[width=\linewidth]{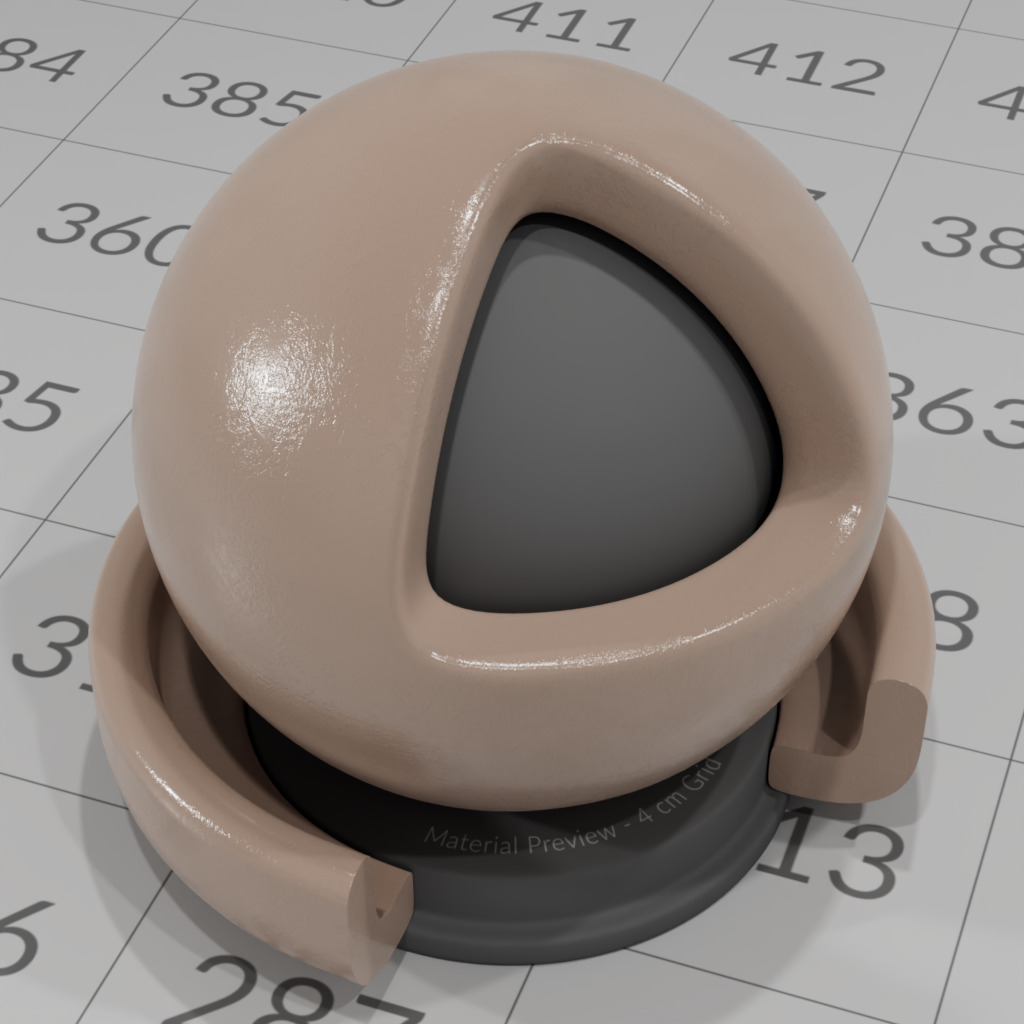}
  \end{subfigure}
  \begin{subfigure}{.22\textwidth}
    \includegraphics[width=\linewidth]{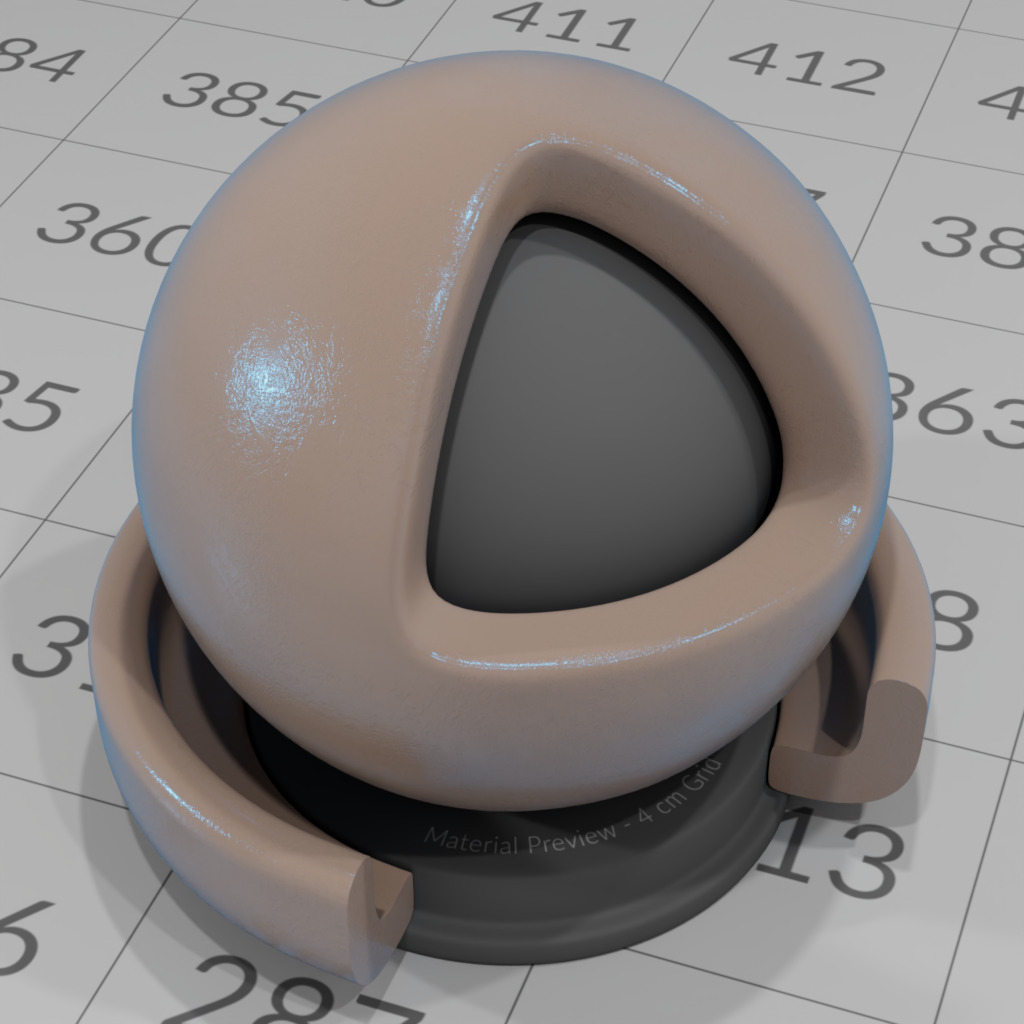}
  \end{subfigure}
  \hspace{0.02\textwidth}
  \begin{subfigure}{.22\textwidth}
    \includegraphics[width=\linewidth]{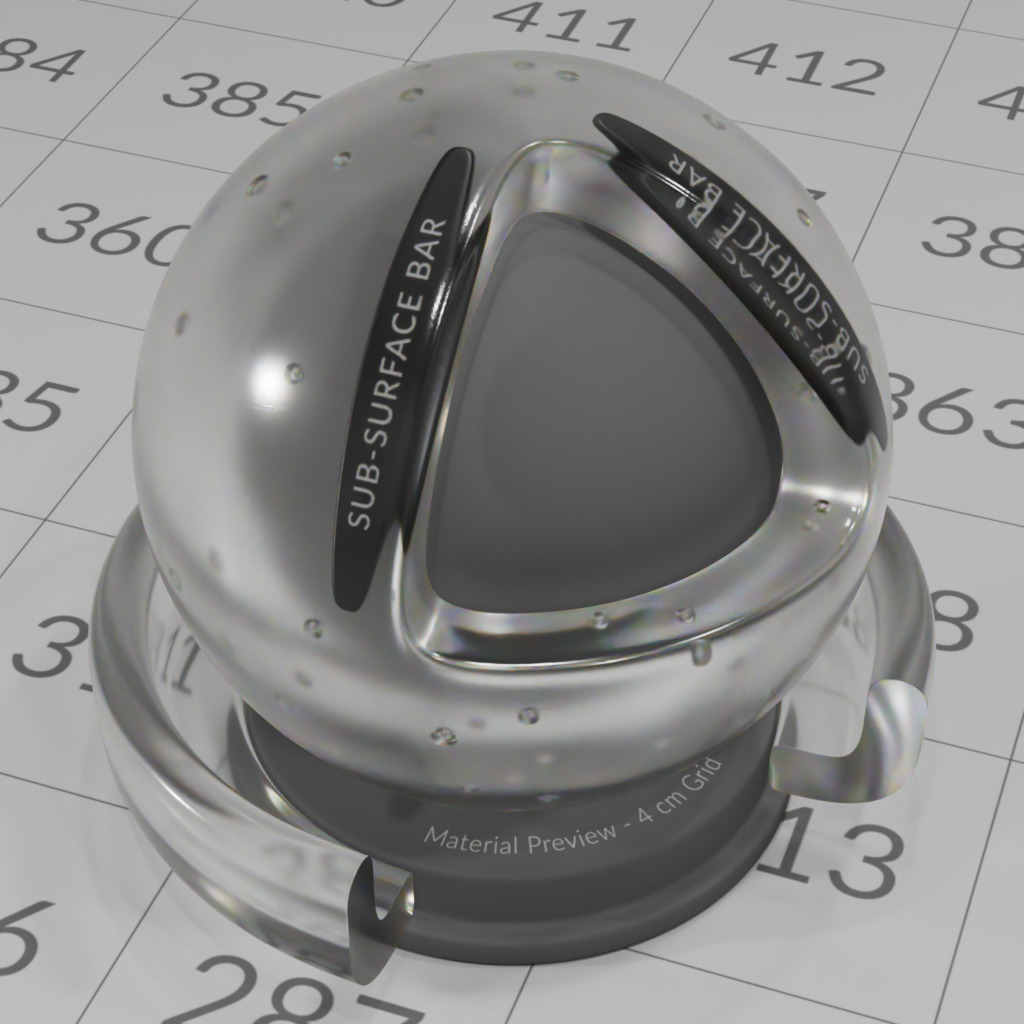}
  \end{subfigure}
  \begin{subfigure}{.22\textwidth}
    \includegraphics[width=\linewidth]{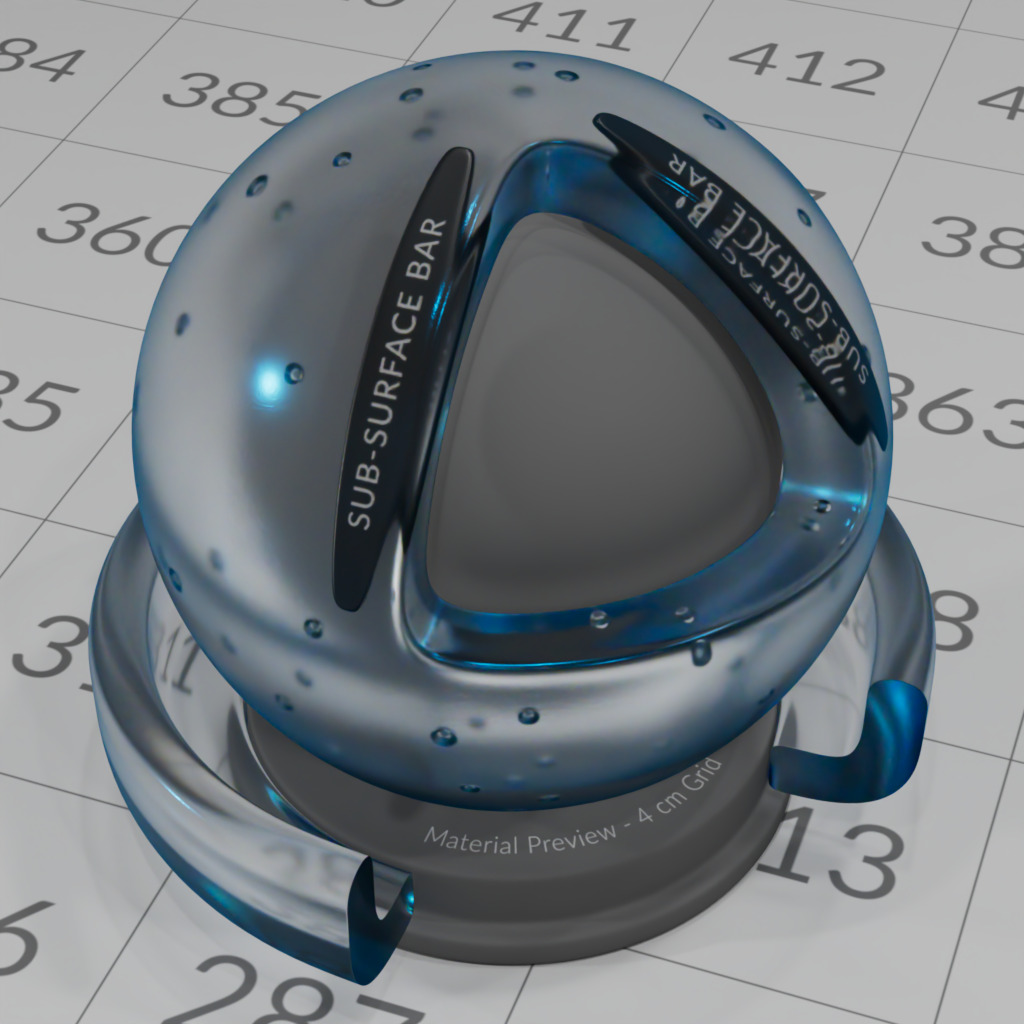}
  \end{subfigure}
\caption{The non-physical behavior of \texttt{specular\_color} of a dielectric, which tints only the Fresnel reflection highlight (shown for both a non-transmissive and transmissive case). \label{fig:specular_color}}
\end{figure}

The \verb|specular_color| parameter tints the Fresnel factor of $f_\mathrm{dielectric}$, but only in the macroscopic BRDF, i.e., the light incident from the upper or lower hemisphere which is reflected back into the same hemisphere (Figure~\ref{fig:specular_color}). The light transmitted from above or below is assumed to be unaffected. This is technically unphysical if altered from the default white color (as real dielectrics have a Fresnel factor dependent only on the index of refraction), but it can be useful in practice to artificially tint the specular highlight in a way that is minimally disruptive to the rest of the model and harmless as it merely multiplies one scattering mode by a factor, effectively removing some energy via an unspecified mechanism.

This was deliberately chosen to differ from the behavior of the \verb|specular_color| parameter in Autodesk Standard Surface, which had the tinting but the light transmitted through the dielectric was then compensated accordingly to preserve the energy balance, thus generating a complementary color if \verb|specular_color| is not white. This was generally considered a bad behavior, hence we opted to fix it in OpenPBR.

%
%

\clearpage

\subsubsection{Glossy-diffuse model: EON}

\label{sec:glossy-diffuse}

The glossy-diffuse slab represents the base dielectric, embedding a semi-infinite bulk of extremely dense scattering material (like the infinitely dense limit of the subsurface). The BRDF of the slab is the combination of a ``glossy'' specular lobe provided by immediate reflection from the dielectric interface, and a diffuse lobe provided by scattering off the embedded substrate. This models, for example, the reflection from shiny, totally opaque surfaces such as dense plastic, rock, and concrete.

\begin{figure}[!b]
  \centering
    \includegraphics[width=0.8\linewidth]{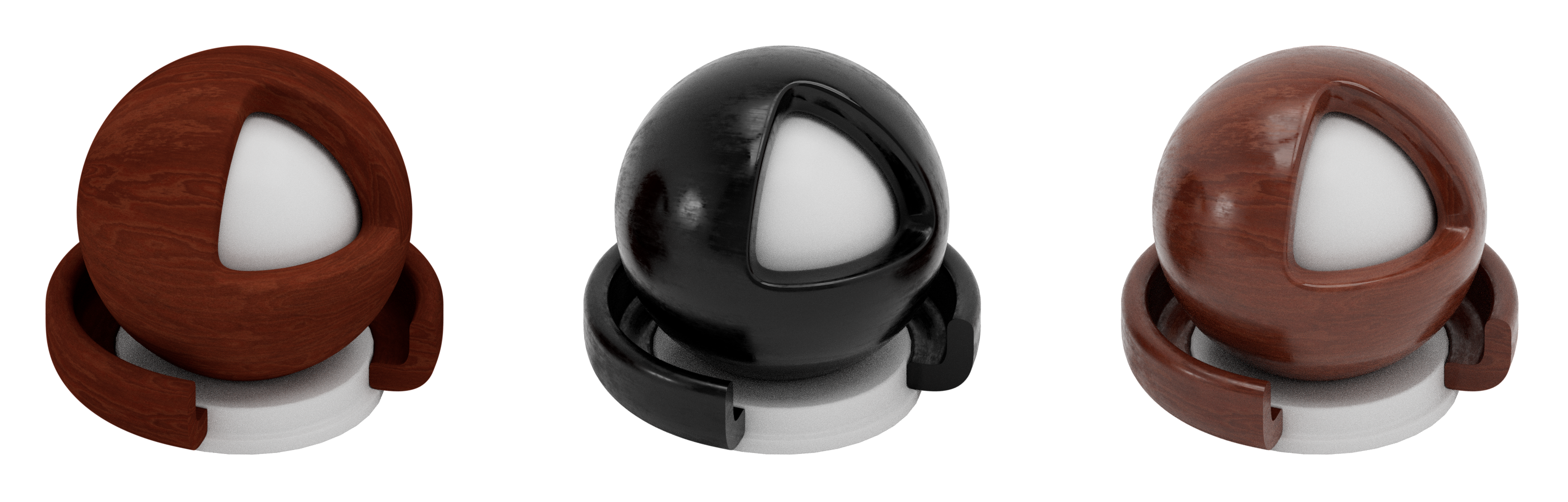}
  \caption{Wood rendered as glossy-diffuse, composed of diffuse lobe (left), specular lobe (middle), and their normalized sum. Near the specular highlight, the diffuse lobe is automatically reduced, since energy is conserved.  \label{fig:glossy-diffuse}}

\end{figure}

There is a long history in computer graphics of modeling such glossy-diffuse, plastic-like materials via a combination of a specular term and diffuse term, with heuristics to ensure energy conservation (Figure~\ref{fig:glossy-diffuse}). The most well-known example is the Ashikhmin--Shirley model \cite{Ashikhmin2000,Pharr2023}, which uses a microfacet specular term and a diffuse term based on the Lambertian BRDF.

Generally speaking, however, these models did not commit to a particular physical interpretation of the underlying material, which for OpenPBR we would like to specify. Since, at the present time, there isn't a standard model of this limit of dense scattering material embedded in a dielectric slab, we choose to model it concretely as a layer of dielectric ``gloss'' on top of an (index-matched) opaque slab with a specified diffuse BRDF:
\begin{eqnarray}
M_\textrm{glossy-diffuse} = \mathrm{\mathbf{layer}}(S_\mathrm{diffuse}, S_\textrm{gloss}) \ ,
\end{eqnarray}
where $S_\textrm{gloss}$ is a thin slab of dielectric with the rough dielectric microfacet BRDF as parameterized in Section~\ref{sec:dielectric-base}, and the same internal medium as the base dielectric except with zero extinction due to infinitesimal layer thickness:
\begin{eqnarray}
S_\textrm{gloss} = \mathrm{Slab}(f_\mathrm{dielectric}, V_\mathrm{dielectric})  \ .
\end{eqnarray}
We opted to describe the glossy-diffuse slab as an explicit layer of dielectric on top of a diffuse base substrate as then it is clear what the base color and roughness mean (i.e., the color and roughness of the Oren--Nayar base).

\begin{figure}[!tb]
  \centering
  \begin{subfigure}{.3\textwidth}
    \includegraphics[width=\linewidth]{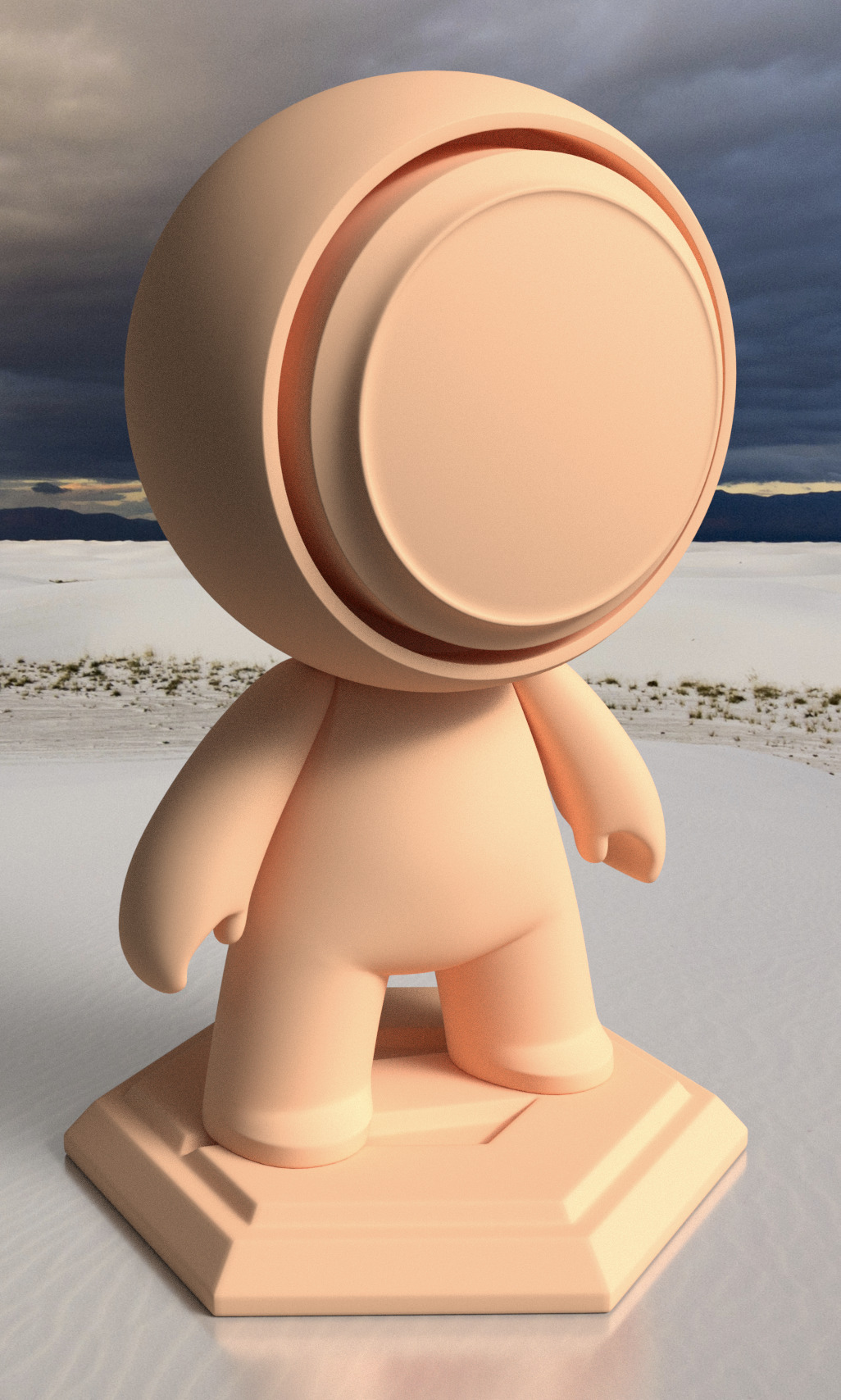}
    \caption{Lambert}
  \end{subfigure}
  \hspace{0.02\textwidth}
  \begin{subfigure}{.3\textwidth}
    \includegraphics[width=\linewidth]{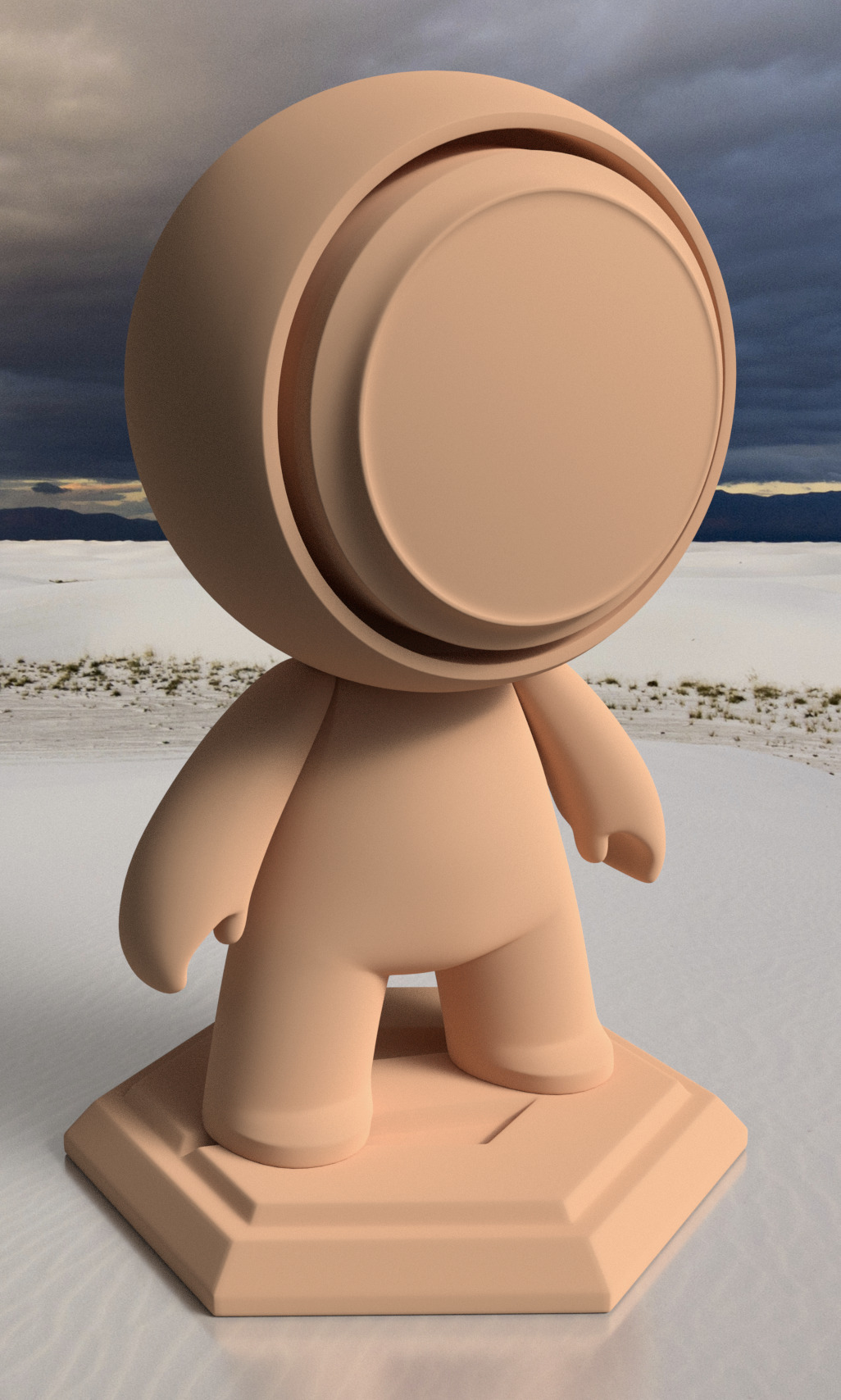}
    \caption{QON}
  \end{subfigure}
  \hspace{0.02\textwidth}
  \begin{subfigure}{.3\textwidth}
    \includegraphics[width=\linewidth]{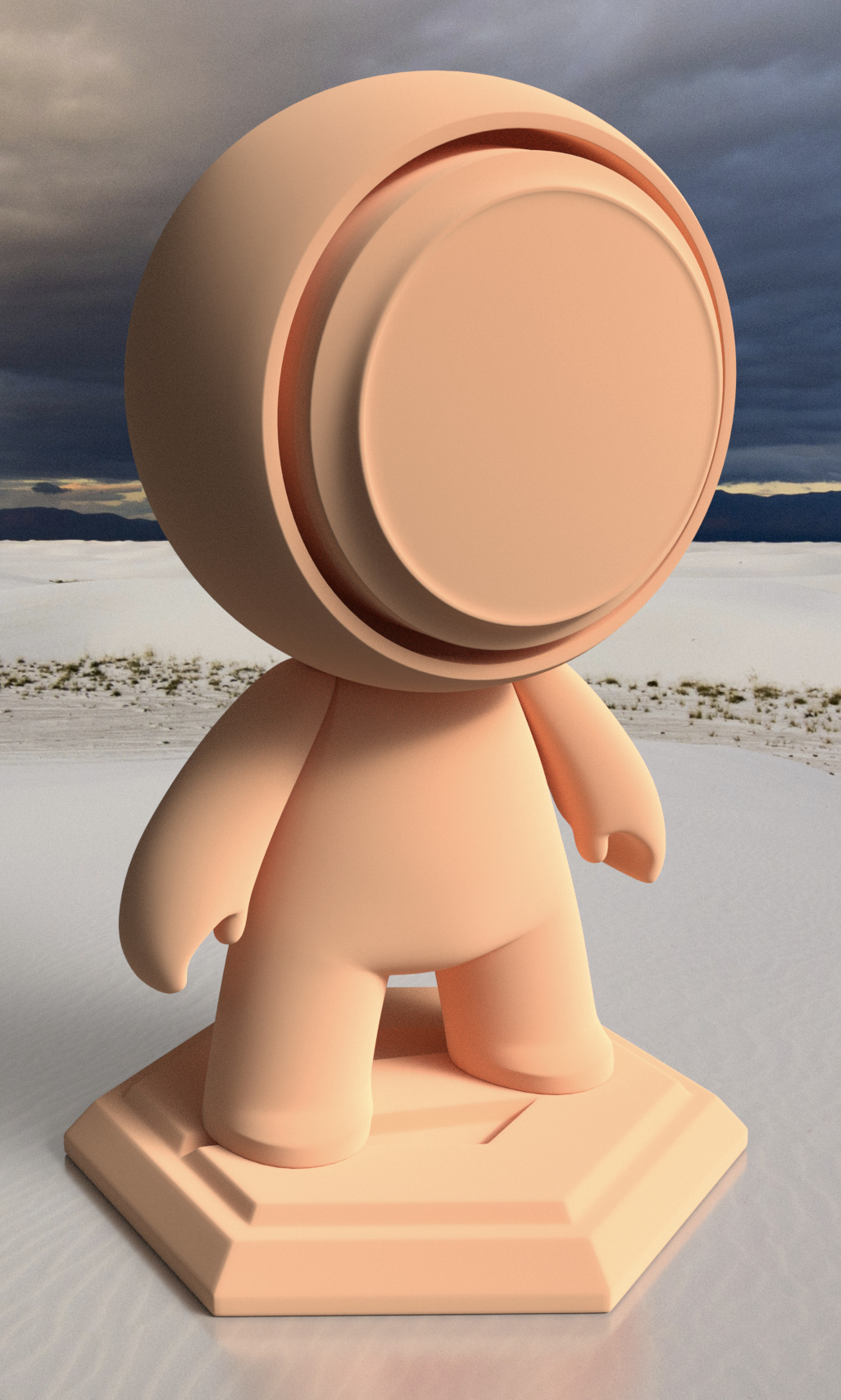}
    \caption{EON}
  \end{subfigure}
  \caption{Comparison of diffuse models at maximum roughess: (left) Lambert, (center) QON, (right) EON. The EON model preserves energy at high roughness, while the QON model loses energy. \label{fig:diffuse_models}}
\end{figure}

Since the diffuse base is index-matched with the gloss, Fresnel reflection is only generated from the top interface of the gloss. The opaque substrate slab has a diffuse BRDF lobe:
\begin{eqnarray}
S_\textrm{diffuse}    = \mathrm{Slab}(f_\mathrm{diffuse}) \ ,
\end{eqnarray}
where $f_\mathrm{diffuse}$ is based on the classic Oren--Nayar model \cite{OrenNayar1994}, which is a V-cavity based microfacet model in which the microfacets are assumed to be individually Lambertian. The Oren--Nayar model is parameterized by roughness $\sigma$ and an overall RGB scale factor $\boldsymbol{\rho}$. The roughness models a ``flattening'' effect (Figure~\ref{fig:diffuse_models}), which is observed in real rough diffuse materials.

During our investigation, we found that the original Oren--Nayar model loses a significant amount of energy when the roughness is high, which is not compatible with the energy-preservation requirements of the OpenPBR model. To address this, we developed a modified version of the Oren--Nayar model (termed the ``energy-preserving Oren--Nayar model'', or EON model) that preserves energy at high roughness values, as described by \textcite{Portsmouth2025}.
This augments the Oren--Nayar model with a simple analytical, reciprocal energy compensation term:
\begin{eqnarray} \label{EON_brdf}
f_\mathrm{diffuse}(\omega_i, \omega_o) = f_\mathrm{ON}(\omega_i, \omega_o) + f^\mathrm{comp}_\mathrm{ON}(\omega_i, \omega_o) \ .
\end{eqnarray}
The Oren--Nayar term $f_\mathrm{ON}$ is given by the form (introduced by Fujii \cite{Portsmouth2025})
\begin{eqnarray} \label{FON_brdf}
f_\mathrm{ON}(\omega_i, \omega_o) = \frac{w_\mathrm{d} \boldsymbol{\rho}}{\pi} \Bigl( A(\sigma) + B(\sigma) \frac{s}{t}
\Bigr) \ .
\end{eqnarray}
The roughness parameter $\sigma \in [0,1]$ is given by \verb|base_diffuse_roughness|. The overall weight is
$w_\mathrm{d}$ = \verb|base_weight|. The RGB $\boldsymbol{\rho}$ parameter is determined by the specified \verb|base_color|, as described below.
The directional albedo $E_\mathrm{ON}(\omega) = w_\mathrm{d} \boldsymbol{\rho}\,\hat{E}_\mathrm{ON}(\omega)$ and corresponding \emph{average albedo} $\langle\hat{E}_\mathrm{ON}\rangle$ of the Oren--Nayar term can be determined analytically. The energy compensation term $f^{\mathrm{comp}}_\mathrm{ON}$ is given in terms of the
albedo $\langle\hat{E}_\mathrm{ON}\rangle$ by
\begin{equation} \label{EON_comp}
f^\mathrm{comp}_\mathrm{ON}(\omega_i, \omega_o) = \frac{w_\mathrm{d} \boldsymbol{\rho}_\mathrm{ms}}{\pi}
\bigl(1 - \hat{E}_\mathrm{ON}(\omega_i)\bigr)
\bigl(1 - \hat{E}_\mathrm{ON}(\omega_o)\bigr) \ ,
\end{equation}
where the factor $\boldsymbol{\rho}_\mathrm{ms}$ accounts approximately for multiple scattering on the microfacet surface:
\begin{equation}
\boldsymbol{\rho}_\mathrm{ms} = \frac{\boldsymbol{\rho}^2}{\pi}
\frac{\langle\hat{E}_\mathrm{ON}\rangle / (1 - \langle \hat{E}_\mathrm{ON}\rangle)}{1 - \boldsymbol{\rho} \bigl(1 - \langle \hat{E}_\mathrm{ON}\rangle\bigr)} \ .
\end{equation}
One can verify that, as $\boldsymbol{\rho} \rightarrow 1$, the total directional albedo of the BRDF of Equation~\ref{EON_brdf}, $E_\mathrm{diffuse}(\omega) \rightarrow 1$ (in the $w_\mathrm{d}=1$ case), thus the compensation term ensures that the white furnace test passes. Note that in the zero roughness ($\sigma \rightarrow 0$) limit, the energy compensation term vanishes, and the parameter $\boldsymbol{\rho}$ is equal to the albedo of $f_\mathrm{diffuse}$. As roughness increases, the albedo of $f_\mathrm{diffuse}$ becomes slightly more dark and saturated than $\boldsymbol{\rho}$ due to the multiple scattering, which is physically realistic.

Although physically the glossy-diffuse slab as described should exhibit darkening and saturation due to the same physics as the coat darkening (i.e., multiple bounces inside the gloss layer losing energy as they hit the colored base), we wanted to prevent this so that the user sees -- to the extent possible -- the base color they specified. To define the meaning of the specified color in terms of the underlying base albedo, we write the normal-direction reflectance of the glossy-diffuse slab, $\mathbf{E}_\textrm{glossy-diffuse}$ in the general form
\begin{eqnarray} \label{glossy_diffuse_albedo}
\mathbf{E}_\textrm{glossy-diffuse} = \mathbf{E}_\textrm{spec} + \mathbf{E}_\textrm{diffuse} \ ,
\end{eqnarray}
where $\mathbf{E}_\mathrm{spec}$ is the normal-direction reflectance of all energy reflected from the dielectric interface without transmission, and $\mathbf{E}_\mathrm{diffuse}$ is the normal-direction reflectance of all energy transmitted through the interface, scattered off the diffuse medium, and transmitted back out.

We then \emph{define} $\mathbf{C} = \mathtt{base\_weight} \times \mathtt{base\_color}$
to be such that the reflectance of the remaining energy transmitted into the slab, $\mathbf{E}_\textrm{diffuse}$ in the \emph{zero} \verb|base_roughness| case (i.e., a Lambertian base) and assuming vacuum exterior, is given by
\begin{equation} \label{glossy_diffuse_albedo_constraint}
\mathbf{E}_\textrm{diffuse} = \bigl( 1 - \mathbf{E}_\textrm{spec} \bigr) \mathbf{C} \ .
\end{equation}
In other words, the selected color $\mathbf{C}$ parameterizes the fraction of the energy transmitted into the dielectric layer that is subsequently transmitted back out due to reflection from a Lambertian interface. Thus, since $0 \le \mathbf{C} \le 1$, $\mathbf{E}_\textrm{glossy-diffuse} \le 1$ so energy is always conserved, and if $\mathbf{C}=1$ then $\mathbf{E}_\textrm{glossy-diffuse}=1$, which guarantees that a white $\mathbf{C}$ passes the furnace test. According to this definition, $\mathbf{C}$ is the observed reflection color (viewed at normal incidence under uniform illumination) in areas where the Fresnel reflection is negligible, and otherwise the observed color is a blend of $\mathbf{C}$ with the gray Fresnel reflection that conserves total reflected energy.

Given the required diffuse albedo $\mathbf{E}_\textrm{diffuse}$ according to the formula above, then in principle the albedo of the diffuse Oren--Nayar lobe $f_\mathrm{diffuse}(\omega_i, \omega_o)$ that generates the required $\mathbf{E}_\textrm{diffuse}$ can be determined. A reasonable, practical (albeit non-reciprocal) approximation to the resulting BRDF of the glossy-diffuse slab that satisfies this requirement is obtained via the non-reciprocal albedo-scaling approximation of Equation~\ref{non-reciprocal-albedo-scaling}:
\begin{equation}
f_\textrm{glossy-diffuse}(\omega_i, \omega_o) \approx f_\mathrm{dielectric}(\omega_i, \omega_o)  +  \bigl(1 - E_\mathrm{dielectric}(\omega_o)\bigr) \,f_\mathrm{diffuse}(\omega_i, \omega_o) \ ,
\end{equation}
where the albedo of the Oren--Nayar lobe $f_\mathrm{diffuse}(\omega_i, \omega_o)$ must be taken to be $\mathbf{C} = \mathtt{base\_weight} \times \mathtt{base\_color}$ in order to satisfy Equation~\ref{glossy_diffuse_albedo_constraint}.
If a reciprocal formulation is desired, the classic Ashikhmin--Shirley or Kelemen model \cite{Ashikhmin2000,Kelemen2001,Kulla2017,Kutz2021} form is
\begin{equation}
f_\textrm{glossy-diffuse}(\omega_i, \omega_o) \approx f_\mathrm{dielectric}(\omega_i, \omega_o)  + \mathcal{N} \bigl(1 - E_\mathrm{dielectric}(\omega_i)\bigr) \bigl(1 - E_\mathrm{dielectric}(\omega_o)\bigr) \,f_\mathrm{diffuse}(\omega_i, \omega_o) \ ,
\end{equation}
where $\mathcal{N}$ is a normalization factor such that Equation~\ref{glossy_diffuse_albedo_constraint} is satisfied. In the case of a zero roughness (i.e., Lambertian) Oren--Nayar base $\mathcal{N}$ can be tabulated in terms of the dielectric IOR and roughness \cite{Kutz2021}, and the required albedo of the Lambertian base is equal to $\mathbf{C}$ as in the non-reciprocal albedo-scaling approximation. Extending this to the more general case of a non-Lambertian rough Oren--Nayar base would require adding the roughness dimension to the tabulation, and the required Oren--Nayar albedo will not simply equal $\mathbf{C}$. We leave the specific choice of model and these details to the implementation.

\begin{figure}[!t]
  \centering
  \begin{subfigure}{.48\textwidth}
    \includegraphics[width=\linewidth]{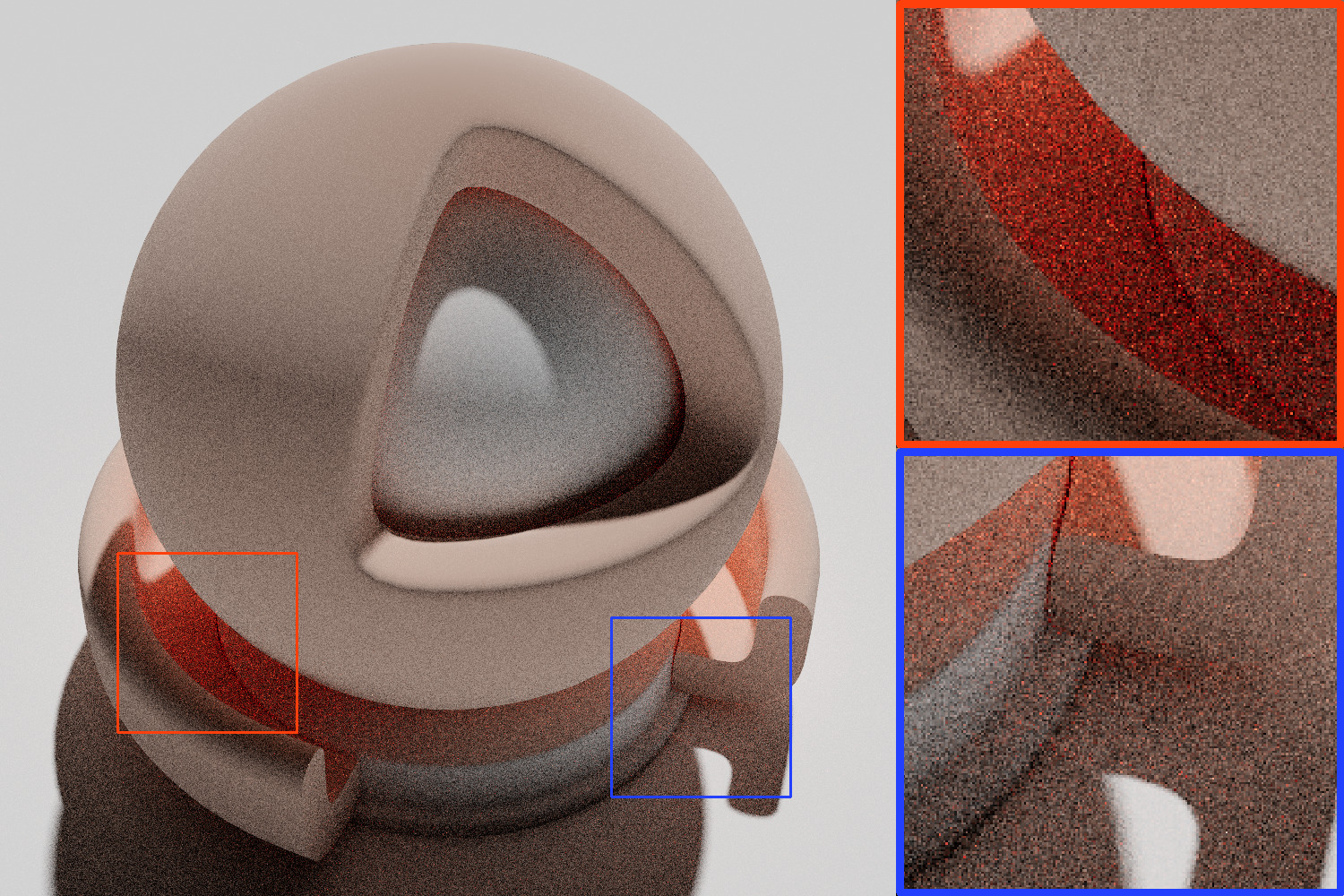}
  \end{subfigure}
  \hspace{0.02\textwidth}
  \begin{subfigure}{.48\textwidth}
    \includegraphics[width=\linewidth]{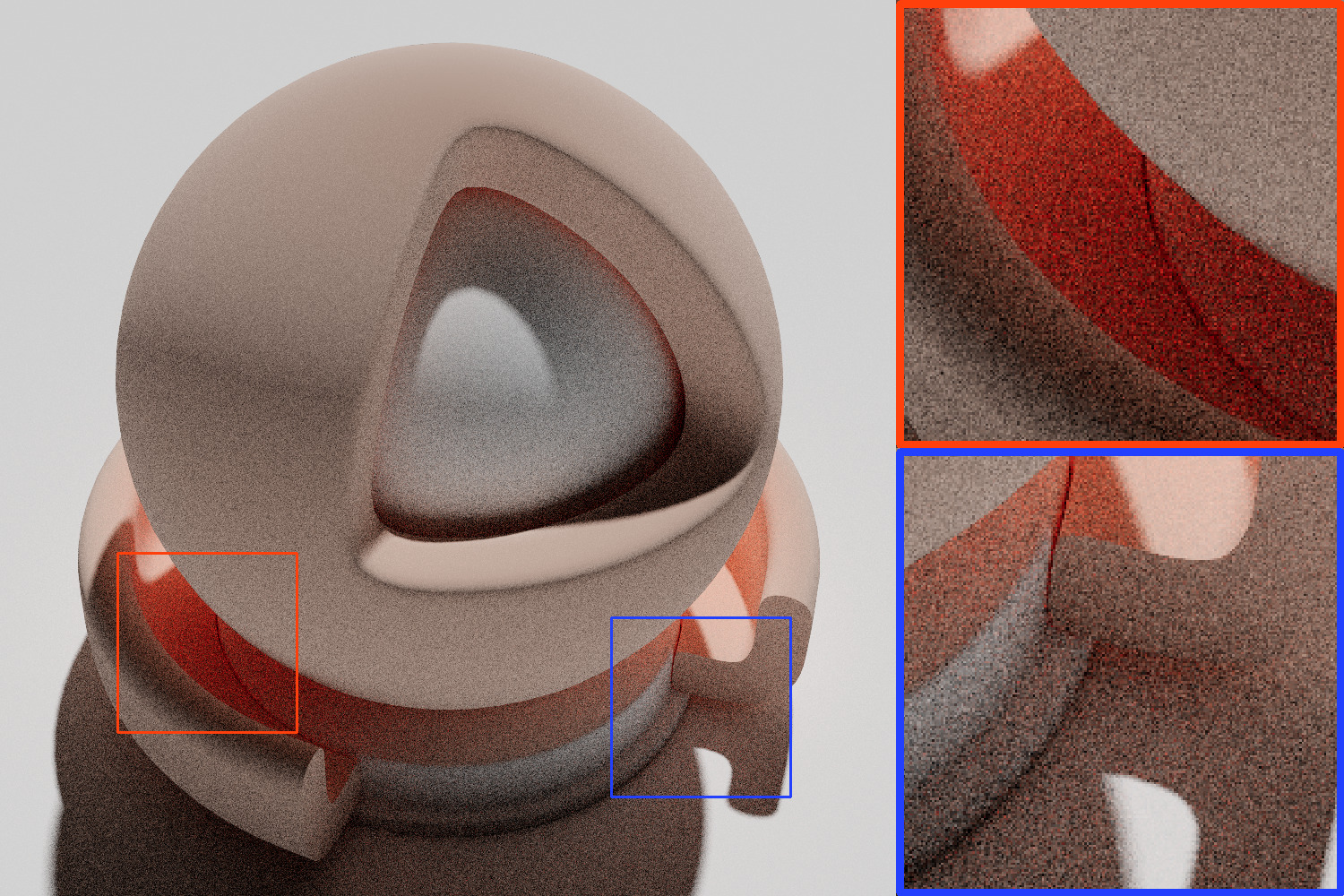}
  \end{subfigure}
\caption{Variance reduction achieved in renders via CLTC importance sampling of the EON model (left panel is cosine sampling, right panel is CLTC sampling). \label{fig:glossy_diffuse_sampling}}
\end{figure}

\begin{figure}[!hb]
  \centering
    \includegraphics[width=0.5\linewidth]{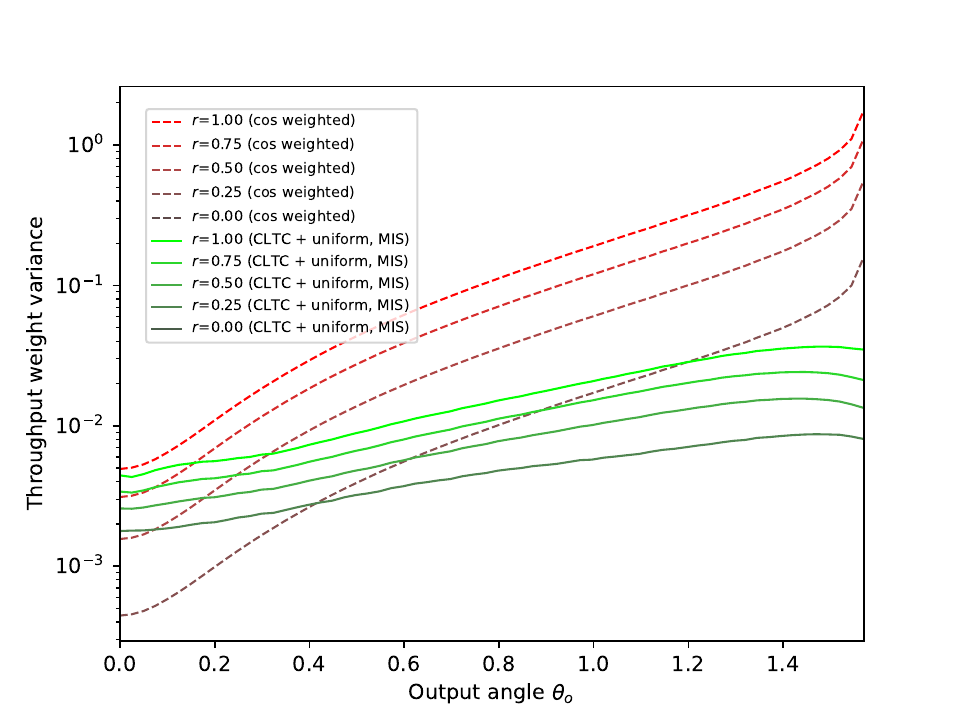}
\caption{A plot of the variance reduction achieved via CLTC importance sampling of the EON model (compared to traditional cosine-weighted sampling), as a function of the incident angle and roughness. Note that the $r=0$ curves (for both uniform and CLTC sampling) are simply zero, so don't show up on the log plot (since in this case, both uniform and CLTC sampling degenerate to perfect importance sampling, with constant throughput weight). \label{fig:glossy_diffuse_sampling_variance}}
\end{figure}

The EON diffuse model also provides more sophisticated importance sampling than standard cosine-weighted sampling typically used for diffuse materials \cite{Portsmouth2025}. The sampling is based on \emph{Linearly Transformed Cosine} (LTC) sampling \cite{Heitz2016b}, with a modification to avoid generating redundant samples under the hemisphere (so-called CLTC sampling). This results in a significant reduction in noise (see Figures~\ref{fig:glossy_diffuse_sampling} and ~\ref{fig:glossy_diffuse_sampling_variance}).

\clearpage

\subsubsection{Subsurface}

\label{sec:subsurface}

\begin{figure}[!b]
  \centering
  \hfill
  \begin{subfigure}{.24\textwidth}
    \includegraphics[width=\linewidth]{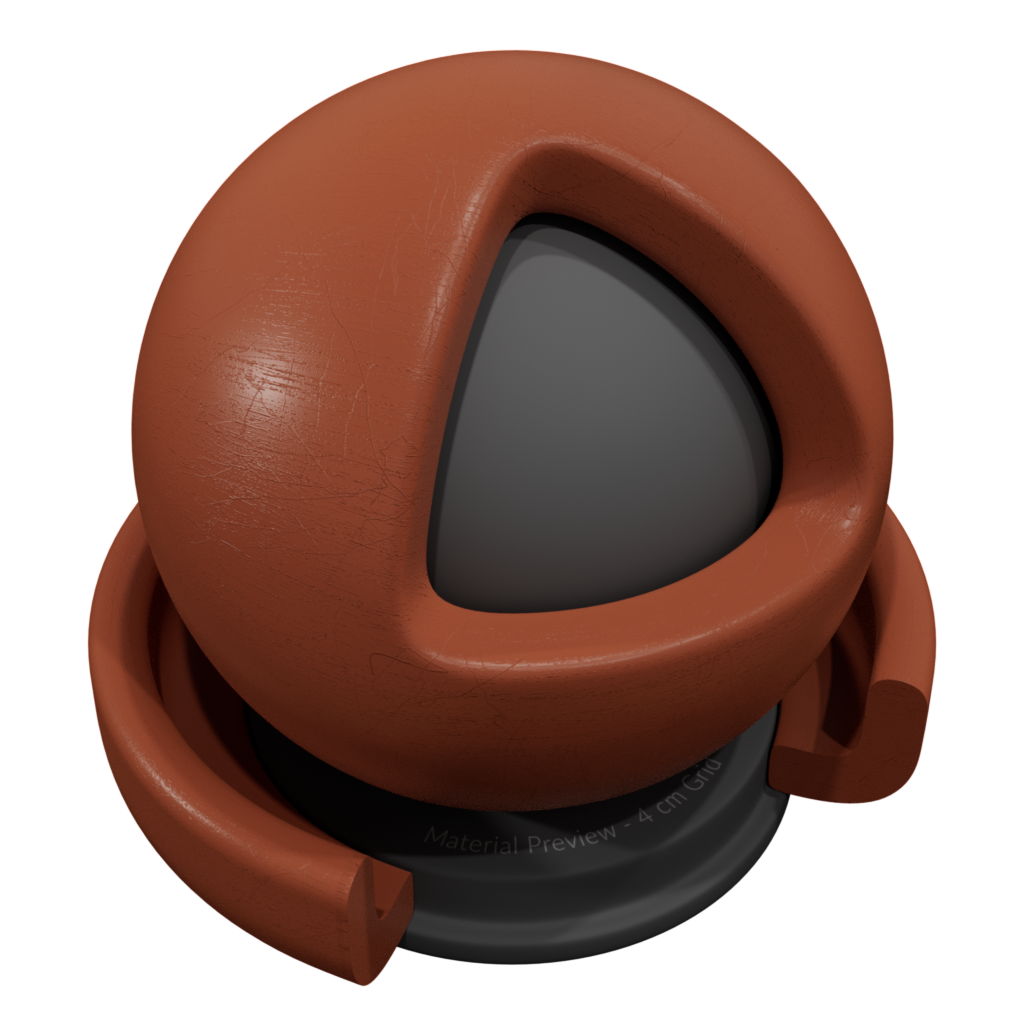}
  \end{subfigure}
  \hfill
  \begin{subfigure}{.24\textwidth}
    \includegraphics[width=\linewidth]{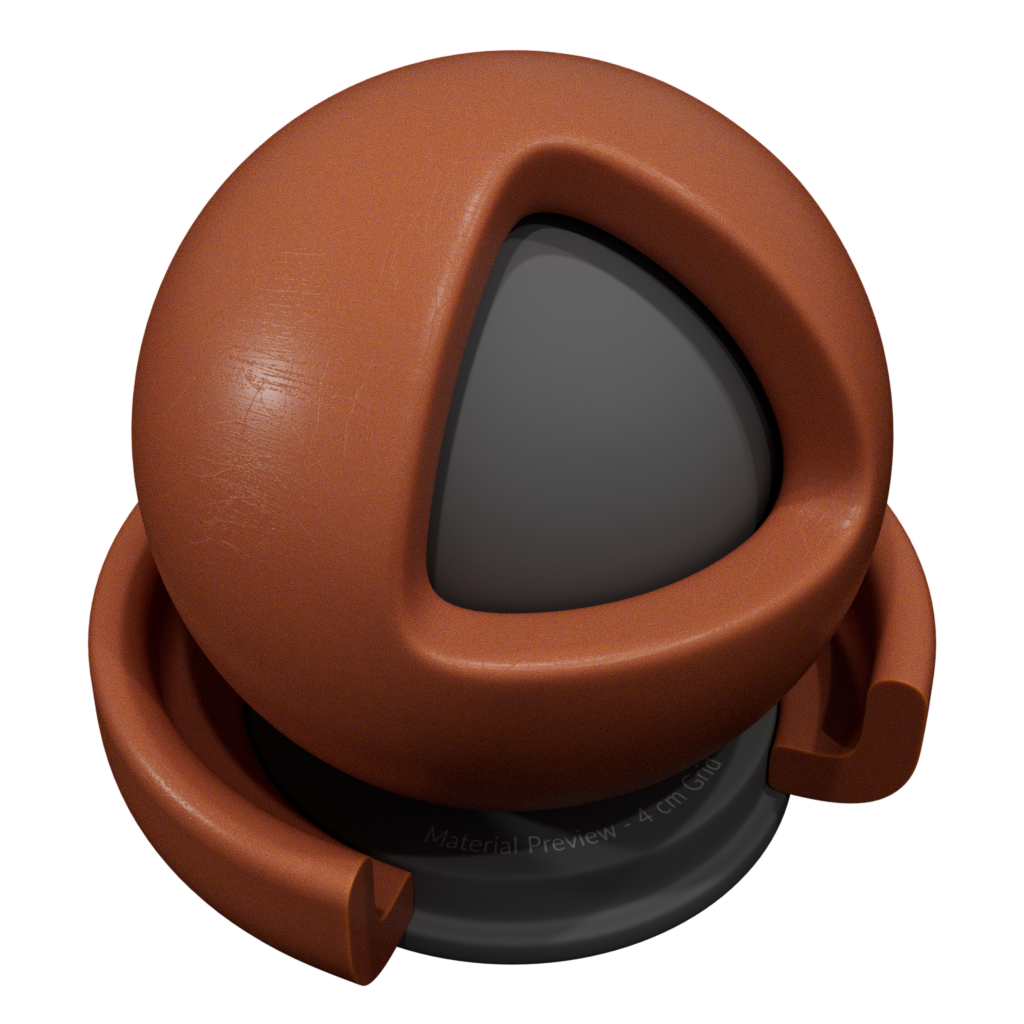}
  \end{subfigure}
  \hfill
  \begin{subfigure}{.24\textwidth}
    \includegraphics[width=\linewidth]{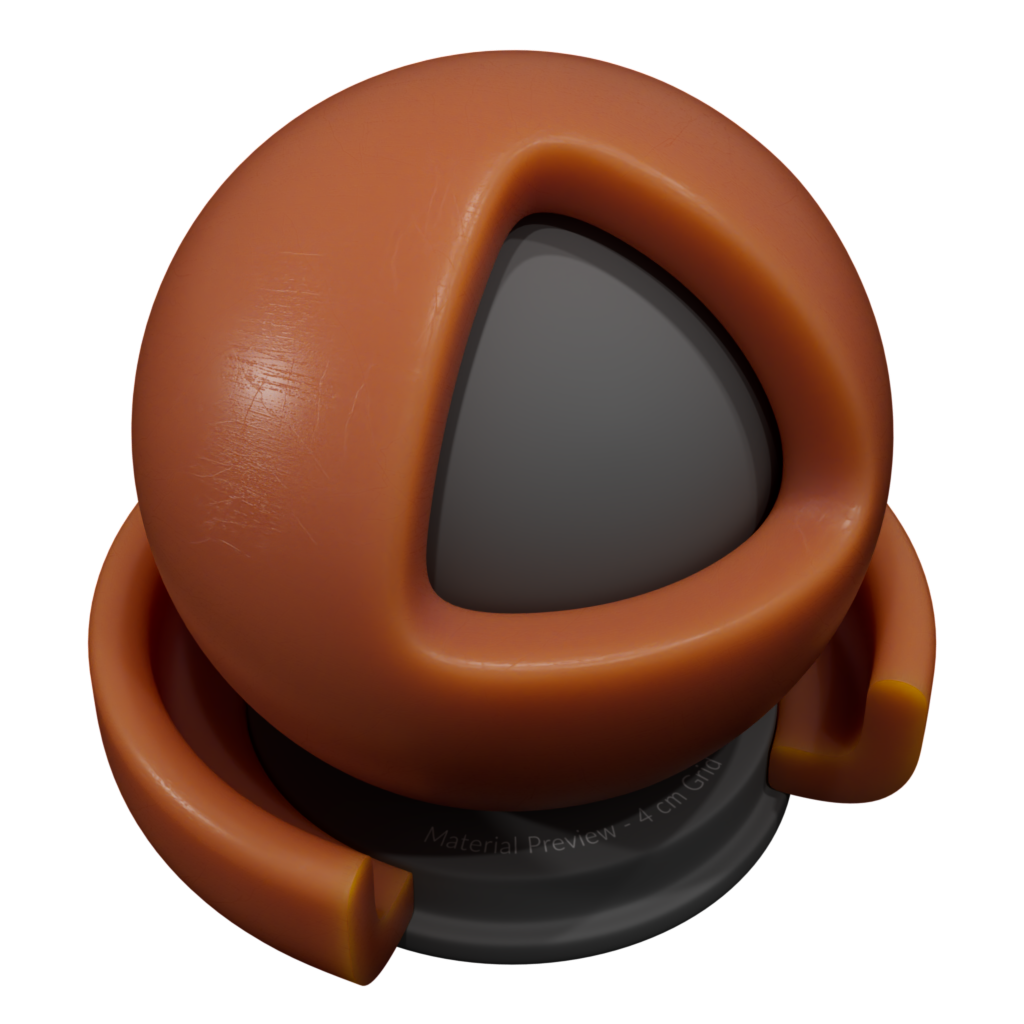}
  \end{subfigure}
  \hfill
  \begin{subfigure}{.24\textwidth}
    \includegraphics[width=\linewidth]{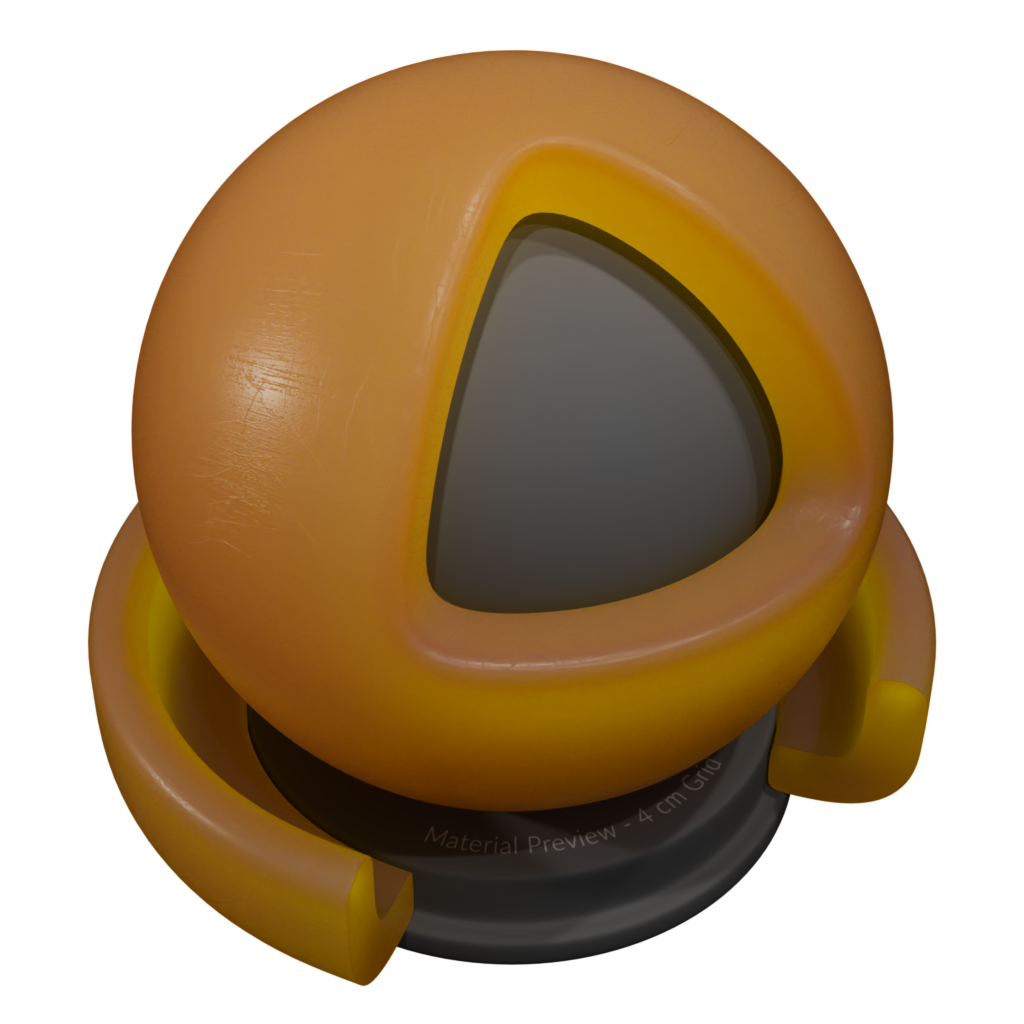}
  \end{subfigure}
  \hfill
  \caption{Comparison of diffuse and subsurface scattering with varying radius. The left image shows a diffuse material with orange \texttt{base\_color}, while the other three images show subsurface scattering with the same orange color \texttt{subsurface\_color} and a yellow \texttt{subsurface\_radius\_scale}, with \texttt{subsurface\_radius} of 0.01, 0.1, and 1.0 respectively. At low radius, the reflection color is a good match to the diffuse case, while at higher radii the yellow color becomes more visible.\label{fig:sss_match_renders}}
\end{figure}

In OpenPBR, we define two separate but related volumetric components: subsurface scattering and the transmission volume (or \hyperref[sec:translucent-base]{translucent base}). While the underlying physics is the same for both, they are parameterized differently and are intended for different use cases. Essentially, subsurface scattering is designed to make an opaque diffuse material more translucent, while the transmission volume is designed to make a clear transparent object less translucent.

While developing OpenPBR, we originally considered a unified volumetric approach, with a continuum from diffuse to subsurface to transmission, since all of these effects are conceptually different scales of the same underlying physics. However, we abandoned that approach in favor of the multi-component model due to its familiarity for artists and the ability to use a different appropriate parameterization for each component.

As in the cases of the \hyperref[sec:glossy-diffuse]{glossy-diffuse} slab and the \hyperref[sec:translucent-base]{translucent base}, the subsurface is bounded by a dielectric interface with BSDF $f_\mathrm{dielectric}$, which generates the primary specular reflection lobe parameterized via the ``specular'' parameters. Combined with this is the reflection generated by light that is transmitted through the dielectric interface into the underlying embedded subsurface medium, where it scatters around and eventually transmits back out. In this case, the subsurface medium $V^\infty_\mathrm{subsurface}$ is given a parameterization which is particularly convenient for controlling the volumetric effect of dense subsurface scattering.

\subparagraph{Subsurface radius}

One of the biggest challenges of parameterizing subsurface scattering is specifying the length scale of the effect.
An ideal scale parameter would be intuitive, expressive, and physically meaningful, exhibit consistent behavior when varied, and not be prone to artifacts. In practice, however, there are many tradeoffs between these properties, and the optimal choice depends on the use case.

For example, for thin or low-albedo volumes, it might be useful to specify the distance traveled by light along a single ray before it first interacts with the volume, but this is not as straightforward as it might sound, since the distance is probabilistic, and since there are two different types of interactions that can occur: scattering and absorption. For dense or high-albedo volumes, it might be useful to specify the distance traveled along the surface before light re-emerges, however this is even less straightforward, since that distance is also probabilistic and varies strongly with the single-scattering albedo of the volume.

While designing OpenPBR, we considered and experimented with a number of different options. One of the options we tentatively finalized involved setting the average distance that light visibly travels along the surface for each color channel after all scattering and absorption has occurred. We based this on the theory presented in \textcite{Christensen2015}. Effectively the user would be controlling the resultant diffusion profile directly (even when simulated using path-traced volumetric scattering). This was conceptually appealing, especially since it aligned well with the perceptual multiple-scattering definition of the subsurface color parameter described below.

However, in practice, it produced inconsistent results and artifacts due to its tight coupling with the albedo.
For example, if the user darkened the albedo, light would travel less far on average due to the increased absorption, so the system would have to increase the translucency of the material to compensate. This was already unintuitive with gray albedos (especially textured ones), but it became even more problematic with colored albedos. Because the albedo affects the shape (peak/tail balance) of the diffusion profile, and because the scale compensation acted independently on each color channel, each color channel would end up with a diffusion profile with both a different shape and a different size.
This could result in situations where one color channel would dominate at short distances while another dominated at long distances, resulting in unexpected and unsightly hue shifts (e.g., a red material might look cyan in shadow regions). We observed that this hue inversion cannot physically occur in real-world situations where either the extinction or scattering coefficient is roughly uniform across the visible spectrum, so we decided to avoid this approach.

Ultimately, we decided to use a simpler and more pragmatic approach where the subsurface radius parameter simply specifies the extinction mean free path. That is,  $\texttt{subsurface\_radius} \times \texttt{subsurface\_radius\_scale}$ defines, per RGB channel, the \emph{mean free path} (MFP) $\mathbf{r}$ -- i.e., the average distance that a ray of light travels through the medium before being absorbed or scattered. The corresponding extinction coefficient $\boldsymbol{\mu}_t = 1/\mathbf{r}$ controls the apparent density of the medium. In the limit of zero MFP, the medium tends towards infinite density, and approaches the look of an opaque diffuse surface. Being a length, \verb|subsurface_radius| can be any value greater than or equal to zero. For convenience, we make the soft range $[0, 1]$, thus covering common cases such as skin where the MFP is lower than the scene length units. The \verb|subsurface_radius_scale| controls the color-channel dependence of the MFP, and thus this color is visible in the light transmitted through thinner regions of the subsurface volume.

In this approach, the user-provided radius doesn't map to anything directly observable, but it does map directly to an underlying physical property of the medium, leading to a parameterization that is straightforward to describe and implement. Most importantly, this approach produces results that are familiar for artists, predictable, and free of artifacts.

The medium's phase function is parameterized by \verb|subsurface_scatter_anisotropy|, giving the scalar anisotropy $g \in [-1, 1]$. In practice, implementations may want to internally clamp the anisotropy as the phase function becomes ill-defined and the rendering unstable near the -1 and 1 limits. Furthermore, path-traced volumetric rendering can be made more efficient by leveraging similarity theory as described in Section~\ref{sec:albedo_inversion}.

\subparagraph{Subsurface color}

For artist directability, it is important that the input parameters produce renders with colors that match to a reasonable approximation.
This corresponds to the requirement that the observed color of the reflected light matches \verb|subsurface_color| given all the other relevant model parameters (i.e., the other volumetric parameters of the subsurface such as anisotropy, and the dielectric IOR and roughness). Renderers should try to achieve that, to the extent possible given their constraints. As shown in Figure~\ref{fig:sss_match_renders}, in the dense limit, the look of a diffuse base with given \verb|base_color| should closely match the look of the subsurface with equal corresponding \verb|subsurface_color|.

To define what it means to say that the observed color ``matches'' the \verb|subsurface_color|, we need to separate the Fresnel reflection from the reflection due to the subsurface scattering. To make this precise, we use a similar mechanism to the definition of the albedo of the \hyperref[sec:glossy-diffuse]{glossy-diffuse} slab. Denoting the directional reflectance along the normal direction by $\mathbf{E}_\mathrm{subsurface}$, this can be broken into two components:
\begin{equation} \label{subsurface_albedo}
\mathbf{E}_\mathrm{subsurface} = \mathbf{E}_\mathrm{spec} + \mathbf{E}_\mathrm{multi-scatter} \ .
\end{equation}
Here, $\mathbf{E}_\mathrm{spec}$ is the normal-direction reflectance of all energy reflected from the dielectric interface \emph{without} macroscopic transmission. While the \emph{multiple-scattering albedo} $\mathbf{E}_\mathrm{multi-scatter}$ is the normal-direction reflectance of all remaining reflected energy due to (macroscopic) transmission through the interface, multiple scattering in the subsurface medium, and transmission back out (assuming external vacuum). We then \emph{define} the \verb|subsurface_color| $\mathbf{C}$ to parameterize, assuming correct physical light transport and semi-infinite homogeneous slab geometry, the fraction of the energy transmitted through the interface that is transmitted back out and observed, i.e.:
\begin{equation} \label{subsurface_albedo_constraint}
\mathbf{E}_\mathrm{multi-scatter} = \left( 1 - \mathbf{E}_\mathrm{spec} \right) \mathbf{C} \ .
\end{equation}
Thus, since $0 \le \mathbf{C} \le 1$, $\mathbf{E}_\mathrm{subsurface} \le 1$, so energy is always conserved, and if $\mathbf{C}=1$ then $\mathbf{E}_\mathrm{subsurface}=1$, which guarantees that a white \verb|subsurface_color| passes the furnace test. According to this definition, the \verb|subsurface_color| $\mathbf{C}$ is the observed reflection color (viewed at normal incidence under uniform illumination) in areas where the Fresnel reflection is negligible, and otherwise the observed color is a blend of $\mathbf{C}$ with the gray Fresnel reflection that conserves total reflected energy.

If the interface is index-matched (i.e., $\mathbf{E}_\mathrm{spec}=0$) then $\mathbf{E}_\mathrm{subsurface} = \mathtt{subsurface\_color}$, so in the index-matched case the color $\mathtt{subsurface\_color}$ drives the base (multi-scatter) albedo directly, as one would expect.
Since $\mathbf{E}_\mathrm{spec}$ is grey, in the general case the final albedo $\mathbf{E}_\mathrm{subsurface}$ has the same hue as $\mathtt{subsurface\_color}$, while satisfying the properties above, so this provides a reasonable match between the desired and resultant observed color given the constraints.

\subparagraph{Albedo inversion}

\label{sec:albedo_inversion}

In principle, given the required \verb|subsurface_color| $\mathbf{C}$, the single-scattering albedo $\boldsymbol{\alpha}$ which generates the required $\mathbf{E}_\mathrm{multi-scatter}$ can be determined (assuming the given extinction $\boldsymbol{\mu}_t$ and anisotropy $g$). We do not require any particular theoretical formula for $\boldsymbol{\alpha}(\mathbf{C})$ be used, but the closer it is to satisfying Equation~\ref{subsurface_albedo_constraint} the closer it is to the ground truth. Ideally, this formula should take all the properties of the medium into account as well as the dielectric interface. A number of such approximations have been derived in the literature.

A well-known approximation is due to van de Hulst \cite{Kulla2017,HHGTMS}, which assumes an index-matched boundary (i.e., $\mathbf{E}_\mathrm{spec} = 0$, thus $\mathbf{E}_\mathrm{multi-scatter} = \mathbf{C}$). According to this, the total reflectance (i.e., multi-scatter albedo) $\mathbf{C}$ from a semi-infinite isotropic medium with no Fresnel (vacuum boundary) and diffusion illumination is given by
\begin{equation} \label{A(s)}
\mathbf{C}(\mathbf{s}) = \frac{(1-\mathbf{s}) (1 - 0.139\mathbf{s})}{1 + 1.17\mathbf{s}} \ .
\end{equation}
Where
\begin{equation}
\mathbf{s} = \sqrt{1 - \boldsymbol{\alpha}^\star}
\end{equation}
and $\boldsymbol{\alpha}^\star$ is the single-scatter albedo for the isotropic case. Inverting Equation~\ref{A(s)} gives (to less than a percent error)
\begin{equation} \label{s(A)}
\mathbf{s}(\mathbf{C}) = 4.09712 + 4.20863 \mathbf{C} - \sqrt{9.59217 + 41.6808 \mathbf{C} + 17.7126 \mathbf{C}^2} \ .
\end{equation}
This provides the desired single-scatter albedo given multi-scatter albedo $\mathbf{C}$ as $\boldsymbol{\alpha}^\star(\mathbf{C}) = 1 - \mathbf{s}^2(\mathbf{C})$.
\textcite{ChiangKutz2016} found their own fitting function for multi-scatter albedo $\boldsymbol{\alpha}^\star(\mathbf{C})$, which was later extended by \textcite{Hyperion}, taking into account TIR from an interface assumed to have IOR 1.4, producing
\begin{equation} \label{hyperion_fits}
\boldsymbol{\alpha}^\star(\mathbf{C}) = 1 - e^{-11.43 \mathbf{C} + 15.38 \mathbf{C}^2 - 13.91 \mathbf{C}^3} \ .
\end{equation}
These formulas are necessarily approximate, as for example they neglect certain effects such as the roughness of the dielectric boundary. Various other alternative analytical formulas which attempt to account more accurately for all the relevant effects are presented by \textcite{HHGTMS}.

To account for anisotropy, the standard approach is to use the following ``similarity relation'' \cite{Zhao2014, Hyperion} between the scattering coefficients in media with different anisotropies $g$ and $g^\star$:
\begin{equation} \label{eq:similarity_relation}
\boldsymbol{\mu}_s (1 - g) = \boldsymbol{\mu}^\star_s (1 - g^\star) \ .
\end{equation}
The absorption coefficients of the two media are assumed to be equal, i.e., $\boldsymbol{\mu}_a = \boldsymbol{\mu}^\star_a$.
The extinction coefficient $\boldsymbol{\mu}_t = 1/\mathbf{r}$ is assumed to set the extinction of the anisotropic medium.  The scattering albedo $\boldsymbol{\alpha}$ of the anisotropic medium can then be shown to be (see Appendix~\ref{sec:similarity_theory_subsurface} for the derivation)
\begin{equation} \label{ss_albedo_anisotropic}
\boldsymbol{\alpha} = \frac{\boldsymbol{\alpha}^\star}{1 - g(1 - \boldsymbol{\alpha}^\star)} \ .
\end{equation}

This formulation for general anisotropy (i.e., scattering albedo given by Equation~\ref{ss_albedo_anisotropic} and Equation~\ref{hyperion_fits}) is used for example in Arnold's OpenPBR subsurface implementation.

\begin{figure}[!tb]
  \centering
  \hfill
  \begin{subfigure}{.48\textwidth}
    \includegraphics[width=\linewidth]{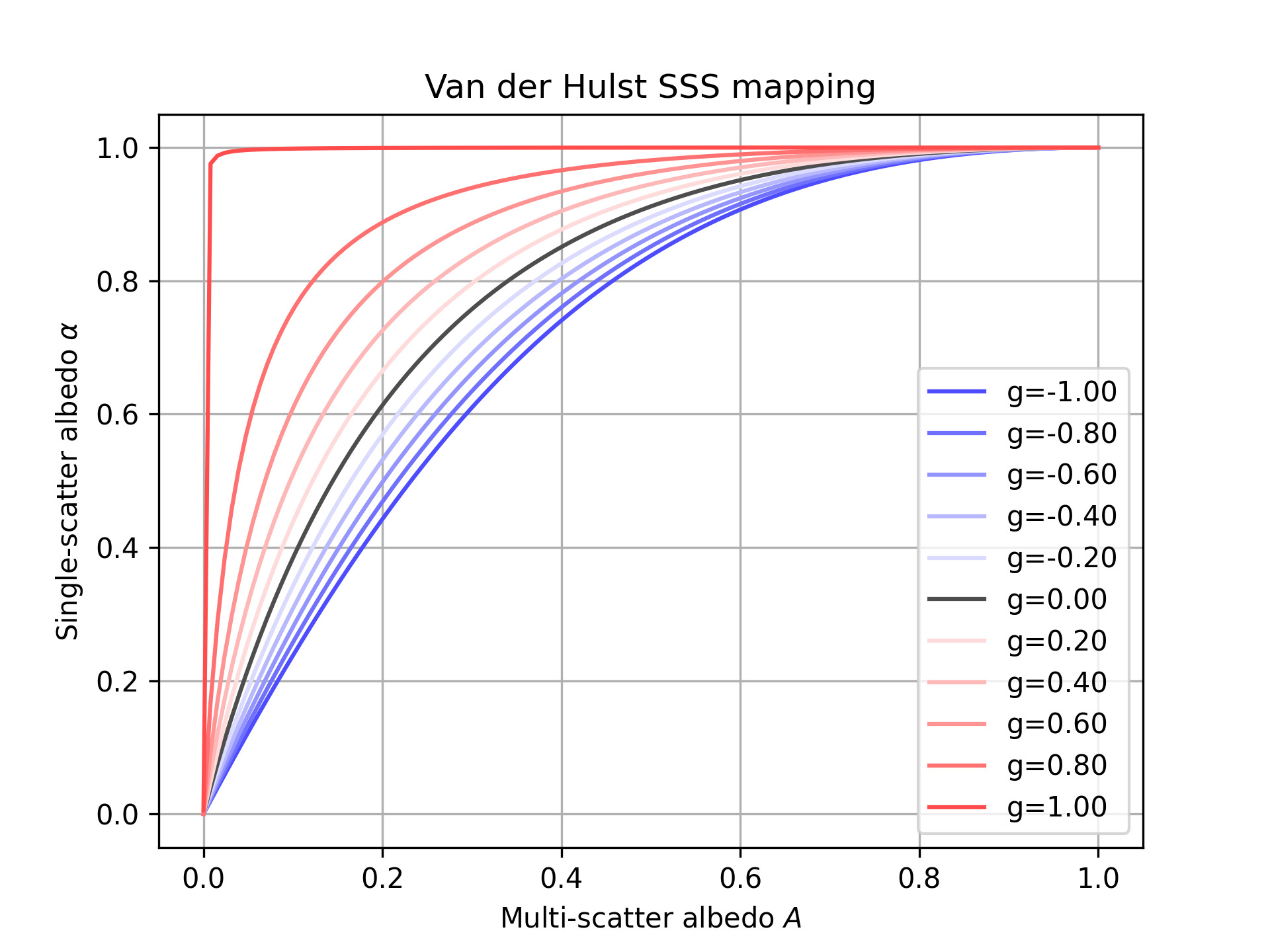}
  \end{subfigure}
  \hfill
  \begin{subfigure}{.48\textwidth}
    \includegraphics[width=\linewidth]{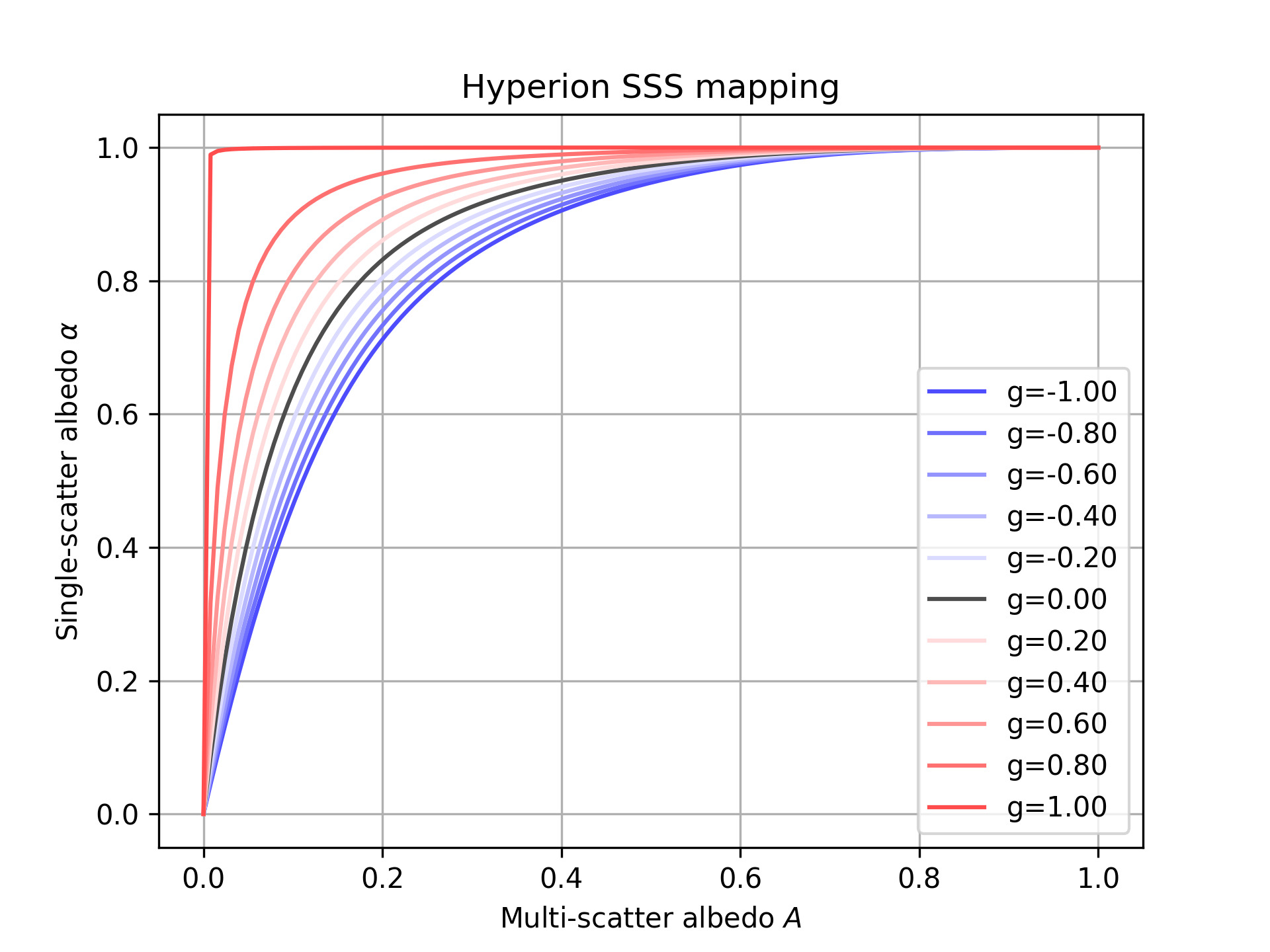}
  \end{subfigure}
  \hfill
  \caption{Comparison of the van de Hulst (left) and Hyperion (right) subsurface scattering mappings from desired multi-scatter albedo $A$ to single-scatter albedo $\alpha$. \label{fig:sss_mappings}}
\end{figure}

Figure~\ref{fig:sss_mappings} shows the form of the van de Hulst and Hyperion mappings as a function of anisotropy $g$. At high anisotropy, a higher single-scatter albedo is required to achieve the same multi-scatter albedo, since light is preferentially scattered in the forward direction so has a higher chance of escaping without scattering again.

Note that the similarity relation described above can also be used to make path-traced volume rendering more efficient by reducing the number of scattering events. This can be done for a volume with anisotropy $g$ by setting $g^\star$ to any value between $g$ and 0 (0 for isotropic scattering), computing $\boldsymbol{\mu}^\star_s$ using Equation~\ref{eq:similarity_relation}, and simply setting $\boldsymbol{\mu}^\star_a = \boldsymbol{\mu}_a$ (since the distance traveled along the path and therefore the amount of absorption is essentially unchanged). As proposed by \textcite{Hyperion}, it is even possible to gradually transition from full anisotropic scattering to reduced isotropic scattering over the first several bounces along a path by setting $g^\star$ differently at each bounce, gradually reducing it from $g$ to 0. This can allow an implementation to achieve the best of both worlds of visuals and performance: view-dependent, low-order scattering and efficient high-order scattering.

\subparagraph{Rayleigh scattering}

\begin{figure}[!tb]
  \centering
  \begin{subfigure}{0.48\textwidth}
    \centering
    \includegraphics[width=\linewidth]{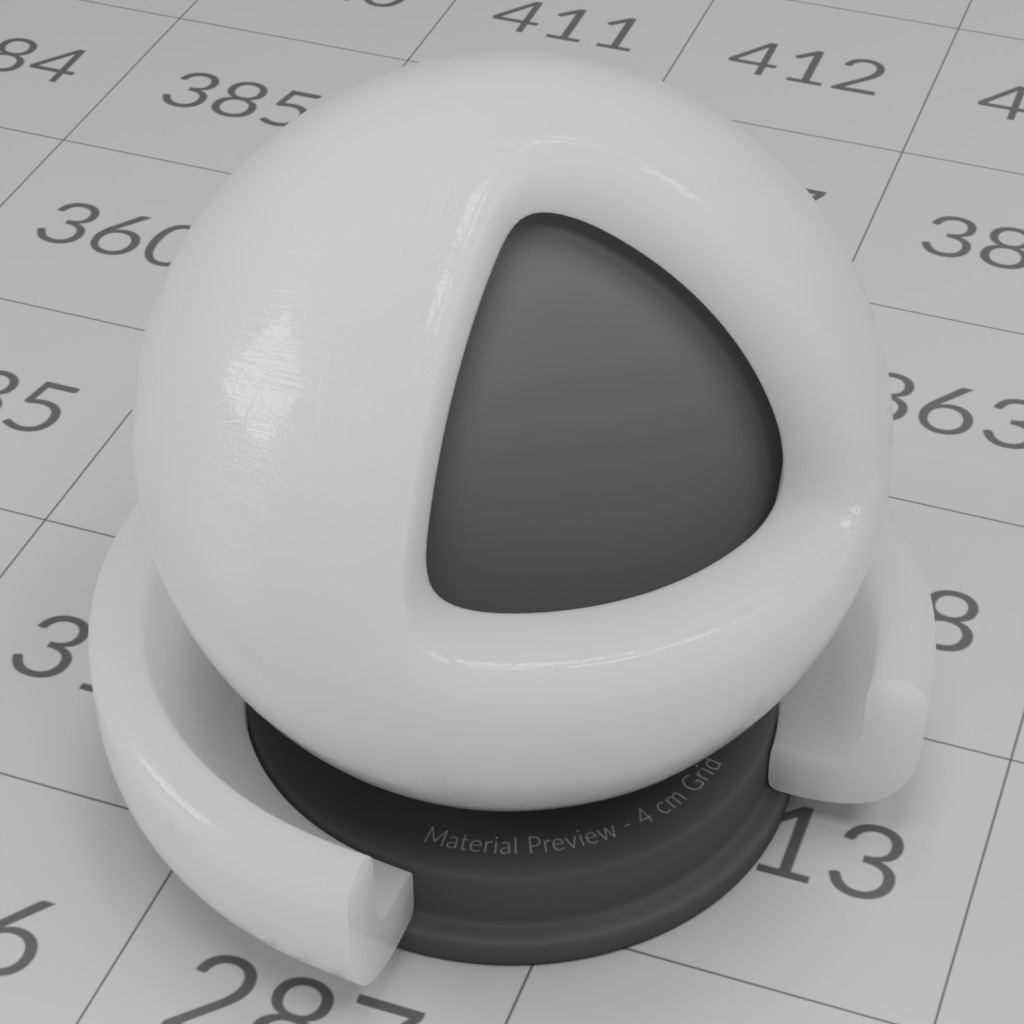}
  \end{subfigure}
  \hspace{0.02\textwidth}
  \begin{subfigure}{0.48\textwidth}
    \centering
    \includegraphics[width=\linewidth]{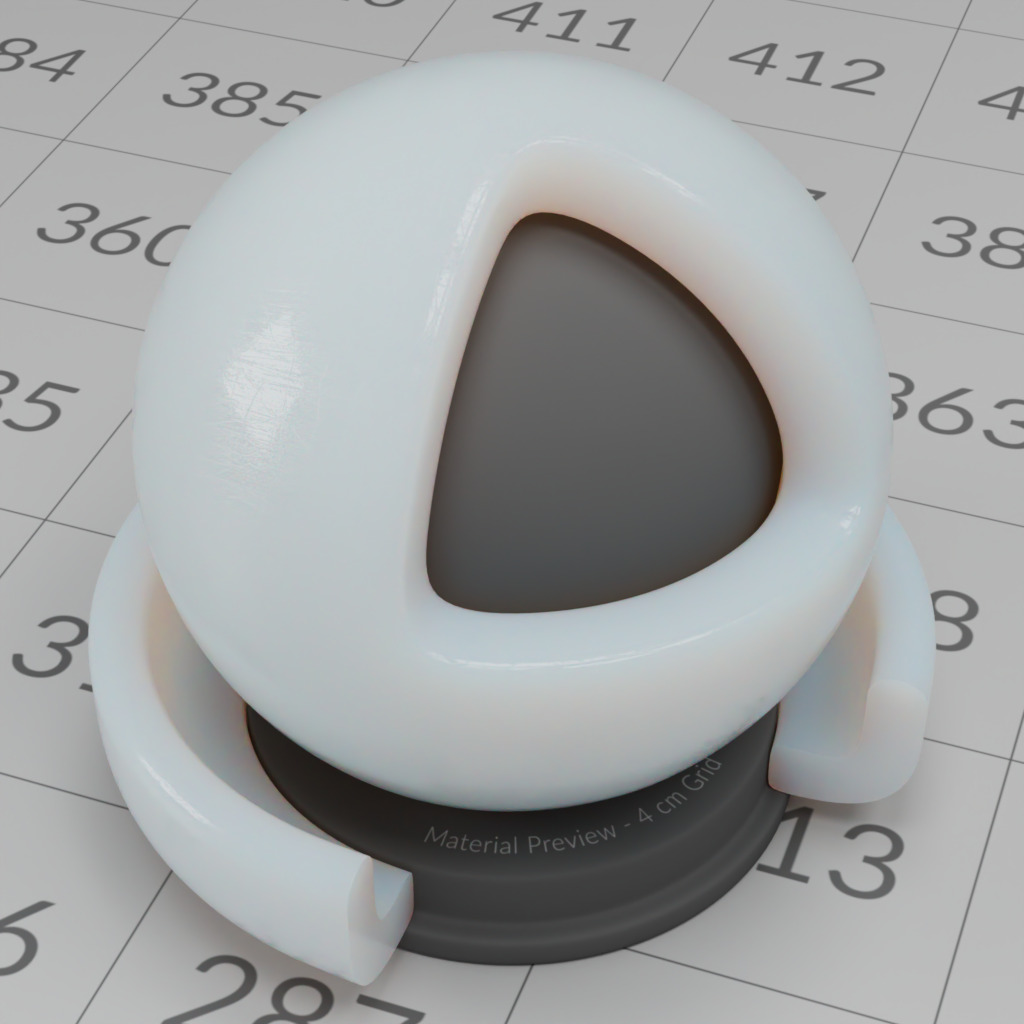}
  \end{subfigure}
  \caption{The subsurface defaults to a \texttt{subsurface\_radius\_scale} of $(1, 0.5, 0.25)$ to approximate the effect of Rayleigh scattering (right), rather than pure white (left). \label{fig:subsurface_rayleigh}}
\end{figure}

Note that the default value of \verb|subsurface_radius_scale| is set at $(1, 0.5, 0.25)$ to approximate the effect of Rayleigh scattering (or Tyndall scattering). If we roughly approximate the wavelength corresponding to each of the RGB channels as $\lambda/\mathrm{nm} \approx 650, 550, 450$ and assume the radii scale like the reciprocal of the extinction, then since in Rayleigh scattering the extinction scales like $\lambda^{-4}$ for light of wavelength $\lambda$, the expected relative magnitudes of the radii are $(1,  (550/650)^4, (450/650)^4) \approx (1, 0.5, 0.25)$, hence the chosen default. This provides a slightly more realistic default look for the subsurface than resulting from gray radii (Figure~\ref{fig:subsurface_rayleigh}). We considered a more explicit control for providing a Rayleigh-scattering effect (such as the model from \textcite{Kutz2021}) but felt that this could always in principle be built of top of the model, and having the default look be reminiscent of Rayleigh scattering was sufficient.

Since the scattered ray may emerge from a different point on the surface than the input point, the material will in general differ there. Thus, on transmitting back out, it will encounter different layer properties. In principle this should be taken into account, but how to model subsurface scattering in the presence of varying subsurface properties is an open question.

\clearpage

\subsubsection{Translucent base}

\label{sec:translucent-base}

As noted earlier, the volume of the base dielectric is parameterized in two ways in OpenPBR, either via \hyperref[sec:subsurface]{subsurface scattering} or using a parameterization more suited to thin volumetric media with visible refraction. We term the latter case the ``translucent base'' dielectric, whose embedded volume is the ``transmission volume'' (being parameterized by the transmission parameters).

As for the dielectric slabs of the \hyperref[sec:glossy-diffuse]{glossy-diffuse} and \hyperref[sec:subsurface]{subsurface} components, the top interface is described by a rough GGX microfacet BSDF $f_\mathrm{dielectric}$ whose ``specular'' parameters are described in the \hyperref[sec:dielectric-base]{Dielectric base} section. The bulk of the dielectric, $V^\infty_\mathrm{dielectric}$ is a volumetric medium supporting absorption and scattering:
\begin{equation}
S_\textrm{translucent-base} = \mathrm{Slab}(f_\mathrm{dielectric}, V^\infty_\mathrm{dielectric})  \ .
\end{equation}
The index of refraction of $V^\infty_\mathrm{dielectric}$ is specified by \verb|specular_ior| (as for the entire dielectric base). The volumetric properties of $V^\infty_\mathrm{dielectric}$ are specified as follows.

The \verb|transmission_depth| $\lambda_T$ is the distance traveled inside the medium by white light before its color becomes exactly \verb|transmission_color| $\mathbf{T}$ by Beer's law, determining the extinction coefficient of the interior medium $\boldsymbol{\mu}_t$:
\begin{equation}
\boldsymbol{\mu}_t = - \frac{\ln \mathbf{T}} {\lambda_T} \ .
\end{equation}
Being a length, \verb|transmission_depth| can be any value greater than or equal to zero. For convenience, we make the soft range $[0, 1]$, thus covering common cases. However, when \verb|transmission_depth| $\lambda_T$ is zero, it is assumed that the interior medium is absent ($\boldsymbol{\mu}_t=0$) and \verb|transmission_color| is used instead to non-physically tint the dielectric refraction Fresnel factor multiplicatively by a constant amount (ignoring the dielectric energy balance).

\begin{figure}[!tb]
  \centering
  \begin{subfigure}{.3\textwidth}
    \includegraphics[width=\linewidth]{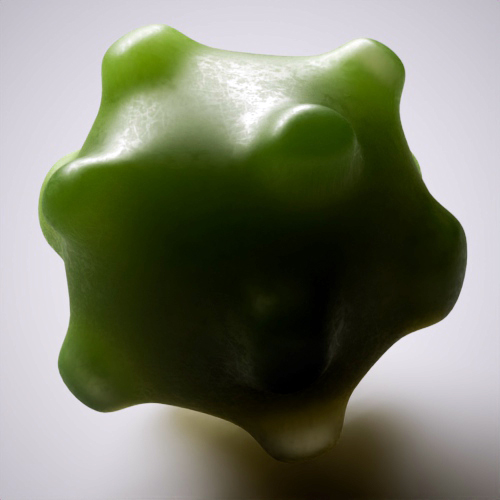}
  \end{subfigure}
  \hspace{0.02\textwidth}
  \begin{subfigure}{.3\textwidth}
    \includegraphics[width=\linewidth]{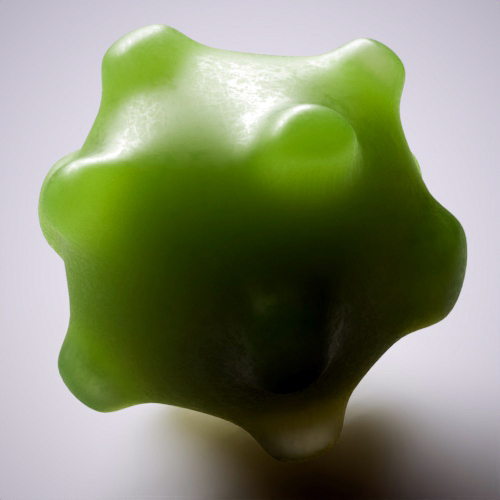}
  \end{subfigure}
  \hspace{0.02\textwidth}
  \begin{subfigure}{.3\textwidth}
    \includegraphics[width=\linewidth]{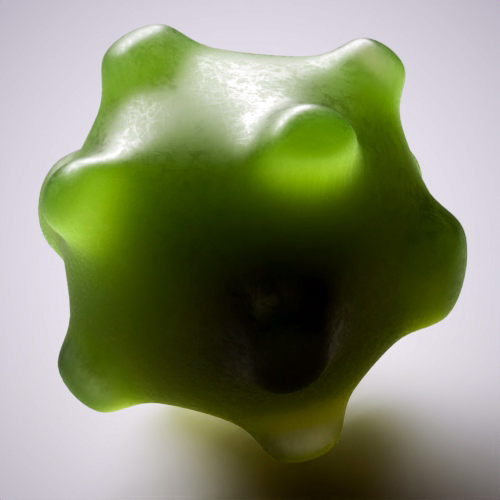}
  \end{subfigure}
  \caption{The effect of \texttt{transmission\_scatter\_anisotropy}. The left image shows a negative anisotropy of -0.5, the center image shows an isotropic phase function, and the right image shows a positive anisotropy of 0.5.\label{fig:transmission_scatter_anisotropy_example}}
\end{figure}

The \verb|transmission_scatter| parameter $\mathbf{S}$ directly sets the medium scattering coefficient $\boldsymbol{\mu}_s$ (as a multiple of the inverse \verb|transmission_depth|):
\begin{equation}
\boldsymbol{\mu}_s = \frac{\mathbf{S}} {\lambda_T} \ .
\end{equation}
The \verb|transmission_scatter| color thus controls the observed color of the single-scattered light. (Note that in the case \verb|transmission_depth| $\lambda_T$ is zero, however, the scattering coefficient is ignored.) The absorption coefficient $\boldsymbol{\mu}_a$ is then computed as
\begin{equation}
\boldsymbol{\mu}_a = \boldsymbol{\mu}_t - \boldsymbol{\mu}_s \ .
\end{equation}
If any component of $\boldsymbol{\mu}_a$ is negative, then $\boldsymbol{\mu}_a$ is shifted by enough gray to make all of the components positive, i.e. (in an obvious notation):
\begin{eqnarray}
&& \mathrm{if \;\; min(\boldsymbol{\mu}_a)} < 0 \nonumber \\
&& \quad\quad \boldsymbol{\mu}_a \leftarrow \boldsymbol{\mu}_a - \mathrm{min}(\boldsymbol{\mu}_a) \ .
\end{eqnarray}
After this shifting, the final extinction coefficient is given by the new $\boldsymbol{\mu}_a + \boldsymbol{\mu}_s$.
This formulation yields volumetric parameters which reproduce reasonably well the independently specified colors of the transmitted and single-scattered light. Finally, the medium phase function anisotropy $g \in [-1, 1]$ is given by \verb|transmission_scatter_anisotropy| (the standard Henyey--Greenstein phase function is assumed). Figure~\ref{fig:transmission_scatter_anisotropy_example} shows the effect of varying the \verb|transmission_scatter_anisotropy| parameter.

\subparagraph{Dispersion}

\label{sec:dispersion}

The phenomenon known as \emph{optical dispersion} (i.e., variation of the IOR with wavelength) produces familiar rainbow-like colors when light refracts through dielectrics such as water, glass and diamond. Control over dispersion is grouped with the transmission parameters since this effect is only significant in highly transparent refractive media. Dispersion is parameterized by the \emph{Abbe number} $V_d$, which is the ratio between differences of refractive indices at short, medium, and long wavelengths as follows:
\begin{equation}
V_d = \frac{n_{\mathrm{medium}} - 1}{n_{\mathrm{short}} - n_{\mathrm{long}}} \ ,
\end{equation}
where in the standard modern definition $n_{\mathrm{long}} = n_C$, $n_{\mathrm{medium}} = n_d$ and $n_{\mathrm{short}} = n_F$ are the IORs of the material at the wavelengths of the Fraunhofer C, d, and F spectral lines (at $\lambda_C = 656.3$~nm, $\lambda_d = 587.6$~nm, and $\lambda_F = 486.1$~nm, respectively). The amount of dispersion (i.e., angular separation of colors) is roughly proportional to the reciprocal of the Abbe number. The IOR at any wavelength can be determined from the Cauchy empirical formula:
\begin{equation}
n(\lambda) = A + \frac{B}{\lambda^2} \ .
\end{equation}
It follows that if the Abbe number $V_d$ and IOR $n_d = n(\lambda_d)$ are given, the coefficients in the Cauchy formula are given by
\begin{equation}
A = n_d - \frac{B}{\lambda^2_d}  \ ,  \quad    B = \frac{n_d - 1}{V_d \; (\lambda^{-2}_F - \lambda^{-2}_C)} \ .
\end{equation}
We assume that \verb|specular_ior| (including any modulation via \verb|specular_weight| as in Equation~\ref{modulated_ior}) defines $n(\lambda_d)$.
Thus the IOR $n$ at any wavelength $\lambda$ is determined, given $V_d$. A renderer can use this known $n(\lambda)$ function to model the effect of dispersion, for example by stochastically choosing a wavelength sample and tracing the refracted ray direction according to the corresponding IOR.

\begin{figure}[!tb]
  \centering
  \hfill
  \begin{subfigure}{.19\textwidth}
    \includegraphics[width=\linewidth]{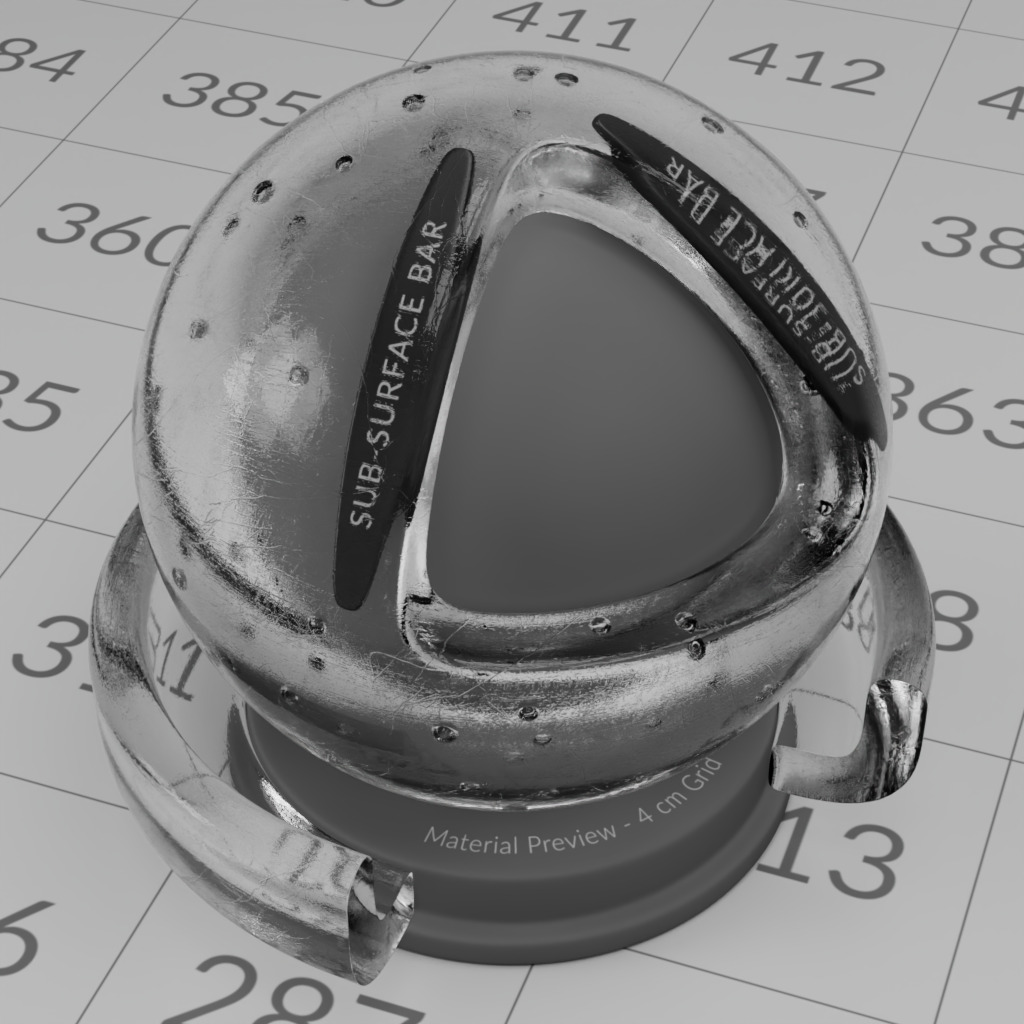}
  \end{subfigure}
  \hfill
  \begin{subfigure}{.19\textwidth}
    \includegraphics[width=\linewidth]{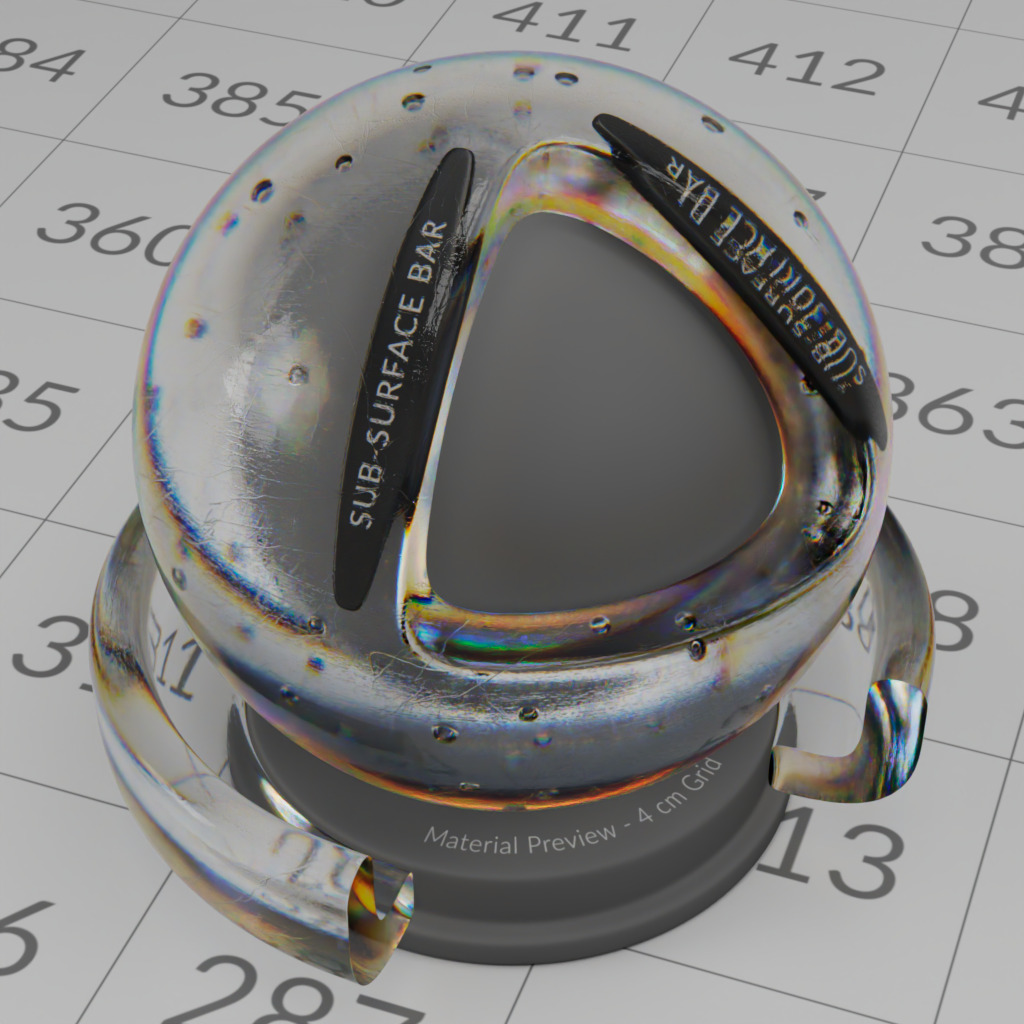}
  \end{subfigure}
  \hfill
  \begin{subfigure}{.19\textwidth}
    \includegraphics[width=\linewidth]{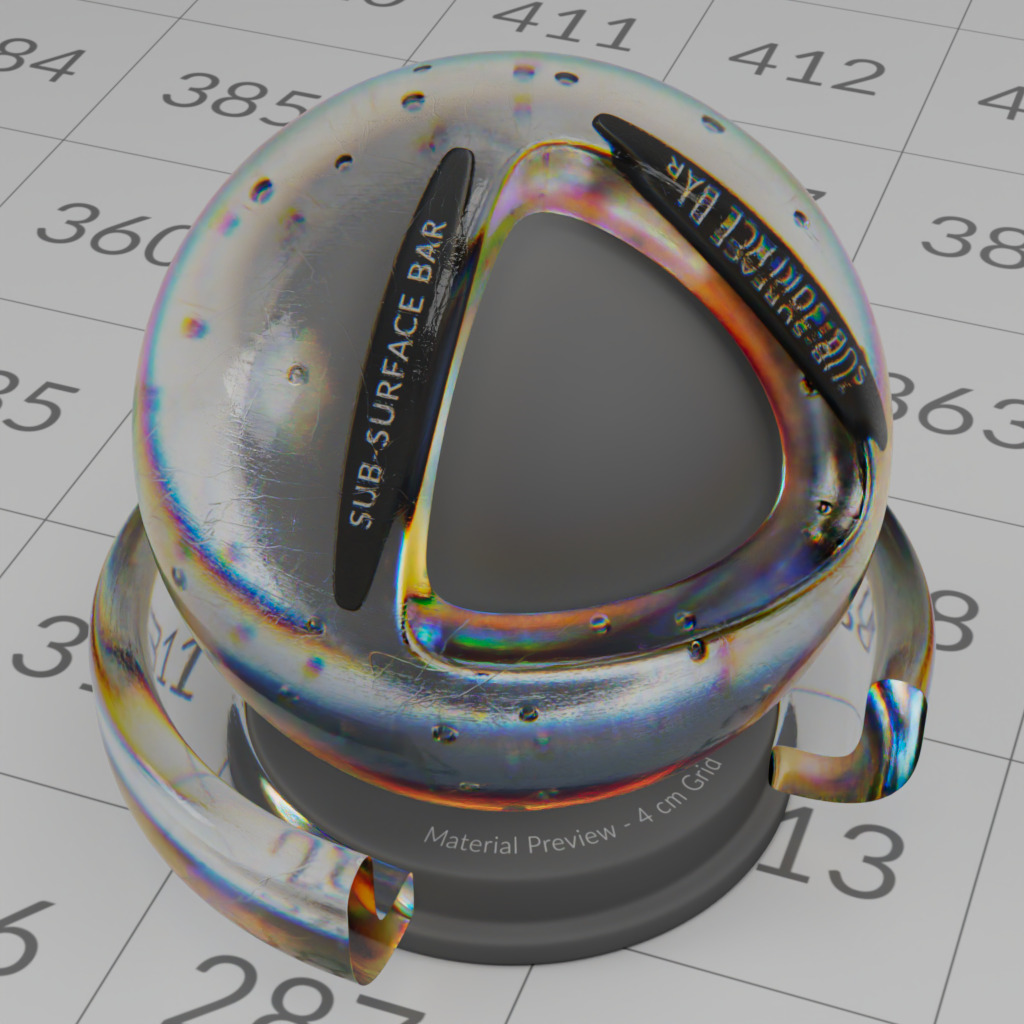}
  \end{subfigure}
  \hfill
  \begin{subfigure}{.19\textwidth}
    \includegraphics[width=\linewidth]{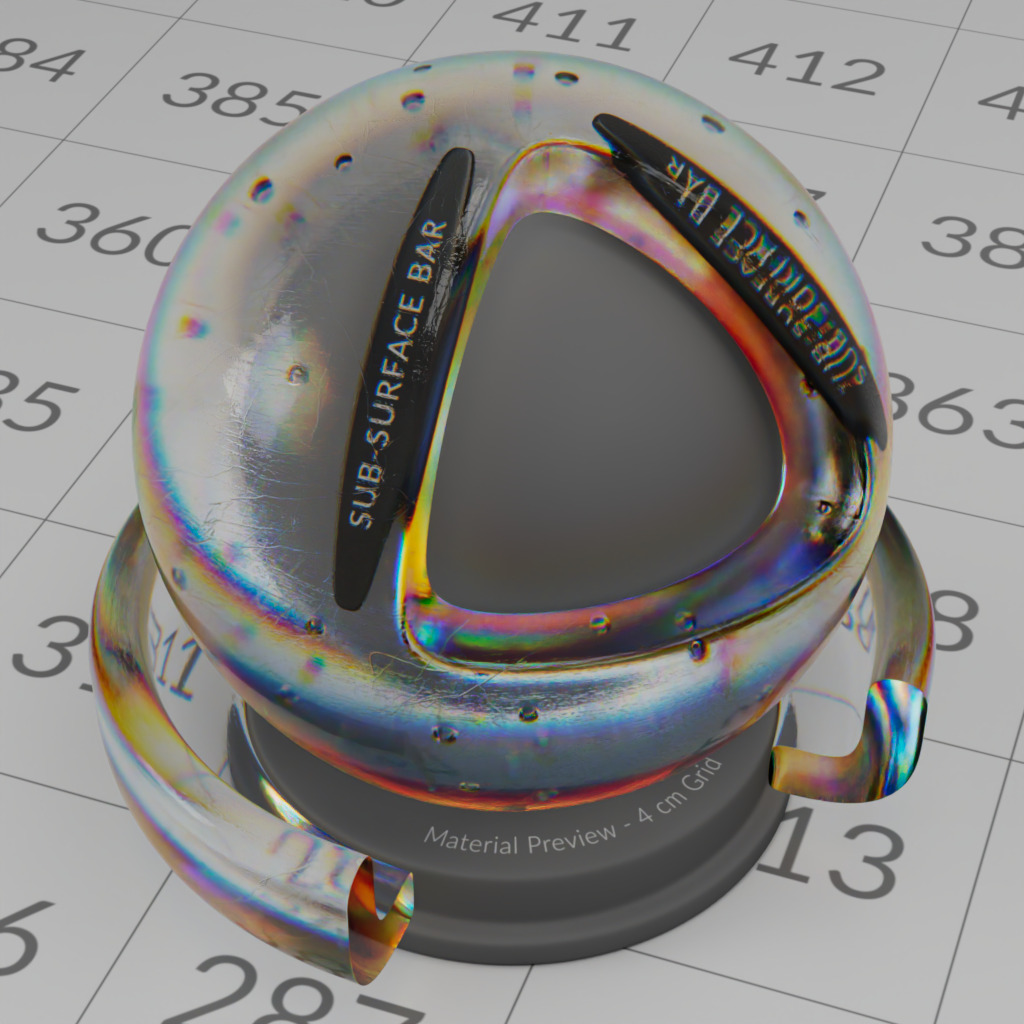}
  \end{subfigure}
  \hfill
    \begin{subfigure}{.19\textwidth}
    \includegraphics[width=\linewidth]{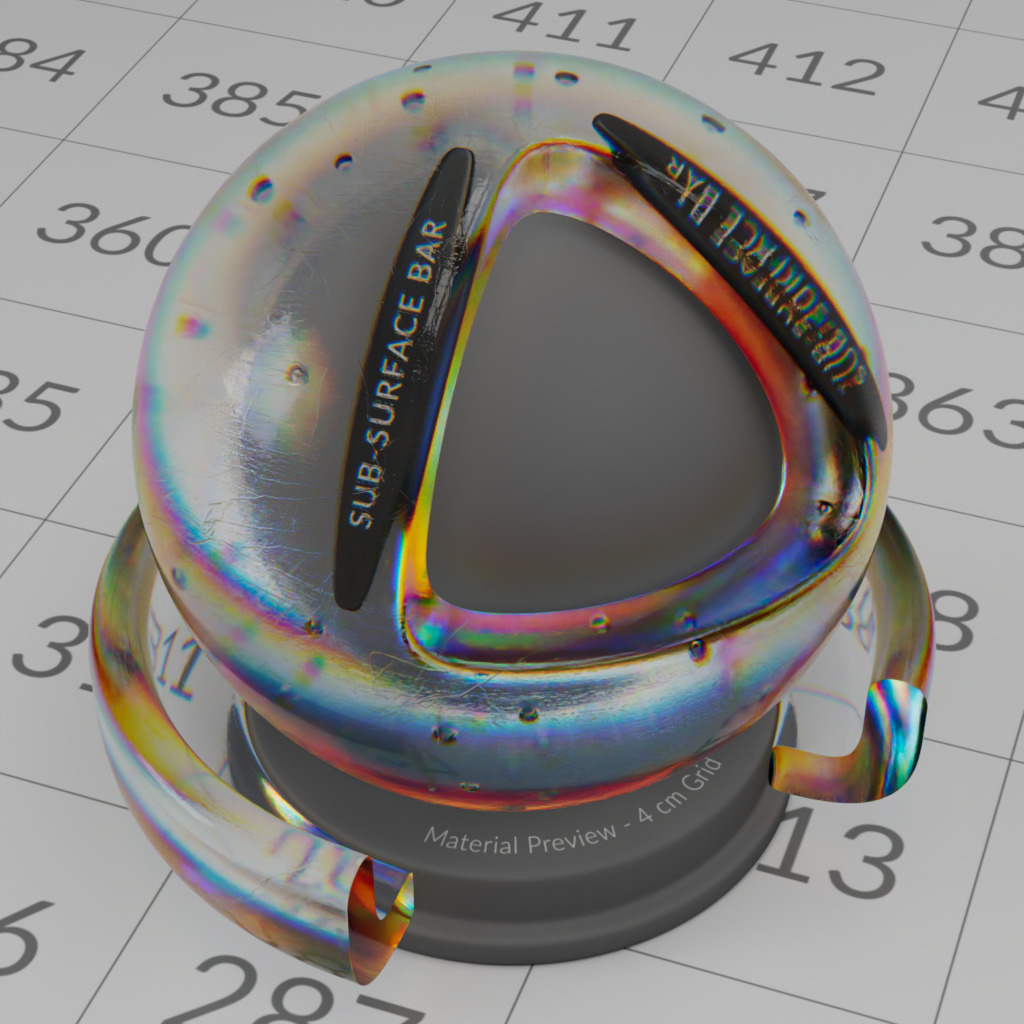}
  \end{subfigure}
  \hfill
  \caption{The effect of \texttt{transmission\_dispersion\_scale}, varying over 0, 0.2, 0.4, 0.6, 1. Here, the Abbe number was set (unrealistically low) to 1, via \texttt{transmission\_dispersion\_abbe\_number}. \label{fig:transmission_dispersion_scale_example}}
\end{figure}

However, the Abbe number itself is not very intuitive to work with, since the dispersion effect increases as the Abbe number decreases (zero dispersion occurs at infinite Abbe number). We therefore prefer to use a more artist-friendly parameterization (see Figure~\ref{fig:transmission_dispersion_scale_example}), where the Abbe number is specified by
\begin{equation}
V_d = \frac{\mathtt{transmission\_dispersion\_abbe\_number}} {\mathtt{transmission\_dispersion\_scale}} \ .
\end{equation}
At the default \verb|transmission_dispersion_scale| of zero, the Abbe number is infinite, which corresponds to no dispersion. Conversely, at the maximum of 1, the Abbe number peaks at $V_d$ = \verb|transmission_dispersion_abbe_number|, which defaults to 20.

The default 20 was chosen since common real materials with the highest dispersion have Abbe numbers roughly greater than or equal to this (for example, different types of glass have Abbe numbers between 20 and 91, water and diamond both have Abbe numbers between 55 and 56, while the oxide mineral rutile, composed of titanium dioxide, exhibits extremely high dispersion with an Abbe number of 9.87 \cite{Polyanskiy2023}).

In most cases, the \verb|transmission_dispersion_scale| therefore functions as a convenient, roughly linear slider from low to high dispersion. For those special cases where the Abbe number of a specific material is required, the Abbe number itself can be changed via \verb|transmission_dispersion_abbe_number|.

Note that dispersion can cause banding artifacts and color noise if not properly managed, however these issues can be mitigated using a careful strategy of wavelength sampling, color-channel selection, and MIS. For sampling wavelengths and incorporating spectral effects inside an RGB renderer, multiple approaches exist, such as the empirical one described by \textcite{Kutz2021}. The wavelengths from such a ``locally spectral'' approach can also be used for rendering thin-film iridescence, as mentioned in Section~\ref{sec:thin-film}. For selecting a color channel to use for the refracted ray direction, it is useful to multiply the path throughput into the probabilities. This way, the channel carrying the most energy along the path so far will be preferentially used for continuing the path. This keeps the magnitude of the path throughput balanced and avoids generating paths that don't contribute to the image.

For example, if one scattering event selects the red channel and generates a refracted path that is only valid for the red channel, the next scattering event will also use the red channel. While careful color-channel selection can reduce variance, it doesn't directly target color noise. Color noise is inevitable when individual paths are only valid for a single wavelength, however when dispersion occurs on a rough microfacet surface, the resultant transmission lobes can overlap significantly, so paths can be shared across color channels using MIS (as described by \textcite{Kutz2021}). Similar approaches can also be leveraged to reduce noise when using path tracing to render volumes with wavelength-varying extinction coefficients: specifically, the path throughput (along with the single-scattering albedo) can be incorporated into the color-channel selection probabilities for distance sampling, and paths can be shared among color channels using MIS (as described by \textcite{ChiangKutz2016} and \textcite{Kutz2017}).



\clearpage

\subsection{Metallic base: the F82-tint model}

\label{sec:metallic-base}

The metallic base is represented as a separate bulk slab consisting of an opaque GGX microfacet conductor BRDF $f_\mathrm{conductor}$ whose NDF is parameterized by \verb|specular_roughness| and \verb|specular_roughness_anisotropy| (overloading the same parameters used for the dielectric BSDF $f_\mathrm{dielectric}$, as noted in the \hyperref[sec:microfacet]{Microfacet} section):
\begin{equation}
S_\mathrm{metal} = \mathrm{Slab}(f_\mathrm{conductor}) \ .
\end{equation}

Metals are completely opaque and have a characteristic and familiar form of specularity due to the Fresnel factor for conductors differing from that of dielectrics. The conductor Fresnel reflection curve is parameterized by the colors at normal and near-grazing incidence (\verb|base_color| and \verb|specular_color| respectively, scaled by the corresponding weights). This allows for art-directable variation in reflectivity toward the grazing edges by directly specifying the (texturable) colors at normal and grazing incidence to simulate the dip in reflectivity observed in real metals, or just for artistic effect \cite{Hoffman2019}. Note that these two color parameters are also used for the non-metallic (i.e., dielectric) \hyperref[sec:dielectric-base]{specular} and \hyperref[sec:glossy-diffuse]{diffuse} BRDFs.

\begin{figure}[!b]
  \centering
  \hfill
  \begin{subfigure}{.19\textwidth}
    \includegraphics[width=\linewidth]{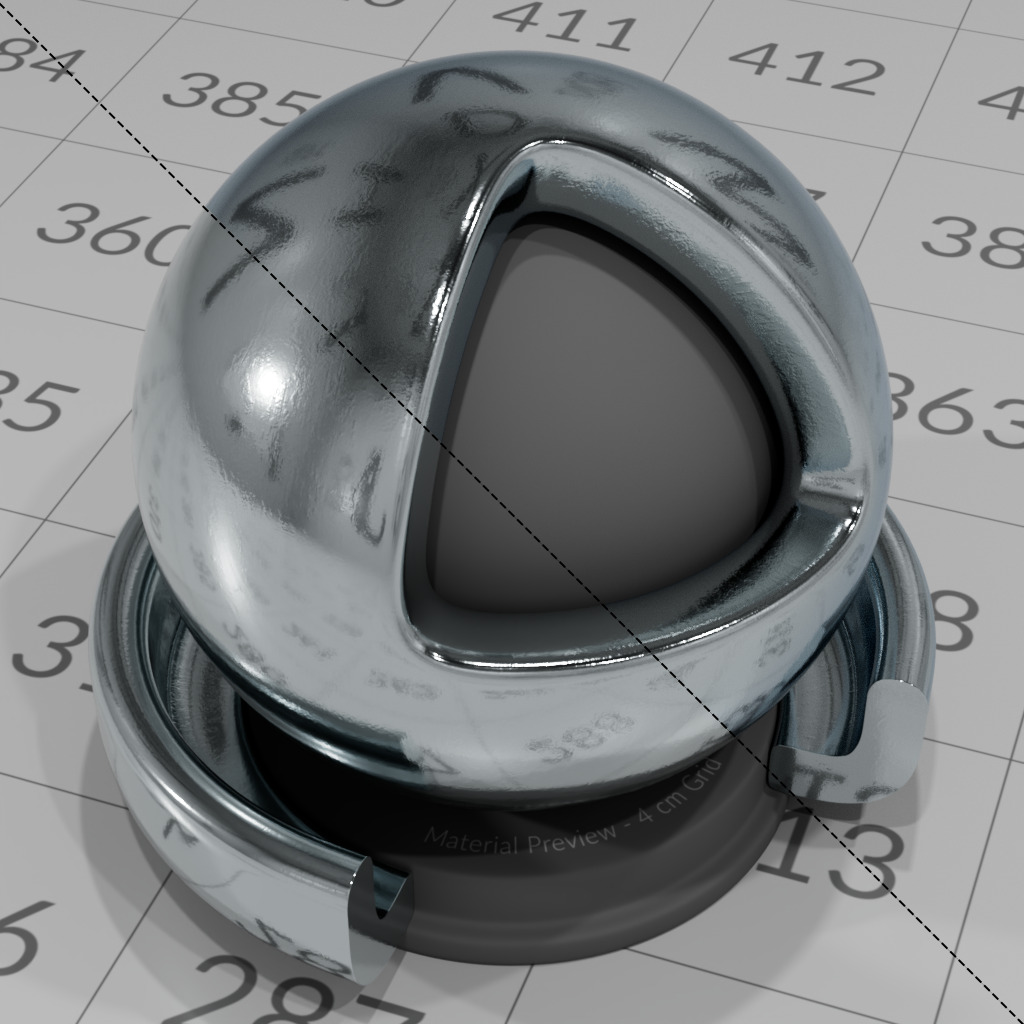}
  \end{subfigure}
  \hfill
  \begin{subfigure}{.19\textwidth}
    \includegraphics[width=\linewidth]{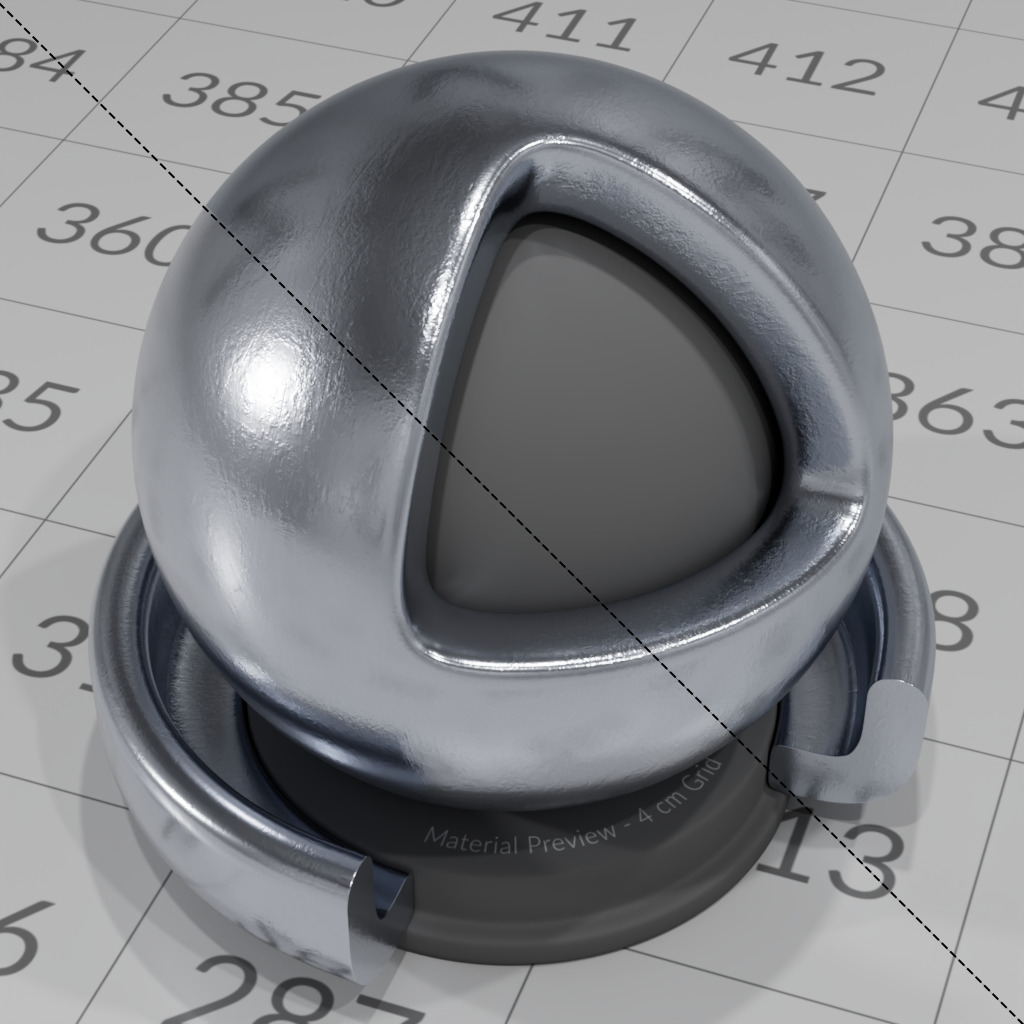}
  \end{subfigure}
  \hfill
  \begin{subfigure}{.19\textwidth}
    \includegraphics[width=\linewidth]{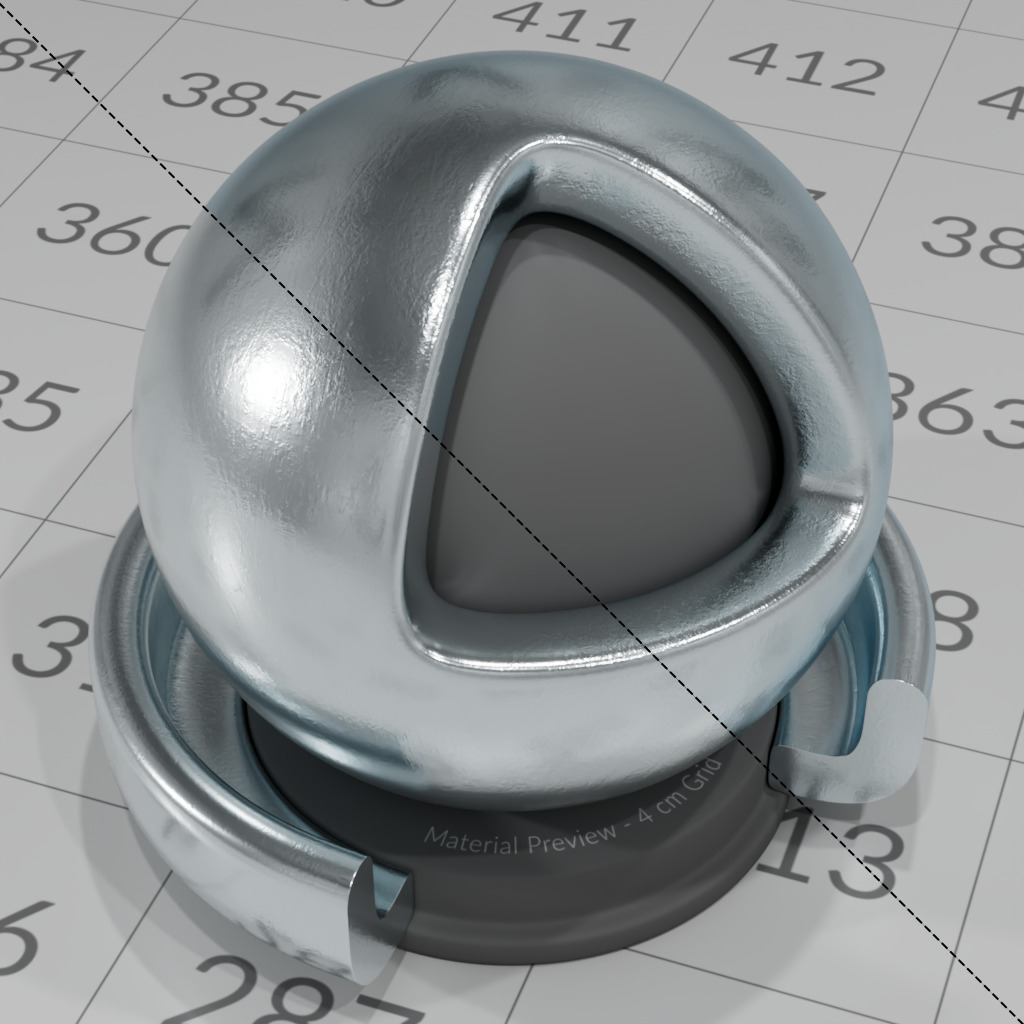}
  \end{subfigure}
  \hfill
  \begin{subfigure}{.19\textwidth}
  \includegraphics[width=\linewidth]{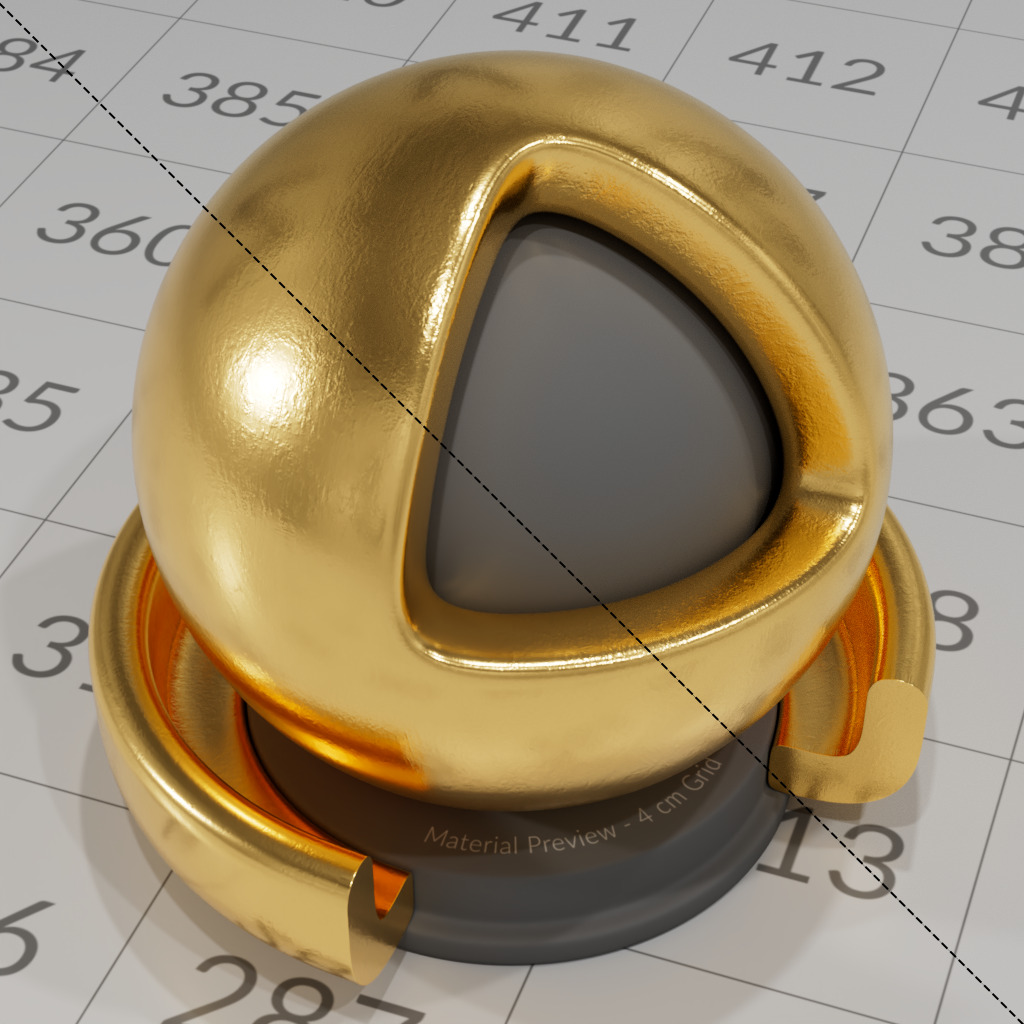}
  \end{subfigure}
  \hfill
  \begin{subfigure}{.19\textwidth}
  \includegraphics[width=\linewidth]{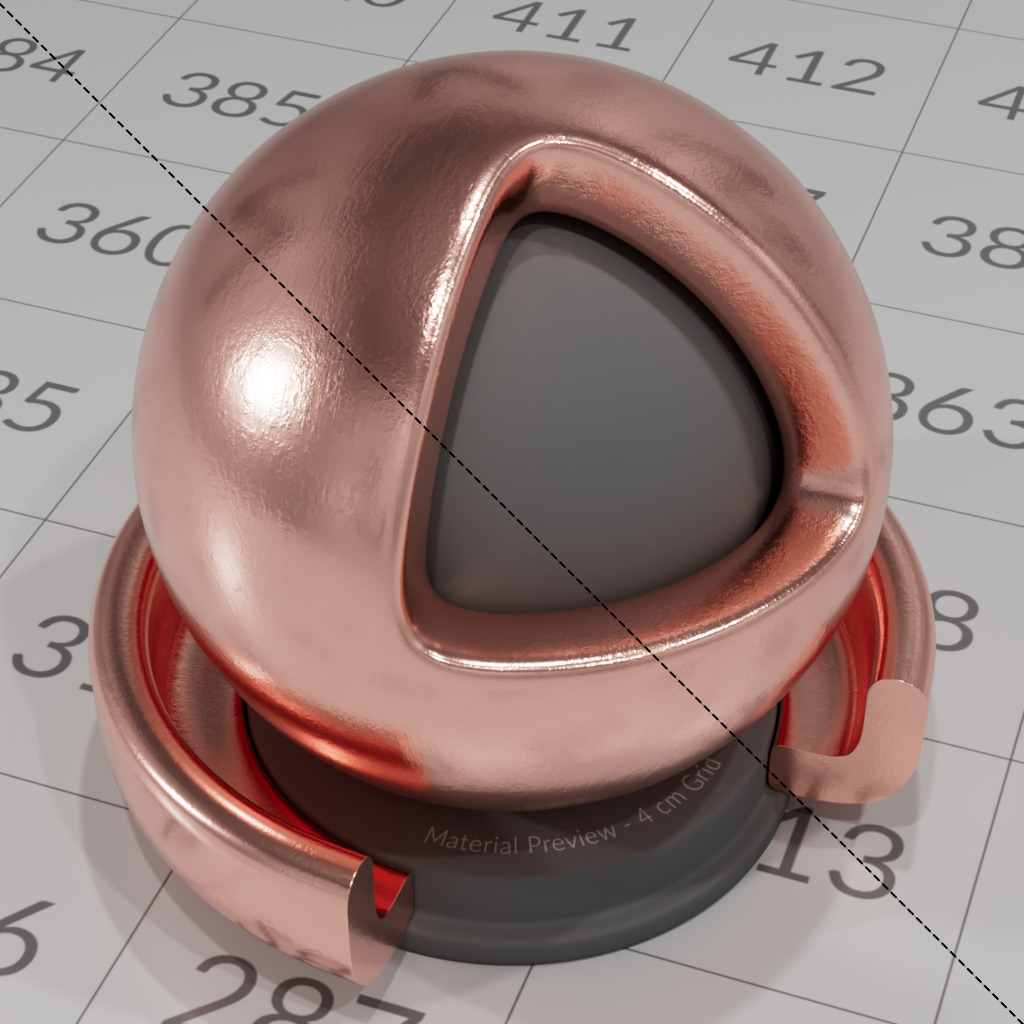}
  \end{subfigure}
  \hfill
  \caption{Renders with the F82-tint model of (from left to right) chromium, lead, zinc, gold and copper. Above the diagonal is the full fit to the measured metal IORs, while below the diagonal has the fitted \texttt{base\_color} but \texttt{specular\_color} is set to white. \label{fig:metal_renders}}
\end{figure}

As noted previously, this non-transmissive metallic base is blended as a statistical mixture with the dielectric base according to the \verb|base_metalness| parameter:
\begin{eqnarray}
M_\textrm{base-substrate} &=& \mathrm{\mathbf{mix}}(M_\textrm{dielectric-base}, S_\mathrm{metal}, \mathtt{M})   \ ,
\end{eqnarray}
where $\mathtt{M}$ = \verb|base_metalness|.

The specific model we stipulate for the metallic Fresnel factor $\mathbf{F}_{\mathrm{metal}}(\mu)$ is the ``F82-tint'' model of \textcite{Kutz2021}, which extends previous work by \textcite{Hoffman2019}.
This is based on the standard Schlick approximation to the metallic Fresnel factor, where $\mathbf{F}_0$ is the RGB reflectivity at normal incidence (i.e., $\mathtt{base\_weight} * \mathtt{base\_color}$), and $\mu$ is the cosine of the incident angle:
\begin{equation}
\mathbf{F}_{\mathrm{Schlick}}(\mu) = \mathbf{F}_0 + (1 - \mathbf{F}_0) (1 - \mu)^5 \ .
\end{equation}
To better approximate the actual Fresnel curve of metals, in the F82-tint model the Schlick approximation is augmented with a correction term:
\begin{equation}
\mathbf{F}_{82}(\mu) = \mathbf{F}_{\mathrm{Schlick}}(\mu) - \mathbf{b} \mu (1 - \mu)^6 \ ,
\end{equation}
where
\begin{equation} \label{f82_b_coeff}
    \mathbf{b} =  \frac{\mathbf{F}_{\mathrm{Schlick}}(\bar{\mu}) - \mathbf{F}(\bar{\mu})}{\bar{\mu}(1 - \bar{\mu})^6} \ .
\end{equation}
Here, $\mathbf{F}(\bar{\mu})$ is the desired metallic reflectivity at the ``grazing edge'' angle cosine $\bar{\mu} = 1/7$ corresponding roughly to $82^\circ$ (i.e., around silhouettes), ensuring $\mathbf{F}_{82}(\bar{\mu}) = \mathbf{F}(\bar{\mu})$. This desired edge reflectivity is user-specified as a fractional tint of the Schlick curve, controlled via $\mathbf{C}_s = \mathtt{specular\_color}$, i.e.
\begin{equation}
\mathbf{F}(\bar{\mu}) = \mathbf{C}_s \, \mathbf{F}_\mathrm{Schlick}(\bar{\mu}) \ .
\end{equation}
The benefit of having \verb|specular_color| function as the tint is that the model reduces to the regular Schlick reflectivity at the default values of \verb|specular_weight| and \verb|specular_color|.

The final metallic Fresnel term we employ is then given by an overall multiplication by $\xi_s = \mathtt{specular\_weight}$, ensuring that the entire metallic lobe is suppressed as the weight goes to zero.
Similar to the dielectric, the weight can exceed one in order to linearly boost the Fresnel, with a clamp put in place to ensure that it remains bounded in $[0,1]$ as $\xi_s \rightarrow \infty$:
\begin{equation}
\mathbf{F}_{\mathrm{metal}}(\mu) = \mathrm{clamp}\bigl(\xi_s \mathbf{F}_{82}(\mu), 0, 1\bigr) \ .
\end{equation}
Note that the clamp at the lower end is applied since $\mathbf{F}_{82}(\mu)$ can become negative for some values of $\mu$, which is not physically meaningful \cite{Hoffman2019}.
Note that the edge cannot be brighter than the standard Schlick term, but this is generally true in real metals. We consider this a benefit of this parameterization, since it makes it impossible to produce physically implausible metals with excessively bright edges.

Figure~\ref{fig:metal_renders} shows renders of chromium, lead, zinc, gold and copper, with and without the physically correct \verb|specular_color| F82-tint color. In the cases of chromium and lead, the effect of the F82-tint color is visually significant near the edge. In the case of zinc it is less obvious, and barely perceptible in the case of gold and copper. In Appendix~\ref{sec:f82_fits}, we provide a table of computed fits to the F82-tint model for a variety of real metals, using publicly available tabulated IOR data \cite{Polyanskiy2023}.

The hemispherical (or average) albedo of the F82-tint model (i.e., $\mathbf{E}_\mathrm{avg} \equiv 2 \int_0^1 \mathbf{F}_{82}(\mu) \,\mu\,\mathrm{d}\mu$) is required in multiple-scattering compensation schemes, for example. It is given exactly by
\begin{equation}
   \mathbf{E}_\mathrm{avg} = \mathbf{F}_0 + (\mathbf{1} - \mathbf{F}_0)/21 - \mathbf{b}/126 \ ,
\end{equation}
where $\mathbf{b}$ is defined in Equation~\ref{f82_b_coeff}. If both $\mathbf{F}_0 = \mathtt{base\_weight} \times \mathtt{base\_color}$ and $\mathbf{C}_s = \mathtt{specular\_color}$ are white, then $\mathbf{F}_0 = \mathbf{1}$ and $\mathbf{b}= \mathbf{0}$ reducing to $\mathbf{E}_\mathrm{avg} = \mathbf{1}$, thus the metal satisfies the furnace test in this case. Note, however, that $\mathbf{F}_{82}(\mu)$ can be slightly negative (for very dark metals), thus it is necessary to clamp it. The albedo of the clamped Fresnel is not exactly given by clamping the formula above, but it is very close (never more than 0.01 incorrect in absolute albedo).

\subparagraph{Future work: Decoupled metallic parameterization}

\label{sec:metal_decoupling}

\begin{figure}[!htb]
  \centering
  \begin{subfigure}{0.35\textwidth}
    \includegraphics[width=\linewidth]{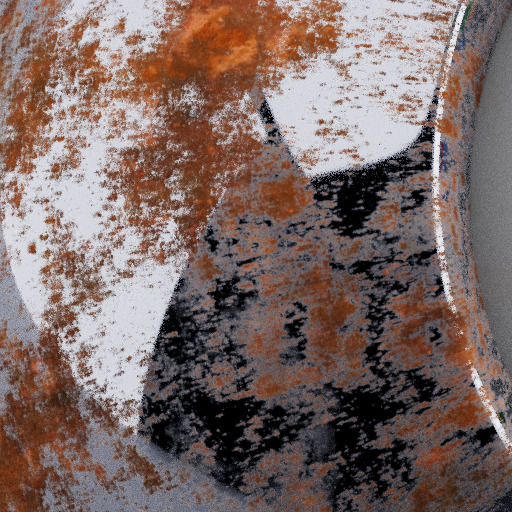}
  \end{subfigure}
  \hspace{0.03\textwidth}
  \begin{subfigure}{0.35\textwidth}
    \includegraphics[width=\linewidth]{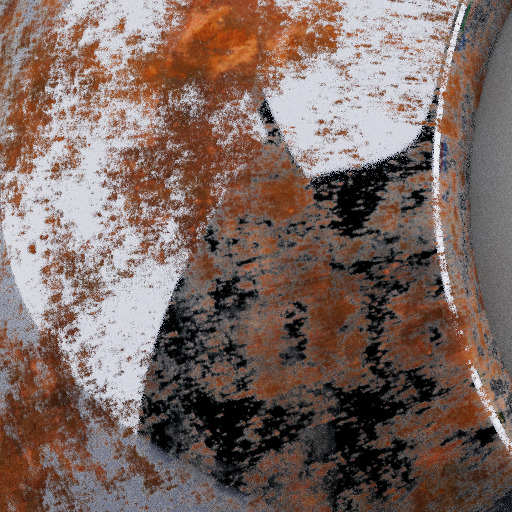}
  \end{subfigure}
  \caption{Metalness blending artifacts with the current model (left), which would be fixed with an expanded parameterization (right). \label{fig:metal_blending_example}}
\end{figure}

The use of \texttt{base\_color} and \texttt{specular\_color} to control the reflectivity of both the metallic and dielectric base is a convenience that originates from the model of \textcite{Burley2012}, and corresponds to the popular ``metalness'' workflow where a single set of textures, in conjunction with a metalness map, can be used to define both metallic and non-metallic regions on the same object.

However, while convenient, this approach does introduce some limitations. Figure~\ref{fig:metal_blending_example} shows an example of the artifacts that can occur when blending between metallic and non-metallic regions, in this case to produce a rusty look. The left image shows a metalness blend done in the current model, while the right image shows the result with the metal and diffuse colors decoupled (made by altering the shader to allow the metal $F_0$ to be defined independently from the diffuse color used for rust). The metalness blend produces a brightening ``halo'' effect in the transition region (slightly blue due to the metal color), while the decoupled result produces a transition that is more plausible for a metal that is partially covered by rust. While this could be avoided (if using a general material layering framework) by adding a completely separate layer on top of the OpenPBR base substrate, it seems preferable to have a single model that can handle blending between metallic and non-metallic regions without artifacts.

To be able to avoid the artifacts, a proposed future extension to the OpenPBR metallic base model is to decouple the metallic and dielectric reflectivities, so that the metalness can be controlled independently of the dielectric reflectivity. This would allow for a more physically plausible transition between metallic and non-metallic regions, as well as more flexibility in defining the material properties.

\begin{figure}[!htb]
  \centering
  \includegraphics[width=0.5\linewidth]{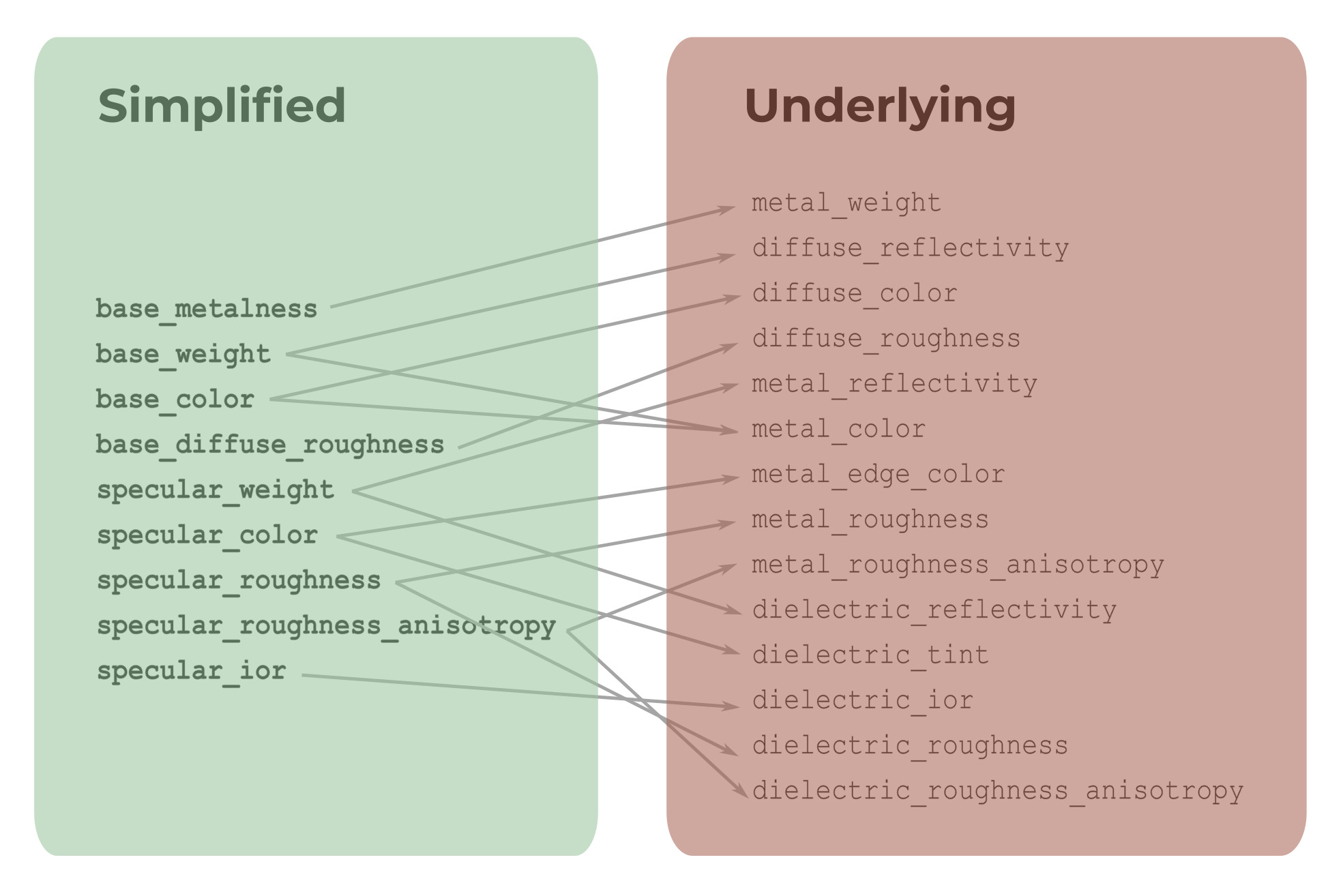}
  \caption{Proposed expanded parameterization for decoupled metalness. \label{fig:metal_decoupling_parameterization}}
\end{figure}

One proposed model for this decoupling is shown in Figure~\ref{fig:metal_decoupling_parameterization}, where the existing convenient controls are retained,
but they are used to drive (as indicated by the arrows) an underlying model which provides full decoupling of the metallic and dielectric properties. So, if necessary, one can drop down to the underlying parameters to control the reflectivity of the metal and dielectric separately.

\clearpage

\subsection{Emission}

\label{sec:emission}

It is assumed that the base substrate below the coat may emit light, with a directionally uniform distribution. One can imagine the light being emitted from the interior of the base substrate with an isotropic luminance given by $L_e$. We put emission below the coating so that emitted light will be tinted due to the absorption in the coat and fuzz layers. This allows for the convenient rendering of low-emission materials that are bounded by a reflective surface (e.g., glow sticks, LEDs, display screens, etc.) without explicit modeling of the emitter and the bounding object.

The intensity of the emission is controlled by a luminance value (in nits) with color and weight multipliers. The color and weight act as multipliers, i.e., the HDR emission in the model color space is defined to have a color given by $\mathtt{emission\_weight} \,\times\, \mathtt{emission\_color} \,\times\,  \mathtt{emission\_luminance}$. The \verb|emission_luminance| parameter thus refers to the luminance the emissive layer would have when the color is white and weight is 1, and in the absence of coat and fuzz. Thus the final resulting luminance may be less than the input parameter, or even zero if the color or weight are zeroed.

Moreover, the overall material luminance may be further reduced in the presence of coat or fuzz, as they can absorb light coming from the emissive layer before it exits the surface. The emission from the top surface should in principle gain a directional dependence due to the combined effects of absorption, total internal reflection (TIR) and multiple bounces in the coat layer, and absorption in the fuzz layer. The combined effect should result mostly in darkening and saturation at grazing angles.

Being an intensity, \verb|emission_luminance| can be any value greater than or equal to zero. For convenience, we make the soft range $[0, 1000]$ nits, corresponding to the typical range of home appliances. (Note that if the renderer does not deal with photometric units internally, a scale factor may need to be applied to bring the emission into a sensible range.)

We found that it is convenient for there to be a simple \verb|emission_weight| control modulating the luminance that can be adjusted via a slider (and also textured with a $[0, 1]$ mask). This is more consistent with the other lobes, which all have an associated weight dialing the strength of the effect. Adding this weight, which defaults to 0, allows \verb|emission_luminance| to default to a reasonable value of 1000 nits.

\begin{figure}[!htb]
  \centering
  \begin{subfigure}{.47\textwidth}
    \centering
    \includegraphics[width=\linewidth]{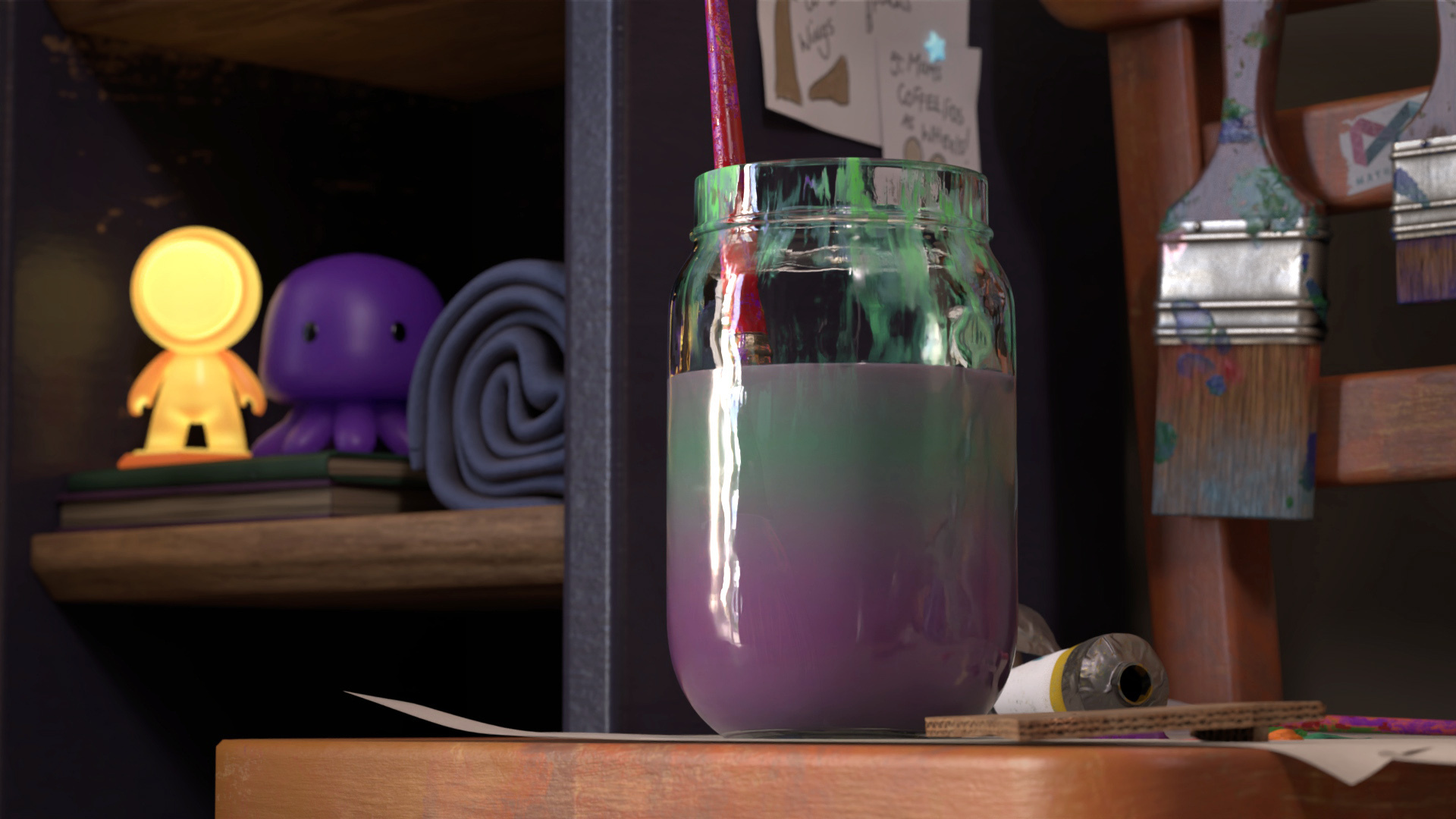}
  \end{subfigure}
  \begin{subfigure}{.47\textwidth}
    \centering
    \includegraphics[width=\linewidth]{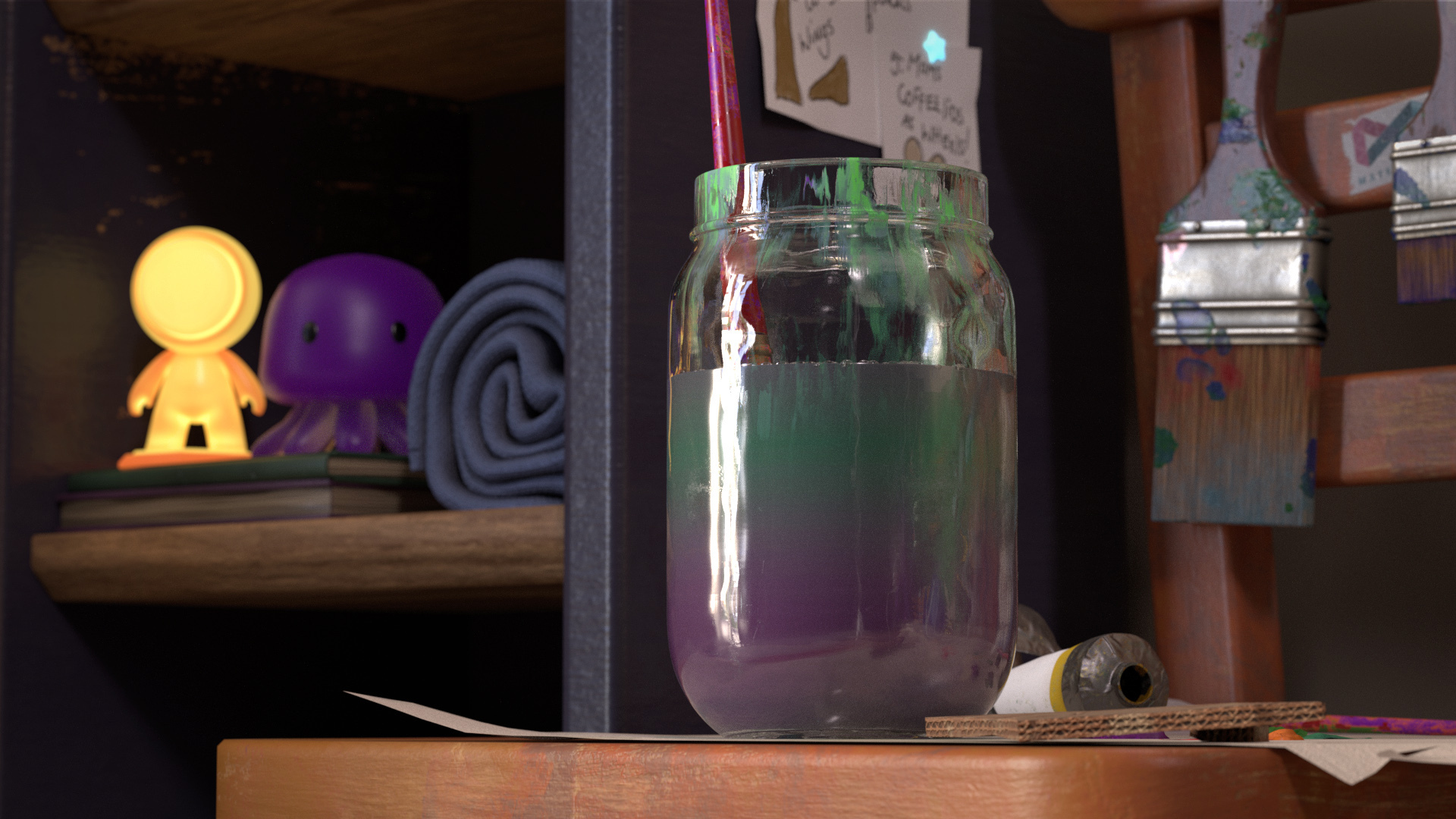}
  \end{subfigure}
  \caption{A scene with an emitting surface (rendered in Arnold on the left, and Adobe's proprietary renderer on the right). \label{fig:emission_example}}
\end{figure}

\clearpage

\enlargethispage{3\baselineskip}

\thispagestyle{empty}  

\subsection{Future work}

\label{sec:base-substrate-future-work}

Here we detail two proposed extensions to the base substrate model: \hyperref[sec:hazy-specular]{hazy-specular} and \hyperref[sec:retro-reflection]{retro-reflection}. These are not yet implemented in OpenPBR (as of version 1.1), but are planned for future releases.

\subsubsection{Hazy specular}

\label{sec:hazy-specular}

A strongly requested feature is a more configurable \hyperref[sec:microfacet]{microfacet} NDF which can support a ``hazy gloss'' look. The \hyperref[sec:coat]{coat} layer does not provide equivalent functionality, since the coat produces a strong Fresnel effect near grazing, roughens the underlying base, and generates darkening due to inter-reflections. A number of models for a generalized microfacet NDF have been proposed. \textcite{Ribardiere2017} introduced a BRDF based on the Student's t-distribution. \textcite{Burley2012} introduced the ``generalized Trowbridge-Reitz'' (GTR) NDF. These add a 1-parameter control over the NDF shape. \textcite{Barla2018} introduced a model with a blend of a primary lobe and secondary lobe differing in roughness, expressed as a sum of a ``specular core'' and a ``surrounding halo'', in such a way that the halo width can be adjusted independently of the core brightness.

An alternative, simpler approach based on \textcite{Barla2018} that we favour is to simply have the BSDF be a blend between microfacet BSDFs with independent roughnesses, the original ``core'' lobe and an added ``haze'' lobe with some mix weight $w_h \in [0,1]$ and effective roughness $r_h$ greater than the core roughness. We noted that this model is actually more expressive and physically plausible than the 1-parameter models, since the mix weight $w_h$ effectively controls the ratio of the energy in the haze relative to the core, while the haze roughness $r_h$ can be varied completely independently of this (supporting weak and strong haze, with independent low or high added roughness).

\begin{figure}[!b]
  \centering
  \hfill
  \begin{subfigure}{.24\textwidth}
    \includegraphics[width=\linewidth]{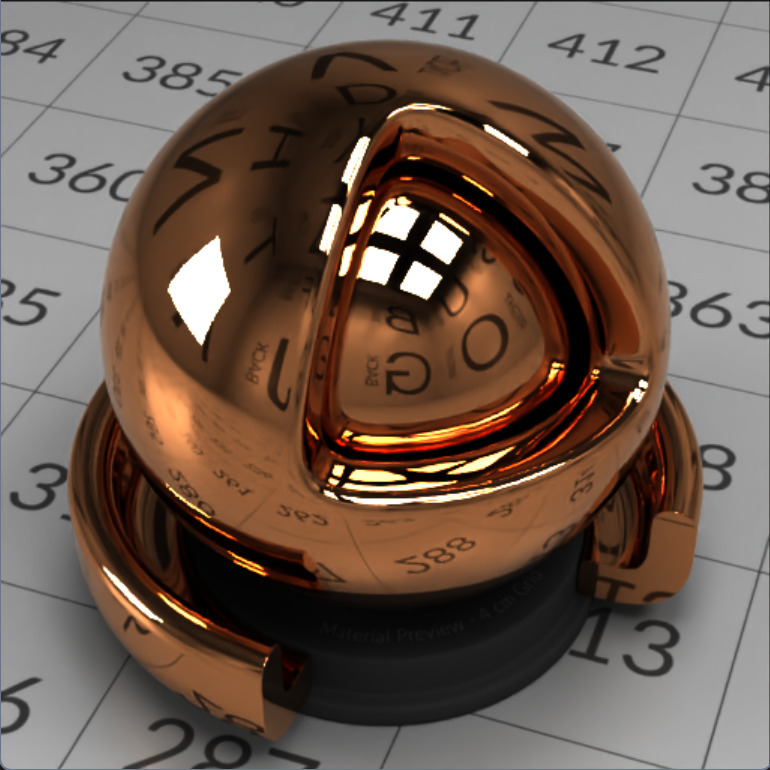}
  \end{subfigure}
  \hfill
  \begin{subfigure}{.24\textwidth}
    \includegraphics[width=\linewidth]{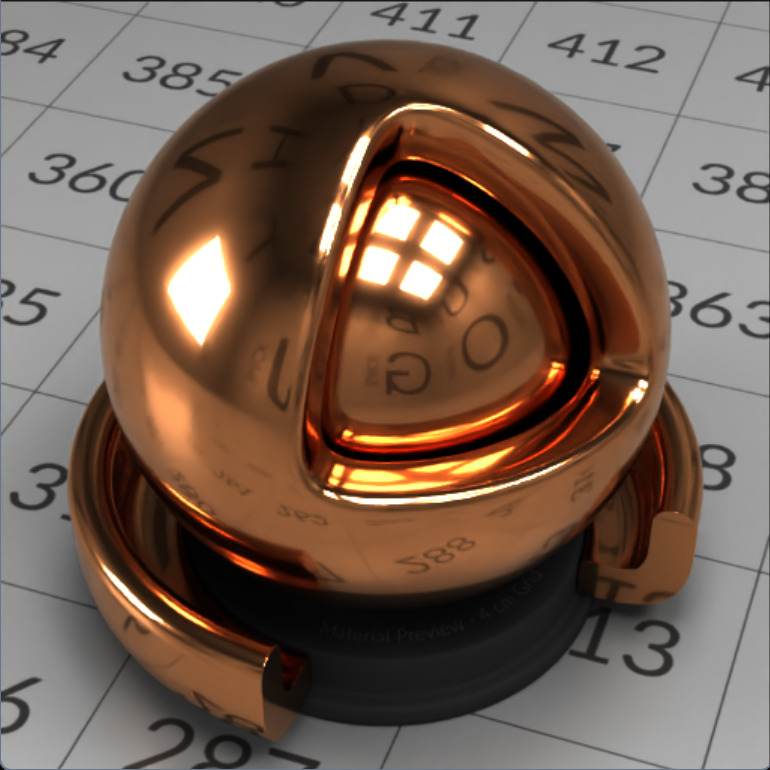}
  \end{subfigure}
  \hfill
  \begin{subfigure}{.24\textwidth}
    \includegraphics[width=\linewidth]{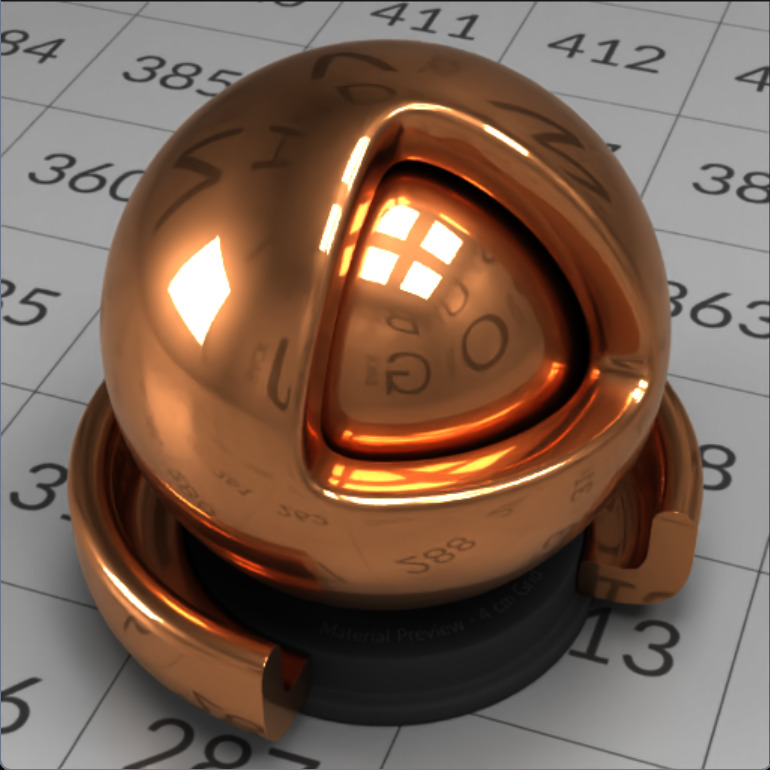}
  \end{subfigure}
  \hfill
    \begin{subfigure}{.24\textwidth}
    \includegraphics[width=\linewidth]{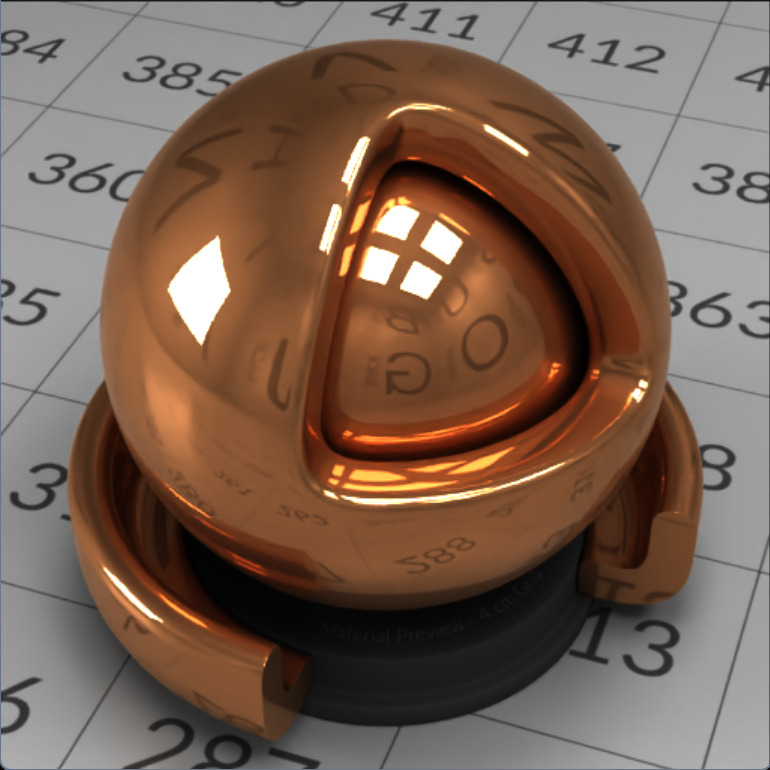}
  \end{subfigure}
  \hfill
  \caption{With \texttt{specular\_haze} $w_h=0.5$, varying the \texttt{specular\_haze\_spread} $\xi_h$ over 0, 0.2, 0.4, 0.6. \label{fig:hazy_specular_example}}
\end{figure}

We plan to provide this via the following parameters:

\begin{itemize}
  \item \verb|specular_haze|, $w_h \in [0,1]$: Provides the blend weight of a secondary specular NDF (defaulting to 0).
  \item \verb|specular_haze_spread|, $\xi_h \in [0,1]$: Specifies the extra roughness of this secondary lobe, in $[0, 1]$. This is mixed with the primary specular lobe roughness (\verb|specular_roughness|, $r$) to produce the roughness value for the secondary lobe $r_h$ (with the sum clamped to the $[0,1]$ range) via $r_h = (1 - \xi_h) r + \xi_h = r_c + \xi_h (1 - r)$. Note that the anisotropy parameters are the same for both primary and secondary specular lobes. Figure~\ref{fig:hazy_specular_example} shows the effect of varying the \verb|specular_haze_spread| parameter, with a fixed \verb|specular_haze| of 0.5.
\end{itemize}

The model then consists simply of two microfacet BSDFS $f_c$ (core) and $f_h$ (haze) with roughnesses $r$ and $r_h$ respectively (and equal anisotropy), blended via $(1 - w_h) f_c + w_h f_h$.

\clearpage

\vspace*{-3.5\baselineskip}

\subsubsection{Retro-reflection}

\label{sec:retro-reflection}

\enlargethispage{1\baselineskip}

Rendering of retro-reflective materials is useful in a variety of contexts (typically for safety applications such as road markings, signs, vehicles and clothing). Such materials are usually designed to be retro-reflective via a substructure of elements that preferentially scatter light backwards. A number of CG models have been proposed (\cite{Belcour2014, Guo2017, Guo2018b}). For practical purposes in visual effects, we are interested in a model which is visually plausible at least, provably energy-conserving, efficient and easy to implement, and ideally a small modification to well-understood microfacet models so it can be implemented as a modification to the specular lobe.

\begin{figure}[!hb]
  \centering
  \begin{subfigure}{0.23\textwidth}
    \includegraphics[width=\linewidth]{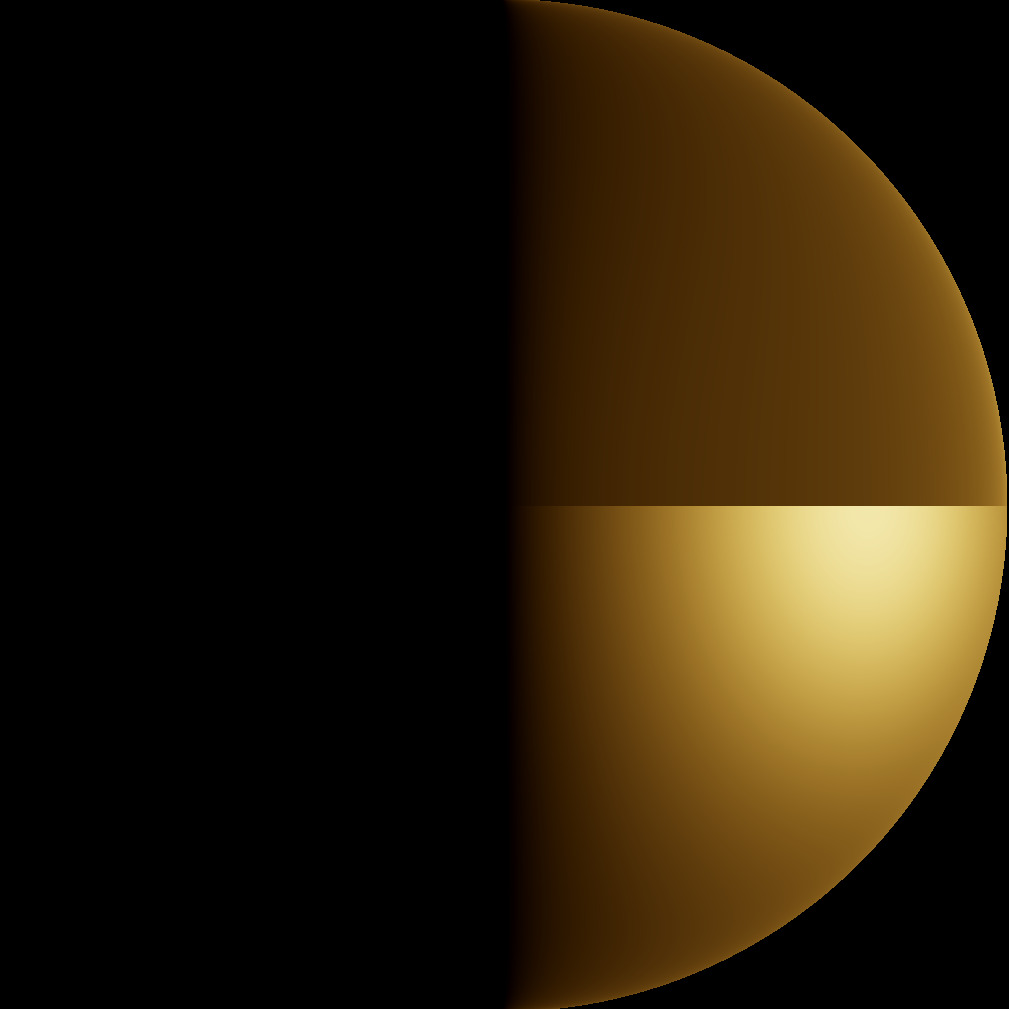}
  \end{subfigure}
  \hspace{0.01\textwidth}
  \begin{subfigure}{0.23\textwidth}
    \includegraphics[width=\linewidth]{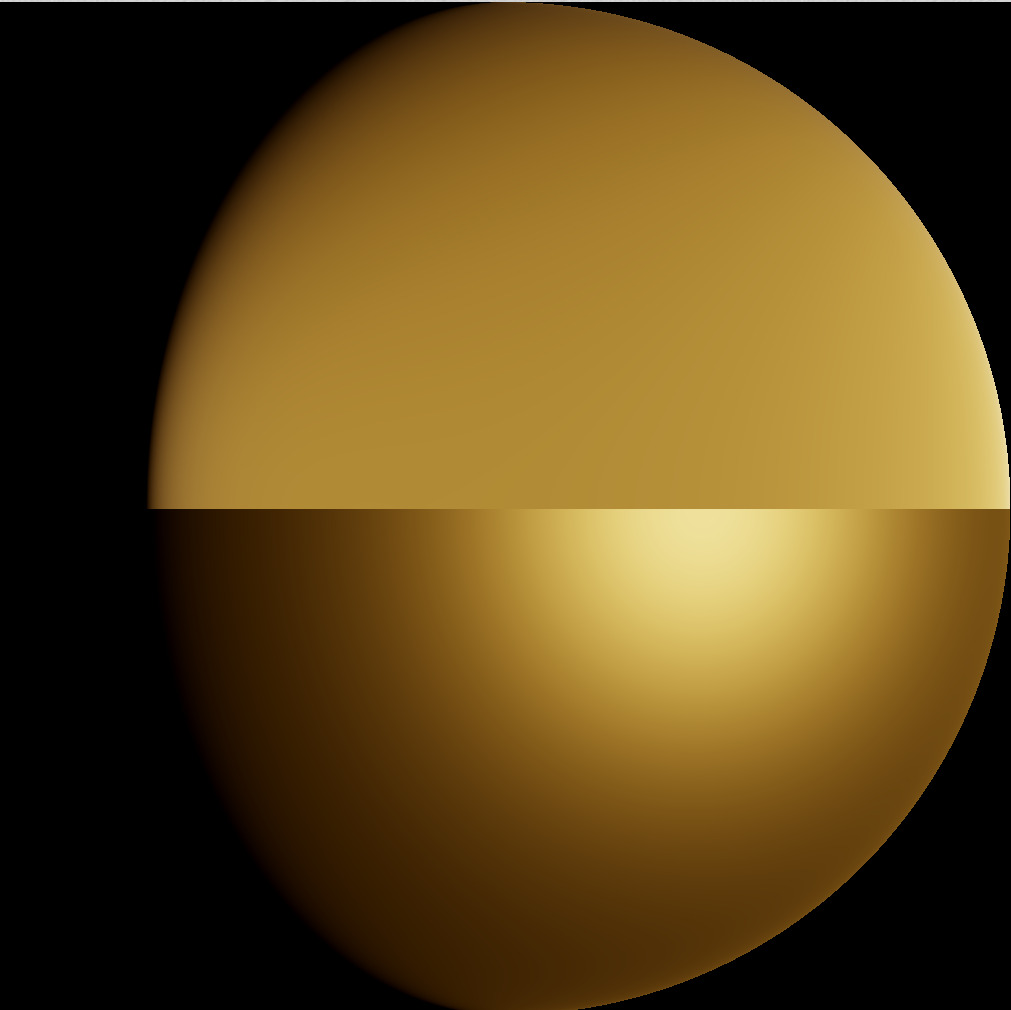}
  \end{subfigure}
  \hspace{0.01\textwidth}
  \begin{subfigure}{0.23\textwidth}
    \includegraphics[width=\linewidth]{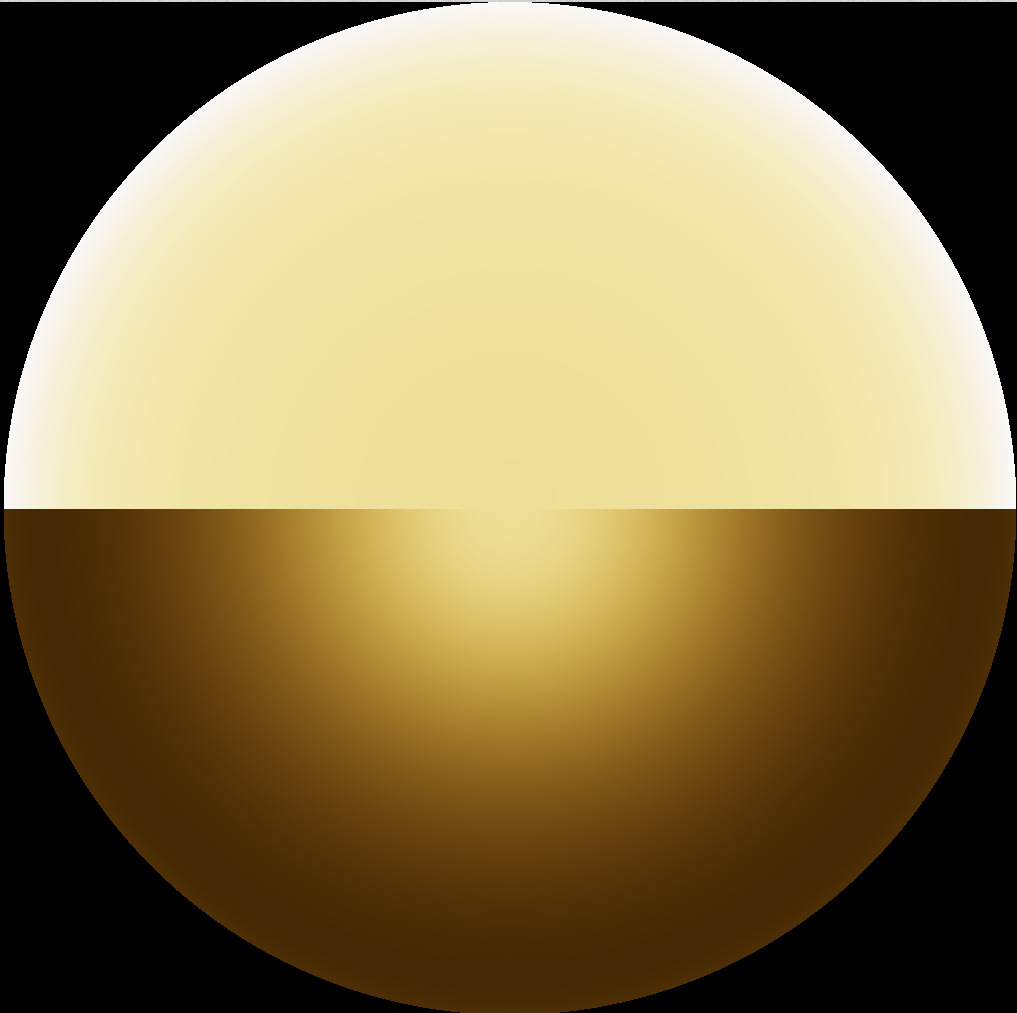}
  \end{subfigure}
  \caption{The top row shows the ``minimal retro-reflection model'' (MRRM) applied to a microfacet conductor, viewed from the side, quarter-on and facing away from the illumination. The bottom row shows the look with the usual non-retro-reflective microfacet conductor model of the same roughness.}
  \label{fig:retroreflection}
\end{figure}

A model we are evaluating is based on the previously published ``back-vector'' formulation~\cite{Belcour2014} which meets these requirements.
This empirically based approach simply replaces the half-vector $\vech$ in the microfacet model with the back-vector $\vecb$ defined as
\begin{equation}
  \vecb(\vecv, \vecl) = \vech(\vecvv, \vecl) = \frac{\vecvv + \vecl}{\lVert \vecvv + \vecl \rVert}, \text{ with } \vecvv = \text{reflect}(\vecv, \vecn) \ .
\end{equation}
\textcite{Belcour2014} demonstrated that this formulation achieves a reasonable match to measured retro-reflective materials.
Given an implementation of a regular microfacet BRDF, extending it to retro-reflection is then extremely straightforward:
\begin{itemize}
\item
  Evaluation merely needs to replace $\vecv$ with $\vecvv$ upfront. Similarly, importance sampling of $\vecl$ given $\vecv$ can be realized by replacing $\vecv$ with $\vecvv$ upfront and then importance sampling the regular microfacet BRDF.
  This may include low variance sampling using the domain of visible microfacets \cite{Heitz2018}.
\item
  As the albedos of the standard BRDF and retro-reflective BRDF are essentially identical, compensating for energy loss in the sense of \textcite{Kelemen2001} can be realized using the same data tables.
\end{itemize}
We tentatively term this model the ``minimal retro-reflective microfacet model'' (MRRM). It can be shown that this leads to a visually plausible, energy-conserving and reciprocal retro-reflective result (Figure~\ref{fig:retroreflection}).
To provide this retro-reflectivity functionality, the BRDF of the base can simply be taken to be a statistical mix of the BRDF with and without the view vector modification, with a mix weight $w_\mathrm{retro}$:
\begin{equation}
f_\mathrm{conductor} \rightarrow (1 - w_\mathrm{retro}) f_\mathrm{conductor} + w_\mathrm{retro}f_\mathrm{retro}
\end{equation}

An open question is whether the back-vector modification should be applied to both the conducting and dielectric microfacet cases, as the behavior of the latter is less plausible as a typical retro-reflective material.

\clearpage

\section{Thin-film iridescence}

\label{sec:thin-film}

\emph{Iridescence} is the occurrence of rainbow-like color fringes in the reflection when a thin dielectric film with thickness on the order of the wavelength of light is placed on top of a material, due to interference between light reflected from the film's top and bottom surfaces, including internal reflections. To model this, we assume such a thin film sits atop the base substrate (whether metal or dielectric), parameterized by

\begin{itemize}
  \item \verb|thin_film_weight|: the coverage (presence) weight of the film.
  \item \verb|thin_film_thickness|: the thickness of the film in micrometers (\SI{}{\micro\meter}). The $[0,1]$ (soft) range of thicknesses, in micrometer units, then corresponds nicely to the typical thickness range needed to observe interference effects.
  \item \verb|thin_film_ior|: the index of refraction (IOR) of the film.
\end{itemize}

\begin{figure}[!b]
  \centering
  \hfill
  \begin{subfigure}{.19\textwidth}
    \includegraphics[width=\linewidth]{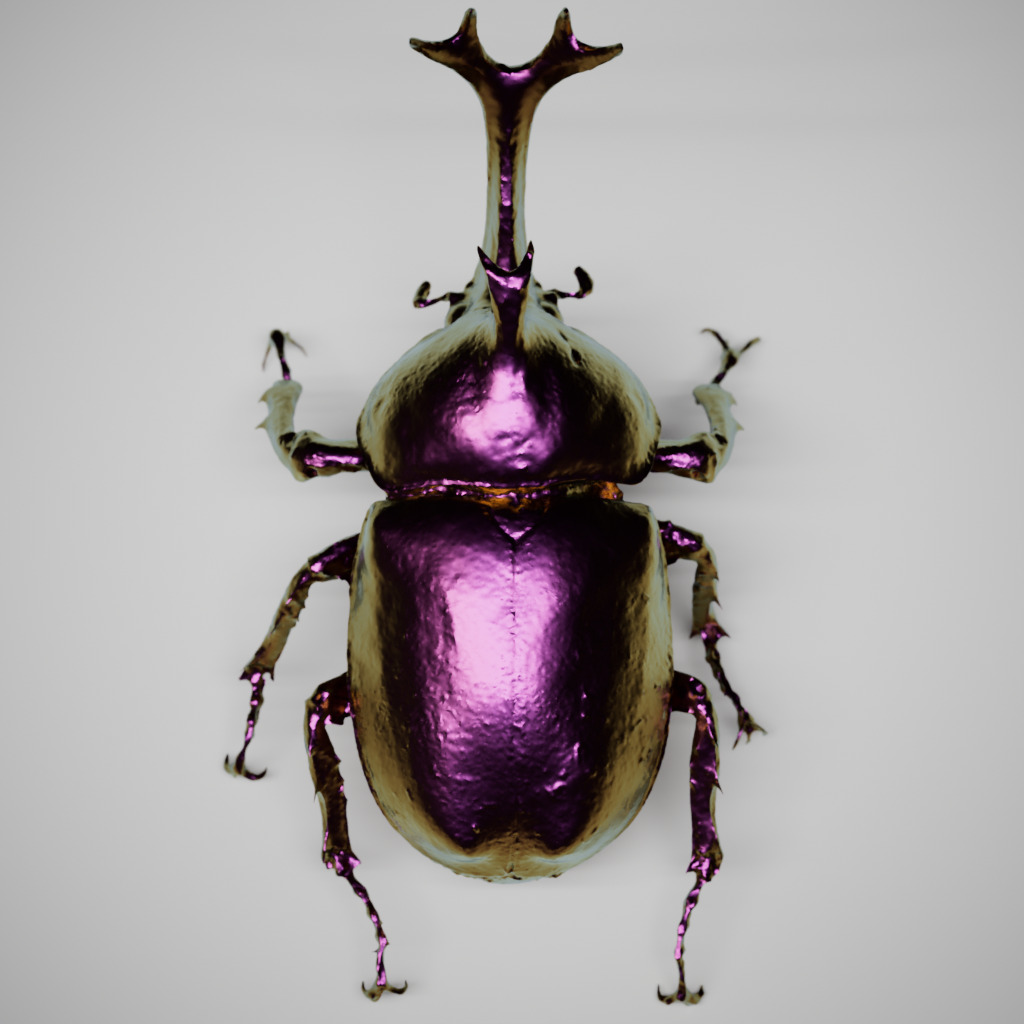}
  \end{subfigure}
  \hfill
  \begin{subfigure}{.19\textwidth}
    \includegraphics[width=\linewidth]{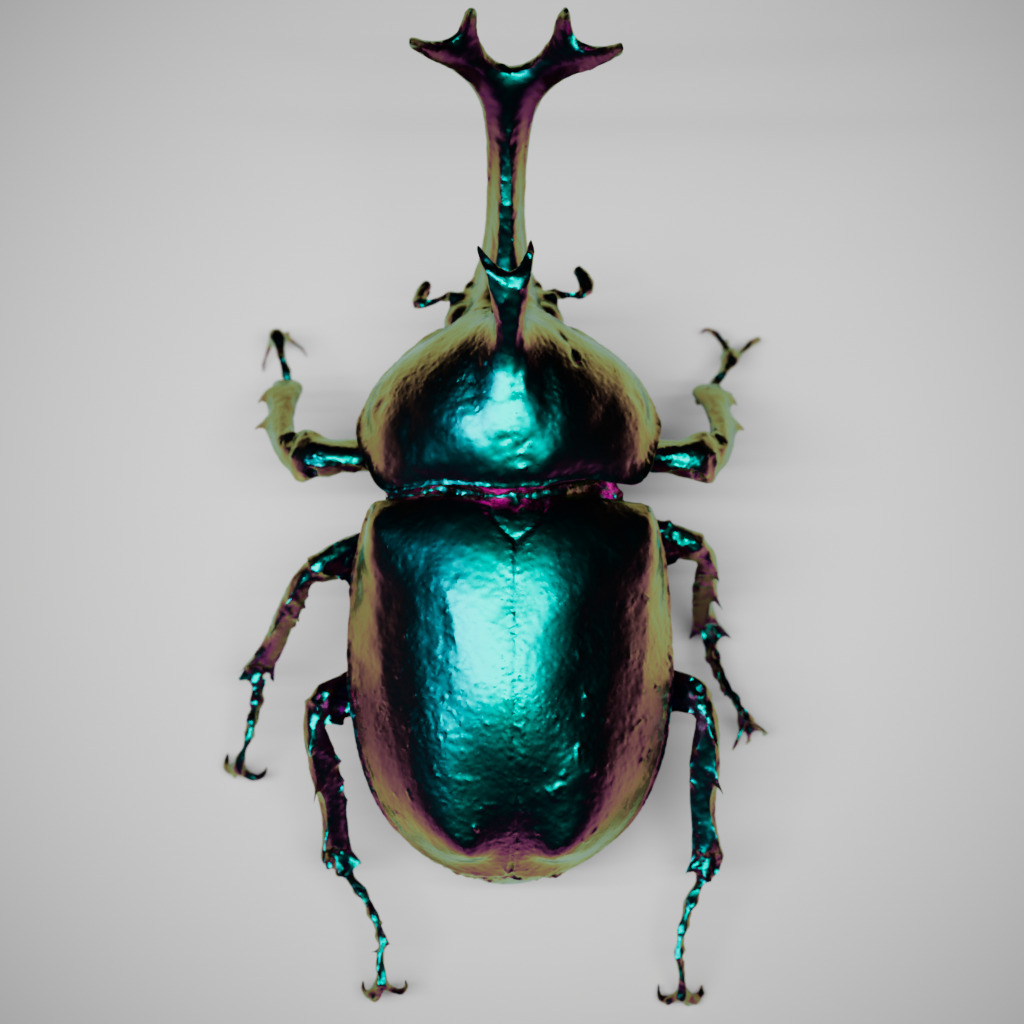}
  \end{subfigure}
  \hfill
  \begin{subfigure}{.19\textwidth}
    \includegraphics[width=\linewidth]{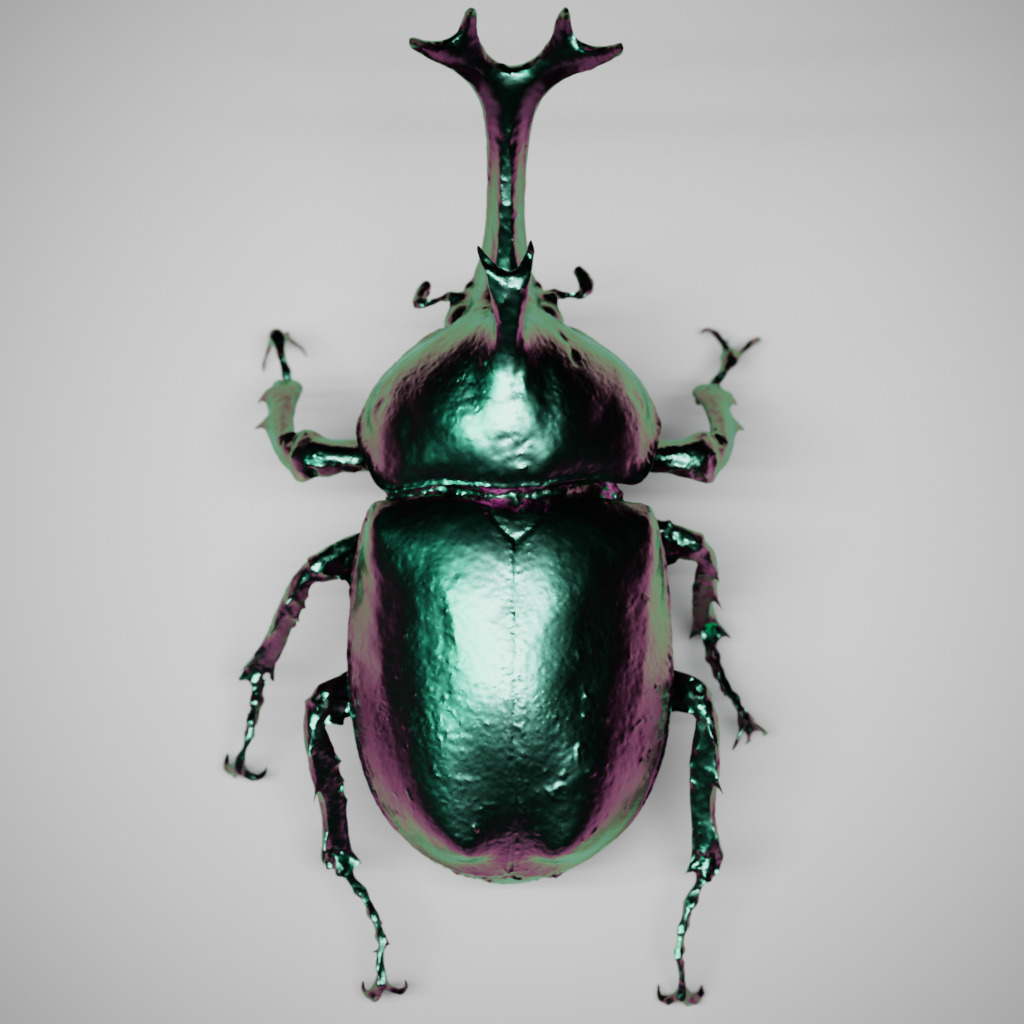}
  \end{subfigure}
  \hfill
  \begin{subfigure}{.19\textwidth}
    \includegraphics[width=\linewidth]{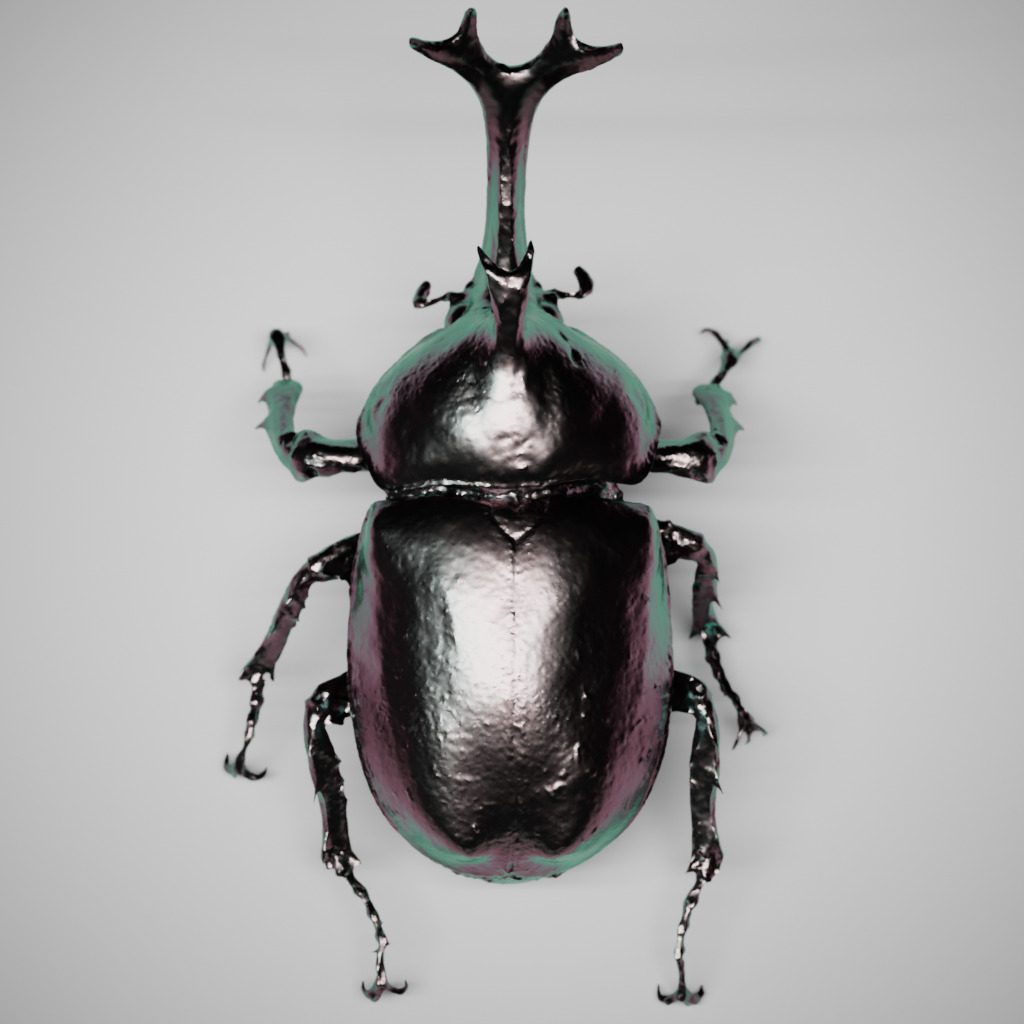}
  \end{subfigure}
  \hfill
 \begin{subfigure}{.19\textwidth}
    \includegraphics[width=\linewidth]{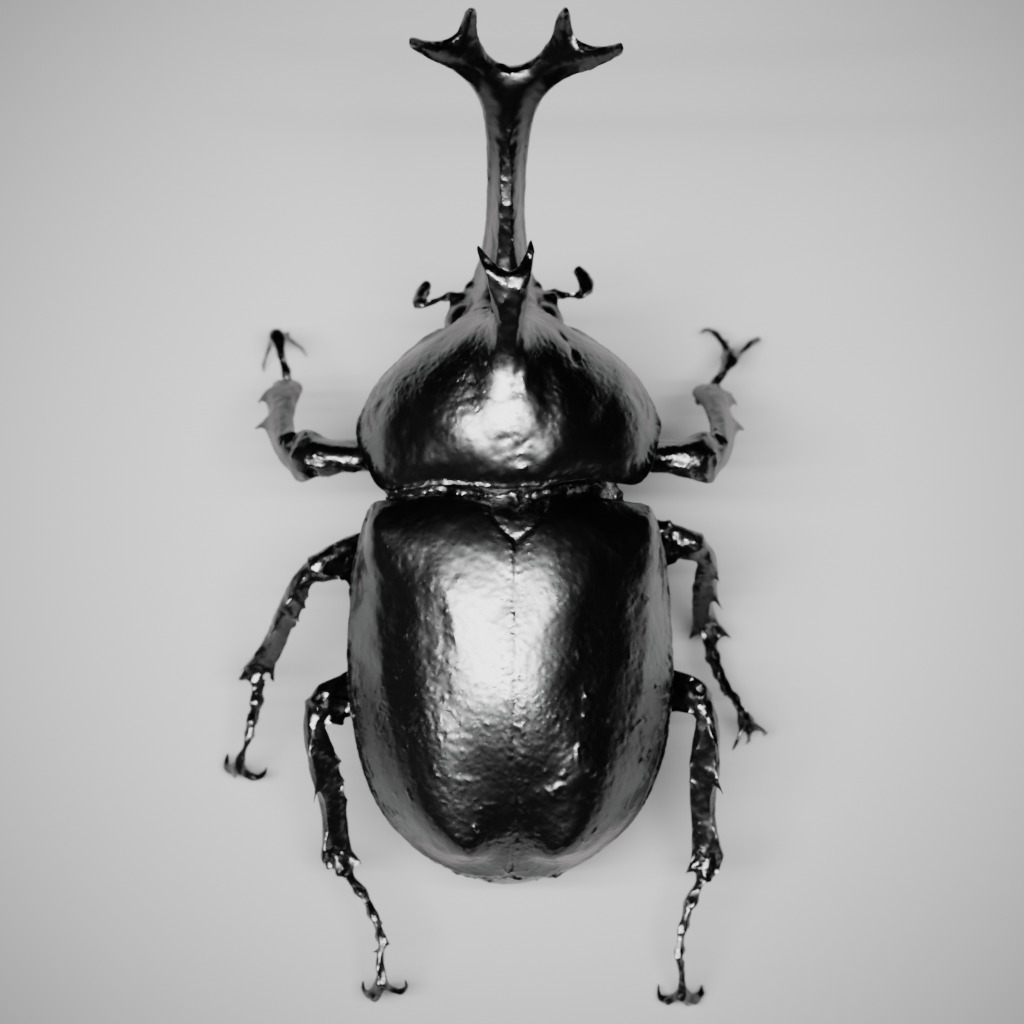}
  \end{subfigure}
  \hfill
  \caption{Varying the thickness of the thin film (from left to right: $0.25, 0.5, 0.75, 1, 2$ $\mu\mathrm{m}$).\label{fig:thin-film-thickness}}
\end{figure}

The thickness and IOR together affect the intensity, spacing, and hue of the color fringes. The coverage weight acts as a blend between the BSDF with and without the presence of the film, allowing the overall strength of the effect to be adjusted without altering its structure or color.
Figure~\ref{fig:thin-film-thickness} shows the effect of varying the thickness of the thin film. As thickness increases, one sees higher-frequency color banding, and the effect eventually converges toward a ``thick film'' look where the colors fade to gray as the interference effects wash away.

The currently recommended thin-film model is that of \textcite{Belcour2017}, which pre-integrates interference effects using Fourier-domain convolutions and Gaussian filtering. This method efficiently produces high-quality fringe patterns in an RGB rendering context, but it can be challenging to implement and may introduce inaccuracies in some cases as it assumes that Fresnel amplitude and phase coefficients remain constant across each spectral band, which limits the model's ability to capture wavelength-dependent dispersion effects.

A more direct alternative that we highlight here (and in more detail in Appendix~\ref{sec:iridescence-from-first-principles}) is a ``locally spectral'' approach that computes reflectance per light path by evaluating the full Fresnel and Airy interference stack -- including complex amplitudes, polarizations, and phase shifts -- at specific wavelengths sampled per path. This can begin with fixed red, green, and blue wavelengths, but better results are achieved by stochastically sampling wavelengths from approximate camera sensitivity curves (as described in Section~\ref{sec:dispersion}). This enables convergence to neutral gray for very thick films and avoids the high-frequency color banding that fixed RGB wavelengths can produce.

The same wavelengths can also be reused to model \hyperref[sec:dispersion]{dispersion}, while all other BSDF components are free to ignore them and operate in RGB as usual. This approach uses only the Airy summation from Belcour and Barla (Equation 3 from \textcite{Belcour2017}) -- a method whose origins trace back to 19th-century studies of thin-plate interference by G.B.~Airy and others -- but requires additional per-wavelength computations and assembling the necessary formulas from multiple sources rather than a single reference. We attempt to collect the necessary equations and implementation instructions in one place in Appendix~\ref{sec:iridescence-from-first-principles}.

\begin{figure}[!hb]
  \centering
\includegraphics[width=0.95\linewidth]{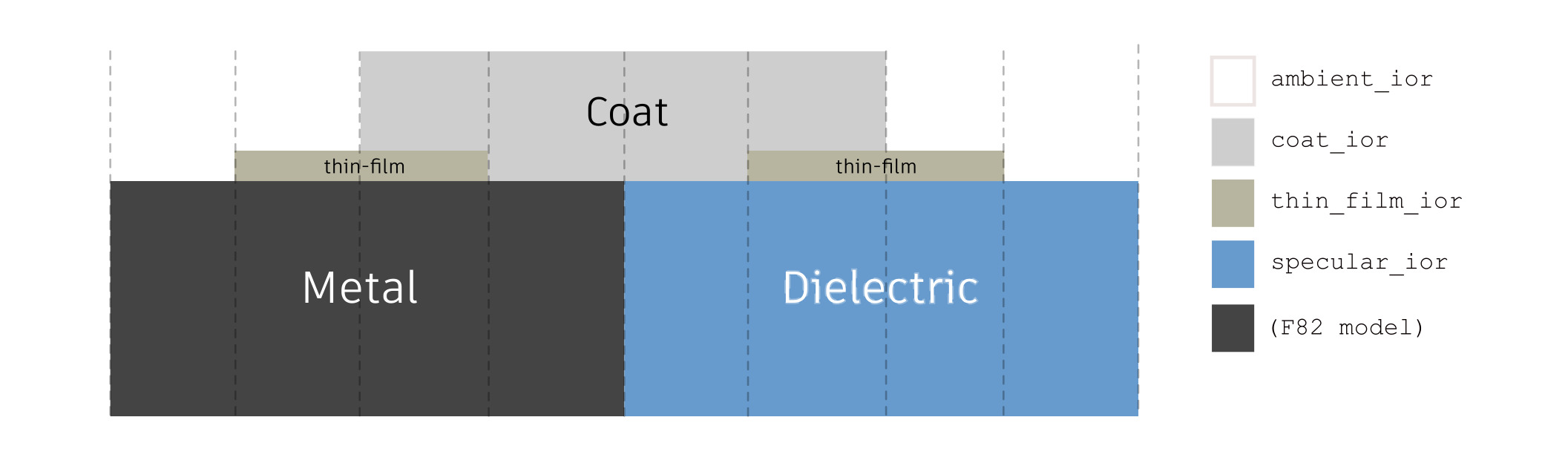}
\caption{Schematic of all 8 possible IOR configurations, including those involving the thin film. \label{fig:ior_configs}}
\end{figure}

Regardless of which approach is chosen, several considerations apply to both:

\begin{itemize}

  \item The shape and color of the fringe patterns in the reflection from the film will be affected by the complex IOR of the adjacent media above and below the film, which in general are a statistical mix of metal and dielectric below, and of coat and ambient medium above (to which the fuzz is index-matched).

  Figure~\ref{fig:ior_configs} illustrates the eight possible different structures depending on the presence of both the film and coat, each of which leads to different Fresnel effects due to the differing IORs at the interfaces. In principle, the implementation should account for all of these configurations accurately, though the precise modeling of these effects is left to the implementation.

  \item In practice, this wave-optics effect is most easily incorporated directly into the Fresnel factor of the microfacet BSDFs of both the \hyperref[sec:metallic-base]{metal} and \hyperref[sec:dielectric-base]{dielectric-base} layers. (For this reason, this effect is not represented by incorporating an explicit thin-film Slab into the model.)

  \item In the case of the dielectric base, the thin film should also generate color fringes in the transmission lobe. This is important for example when rendering soap bubbles (see \textcite{Belcour2017}).

  \item In the case of the metallic base, the physics is somewhat ambiguous since the Fresnel factor for metal is defined according to the Schlick-based ``F82-tint'' parameterization (as described in Section~\ref{sec:metallic-base}), which does not specify the underlying physical complex IOR. We suggest here that some reasonable approximation is employed to map the Fresnel factor to the best matching effective complex IOR, for example that described by \textcite{Gulbrandsen2014}.

  \item Because the thin film is non-absorbing and interference based, it only redistributes the probabilities of reflection and transmission; therefore, it should not violate energy conservation. In the limit, where the base metal or dielectric has a Fresnel factor of 1, the thin film has no effect and passes a white furnace test.

\end{itemize}

\clearpage

\section{Coat}

\label{sec:coat}

The coat slab is a layer of dielectric that transmits light without scattering, but with possible absorption. This is intended to support the appearance of objects with a coat of colored varnish or lacquer:
\begin{equation}
S_\mathrm{coat} = \mathrm{Slab}(f_\mathrm{coat}, V_\mathrm{coat}) \ .
\end{equation}
The BSDF of the interface $f_\mathrm{coat}$ is that of a GGX \hyperref[sec:microfacet]{microfacet} dielectric parameterized by \verb|coat_roughness| and \verb|coat_roughness_anisotropy|. The IOR \verb|coat_ior| of this dielectric layer is distinct from that of the base dielectric, as described below. There is also assumed to be an embedded \emph{purely absorbing} medium $V_\mathrm{coat}$.
\begin{figure}[!b]
  \centering
  \begin{subfigure}{.3\textwidth}
    \includegraphics[width=\linewidth]{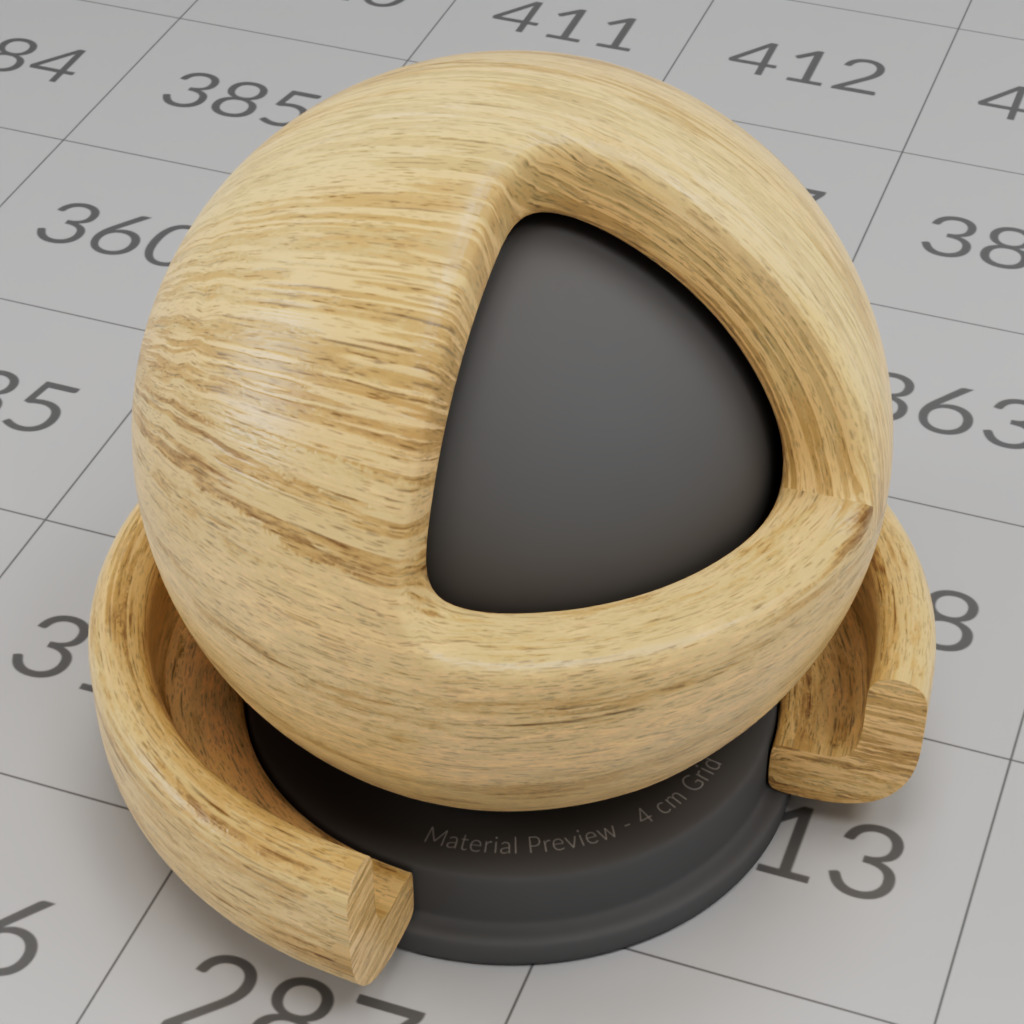}
  \end{subfigure}
  \hspace{0.02\textwidth}
  \begin{subfigure}{.3\textwidth}
    \includegraphics[width=\linewidth]{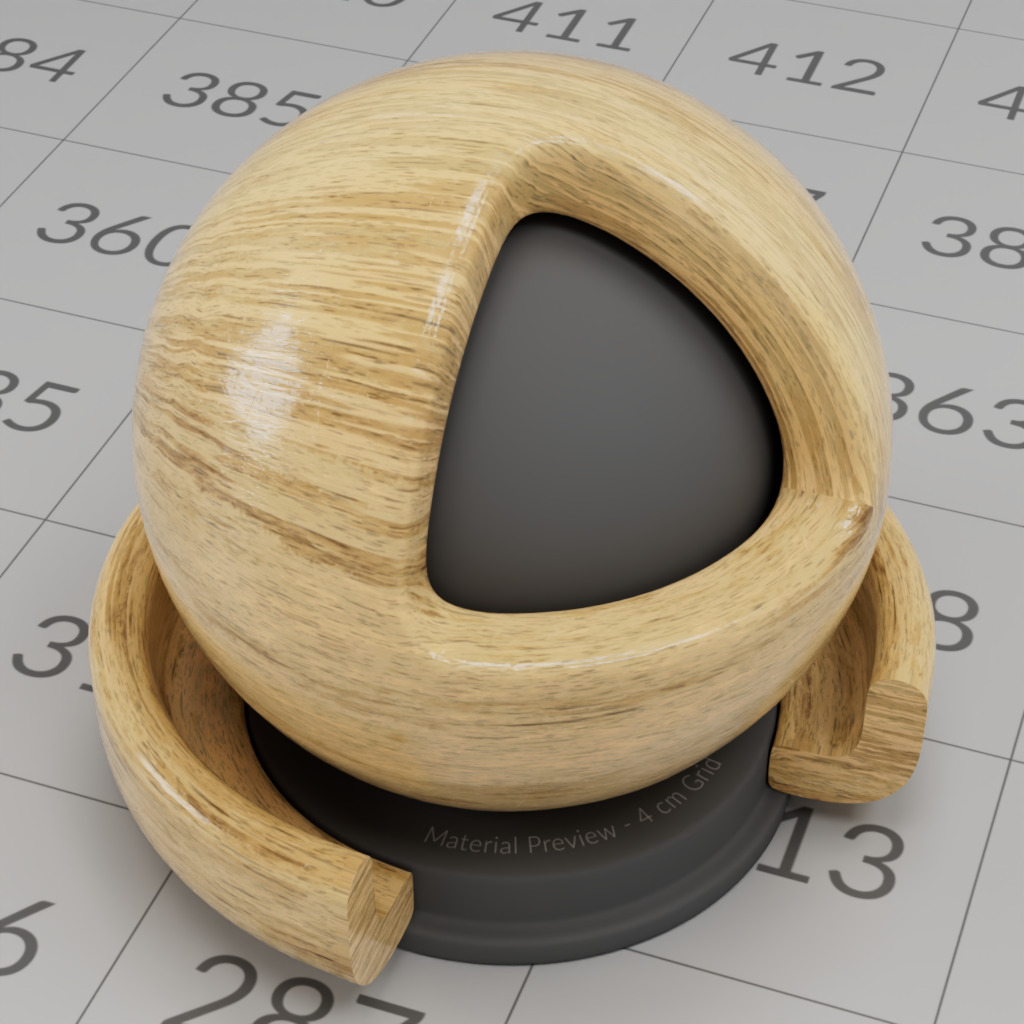}
  \end{subfigure}
  \hspace{0.02\textwidth}
  \begin{subfigure}{.3\textwidth}
    \includegraphics[width=\linewidth]{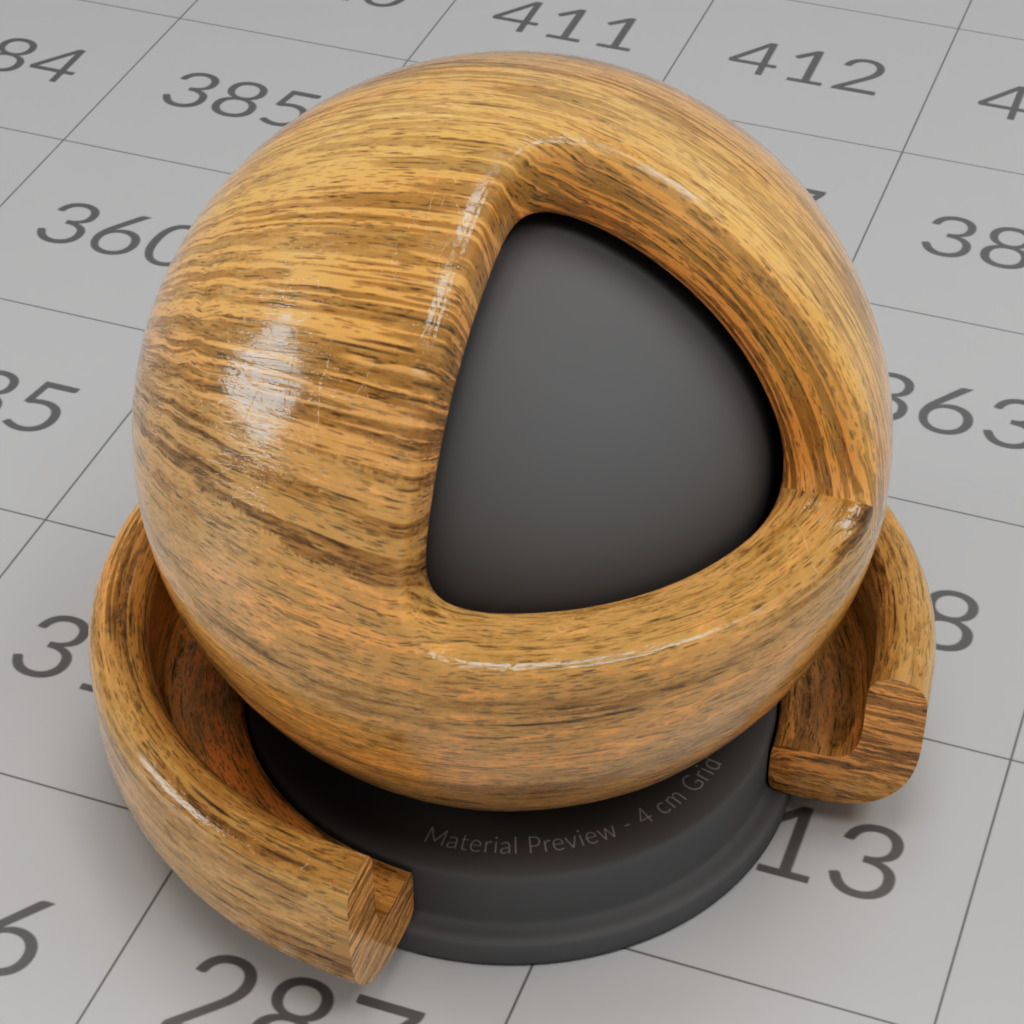}
  \end{subfigure}
  \caption{A wood-textured base with no coat (left), and with a clear-coat (IOR 1.5) with \texttt{coat\_darkening} $\delta=0$ (middle) and $\delta=1$ (right).}
  \label{fig:coat_darkening_example}
\end{figure}
The coat is applied on top of the base substrate, with a coverage weight $\mathtt{C} = \mathtt{coat\_weight}$ as follows:
\begin{equation}
M_\textrm{coated-base} = \mathrm{\mathbf{layer}}(M_\textrm{base-substrate}, S_\textrm{coat}, \mathtt{C}) \ .
\end{equation}

The IOR $n_c = \mathtt{coat\_ior}$ of the coat medium $V_\mathrm{coat}$ will alter the Fresnel factor of both the coat top interface and the underlying metal or dielectric. If there is a fractional $\mathtt{coat\_weight}$ $\mathtt{C}$, then the surrounding IOR of the base dielectric or metal varies statistically across the surface depending on whether the coat is locally present (and the fuzz layer can be assumed to have the ambient IOR $n_a$). The ratio between the specular IOR $n_b = \mathtt{specular\_ior}$ and the surrounding medium can thus be reasonably approximated as
\begin{equation} \label{specular_ior_ratio}
\eta_s = \mathrm{lerp}(n_b/n_a, n_b/n_c, \mathtt{C}) \ .
\end{equation}
This ratio then determines the specular Fresnel factor, as in Equation~\ref{modulated_ior}. As discussed \hyperref[sec:coat_TIR]{below}, evaluation of the specular Fresnel factor may need to be further modified to model the refraction of the ray inside the coat.

The absorption of the medium $V_\mathrm{coat}$ is parameterized by \verb|coat_color|, which is assumed to specify the \emph{square} of the transmittance of the coat at normal incidence (i.e., $\mathbf{T}^2_\mathrm{coat} = \mathtt{coat\_color}$ in the notation of Equation~\ref{non-reciprocal-albedo-scaling-with-T}). Thus the observed tint color, at normal incidence, of the underlying base due to absorption in the coat is approximately given by \verb|coat_color| due to the absorption along the incident and outgoing rays. (Note that the specular reflection from the coat itself is \emph{not} tinted.)

In the full light transport within the coat, various physical effects occur which we assume are accounted for in the ground truth appearance:

\begin{itemize}

  \item \hyperref[sec:coat_darkening]{\textbf{Darkening}}: The observed color of the coated base is darkened and saturated due to multiple internal reflections from the inside of the coat, which causes light to strike the underlying material multiple times and undergo more absorption. This effect is controlled via the \verb|coat_darkening| parameter. Figure~\ref{fig:coat_darkening_example} shows an example of this effect, where the middle image shows the effect of removing the darkening due to internal reflections in the coat via \texttt{coat\_darkening} $\delta=0$, while the right image shows the physically correct darkening effect.

  \item \hyperref[sec:coat_absorption]{\textbf{View-dependent absorption}}: The observed \verb|coat_color| tint also darkens as the incidence angle changes due to the change in path length in the medium.

  \item \hyperref[sec:coat_roughening]{\textbf{Base roughening}}: The presence of a rough coat will increase the apparent roughness of the BSDF lobes of the underlying base.

  \item \hyperref[sec:coat_TIR]{\textbf{TIR}}: Care needs to be taken in the implementation to account for the refraction of the ray direction inside the coat.

\end{itemize}

In reality, coats can also darken the underlying surface due to a different mechanism where the coat modifies the Fresnel factor of the base due to the coat material filling in air gaps between granules or threads of a porous base material, which reduces the relative IORs at the internal interfaces. This occurs, for instance, when adding water to sand or fabric, or adding a penetrating wood finish. We assume here that this effect explicitly does \emph{not} occur, at present, since we do not have enough knowledge about the properties of the underlying substance to model it. We can only safely assume that the first mechanism of darkening (i.e., internal reflections) occurs.

\subsection{Darkening}

\label{sec:coat_darkening}
\enlargethispage{1\baselineskip}

In Appendix~\ref{sec:coat_physics}, we provide the formulas for the physical darkening effect in the case of a smooth coat.

However, this darkening may not always be desirable artistically, as in some applications it is beneficial for the observed color of the coated base to ``match'' the input base color (in a sense defined more precisely below in Equation~\ref{undarkened_coat_albedo}). We allow for this by introducing a \verb|coat_darkening| parameter, $\delta$. By default $\delta = 1$, in which case the physically correct darkening effect due to internal reflections occurs as normal. In the case $\delta = 0$, however, the base albedo is modified to counteract the darkening effect (see Figure~\ref{fig:coat_darkening_example}).

To define what we mean by ``counteract the darkening'', we write the (physically darkened) coat albedo at normal incidence in the general form, similar to Equation~\ref{glossy_diffuse_albedo} and Equation~\ref{subsurface_albedo}:
\begin{equation} \label{general_coat_albedo}
\mathbf{E}_c = \mathbf{E}_\mathrm{spec} + (1 - \mathbf{E}_\mathrm{spec}) \mathbf{E}^\prime_c
\end{equation}
where $\mathbf{E}_\mathrm{spec}$ is the normal-direction reflectance of all energy reflected from the coat's dielectric interface \emph{without} macroscopic transmission, and $\mathbf{E}^\prime_c$ represents the albedo due to transmission into the coat medium (and scattering off the base substrate, potentially multiple internal reflections off the coat interface, and re-transmission back out).
We then require that the \emph{effect} of $\delta = \mathtt{coat\_darkening}$ is to multiply $\mathbf{E}^\prime_c$ by an RGB boost factor
\begin{equation} \label{undarkened_coat_albedo_scaling}
\mathbf{B}(\delta) = \mathrm{lerp}\biggl(\frac{\mathbf{T}_\mathrm{coat}^2 \mathbf{E}_b}{\mathbf{E}^\prime_c}, 1, \delta \biggr)
\end{equation}
where $\mathbf{T}_\mathrm{coat}$ is the coat absorption transmittance, and $\mathbf{E}_b$ represents the normal-incidence albedo of the entire base beneath the coat (which can be approximated as the normal-incidence albedos of the individual slabs of the base, blended according to their mix weights).

This is a straightforward implementation of the requirement for the albedo of the ``un-darkened'' coat to be equal to the usual albedo scaling formula of Equation~\ref{non-reciprocal-albedo-scaling-with-T} (which involves no color shift other than that due to the absorption, and the combination with the Fresnel factor), i.e.
\begin{equation} \label{undarkened_coat_albedo}
\mathbf{E}_c = \mathbf{E}_\mathrm{spec} + (1 - \mathbf{E}_\mathrm{spec}) \mathbf{E}^\prime_c \mathbf{B}(\delta) \; \rightarrow \; \mathbf{E}_\mathrm{spec} + \mathbf{T}^2_\mathrm{coat} \, \mathbf{E}_b (1 - \mathbf{E}_\mathrm{spec})
\end{equation}
as $\delta \rightarrow 0$.

Note that we have not specified the detailed physical mechanism by which the boosting of $\mathbf{E}^\prime_c$ occurs. Instead, we have only defined what the resultant effective albedo modification (at normal incidence) has to be as \verb|coat_darkening| is varied. While this defines the behavior at the level of the required albedos, in practice implementations will need to develop some specific approximation for the coat darkening effect consistent with their physical approximations. Due to the physical effect of the darkening due to the internal reflections, the base BSDF would generally need to be altered (most simply by multiplying by a boost factor) in order for the coated base to achieve the un-darkened albedo. While this can be implementation dependent, we suggest here a reasonably simple, efficient scheme that captures the essential behavior.

According to Appendix~\ref{sec:coat_physics}, the physical effect of the inter-reflections in the coat can be modeled by multiplying the naive albedo-scaling formula for the coat layering by a darkening factor (Equation~\ref{coat_darkening_factor_rough_case}):
\begin{equation}  \label{albedo_scaling_darkening2}
  \Delta(\mathbf{E}_b, \eta_c) = \frac{1 - K_0} {1 - \mathbf{E}_b K_0 \mathbf{T}^2_\mathrm{coat}} \ .
\end{equation}

\begin{figure}[!tb]
\begin{inputcode}
\begin{lstlisting}[label=listing:coat_darkening, frame=trBL, caption={Example implementation of coat darkening.}]

  void AddCoat(const ShaderGlobals& sg, const Params& P, BSDFs& substrate_bsdfs)
  {
      float C         = P.coat_weight; if (C <= 0.f) return;
      float delta     = P.coat_darkening;
      vec3 CC         = P.coat_color;
      float n_coat    = P.coat_ior;
      vec3 base_color = P.base_weight * P.base_color;
      float M         = P.base_metalness;
      float xi        = P.specular_weight;
      float n_spec    = P.specular_ior;
      float r_spec    = P.specular_roughness;
      float T         = P.transmission_weight;
      float S         = P.subsurface_weight;
      vec3 Sc         = P.subsurface_color;

      // Compute the base_darkening factor based on the coat properties:
      vec3 base_darkening = WHITE;
      if (delta > 0.f)
      {
          float EF = DielectricFresnelAvg(n_coat); // Hemispherical Fresnel albedo, Eqn./*!\ref{hemispherical_fresnel_albedo}!*/)
          float Kr = 1.f - (1.f - EF)/sqr(n_coat); // (rough case, Eqn./*!\ref{internal_diffuse_reflection_coefficient_for_rough_base}!*/)
          float Ks = DielectricFresnel(abs(dot(sg.wi, sg.N)), n_coat); // (smooth case, Eqn./*!\ref{internal_diffuse_reflection_coefficient_for_smooth_base2}!*/)
          float eta_s = mix(n_spec, n_spec/n_coat, C); // specular/coat IOR ratio
          float F0 = DielectricFresnel(1.f, eta_s); // coat Fresnel at normal incidence
          float Fs = clamp(xi * F0, 0.f, 1.f); // modulated dielectric Fresnel
          float rd = mix(1.f, r_spec, Fs); // estimate roughness of dielectric base
          float rb = mix(rd, r_spec, M); // roughness of entire base, Eqn./*!\ref{base_roughness_estimate}!*/
          float K = mix(Ks, Kr, rb); // internal diffuse reflection coefficient K, Eqn./*!\ref{internal_diffuse_reflection_coefficient_for_general_base}!*/
          vec3 E_dielec = mix(mix(base_color, Sc, S), vec3(1.f-F0), T); // dielectric base albedo
          vec3 E_metal = clamp(base_color * xi, 0.f, 1.f); // metallic base albedo
          vec3 E_base = mix(E_dielec, E_metal, M); // entire base albedo
          vec3 Delta = max(1.f - K, 0.f) / (WHITE - E_base*K*CC); // darkening factor, Eqn./*!\ref{albedo_scaling_darkening2}!*/
          base_darkening *= mix(WHITE, Delta, delta); // coat_darkening modulation, Eqn./*!\ref{modulated_darkening_factor}!*/
      }

      // Apply albedo scaling to model the layering of the coat on the base.
      // The coat BSDF has the coat_weight presence weight functioning as a multiplier:
      BSDF* coat_bsdf = C * DielectricBSDF(sg, n_coat, P.coat_roughness);

      // Account for absorption along the refracted entry/exit paths in the coat medium:
      float mu_o = dot(sg.wo, P.coat_normal); mu_o = sqrt(1 - (1 - sqr(mu_o))/sqr(n_coat));
      float mu_i = dot(sg.wi, P.coat_normal); mu_i = sqrt(1 - (1 - sqr(mu_i))/sqr(n_coat));
      vec3 absorb_tint = pow(CC, 0.5f*(1.f/mu_i + 1.f/mu_o));

      // The base BSDFs are then multiplied by the coat_throughput factor,
      // which includes the base_darkening effect and the absorption tinting:
      vec3 coat_albedo = Albedo(coat_bsdf);
      vec3 coat_throughput = mix(WHITE, base_darkening * absorb_tint * (WHITE - coat_albedo), C);
      substrate_bsdfs *= coat_throughput;
      substrate_bsdfs.add(coat_bsdf);
  }

\end{lstlisting}
\end{inputcode}
\end{figure}

Here, $K_0\in [0,1]$ is the \emph{internal diffuse reflection coefficient}, which corresponds to the fraction of the energy leaving the base that returns, in the case of a clear coat. For a Lambertian base (which should be a reasonable approximation to the rough metal, dielectric, or diffuse cases), $K_0 = K_r$, where
\begin{equation} \label{internal_diffuse_reflection_coefficient_for_rough_base}
K_r = 1 - \Bigl(1 - E_F(\eta_c)\Bigr)/\eta_c^2
\end{equation}
with relative coat IOR $\eta_c$. As discussed at the end of Appendix~\ref{sec:coat_physics}, if the base has specular reflection (due to a smooth metallic or dielectric interface), the appropriate value of the internal diffuse reflection coefficient $K_0$ should be closer to
\begin{equation} \label{internal_diffuse_reflection_coefficient_for_smooth_base2}
K_s = F(\omega_o, \eta_c) \ .
\end{equation}
We thus recommend taking
\begin{equation} \label{internal_diffuse_reflection_coefficient_for_general_base}
K_0 = \mathrm{lerp}(K_s, K_r, r_b) \ ,
\end{equation}
where $r_b$ is an estimate of the effective base roughness, a blend between dielectric $r_d$ and metal $r_m$ roughness estimates according to the \verb|base_metalness| $\mathtt{M}$:
\begin{equation} \label{base_roughness_estimate}
r_b = \mathrm{lerp}(r_d, r_m, \mathtt{M}) \ .
\end{equation}
The base dielectric roughness $r_d$ can be reasonably estimated as a mix between the high roughness of an assumed underlying base and the microfacet dielectric roughness $r = \mathtt{specular\_roughness}$, according to the base dielectric Fresnel factor modulated via \verb|specular_weight| $\xi_s$ (see Equation~\ref{unmodulated_fresnel} and Equation~\ref{modulated_ior}):
\begin{equation} \label{dielectric_roughness_estimate}
r_d = \mathrm{lerp}(1, r, \xi_s F_s) \ ,
\end{equation}
while the metallic roughness can be taken to be $r_m = r$. (Note that, in this formula for $r_d$, a clamp must be applied to ensure that $\xi_s F_s \in [0, 1]$.)
\enlargethispage{1\baselineskip}

We can thus approximate $\mathbf{E}^\prime_c \approx \mathbf{E}_b \mathbf{T}^2_\mathrm{coat} \mathbf{\Delta}$. The boost factor of Equation~\ref{undarkened_coat_albedo_scaling} then reduces to
\begin{equation} \label{B_approx}
\mathbf{B}(\delta) = \mathrm{lerp}\biggl(\frac{1}{\mathbf{\Delta}}, 1, \delta\biggr) \ .
\end{equation}

A reasonable approximate scheme, assuming no other compensation is made to approximate the effect, is to multiply the base BSDF by the uniform \emph{modulated darkening factor} (depending on the darkening parameter $\delta = \mathtt{coat\_darkening}$):
\begin{equation} \label{modulated_darkening_factor}
\mathbf{B}(\delta) \,\mathbf{\Delta} \mathbf{T}^2_\mathrm{coat} = \mathrm{lerp}\left(1, \mathbf{\Delta}, \delta\right) \,\mathbf{T}^2_\mathrm{coat}
\end{equation}
with $\mathbf{\Delta}$ evaluated via Equation~\ref{albedo_scaling_darkening2}  with the internal diffuse reflection coefficient, accounting for base roughness, given by Equation~\ref{internal_diffuse_reflection_coefficient_for_general_base}.
Listing~\ref{listing:coat_darkening} gives pseudocode of an example implementation in an albedo-scaling based layer framework.

\subsection{View-dependent absorption}

\label{sec:coat_absorption}

\begin{figure}[!tb]
  \centering
  \hfill
  \begin{subfigure}{.24\textwidth}
    \includegraphics[width=\linewidth]{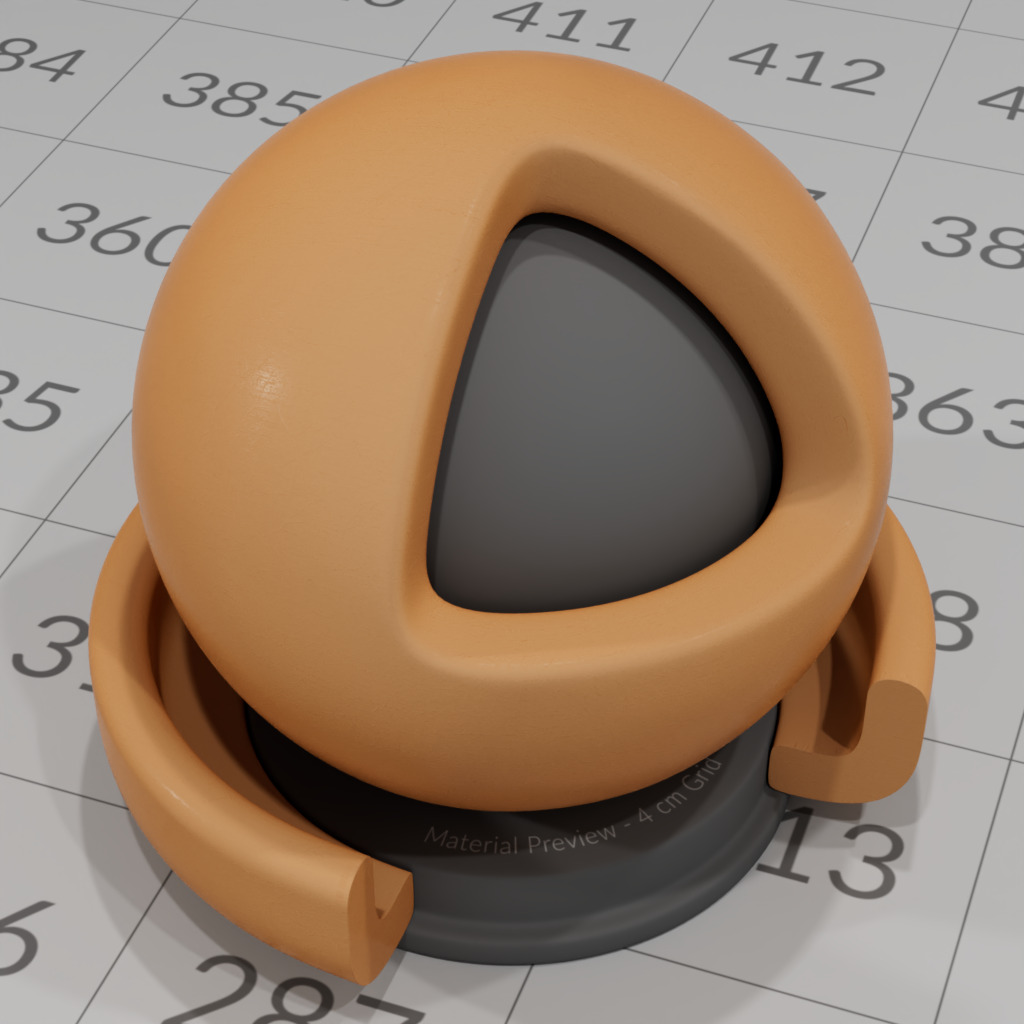}
    \caption{\texttt{coat\_ior} 1.1}
  \end{subfigure}
  \hfill
  \begin{subfigure}{.24\textwidth}
    \includegraphics[width=\linewidth]{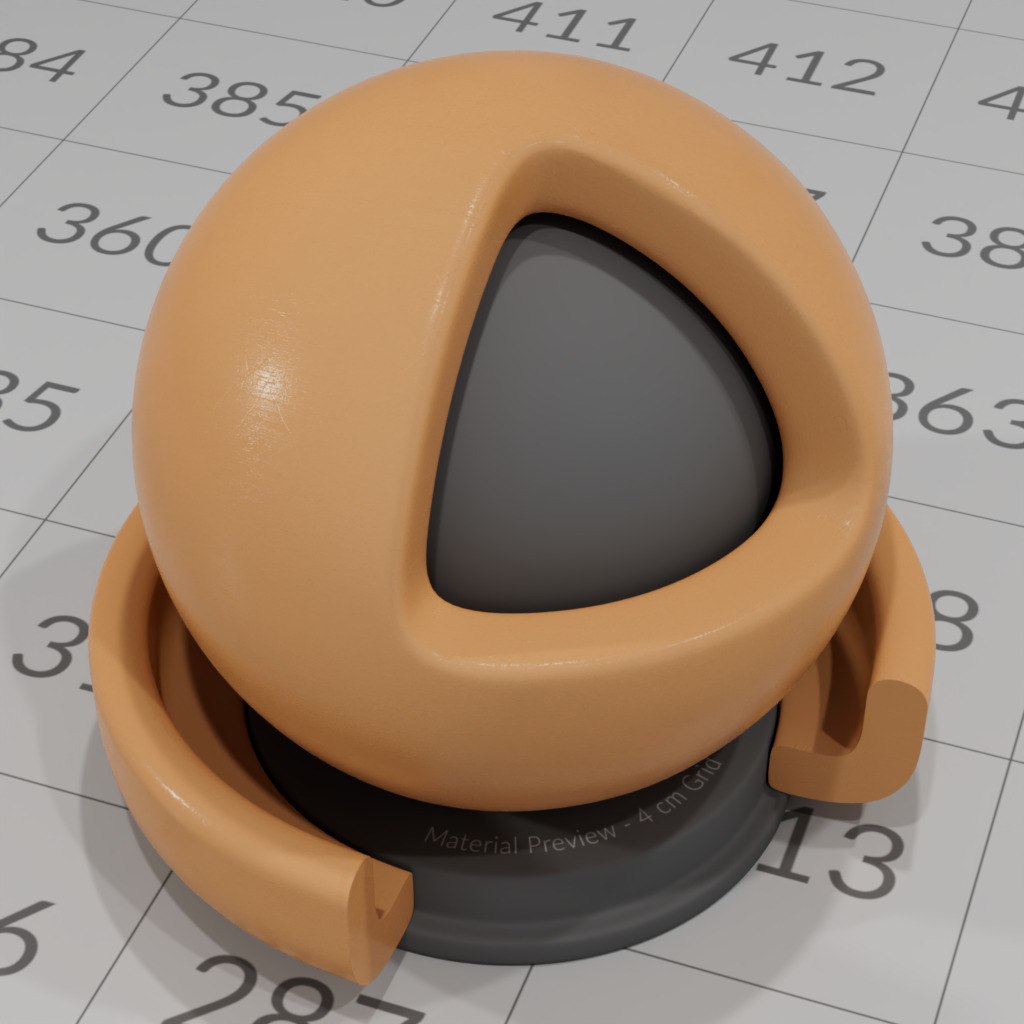}
    \caption{\texttt{coat\_ior} 1.2}
  \end{subfigure}
  \hfill
  \begin{subfigure}{.24\textwidth}
    \includegraphics[width=\linewidth]{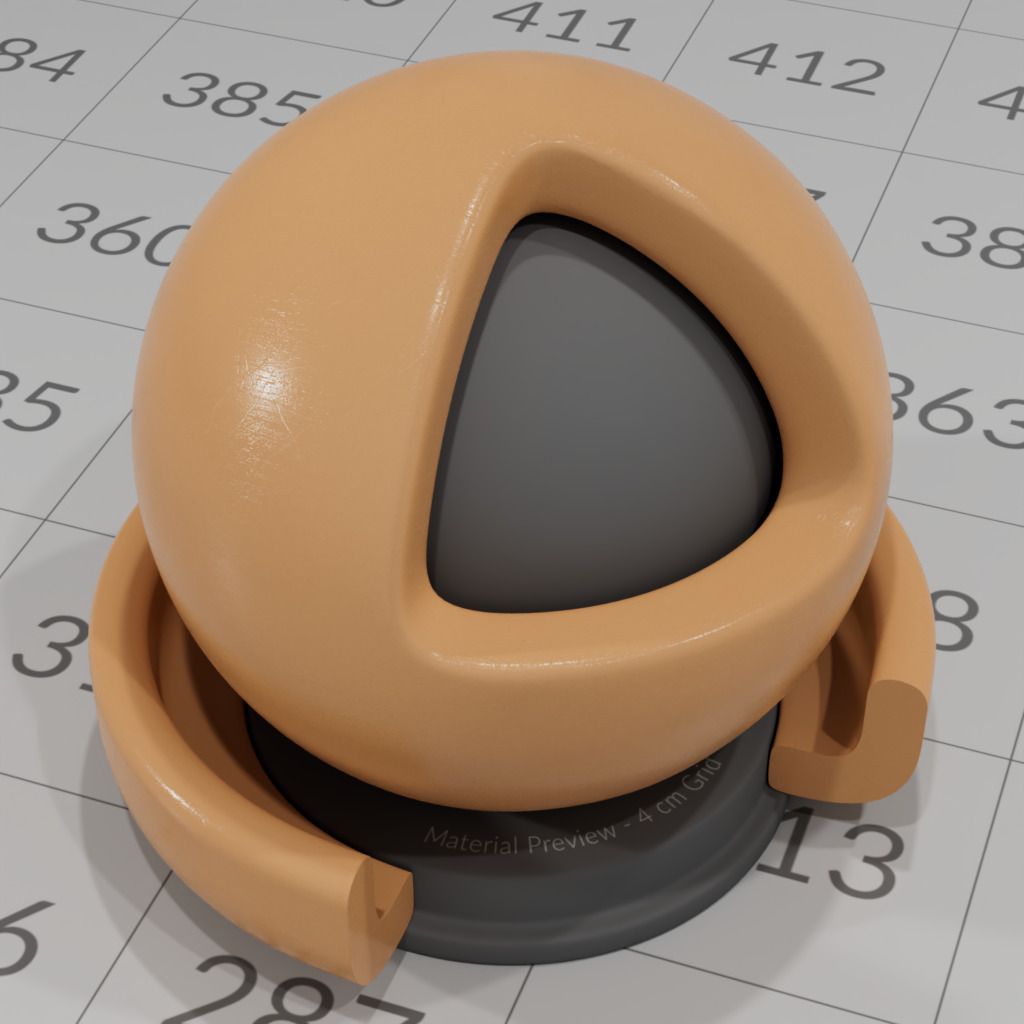}
    \caption{\texttt{coat\_ior} 1.3}
  \end{subfigure}
  \hfill
  \begin{subfigure}{.24\textwidth}
    \includegraphics[width=\linewidth]{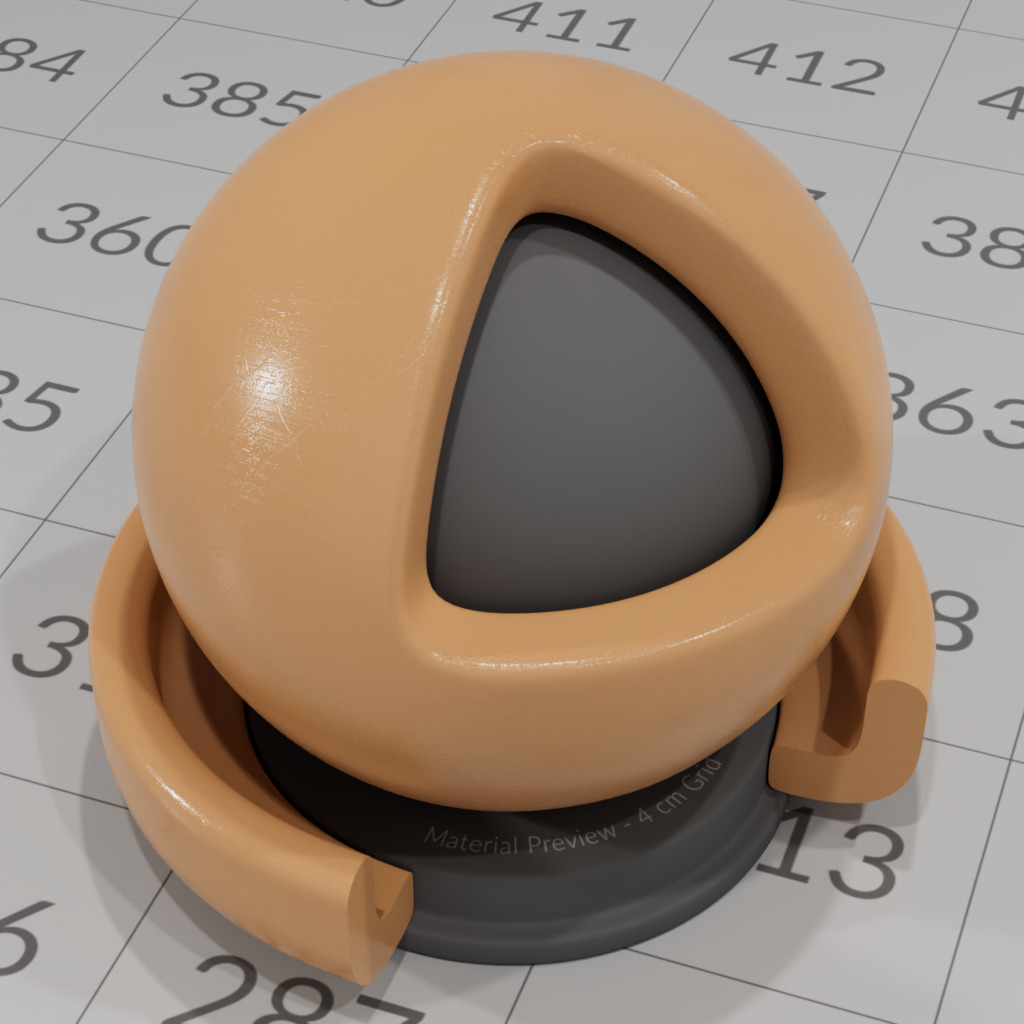}
    \caption{\texttt{coat\_ior} 1.4}
  \end{subfigure}
  \hfill
  \caption{The color of an absorbing coat becomes slightly darker and more saturated at grazing angles, with a stronger effect for low \texttt{coat\_ior}. \label{fig:coat_absorption}}
\end{figure}

In the case of an absorbing coat, there is also enhanced darkening and saturation at grazing angles due to increased path length within the coat medium. The effect of this can be modeled via a factor in the coat BRDF of
\begin{equation}
T_\mathrm{coat}^{1/\mu^t_i + 1/\mu^t_o} \ ,
\end{equation}
where $\mu^t_i$ is the angle cosine of the incident ray refracted into the coat, i.e.
\begin{equation}
\mu^t_i =\sqrt{1 - \frac{1 - \mu^2_i}{\eta^2_c}} \ ,
\end{equation}
and similarly for $\mu^t_o$. Note that, at normal incidence, this factor reduces to $T^2_\mathrm{coat}$, which we defined to equal \verb|coat_color|.
The effect of the absorption is more apparent for low \verb|coat_ior|, as the path length within the coat layer increases (see Figure~\ref{fig:coat_absorption}).

\subsection{Roughening}

\label{sec:coat_roughening}

\begin{figure}[!tb]
  \centering
  \hfill
  \begin{subfigure}{.24\textwidth}
    \includegraphics[width=\linewidth]{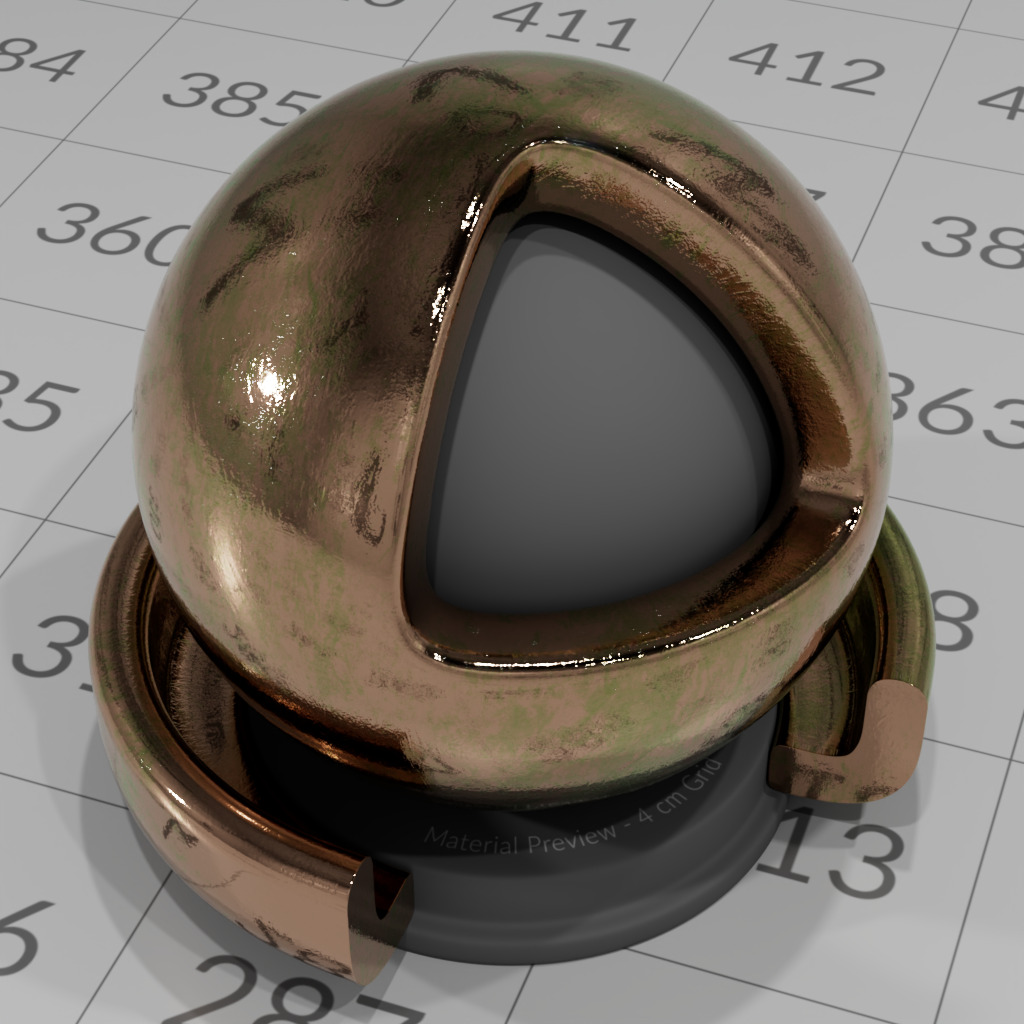}
  \end{subfigure}
  \hfill
  \begin{subfigure}{.24\textwidth}
    \includegraphics[width=\linewidth]{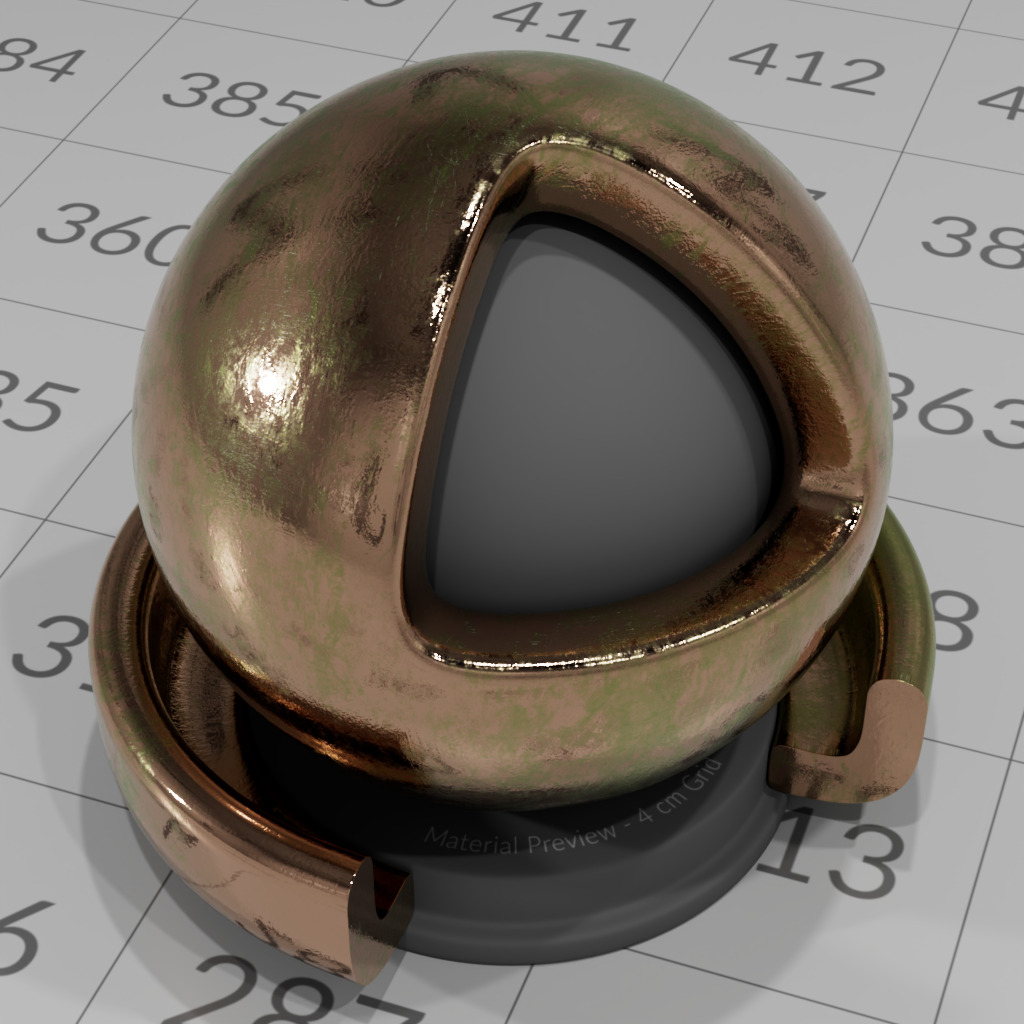}
  \end{subfigure}
  \hfill
  \begin{subfigure}{.24\textwidth}
    \includegraphics[width=\linewidth]{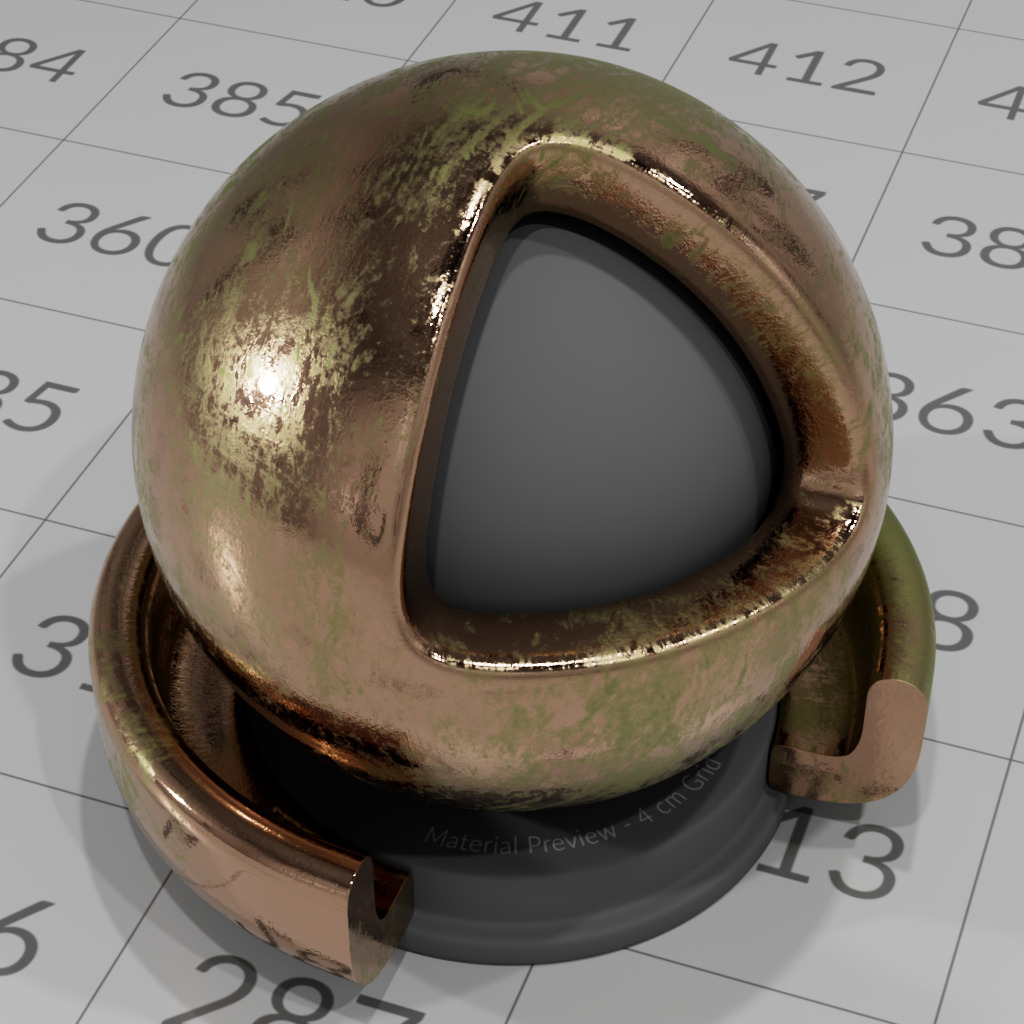}
  \end{subfigure}
  \hfill
    \begin{subfigure}{.24\textwidth}
    \includegraphics[width=\linewidth]{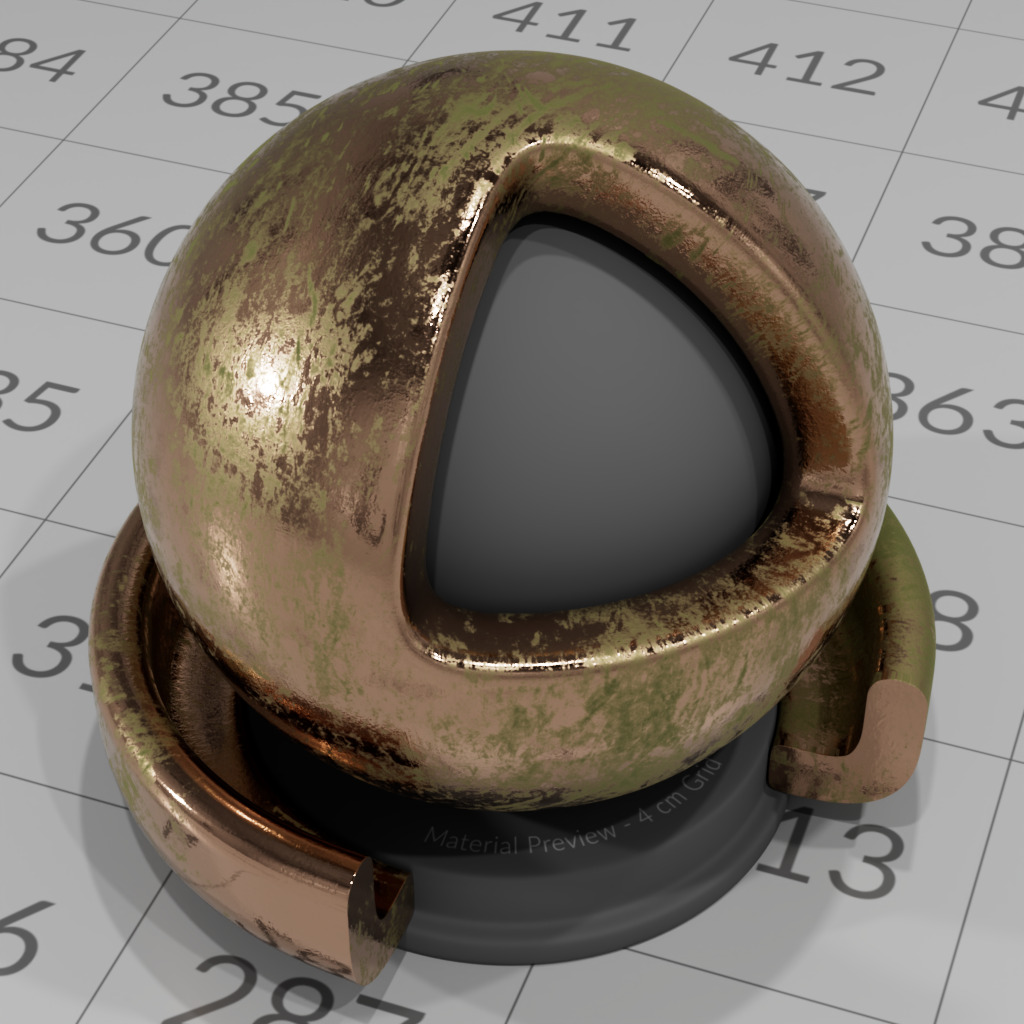}
  \end{subfigure}
  \hfill
  \caption{Appearance of a smooth metallic base with a coat ``patina'' (with textured weight) of varying roughness, for $r=0$ (left), $r=0.25$ (middle left), $r=0.5$ (middle right), and $r=0.75$ (right).}
  \label{fig:coat_roughening_example}
\end{figure}

If the coat is rough, the microfacet BSDF lobes of the underlying base substrate (metal and dielectric) are also effectively roughened. If this is not otherwise accounted for by the light transport, it can instead be reasonably approximated by directly altering the NDF of the base BSDFs. Figure \ref{fig:coat_roughening_example} illustrates the (approximate) effect of varying the coat roughness on the appearance of the base material.

A heuristic we recommend for this is obtained by identifying the NDF of each microfacet lobe as corresponding approximately to a Gaussian in slope space with variance given by $\alpha_t^2 + \alpha_b^2 = r^4$ (in the notation of Section~\ref{sec:microfacet}).
The effect of the coat roughening can then be modeled as the convolution of these Gaussian NDFs, which corresponds to adding the variances (double counting the coat variance, since the reflection passes through the coat boundary twice). The IOR ratio of the coat and ambient medium $\eta_\mathrm{ca}$ also needs to be accounted for, since as $\eta_\mathrm{ca} \rightarrow 1$, the roughening due to the coat goes to zero.
This leads to the following suggested approximation for the modified roughness $r'_\mathrm{B}$  of the base due to the coat:
\begin{equation} \label{coat_roughening_heuristic}
r'_\mathrm{B} = \mathrm{min} \bigl(1, r^4_\mathrm{B} + 2 x_C r^4_\mathrm{C} \bigr)^\frac{1}{4} \quad \textrm{with  } x_C  = 1 - \mathrm{min}(\eta_\mathrm{ca}, 1/\eta_\mathrm{ca}) \ ,
\end{equation}
where $r_\mathrm{B}=\mathtt{specular\_roughness}$ and $r_\mathrm{C}=\mathtt{coat\_roughness}$.
Of course, the presence weight of the coat ($\mathtt{C}= \mathtt{coat\_weight}$) also needs to be taken into account, ideally by blending between the effect with and without the coat present. Alternatively, a cruder approximation would be to just set the roughness of the base to $\mathrm{lerp}(r_\mathrm{B}, r'_\mathrm{B}, \mathtt{C})$.

We intend to provide better approximations in the future. For example, a derivation of the roughening due to transmission through a rough GGX dielectric interface was provided by \textcite{Belcour2018}, which could provide a more physically accurate result.

\subsection{Total internal reflection}

\label{sec:coat_TIR}

\begin{figure}[!tb]
  \centering
\includegraphics[width=0.6\linewidth]{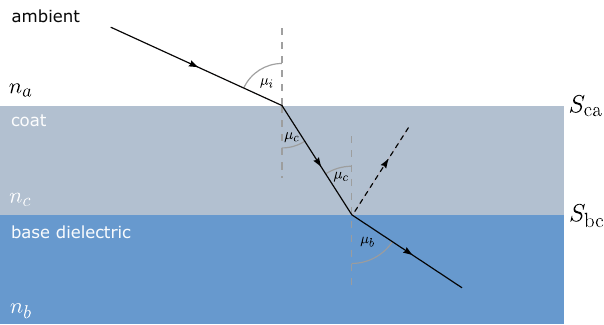}
\caption{Ray configuration at coat-base boundary.}
\label{fig:coat_base_schematic}
\end{figure}

A technical issue which can cause difficulties in the implementation of the BRDF of the coated dielectric base should also be mentioned.
See Figure~\ref{fig:coat_base_schematic} for a schematic of a ray refracting from the ambient exterior medium (with IOR $n_a$) into the coat (with IOR $n_c$), and then into the base dielectric (with IOR $n_b$).

If the coat IOR $n_c$ exceeds the IOR of the base dielectric $n_b$, then rays incident to the base can in general undergo total internal reflection (TIR) at the base-coat boundary surface $S_\mathrm{bc}$ (with IOR ratio $\eta_\mathrm{bc} = n_b/n_c$). In this case, the reflection from the base (the dashed line) is enhanced since no refraction into the base along the $\mu_b$ direction occurs.

However, if the surfaces are smooth and parallel, it is easy to show that \emph{no such TIR is possible} for a ray incident from the exterior due to the refraction of the ray at the upper coat-ambient boundary $S_\mathrm{ca}$ (with IOR ratio $\eta_\mathrm{ca} = n_c/n_a$). In the general case of rough boundaries, TIR can occur though it will be increasingly suppressed as the roughness decreases. Thus if the implementation of the \hyperref[sec:dielectric-base]{dielectric-base} BSDF does not account for the refraction of the incident ray in the coat, spurious TIR will be generated, which produces an obviously incorrect appearance manifesting as a bright ``ring'' near grazing angles. In an implementation which reduces the model to a mixture of independent lobes sampled as a function of incident direction, it is non-obvious how to account for this.

We suggest as a reasonable practical approximation (following \textcite{Kutz2021}) to simply invert the IOR ratio at the coat-base boundary, i.e., replace $\eta_\mathrm{bc}$ as follows:
\begin{equation} \label{specular_ior_ratio_with_tir_fix}
\eta_\mathrm{bc} \rightarrow
\begin{cases}
   \eta_\mathrm{cb} = n_c/n_b & \quad\mathrm{if} \;\; n_c >   n_b \\
   \eta_\mathrm{bc}           & \quad\mathrm{if} \;\; n_c \le n_b \ .
\end{cases}
\end{equation}
This eliminates the TIR while plausibly maintaining the specular reflection. Then, to account for partial coat coverage, this modified $\eta_\mathrm{bc}$ is used in place of the $n_b/n_c$ term in Equation~\ref{specular_ior_ratio}.

An alternative approach is to explicitly modify the refraction angle cosine of the incident direction to the base. Assuming, for simplicity, that in the local space of the base microfacet the overlying coat microfacet has matching micronormal, the angle cosine $\mu_c$ to the base micronormal is given by refraction of the incident angle cosine $\mu_i$ as follows:
\begin{equation} \label{refracted_coat_dir}
\mu_c^2 = 1 - (1 - \mu_i^2) / \eta^2_\mathrm{ca} \ .
\end{equation}
The resultant Fresnel factor can then be approximated as a blend of the Fresnel factors evaluated at the original and refracted incident directions, according to the presence weight of the coat:
\begin{equation}
\mathrm{lerp}(F(\mu_i, \eta_\mathrm{ba}), F(\mu_c, \eta_\mathrm{bc}), \mathtt{C}) \ ,
\end{equation}
where $\eta_\mathrm{ba} = n_b/n_a$, and this blended Fresnel factor is then substituted into the microfacet BRDF (this replaces the Fresnel calculation of Equation~\ref{specular_ior_ratio}, which ignores the coat refraction). If $\eta_\mathrm{ca} < 0$ then TIR at $S_\mathrm{ca}$ may occur (i.e., $\mu_c^2 \le 0$ in Equation~\ref{refracted_coat_dir}), in which case the Fresnel factor $F(\mu_c, \eta_\mathrm{bc})$ can be assumed to be zero.

Both of these schemes produce a plausible appearance that eliminates the spurious TIR. A more physically correct approach would require something similar to the Weidlich--Wilkie layering model \cite{Weidlich2007}, or the layered material model described in \textcite{Pharr2023}, where the scattering through the layers is explicitly modeled via Monte Carlo simulation.

\clearpage

\section{Fuzz}

\label{sec:fuzz}

The \emph{fuzz} BRDF \(f_\mathrm{fuzz}\) and corresponding volumetric distribution function (VDF) \(V_\mathrm{fuzz}\) are modeled as a homogeneous scattering layer composed of anisotropic microflakes with a fiber-like distribution -- often referred to as \emph{sheen} in the literature.

Previous approaches include the microfacet-based sheen of \textcite{Estevez2017} as used by the Autodesk Standard Surface model \cite{Georgiev2019}, and the single-scattering microflake model of Adobe Standard Material \cite{Kutz2021}. In OpenPBR, we adopt the multi-scattering volumetric microflake model of \textcite{Zeltner2022}, which uses the SGGX microflake representation of \textcite{Heitz2015} and provides efficient evaluation and importance sampling via linearly transformed cosines (LTCs) \cite{Heitz2016b}.

While structurally similar to Adobe Standard Material's single-scattering approach, Zeltner's model captures energy-conserving multiple scattering and supports principled importance sampling and layering.

\begin{figure}[!b]
  \centering
  \begin{subfigure}{.3\textwidth}
    \includegraphics[width=\linewidth]{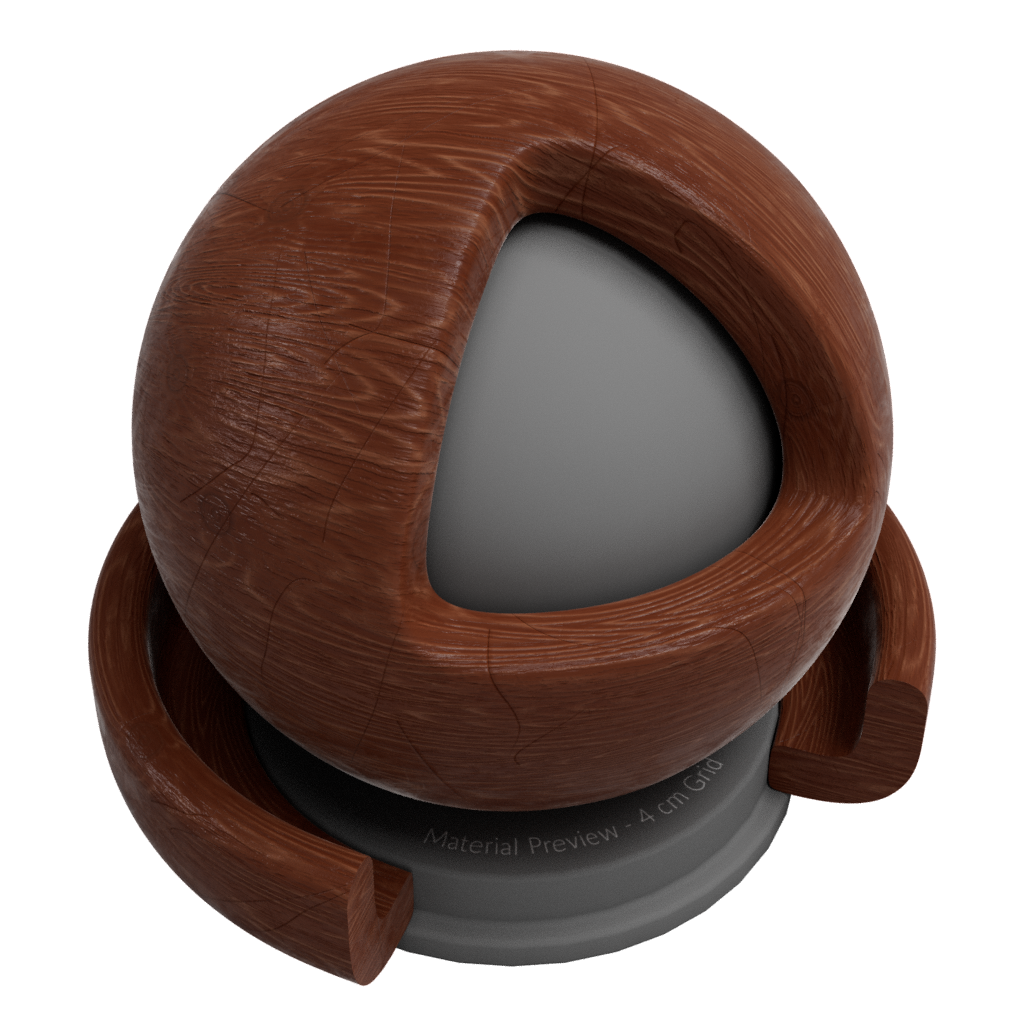}
  \end{subfigure}
  \hspace{0.02\textwidth}
  \begin{subfigure}{.3\textwidth}
    \includegraphics[width=\linewidth]{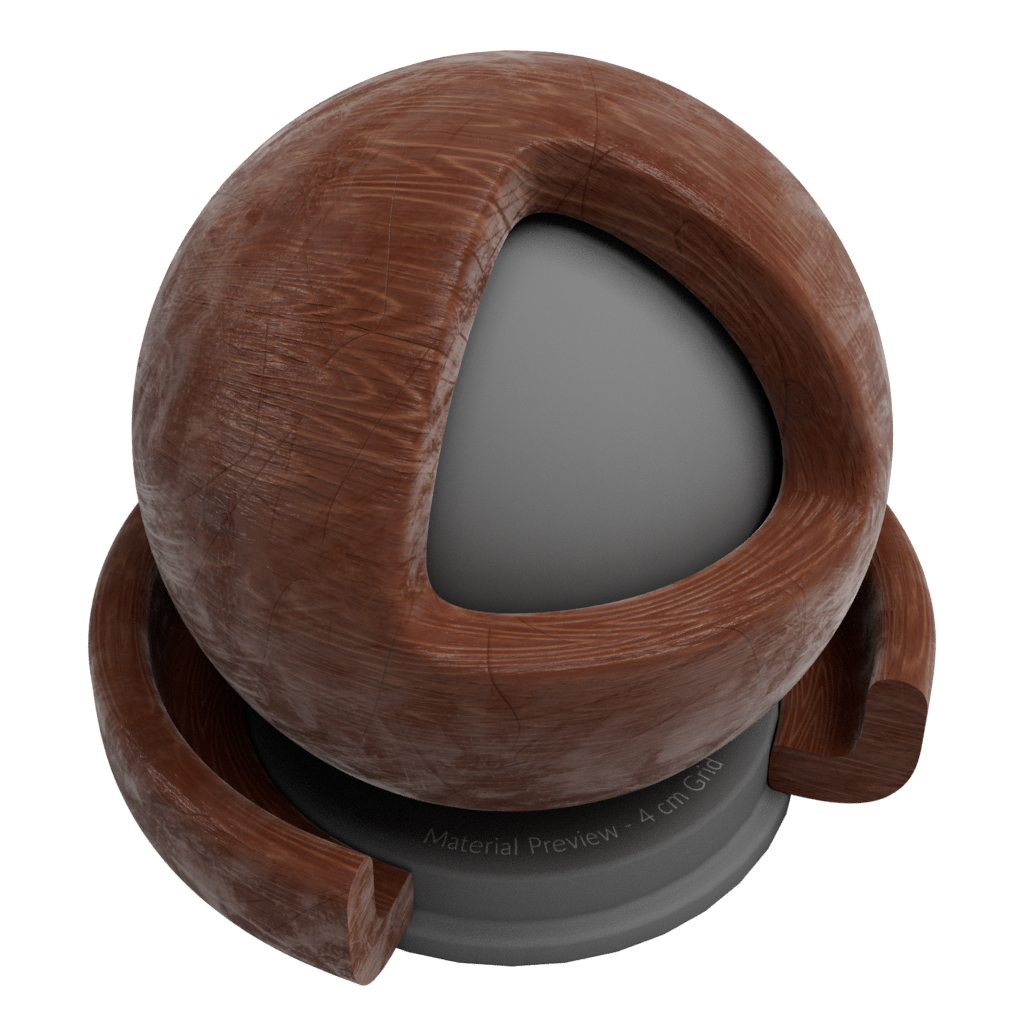}
  \end{subfigure}
  \hspace{0.02\textwidth}
  \begin{subfigure}{.3\textwidth}
    \includegraphics[width=\linewidth]{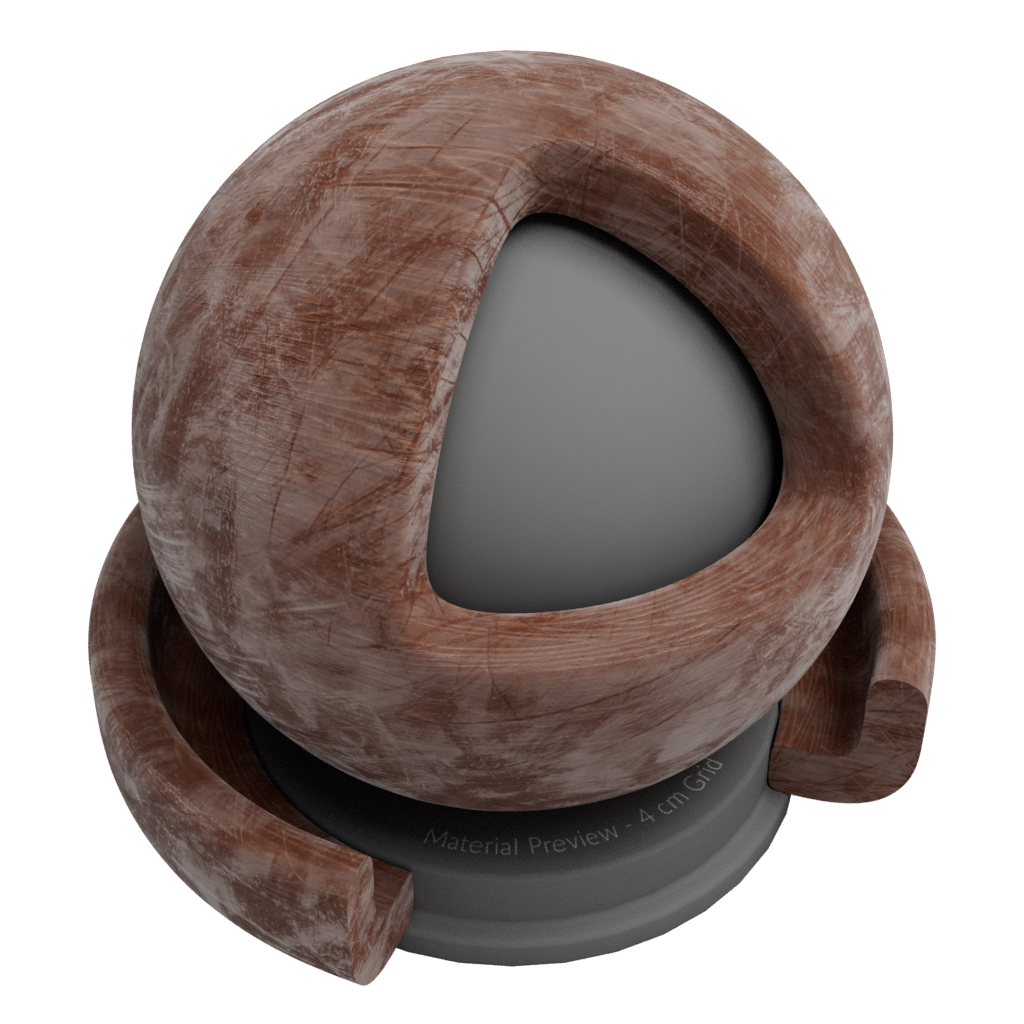}
  \end{subfigure}
  \caption{From left to right, varying the \texttt{fuzz\_roughness} over 0.25, 0.5 (default), 0.75 produces a progressively more dusty appearance. \label{fig:fuzz_roughness}}
\end{figure}

Key features of the fuzz/sheen model in OpenPBR:
\begin{itemize}
  \item \textbf{Volumetric foundation:}
    A homogeneous, purely scattering layer with a fiber-like SGGX microflake phase function \cite{Heitz2015}, providing a physically plausible appearance.
  \item \textbf{Multiple scattering and LTC fitting:}
    Reflection is approximated by an LTC fitted to full volumetric multi-scattering simulations, enabling efficient evaluation and importance sampling.
  \item \textbf{Layering model:}
    The layer has unit optical thickness, is index-matched to the exterior medium (eliminating Fresnel at the interface), and conserves energy -- all unreflected light transmits uncolored, as there is no absorption, which allows for principled layering (unlike the \textcite{Estevez2017} model).
  \item \textbf{Control parameters:}
    \begin{itemize}
      \item \verb|fuzz_weight|: Controls the layer's coverage weight.
      \item \verb|fuzz_roughness|: Adjusts microflake shape -- low values yield tight fiber-like sheen (e.g., silk), while high values yield diffuse, dust-like scattering (see Figure~\ref{fig:fuzz_roughness}).
      \item \verb|fuzz_color|: Tints reflected light, allowing the sheen to lighten or darken.
    \end{itemize}
\end{itemize}

The mathematical form of this model is the following (with $\mu_i, \mu_o$ the angle cosines to the normal of $\omega_i, \omega_o$):
\begin{equation}
\mu_i \, f_\mathrm{fuzz}(\omega_i, \omega_o) = \mathbf{F} \, E_\mathrm{fuzz}(\mu_o, \alpha) \, D(\mu_i | \mu_o, \alpha) \ ,
\end{equation}
where $\mathbf{F}$ = \verb|fuzz_color|, $E_\mathrm{fuzz}(\mu_o, \alpha)$ (termed $R$ by \textcite{Zeltner2022}) is the reflectance at angle cosine $\mu_o$ given roughness $\alpha$ = \verb|fuzz_roughness| $\in [0,1]$, and $D(\mu_i | \mu_o, \alpha)$ is a lobe defined by linear transformations of a cosine lobe (LTCs), where the transformation matrices (and $E_\mathrm{fuzz}$) are fitted (over $\mu_o$ and $\alpha$) to a simulation of the scattering in the volumetric fuzz microflake layer. Since the LTC lobe $D$ is a normalized PDF over the hemisphere, the resulting albedo of $f_\mathrm{fuzz}$ is $\mathbf{F} \, E_\mathrm{fuzz}(\mu_o, \alpha)$.

If using the albedo-scaling interpretation of layering, a reasonable approximation of the reflection from the fuzz layer combined with the reflection from the base is to take
\begin{eqnarray}
\mathrm{\mathbf{layer}}(M_\textrm{coated-base}, S_\mathrm{fuzz}) &\rightarrow& f_\mathrm{fuzz} +  \bigl(1 - E_\mathrm{fuzz} \bigr) \,f_\textrm{coated-base} \ ,
\end{eqnarray}
where the albedo-scaling is explicitly modified to not tint the base, since the tint $\mathbf{F}$ appears only in the first term via $f_\mathrm{fuzz}$.
Then, accounting for the coverage weight of the fuzz layer, $\mathtt{F}$ = \verb|fuzz_weight|, gives
\begin{eqnarray}
\mathrm{\mathbf{layer}}(M_\textrm{coated-base}, S_\mathrm{fuzz}, \mathtt{F})  &\rightarrow&  \mathtt{F} \,f_\mathrm{fuzz} +  \mathrm{lerp}\bigl(1, 1 - E_\mathrm{fuzz}, \mathtt{F}\bigr) \,f_\textrm{coated-base} \ . \label{fuzz-layering-approx}
\end{eqnarray}

The fuzz shading normal is assumed to inherit from that of the substrate layer, the physical picture being that the fuzz volume settles and conforms to the geometry of the substrate. The substrate is generally a mixture of coat and uncoated base. Thus physically the fuzz model should be evaluated with each of the \verb|geometry_coat_normal| and \verb|geometry_normal| separately (if they differ), and the final result blended according to the \verb|coat_weight|. As a practical consideration, it may be more convenient and efficient to instead approximate the fuzz normal by interpolating the coat and base normal according to \verb|coat_weight|.

\subparagraph{Future work: roughening heuristics}

\label{sec:fuzz:roughening}

\begin{figure}[!tb]
  \centering
  \hfill
  \begin{subfigure}{.24\textwidth}
    \includegraphics[width=\linewidth]{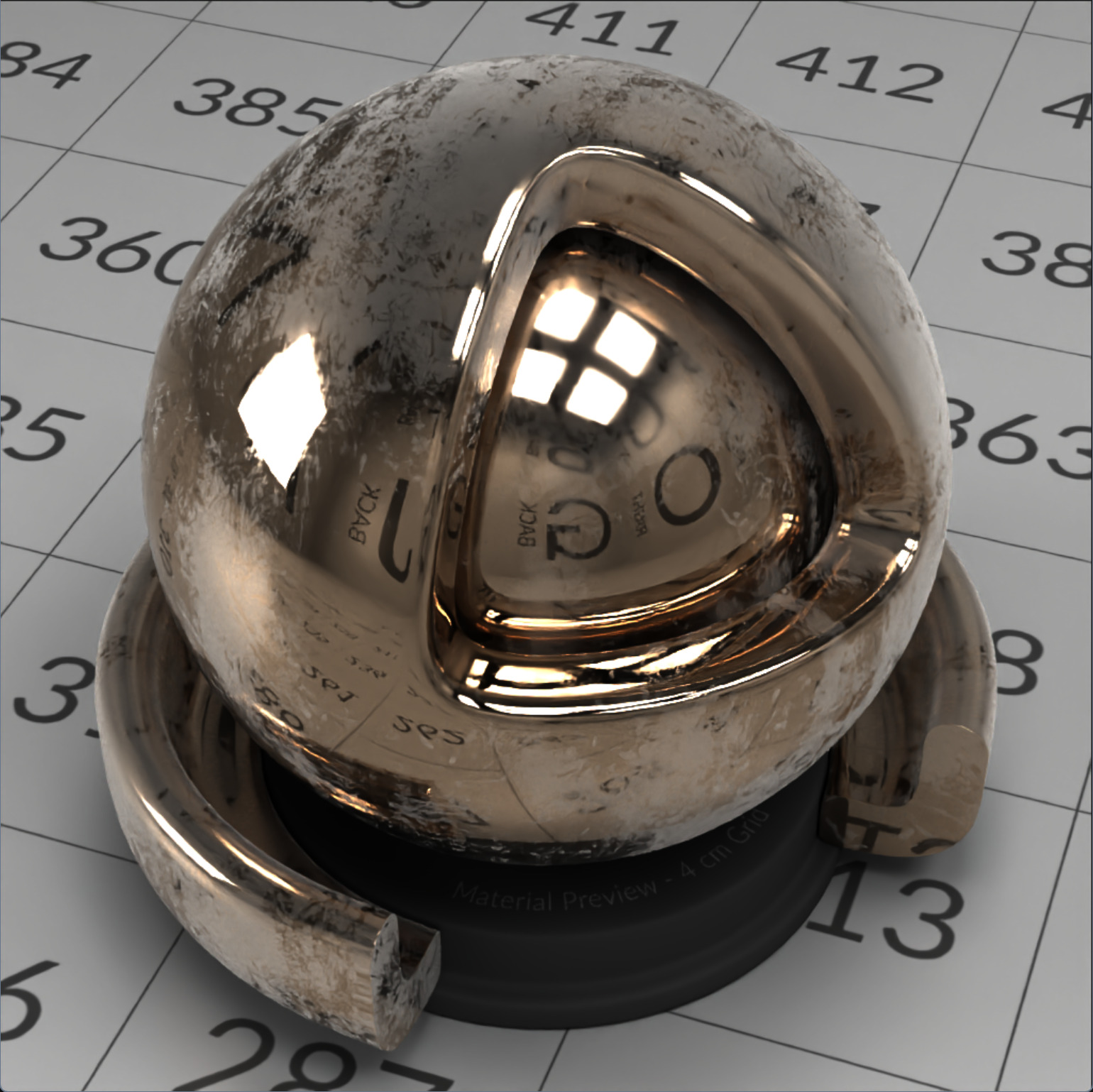}
  \end{subfigure}
  \hfill
  \begin{subfigure}{.24\textwidth}
    \includegraphics[width=\linewidth]{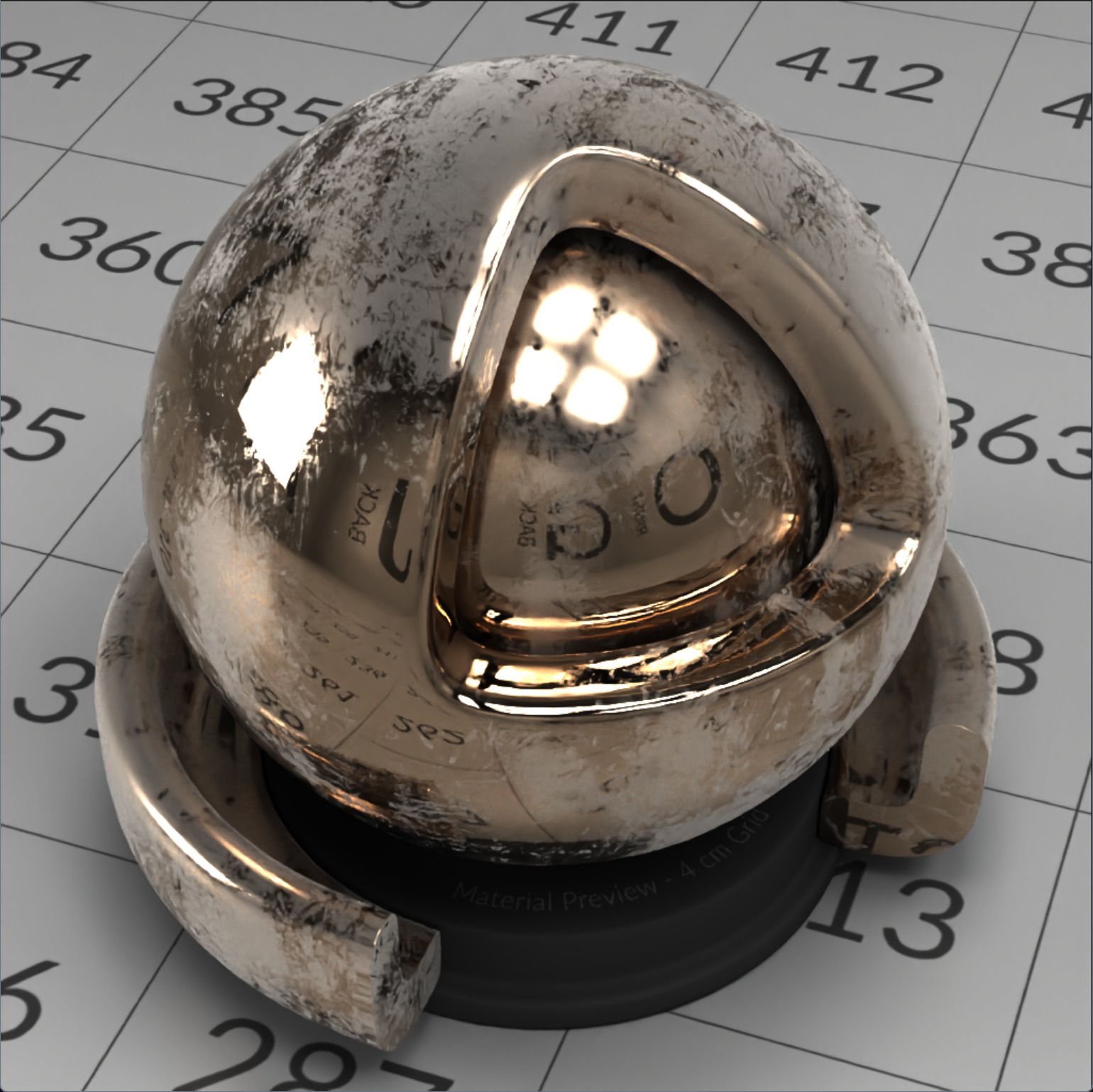}
  \end{subfigure}
  \hfill
  \begin{subfigure}{.24\textwidth}
    \includegraphics[width=\linewidth]{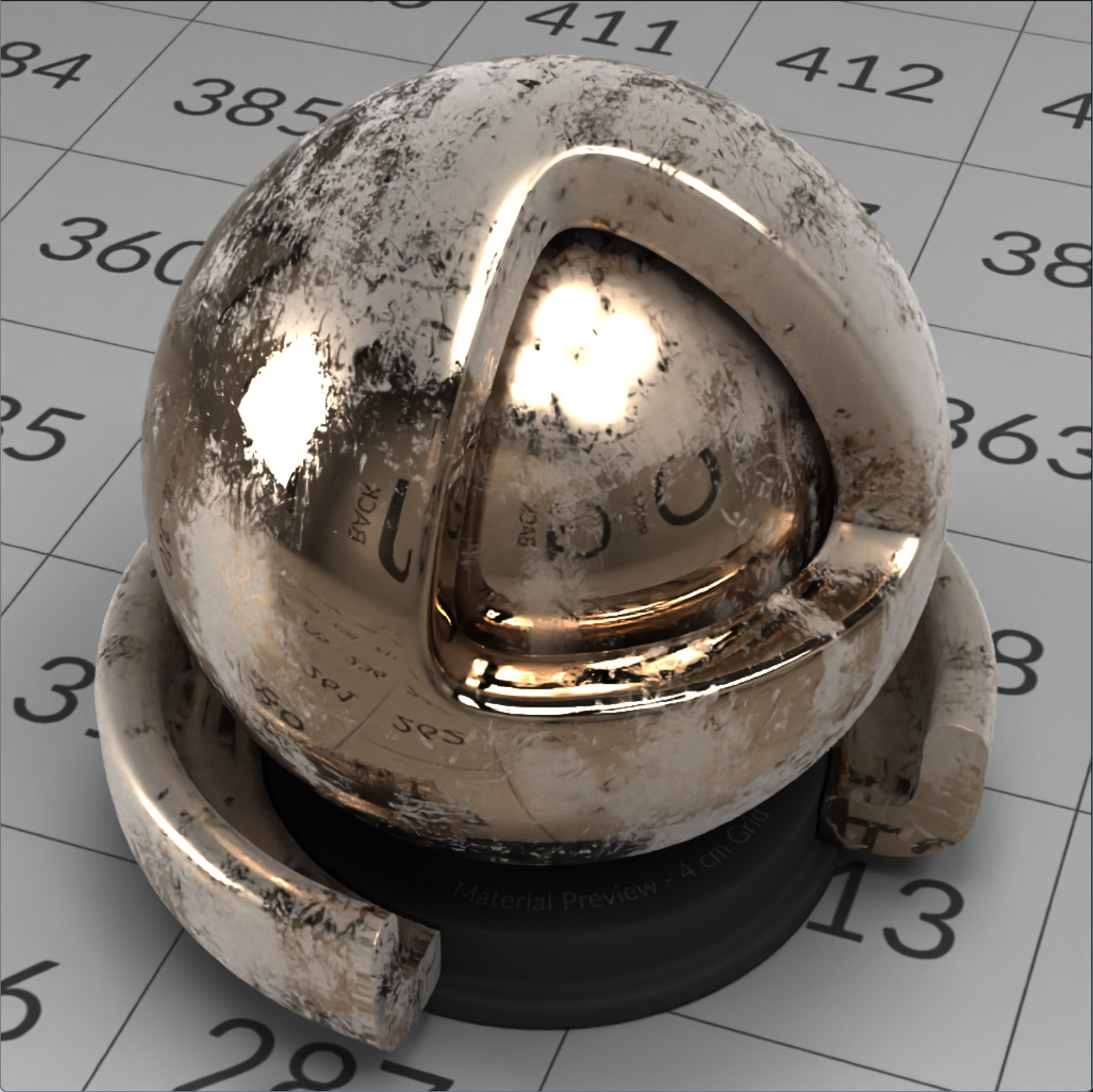}
  \end{subfigure}
  \hfill
    \begin{subfigure}{.24\textwidth}
    \includegraphics[width=\linewidth]{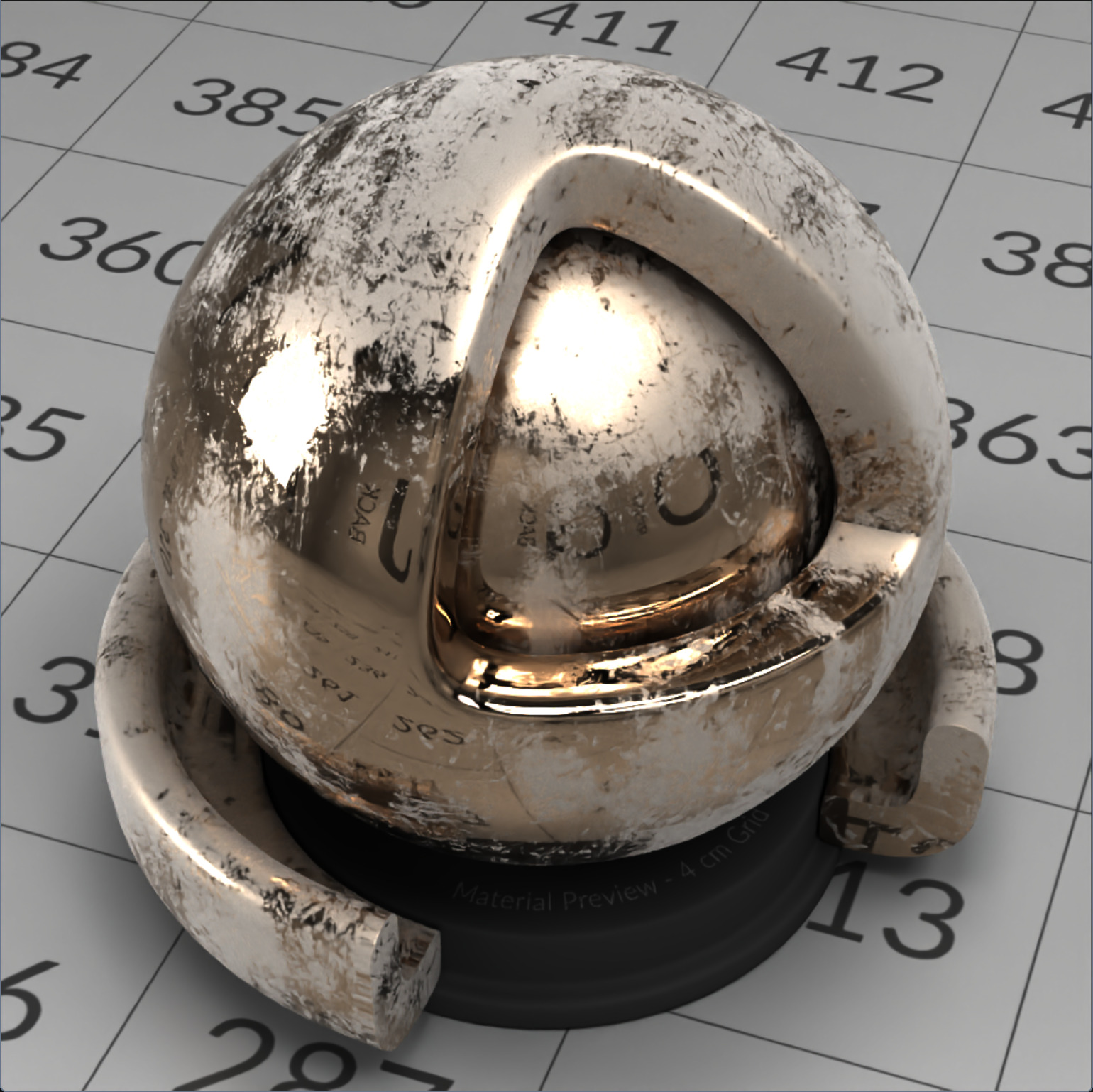}
  \end{subfigure}
  \hfill
  \caption{Appearance of a smooth metallic base with a fuzz layer (with textured weight) modeling dust of varying roughness, for $r=0.7$ (left), $r=0.8$ (middle left), $r=0.9$ (middle right), and $r=1.0$ (right).}
  \label{fig:fuzz_roughening_example}
\end{figure}

The scattering within the fuzz layer will have the effect of roughening the appearance of the substrate beneath it. A simple suggested approximation for this can be adapted from the formula used to model the coat roughening, in Equation~\ref{coat_roughening_heuristic}. If we consider the fuzz layer to generate roughening by scattering, we can approximate its effective roughness as being proportional to the albedo of the layer, as well as to the tint color (since darker fuzz will physically scatter less and absorb more). This leads to the following heuristic for the modified roughness $r'_\mathrm{B}$ of the substrate lobe:
\begin{equation} \label{fuzz_roughening_heuristic}
r'_\mathrm{B} = \mathrm{min} \bigl(1, r^4_\mathrm{B} + 2 R^4_\mathrm{F} \bigr)^\frac{1}{4} \quad \textrm{with $R_F = \mathrm{lum}(\mathbf{F} E_\mathrm{fuzz})$} \ ,
\end{equation}
where $r_\mathrm{B}$ is the original substrate roughness, and $\mathrm{lum}(\cdots)$ computes the luminance of the RGB argument. This should be applied to both the coat (if present) and the base lobes. If both the fuzz and the coat are present, then the base lobe roughness will be broadened by both the coat and fuzz formulas successively. The presence weights of the fuzz and coat should be accounted for appropriately.
Figure~\ref{fig:fuzz_roughening_example} shows the approximate effect of varying the fuzz roughness on the appearance of a smooth \hyperref[sec:metallic-base]{metallic base} with a fuzz layer using this heuristic.

The described fuzz roughening heuristic above ignores the fact that, in reality or in a full simulation, the directional distribution of the transmitted light is more complex: some light transmits completely through the fuzz layer without interacting, and the light that is scattered is scattered anisotropically (i.e., preferentially in the backward direction). This should be taken into account when combining with the base lobe.

As for the \hyperref[sec:coat_roughening]{coat roughening} effect, we would like to provide such improved heuristics in the future.

\subparagraph{Future work: generalized fuzz}

\label{sec:fuzz:generalized}

We would find a slightly more generalized model of fuzz a useful improvement, though this is not presently available.

In the model of \textcite{Zeltner2022}, the optical depth is at most 1, which means the opacity at normal incidence is at most $1-\exp(-1) = 0.63$, i.e., at least 36\% of the light gets through. In really thick dust, the opacity could be arbitrarily close to 1. If the model were to set the optical depth to a higher value (e.g., 3), thicker fuzz could be represented, though that would make the fuzz opacity more uniform across all viewing angles.
Ideally, the optical depth would be exposed as a parameter. In other words, it would be helpful to be able to control the density and thickness in order to make a fully opaque layer of dust.

Ideally the weight parameter $\mathtt{F} = \mathtt{fuzz\_weight}$  would control the total optical depth $\tau$ of the slab via $\tau= -\ln(1-\mathtt{F})$, so weight 1 would be like scattering from a half-space of fuzz of the given microflake roughness. That would increase the brightness of the low-roughness fuzz, compared to the current look which assumes an optical depth 1 in all cases. The weight then also functions naturally as the main control for the fuzz effect, instead of having to treat it as a coverage weight.

Ideally the fuzz color would also correspond to the true multi-scatter albedo (at, say, normal incidence), from which the single-scatter albedo of the flakes is derived, rather than an ad hoc tint multiplier applied to a purely scattering volume.

\clearpage

\section{Thin-walled mode}

\label{sec:thin-walled}

If the \verb|geometry_thin_walled| boolean is enabled, then the surface is assumed to be in a ``thin-walled'' mode.
In this case, we make the assumption that the surface is essentially the same as the bulk structure but mirrored around the base, with the slabs at the base assumed to be thin enough that macroscopically the material can be treated as a two-dimensional sheet with no interior. This sheet thus appears identical viewed from either side.

A convenient rough approximation of this is to ignore the coat and fuzz layers on the underside, and assume that the surface is automatically oriented so that the incident ray is always entering the surface from the top, so the surface appears the same viewed from above or below. Though a crude approximation, this is quite convenient as the light transport can always be dealt with essentially the same as in the bulk case except with a modified base layer (i.e., incident rays enter top-down, and leave from the base layer without further interaction).

As noted earlier, \verb|geometry_opacity| makes physical sense as a fractional value in the thin-walled case (unlike the bulk case), controlling the overall presence weight of the thin wall.

The metal and glossy-diffuse slabs remain but are considered double-sided, with the top BRDF mirrored to the bottom (and a totally opaque, albeit infinitesimally thin, interior). Emission should apply to both sides of a thin-walled surface. This needs to be taken into account in importance sampling of direct illumination; we can no longer assume that emissive geometry only emits from the front.
The thin-walled interpretation of the \hyperref[sec:translucent-base]{translucent base} and \hyperref[sec:subsurface]{subsurface} differ.

\begin{figure}[!b]
  \centering
  \hfill
  \begin{subfigure}{.19\textwidth}
    \includegraphics[width=\linewidth]{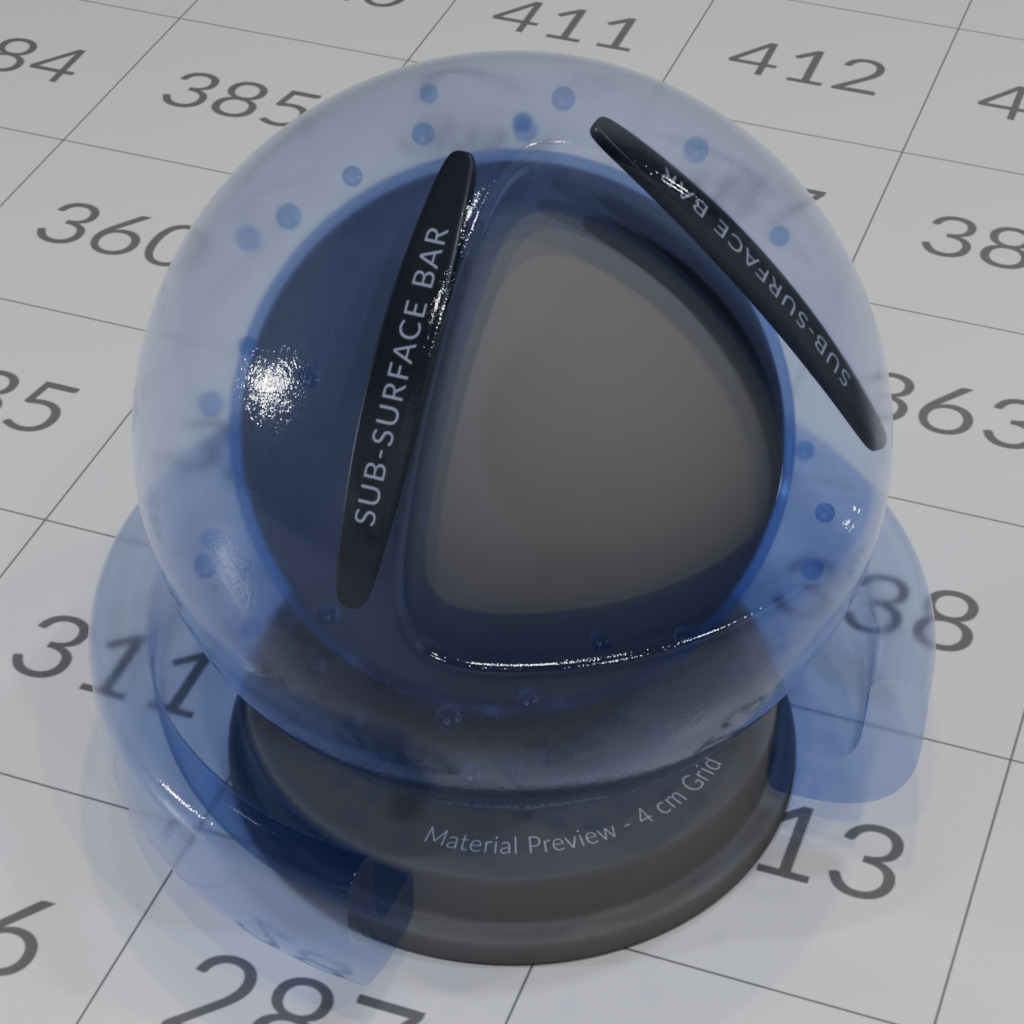}
  \end{subfigure}
  \hfill
    \begin{subfigure}{.19\textwidth}
    \includegraphics[width=\linewidth]{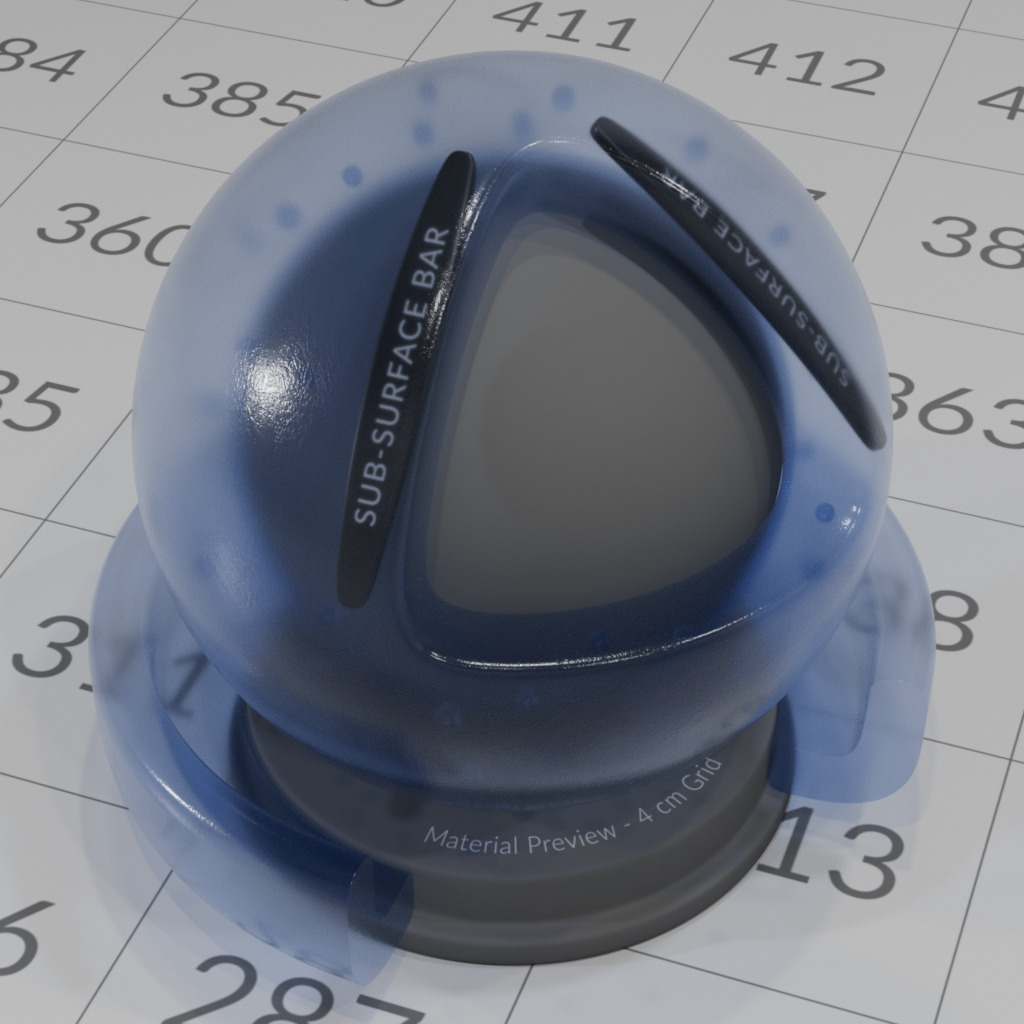}
  \end{subfigure}
  \hfill
    \begin{subfigure}{.19\textwidth}
    \includegraphics[width=\linewidth]{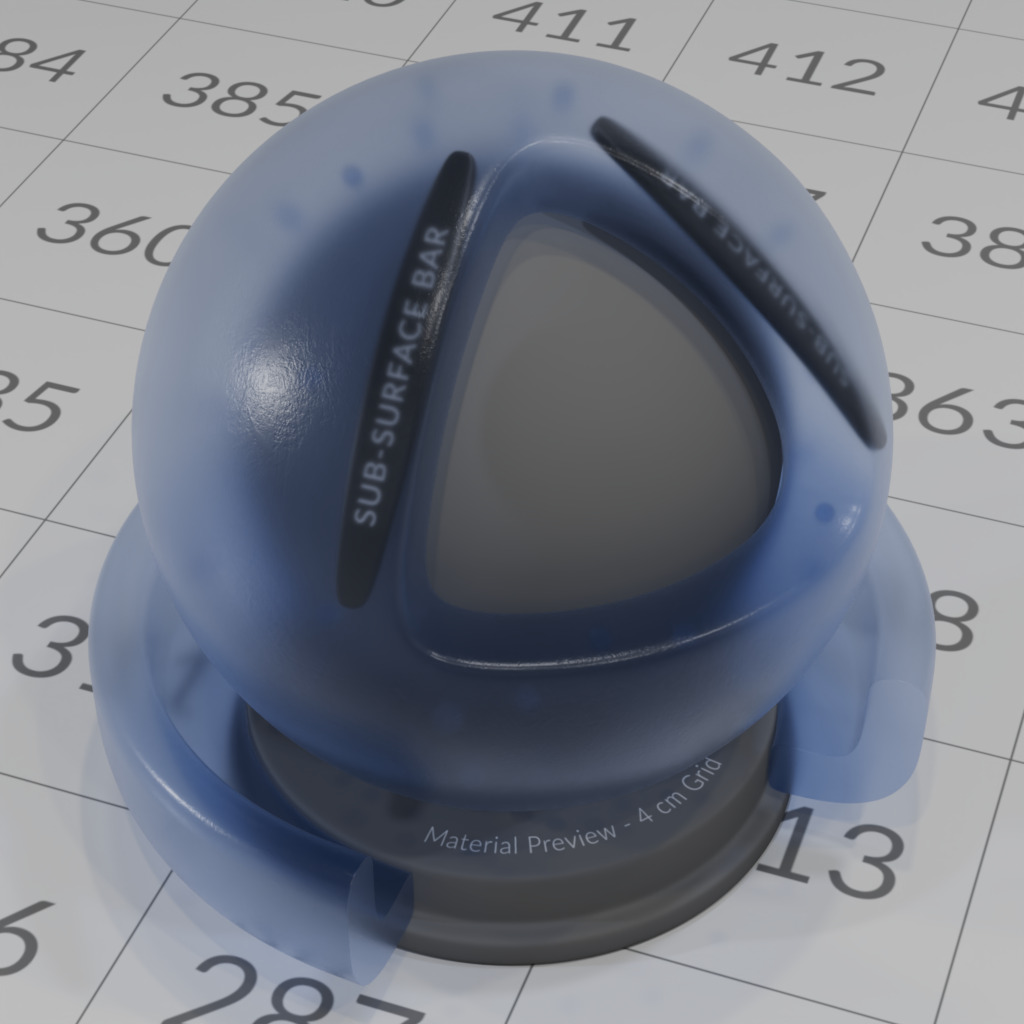}
  \end{subfigure}
  \hfill
    \begin{subfigure}{.19\textwidth}
    \includegraphics[width=\linewidth]{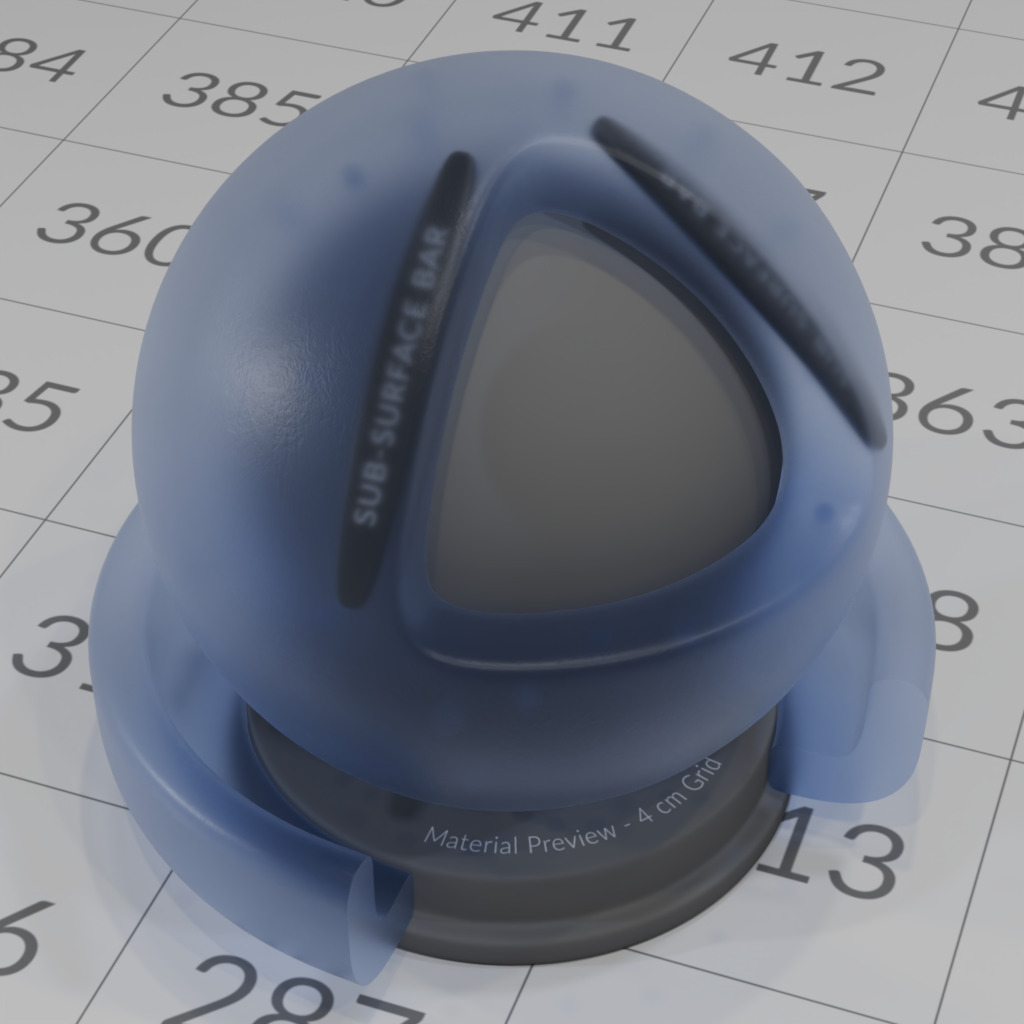}
  \end{subfigure}
  \hfill
    \begin{subfigure}{.19\textwidth}
    \includegraphics[width=\linewidth]{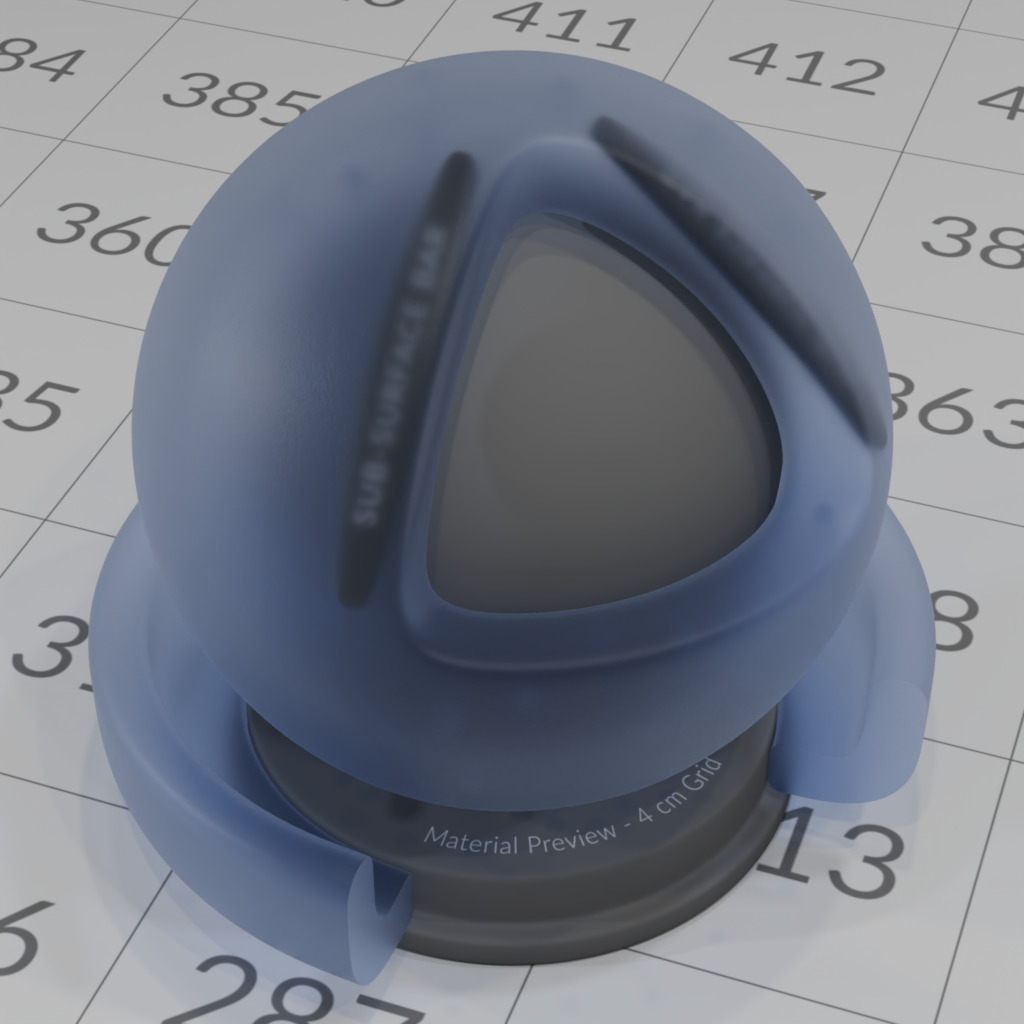}
  \end{subfigure}
  \hfill
  \caption{Thin-wall glass, with \texttt{specular\_roughness} varying over 0.1--0.5. Notice the absence of refraction. \label{fig:thin-wall-glass}}
\end{figure}

\subsection{Thin-wall dielectric}

\label{sec:thin-walled-dielectric}

In thin-walled mode, the \hyperref[sec:translucent-base]{translucent-base} slab can be considered an infinitesimally thin sheet of dielectric (with an embedded absorbing but non-scattering medium), with the BSDF $f_\mathrm{dielectric}$ on both sides. A ladder of inter-reflections occurs inside this slab, producing a reflected lobe and undeflected refracted lobe. The \verb|transmission_color| can be assumed to give the transmittance through the thin sheet at normal incidence due to absorption (with \verb|transmission_depth| ignored). In the smooth case, the BRDF and BTDF of this sheet can be solved exactly by summing over a geometrical series of terms containing Fresnel and absorption factors, and this can be extended to a good approximation of the rough case by appropriately roughening the transmission lobe (as described by \textcite{Kulla2017}). This model of thin-walled glass is a cheaper, much more convenient way to render windows than a finite thickness non-thin-walled mesh.

This model is similar to the \verb|ThinDielectricBxDF| of PBRT, except with absorption and roughness accounted for \cite{Kulla2017, Pharr2023}. The roughening effect of the thin wall can be modeled by considering the broadening of the incident and transmitted rays. \textcite{Belcour2018} provides a useful heuristic for this, but a more complete model would be a welcome advancement.
Figure~\ref{fig:thin-wall-glass} shows the effect of varying the \verb|specular_roughness| parameter on the appearance of a thin-wall glass material.

\subsection{Thin-wall subsurface}

\label{sec:thin-walled-subsurface}

\begin{figure}[!tb]
  \centering
  \hfill
  \begin{subfigure}{.19\textwidth}
    \includegraphics[width=\linewidth]{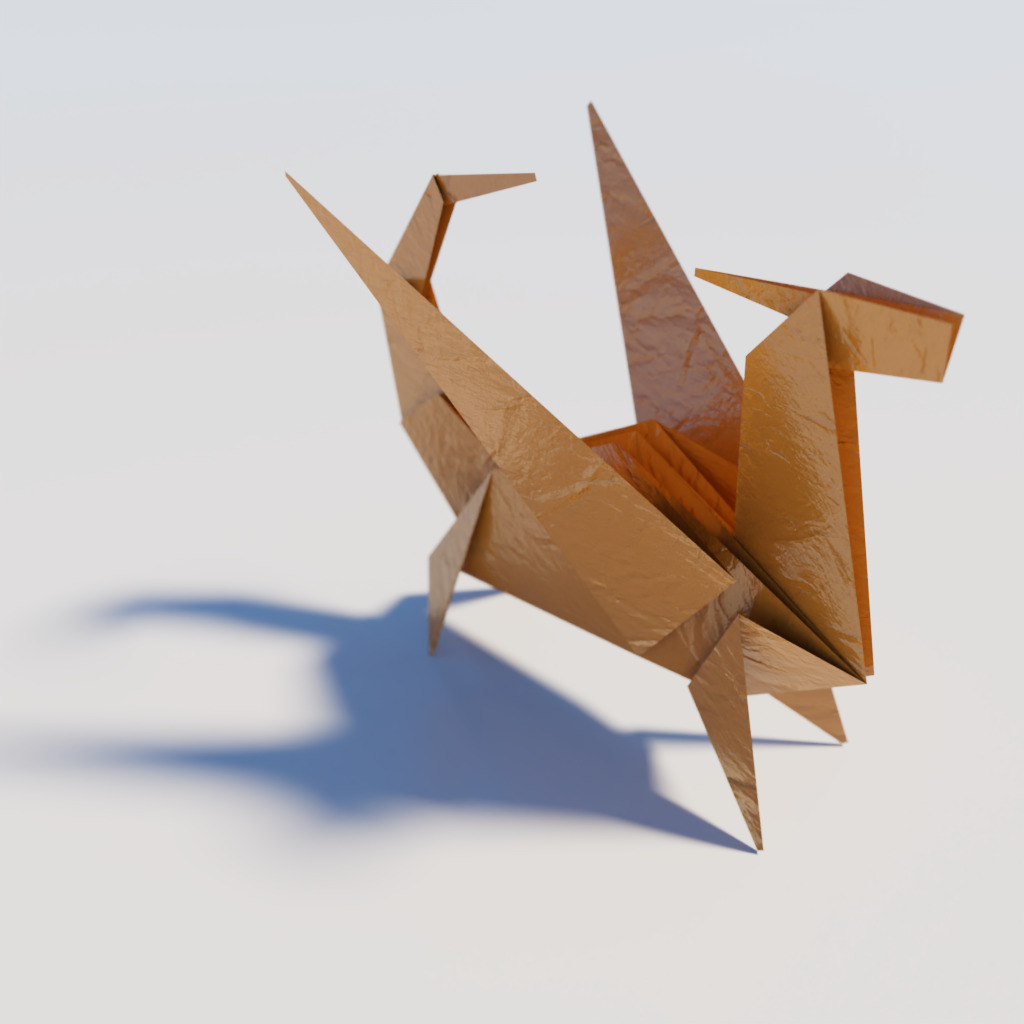}
  \end{subfigure}
  \hfill
  \begin{subfigure}{.19\textwidth}
    \includegraphics[width=\linewidth]{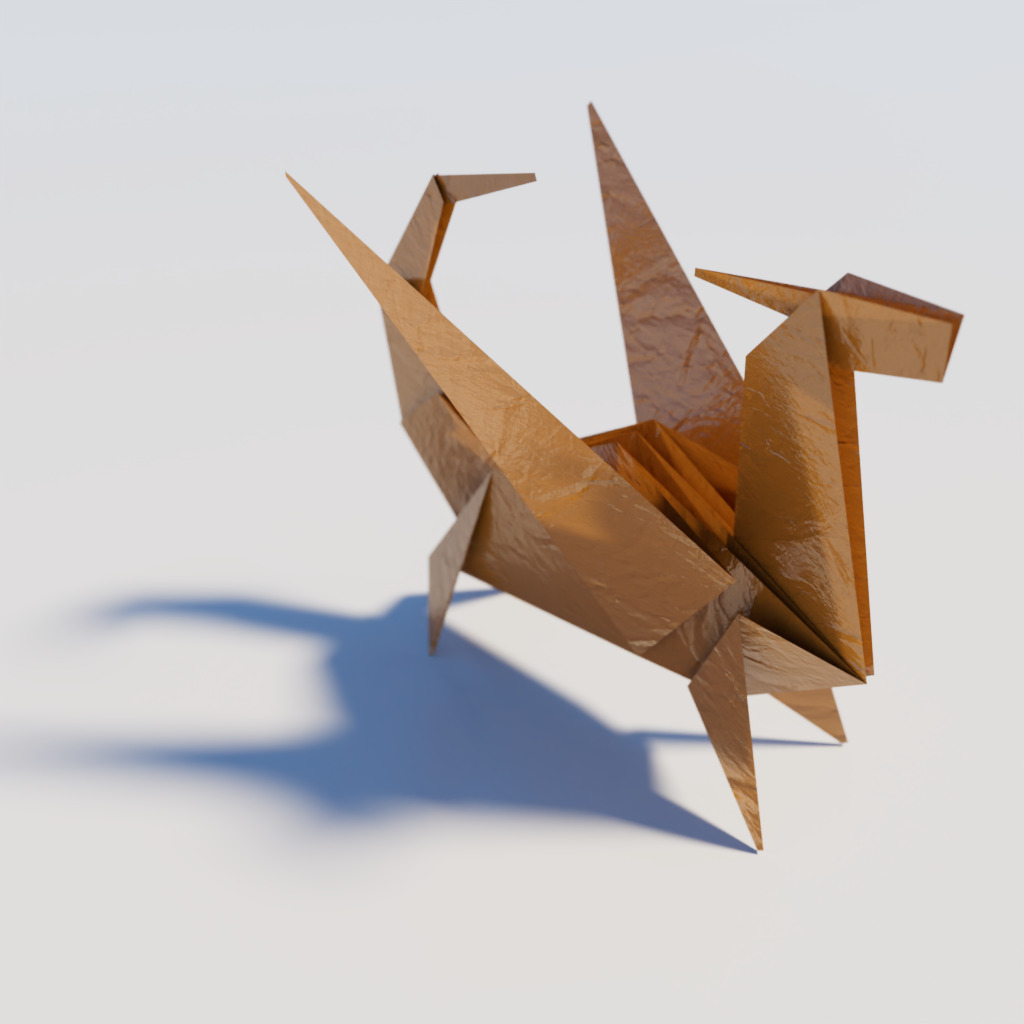}
  \end{subfigure}
  \hfill
  \begin{subfigure}{.19\textwidth}
    \includegraphics[width=\linewidth]{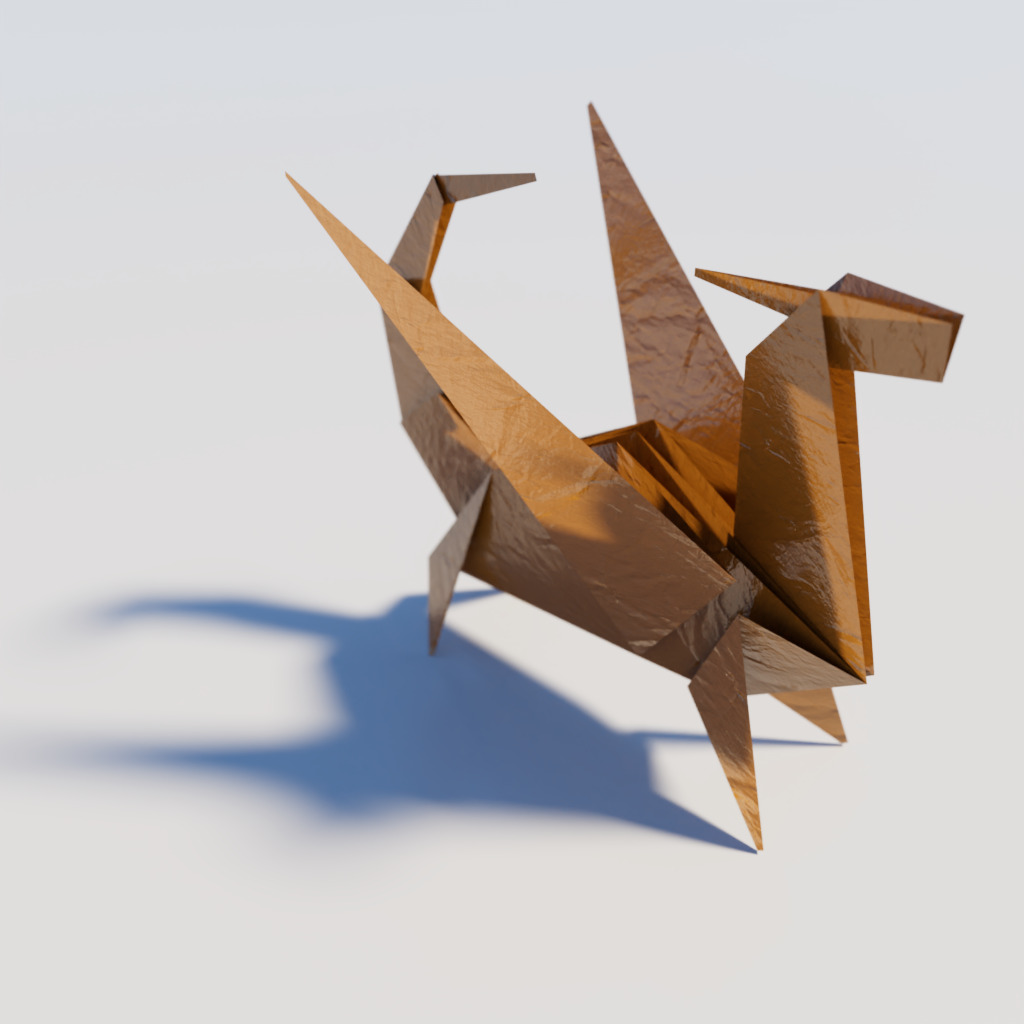}
  \end{subfigure}
  \hfill
    \begin{subfigure}{.19\textwidth}
    \includegraphics[width=\linewidth]{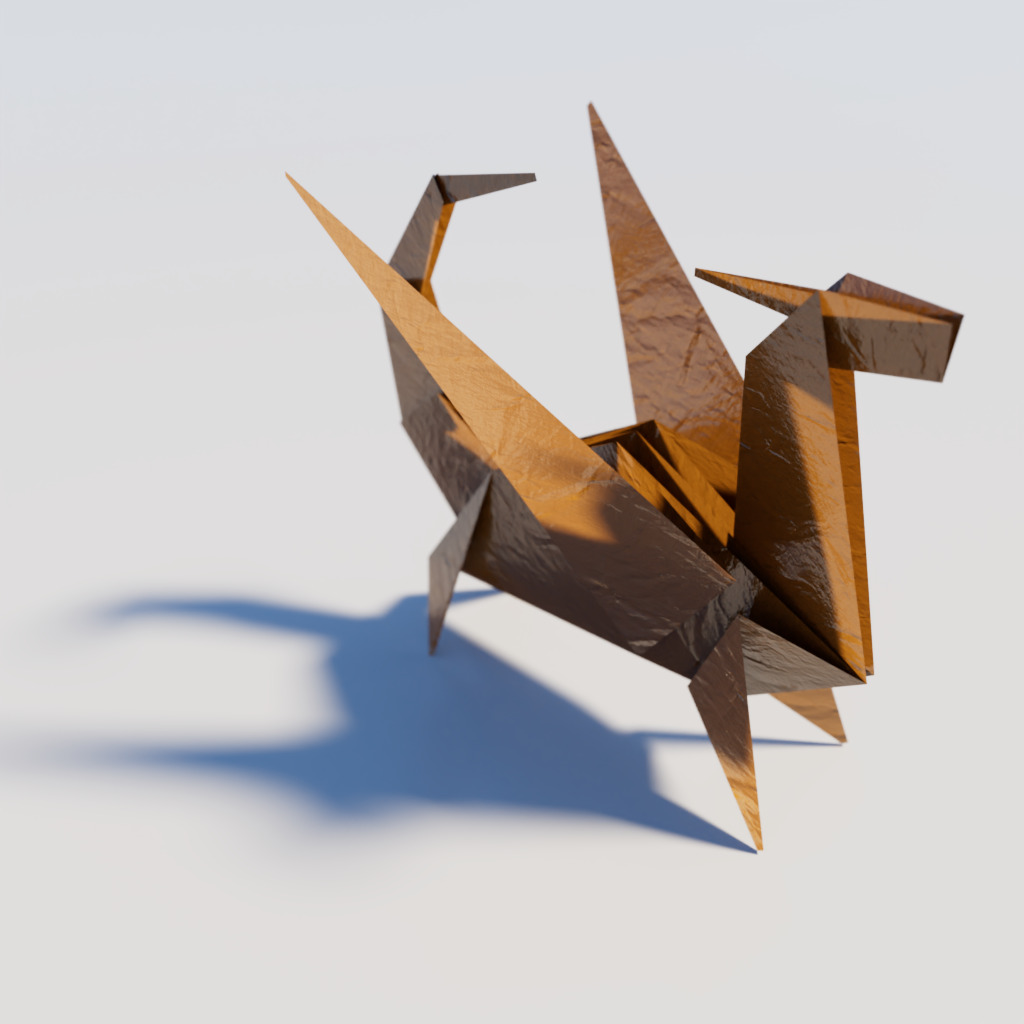}
  \end{subfigure}
  \hfill
  \begin{subfigure}{.19\textwidth}
    \includegraphics[width=\linewidth]{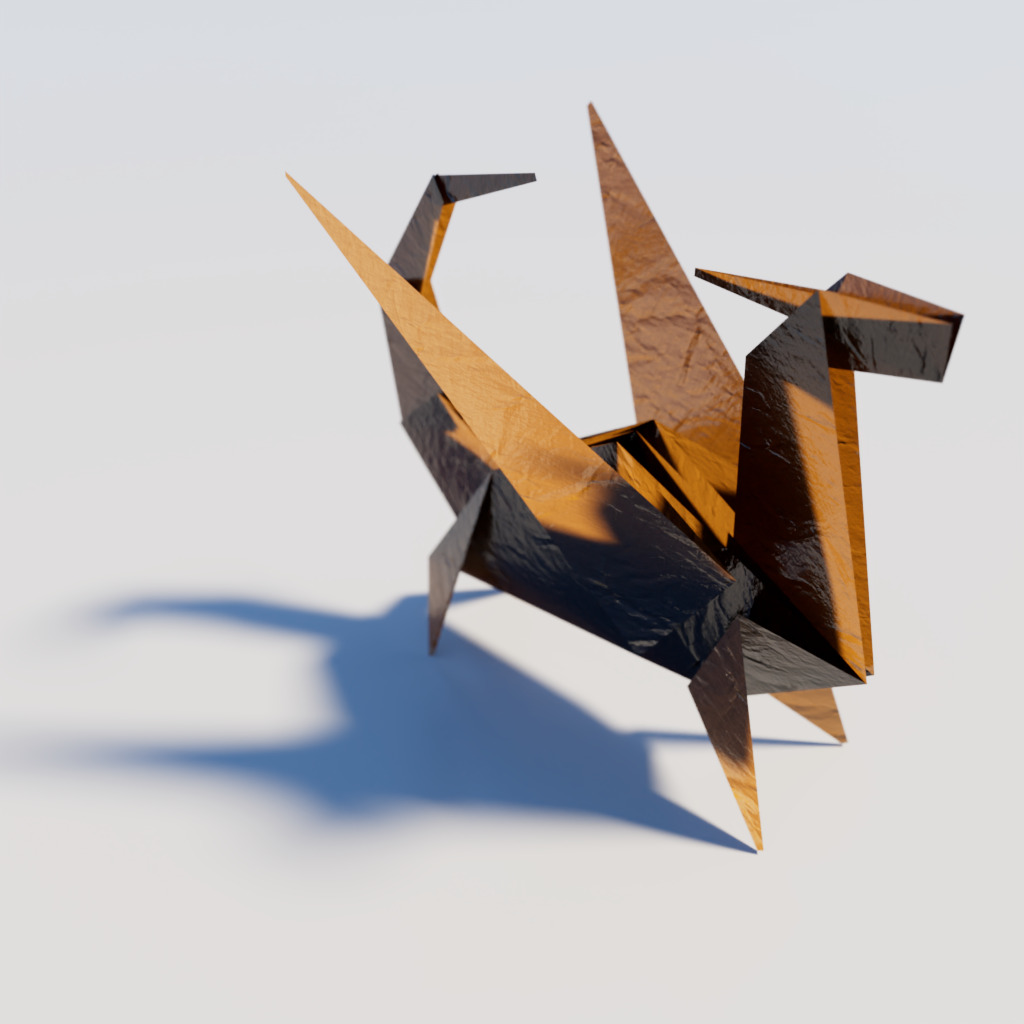}
  \end{subfigure}
  \hfill
  \caption{Varying the \texttt{subsurface\_anisotropy} over -1, -0.5, 0 (default), 0.5, 1. Note that the specular reflection of the embedding dielectric slab is present. \label{fig:thin-wall-subsurface}}
\end{figure}

The $\textrm{subsurface}$ slab is considered to degenerate into an infinitesimally thin sheet of dense scattering material (bounded by dielectric interfaces $f_\mathrm{dielectric}$), which scatters a fraction $S = \texttt{subsurface\_color}$ of the incident light, split between a diffuse reflection lobe $f^R_\mathrm{diffuse}$ and diffuse transmission lobe $f^T_\mathrm{diffuse}$ according to $g = \texttt{subsurface\_scatter\_anisotropy} \in [-1, 1]$. That is, where $f_+$, $f_-$ are albedo 1 diffuse lobes in the positive and negative hemisphere, respectively:
\begin{eqnarray}
      f^R_\mathrm{diffuse} &=& \frac{1}{2} S (1 - g)\, f_+ \ , \nonumber \\
      f^T_\mathrm{diffuse} &=& \frac{1}{2} S (1 + g)\, f_- \ . \label{thin_wall_subsurface}
\end{eqnarray}
This ensures total energy conservation, i.e., the sum of the reflection and transmission albedos is less than 1:
\begin{equation}
    E_R[f^R_\mathrm{diffuse}] + E_T[f^T_\mathrm{diffuse}] = S \le 1 \ .
\end{equation}
At the default of zero anisotropy ($g=0$) the energy is balanced equally between diffuse reflection and transmission.
The diffuse transmission lobe shape (in both hemispheres) is assumed to be controlled by the \verb|base_diffuse_roughness| parameter. Typically the diffuse lobes $f_+$, $f_-$ will be represented by an Oren--Nayar (EON) lobe flipped into the appropriate hemisphere. This model is useful for rendering cases such as light scattering through a thin sheet of paper.

The thin-wall subsurface is interpreted as being embedded with a finite (but small) thickness dielectric slab. Thus a specular lobe is generated as usual due to reflection from this dielectric interface, with the IOR of the slab given by \verb|subsurface_ior|. Similarly to the glossy-diffuse slab, the effective albedo of diffuse base should be unaffected by the physical effect of inter-reflections in the dielectric slab. (This detail needs clarification in the existing OpenPBR specification, and is being improved.)

Figure~\ref{fig:thin-wall-subsurface} shows the effect of varying the \verb|subsurface_anisotropy| parameter on the appearance of a thin-wall subsurface material. The specular reflection from the dielectric slab is visible in all cases, and the diffuse lobes are anisotropic, with the anisotropy increasing from left to right.

\paragraph{Future work: generalized thin-wall model}

\label{sec:thin-walled-generalized}

Currently we treat the thin-wall limit of the translucent volume and the subsurface as two distinct cases (essentially absorbing only and scattering-dominated, respectively).
However, it would be useful to have a more general model where the thin-wall interior medium is a scattering volume with configurable transmittance and albedo. If the scattering is not totally dominant then there is an intermediate regime between ``thin-wall dielectric'' (which roughens due to the refractive boundaries but does not scatter), and ``thin-wall subsurface'' that produces essentially diffuse reflection and transmission lobes. However, at present, there isn't a well-defined model for this intermediate case. Furthermore,  this generalized thin-wall model would make more sense in the context of a unified base volume, which is also considered future work.

It should also be noted that there currently isn't a completely satisfactory model for the case of the described absorbing only thin-wall dielectric with \emph{rough} boundaries, for example the \verb|ThinDielectricBxDF| of PBRT \cite{Pharr2023} simplified to the case of smooth boundaries. There are suggested heuristics \cite{Kulla2017, Belcour2018} but a fully worked out model would be beneficial.

\paragraph{Future work: Two-sided thin-walled materials}

\label{sec:thin-walled-two-sided}

We do not currently explicitly support the case of a thin-walled material where the properties differ on each side, for example a leaf represented by thin-walled SSS with fuzz on the bottom and a coat on top (but no coat on the bottom, and no fuzz on top).

A more general interpretation would be to allow each side of the thin wall to be defined by a distinct thin-walled OpenPBR Surface (one associated with the top, in the direction of the normal, and one with the bottom in the opposite direction) so the material consists of two thin walls sandwiched together, and the light propagates through the whole structure. Then if there are different \hyperref[sec:translucent-base]{translucent-base} or \hyperref[sec:subsurface]{subsurface} parameters on each side, these need to be resolved somehow (e.g., the parameters could be blended, or the light transport could model the presence of the two distinct thin layers of dielectric).

Defining this more general thin-walled model is a topic for future work, and we would like to hear from the community about whether such a more general thin-walled model would be useful for them, or perhaps it is niche enough to not be needed in an uber-shader and a bespoke shader would instead be sufficient.

\clearpage

\enlargethispage{2\baselineskip}

\vspace*{-4\baselineskip}

\thispagestyle{empty}  

\section{Conclusion and Future Work}

In this report, we have given a detailed technical overview of the OpenPBR model, complementary to the official specification available at \url{https://academysoftwarefoundation.github.io/OpenPBR/}. The following future investigations are planned for the OpenPBR model:

\begin{itemize}

  \item Continued interaction with the CG community to help adoption, and gather feedback and suggestions for improvements to the model, ensuring it meets the needs of artists and developers in various industries.

  \item Further alignment and synchronization of the OpenPBR model with the MaterialX standard, ensuring the features required to implement OpenPBR are fully supported in MaterialX.

  \item Completion of the \hyperref[sec:hazy-specular]{hazy-specular} and \hyperref[sec:retro-reflection]{retro-reflection} parameterizations.

  \item Further investigation of the proposed \hyperref[sec:metal_decoupling]{decoupling} of the metallic and dielectric parameters to avoid certain artifacts.

  \item Development of a more comprehensive \hyperref[sec:fuzz:generalized]{generalized} fuzz model that accounts for varying optical depths and anisotropic scattering. Also, further work on developing more accurate physically based heuristics for \hyperref[sec:fuzz:roughening]{roughening} effects (for both coat and fuzz).

  \item Thin-walled materials: further investigation of a \hyperref[sec:thin-walled-generalized]{generalized thin-wall} model, and possible support for \hyperref[sec:thin-walled-two-sided]{two-sided thin-walled} materials.

  \item There is interest in adding explicit support for a model for rendering glints \cite{Jakob2014b,Yan2014} or flakes \cite{Atanasov2016}, which are important effects for certain uses cases, such as car paint.

  \item Development (currently in collaboration with \textcite{Palmqvist2025}) of a set of example materials suitable as presets, and for demonstration of the model's capabilities.

  \item More work on facilitating translation of the OpenPBR model to and from other shading models (to provide an upgrade path, or to map the model to existing shading models more suitable for certain use cases).

  \item Further elucidation and exploration of the use of the model in real-time rendering applications, such as video games or interactive simulations, to achieve high-quality visuals while maintaining performance.

  \item Potential extension of the effort to include non-surface material models in need of standardization, such as volumetric materials or hair.

\end{itemize}

\section{Acknowledgements}

We thank our fellow co-authors of the OpenPBR specification for extensive discussions over the past three years: Zap Andersson, Paul Edmondson, Julien Guertault, Adrien Herubel, Alan King, Andréa Machizaud, Frédéric Servant and Jonathan Stone.

We would also like to thank the following individuals for their valuable feedback and contributions: François Beaune, Henrik Edstrom, Eugene d'Eon, Jerry Gamache, Iliyan Georgiev, Dhruv Govil, Lee Griggs, Larry Gritz, Niklas Harrysson, Miloš Hašan, Thiago Ize, Lee Kerley, Chris Kulla, Vladimir Koylazov, Bernard Kwok, Anders Langlands, Frankie Liu, Thomas Makryniotis, Thomas Mansencal, Arnon Marcus, André Mazzone, Nikie Monteleone, Michael Nickelsky, Anton Palmqvist, Nick Porcino, Guido Quaroni, Nathan Reed, Anthony Salvi, Brian Sharpe, Rob Slater, Lukas Stockner, Masuo Suzuki, Brecht van Lommel, Andrea Weidlich and Nicolas Wirrmann.

The ``shader playground'' renders shown in Figure~\ref{fig:render_comparison_teaser} and Figure~\ref{fig:emission_example} were created by Nikie Monteleone.

\clearpage


\nocite{*}
\printbibliography

\appendix

\clearpage

\vspace*{-5\baselineskip}

\section{Parameterization}

\thispagestyle{plain} 

\label{sec:parameters}

Here we provide a summary of the parameters of the OpenPBR layered material model, including their types, ranges, and default values.

\rowcolors{2}{gray!10}{white}
\renewcommand{\arraystretch}{1.2}

\begin{footnotesize}
\begin{longtable*}[!t]{@{}>{\raggedright\arraybackslash}p{5.5cm} >{\raggedright\arraybackslash}p{1.5cm} >{\raggedright\arraybackslash}p{1.5cm} >{\raggedright\arraybackslash}p{1.5cm} >{\raggedright\arraybackslash}p{3cm}@{}}
\toprule
\textbf{Parameter} & \textbf{Type} & \textbf{Range} & \textbf{Default} & \textbf{Units} \\
\midrule
\endhead
\vspace{-\normalbaselineskip} 
\label{table:parameters}

\verb|base_weight| & float & $[0,1]$ & $1$ & - \\
\verb|base_color| & color3 & $[0,1]^3$ & $(0.8,\ 0.8,\ 0.8)$ & - \\
\verb|base_metalness| & float & $[0,1]$ & $0$ & - \\
\verb|base_diffuse_roughness| & float & $[0,1]$ & $0$ & - \\

\verb|specular_weight| & float & $[0,\infty]$ & $1$ & - \\
\verb|specular_color| & color3 & $[0,1]^3$ & $(1,\ 1,\ 1)$ & - \\
\verb|specular_roughness| & float & $[0,1]$ & $0.3$ & - \\
\verb|specular_roughness_anisotropy| & float & $[0,1]$ & $0$ & - \\
\verb|specular_ior| & float & $[1,\ 3]$ & $1.5$ & - \\

\verb|transmission_weight| & float & $[0,1]$ & $0.0$ & - \\
\verb|transmission_color| & color3 & $[0,1]^3$ & $(1,\ 1,\ 1)$ & - \\
\verb|transmission_depth| & float & $[0,\infty)$ & $0$ & scene units (e.g., meters) \\
\verb|transmission_scatter| & color3 & $[0,1]^3$ & $(0,\ 0,\ 0)$ & - \\
\verb|transmission_scatter_anisotropy| & float & $[-1,1]$ & $0$ & - \\
\verb|transmission_dispersion_scale| & float & $[0,1]$ & $0$ & - \\
\verb|transmission_dispersion_abbe_number| & float & $[0,\infty)$ & $20$ & - \\

\verb|subsurface_weight| & float & $[0,1]$ & $0$ & - \\
\verb|subsurface_color| & color3 & $[0,1]^3$ & $(0.8,\ 0.8,\ 0.8)$ & - \\
\verb|subsurface_radius| & float & $[0,\infty)$ & $1$ & scene units (e.g., meters) \\
\verb|subsurface_radius_scale| & color3 & $[0,1]^3$ & $(1,\ 1,\ 1)$ & - \\
\verb|subsurface_anisotropy| & float & $[-1,1]$ & $0$ & - \\

\verb|coat_weight| & float & $[0,1]$ & $0$ & - \\
\verb|coat_color| & color3 & $[0,1]$ & $(1,\ 1,\ 1)$ & - \\
\verb|coat_roughness| & float & $[0,1]$ & $0$ & - \\
\verb|coat_ior| & float & $[1,\ 3]$ & $1.6$ & - \\
\verb|coat_darkening| & float & $[0,1]$ & $1$ & - \\

\verb|fuzz_weight| & float & $[0,1]$ & $0.0$ & - \\
\verb|fuzz_color| & color3 & $[0,1]$ & $(1.0,\ 1.0,\ 1.0)$ & - \\
\verb|fuzz_normal| & normal & — & — & - \\

\verb|emission_weight| & float & $[0,1]$ & $0$ & - \\
\verb|emission_color| & color3 & $[0,\infty)^3$ & $(1, 1, 1)$ & - \\
\verb|emission_luminance| & float & $[0,\infty)$ & $1000$ & nits (cd/m$^2$) \\

\verb|thin_film_weight| & float & $[0,1]$ & $0$ & - \\
\verb|thin_film_thickness| & float & $[0,1]$ & $0.5$ & micrometers ($\mu$m) \\
\verb|thin_film_ior| & float & $[1.0,\ 3.0]$ & $1.4$ & - \\

\verb|geometry_opacity| & float & $[0,1]$ & $1.0$ & - \\
\verb|geometry_thin_walled| & bool & \{true, false\} & false & - \\
\verb|geometry_normal| & vector3 & — & — & - \\
\verb|geometry_tangent| & vector3 & — & — & - \\
\verb|geometry_coat_normal| & vector3 & — & — & - \\
\verb|geometry_coat_tangent| & vector3 & — & — & - \\

\bottomrule
\end{longtable*}
\end{footnotesize}


\clearpage

\section{A mixture model example}

\label{sec:mixture_model}

In various practical models such as Disney's ``Principled'' shader \cite{Burley2012}, Autodesk 3ds Max's Physical Material \cite{Andersson2016} and Autodesk Standard Surface \cite{Georgiev2019}, the model is defined as a mixture (i.e., linear combination) of BSDF/BSSRDFs corresponding to the constituent BSDF lobes of the reflection and transmission. The light transport between the layers and overall energy balance is approximated via the mix weights. Given such a representation, the integration into a renderer is relatively straightforward as the individual lobes can be importance sampled according to their albedos and combined with direct lighting estimates via standard techniques such as multiple importance sampling (MIS) \cite{Veach1998}.

As an example, we will give a brief derivation of a mixture model representation analogous to Autodesk Standard Surface, from the stated material structure of OpenPBR. Following Autodesk Standard Surface, we assume here that layering is implemented via the non-reciprocal albedo-scaling of Equation~\ref{non-reciprocal-albedo-scaling}.

Consider first the non-thin-walled case (i.e., \verb|geometry_thin_walled| is false).
For brevity, in the following we suppress all the direction arguments, and use the notation of the tree diagram in the \hyperref[sec:model-structure]{Model} section for the weight factors, i.e.:
\begin{eqnarray}
\mathtt{\alpha} &=& \mathtt{geometry\_opacity}     \nonumber \\
\mathtt{F}      &=& \mathtt{fuzz\_weight}          \nonumber \\
\mathtt{C}      &=& \mathtt{coat\_weight}          \nonumber \\
\mathtt{M}      &=& \mathtt{base\_metalness}       \nonumber \\
\mathtt{T}      &=& \mathtt{transmission\_weight}  \nonumber \\
\mathtt{S}      &=& \mathtt{subsurface\_weight}    \nonumber \\
\end{eqnarray}

The base substrate is a mix, which can be mapped to a BSDF as follows (where the $\mathcolor{Blue}{\mathrm{blue}}$ color indicates a primitive BSDF lobe):
\begin{equation}
f_\textrm{base-substrate} = \mathrm{lerp}\left(f_\textrm{dielectric-base}, \mathcolor{Blue}{f_\mathrm{conductor}}, \mathtt{M}\right) \ .
\end{equation}
Similarly, the dielectric-base mix can be written as
\begin{equation}
f_\textrm{dielectric-base} = \mathrm{lerp}\left(f_\textrm{opaque-base}, f_\textrm{translucent-base}, \mathtt{T}\right) \ ,
\end{equation}
and the opaque-base mix as
\begin{equation}
f_\textrm{opaque-base} = \mathrm{lerp}\left(f_\textrm{glossy-diffuse}, f_\textrm{subsurface}, \mathtt{S}\right) \ .
\end{equation}
All of $f_\textrm{translucent-base}$, $f_\textrm{subsurface}$ and $f_\textrm{glossy-diffuse}$ represent the BSDF of a microfacet dielectric interface bounding the dielectric interior and its volumetric media.
In each case, the BSDF can be represented as the sum of a ``primary specular'' BRDF $\mathcolor{Blue}{f^R_\textrm{specular}}$ corresponding to reflection from the dielectric surface without interaction with the internal medium, and a substrate lobe corresponding to the effect of transmission into and scattering within the medium:
\begin{align}
f_\textrm{translucent-base} &= \mathcolor{Blue}{f^R_\textrm{specular}}  + (1 - E[\mathcolor{Blue}{f^R_\textrm{specular}}]) \,\mathcolor{Blue}{f^T_\textrm{specular}}     \ , \nonumber \\
f_\textrm{subsurface}       &= \mathcolor{Blue}{f^R_\textrm{specular}}  + (1 - E[\mathcolor{Blue}{f^R_\textrm{specular}}]) \,\mathcolor{Blue}{f_\textrm{SSS}}            \ , \nonumber \\
f_\textrm{glossy-diffuse}   &= \mathcolor{Blue}{f^R_\textrm{specular}}  + (1 - E[\mathcolor{Blue}{f^R_\textrm{specular}}]) \,\mathcolor{Blue}{f_\mathrm{diffuse}}        \ ,
\end{align}
where, as described in the \hyperref[sec:layer_formalism]{Slabs} section, $E[f_X]$ denotes the directional albedo of $f_X$.

Here, the substrate lobes $\mathcolor{Blue}{f^T_\textrm{specular}}$ and $\mathcolor{Blue}{f_\textrm{SSS}}$ are technically BSSRDFs, which model the entry into the internal medium via the dielectric interface, transport of light from entry point to exit points including absorption and scattering processes, and exit from the medium back though the interface, generating both a reflection and a transmission component.
The ``specular'' BTDF/BSSRDF $\mathcolor{Blue}{f^T_\textrm{specular}}$ corresponds to transmission into the medium parameterized in the \hyperref[sec:translucent-base]{Translucent base} section, and BSSRDF $\mathcolor{Blue}{f_\textrm{SSS}}$ corresponds to transmission into the medium parameterized in the \hyperref[sec:subsurface]{Subsurface} section. In the case of $f_\textrm{glossy-diffuse}$, the BSSRDF degenerates into the BRDF $\mathcolor{Blue}{f_\mathrm{diffuse}}$ as described in the \hyperref[sec:glossy-diffuse]{Glossy-diffuse} section.

Note that in this albedo-scaling approximation, the transmission Fresnel factor associated with $\mathcolor{Blue}{f^T_\textrm{specular}}$ and $f_\textrm{SSS}$ can be \emph{omitted} as the energy conservation of the dielectric BSDF as a whole is maintained automatically, even without explicit multiple-scattering compensation or in the presence of modifications to the reflection Fresnel factor via \verb|specular_color|.

Since $\mathcolor{Blue}{f^R_\textrm{specular}}$ appears in each of the three component slabs of the \hyperref[sec:dielectric-base]{dielectric base}, it follows that on collecting terms, $f_\textrm{dielectric-base}$ reduces to
\begin{equation}
f_\textrm{dielectric-base} = \mathcolor{Blue}{f^R_\textrm{specular}}  + (1 - E[\mathcolor{Blue}{f^R_\textrm{specular}}]) \,f^T_\mathrm{dielectric-base} \ ,
\end{equation}
where $f^T_\mathrm{dielectric-base}$, the total effective transmission lobe of the dielectric base, is
\begin{align}
f^T_\mathrm{dielectric-base} &= \mathtt{T} \mathcolor{Blue}{f^T_\textrm{specular}} + (1 - \mathtt{T}) \left(\mathtt{S} \,\mathcolor{Blue}{f_\textrm{SSS}}  + (1 - \mathtt{S}) \,\mathcolor{Blue}{f_\mathrm{diffuse}}\right) \nonumber \\
                              &= \mathrm{lerp}\left(\mathrm{lerp}(\mathcolor{Blue}{f_\mathrm{diffuse}}, \mathcolor{Blue}{f_\textrm{SSS}}, \mathtt{S}), \mathcolor{Blue}{f^T_\textrm{specular}}, \mathtt{T}\right) \ .
\end{align}

Next, the coat is layered on top of the base substrate with the coverage weight $\mathtt{C}$, where the BRDF of the coat dielectric interface is taken to be $\mathcolor{Blue}{f_\mathrm{coat}}$, with a transmittance $\mathbf{T}_\mathrm{coat}$ (applied both on entry and exit). As in Equation~\ref{coat_layering_formula_with_albedo_scaling}, this can be expressed as
\begin{equation}
f_\mathrm{coated-base} = \mathtt{C} \,\mathcolor{Blue}{f_\mathrm{coat}} +  \mathrm{lerp}\left(1, \mathbf{T}^2_\mathrm{coat} (1 - E[\mathcolor{Blue}{f_\mathrm{coat}}]), \mathtt{C}\right) f_\textrm{base-substrate} \ .
\end{equation}

Similarly, the fuzz layer is applied with coverage weight $\mathtt{F}$ (with albedo-scaling adjusted to account for the gray transmission according to Equation~\ref{fuzz-layering-approx}), producing
\begin{equation}
f_\mathrm{surface} = \mathtt{F} \,\mathcolor{Blue}{f_\mathrm{fuzz}} +  \mathrm{lerp}\left(1, 1 - E[\mathcolor{Blue}{\overline{f_\mathrm{fuzz}}}], \mathtt{F}\right) f_\textrm{coated-base} \ ,
\end{equation}
where $\mathcolor{Blue}{\overline{f_\mathrm{fuzz}}}$ is the fuzz BRDF with the tint color set to white.

Finally, the opacity mix operation is applied, producing
\begin{equation}
f_\mathrm{PBR} = \mathtt{\alpha} \,f_\mathrm{surface}  + (1 - \mathtt{\alpha}) \,\mathcolor{Blue}{f_\textrm{transparent}} \ ,
\end{equation}
where $\mathcolor{Blue}{f_\textrm{transparent}}$ is understood to denote a delta-function BSDF corresponding to the absence of any surface interaction.

If $\mathtt{E}$ represents the isotropic emission luminance from the base, then the total EDF lobe $\mathcolor{Blue}{L_e}$ can be modeled according to the absorption in the coat layer as
\begin{eqnarray}
\mathcolor{Blue}{L_e}  &=& (1 - \mathtt{C}) \,\mathtt{E} + \mathtt{C} \,\mathbf{T}^2_\mathrm{coat} \mathtt{E}  \nonumber \\
                        &=& \mathrm{lerp}\left(1, \mathbf{T}^2_\mathrm{coat}, \mathtt{C}\right) \mathtt{E} \ ,
\end{eqnarray}
which, in principle, is a function of direction due to the varying transmittance. This can also be thought of as a lobe in its own right (representing light self-emitted, rather than reflected or transmitted).

To summarize, we have thus expressed the model as the following linear combination of component BRDF/BTDF/BSSRDF lobes (and a separate EDF lobe):
\begin{equation}
\begin{aligned}
            \; f_\mathrm{PBR}               &=&                                                     &\mathrm{lerp}\left(\mathcolor{Blue}{f_\textrm{transparent}},                              f_\mathrm{surface}          ,      \mathtt{\alpha}\right)                           \ ,   \; \nonumber \\
            \; f_\mathrm{surface}           &=& \mathtt{F} \,\mathcolor{Blue}{f_\mathrm{fuzz}}    + &\mathrm{lerp}\left(1,                              \quad\quad\; 1 - E[\mathcolor{Blue}{\overline{f_\mathrm{fuzz}}}]   ,    \;\mathtt{F}     \right) f_\textrm{coated-base}    \ ,   \; \nonumber \\
            \; f_\mathrm{coated-base}       &=& \mathtt{C} \,\mathcolor{Blue}{f_\mathrm{coat}}    + &\mathrm{lerp}\left(1,               \mathbf{T}^2_\mathrm{coat} (1 - E[\mathcolor{Blue}{f_\mathrm{coat}}])             ,      \mathtt{C}     \right) f_\textrm{base-substrate} \ ,   \; \nonumber \\
            \; \mathcolor{Blue}{L_e}        &=&                                                     &\mathrm{lerp}\left(1,                                                                        T_\mathrm{coat}          ,      \mathtt{C}     \right) \mathtt{E}                \ ,   \; \nonumber \\
            \; f_\textrm{base-substrate}    &=&                                                     &\mathrm{lerp}\left(f_\textrm{dielectric-base}, \mathcolor{Blue}{f_\mathrm{conductor}}                                 ,      \mathtt{M}     \right)                           \ ,   \; \nonumber \\
            \; f_\textrm{dielectric-base}   &=& \mathcolor{Blue}{f^R_\textrm{specular}}           + &(1 - E[\mathcolor{Blue}{f^R_\textrm{specular}}]) f^T_\mathrm{dielectric-base}                                                                                                 \ ,   \; \nonumber \\
            \; f^T_\mathrm{dielectric-base} &=&                                                    &\mathrm{lerp}(\mathrm{lerp}(\mathcolor{Blue}{f_\mathrm{diffuse}}, \mathcolor{Blue}{f_\textrm{SSS}}, \mathtt{S}), \mathcolor{Blue}{f^T_\textrm{specular}}, \mathtt{T})          \ .   \;
\end{aligned}
\end{equation}

Where the component lobes are listed below:

\begin{table}[h]
\centering
\begin{tabular}{|l|c|l|l|}
\hline
\textbf{Lobe name} & \textbf{Lobe symbol} & \textbf{Description} & \textbf{Parameters} \\
\hline
Transparency           & $\color{darkblue}{f_\textrm{transparent}}$  & pass-through (delta BTDF)  & as in the \hyperref[sec:geometry]{Geometry} section   \\
Coating                & $\color{darkblue}{f_\mathrm{coat}}$         & coat BRDF                  & as in the \hyperref[sec:coat]{Coat} section                        \\
Emission               & $\color{darkblue}{L_e}$                     & emission EDF               & as in the \hyperref[sec:emission]{Emission} section         \\
Metal                  & $\color{darkblue}{f_\mathrm{conductor}}$    & metal BRDF                 & as in the \hyperref[sec:metallic-base]{Metal} section              \\
Specular reflection    & $\color{darkblue}{f^R_\textrm{specular}}$   & specular BRDF              & as in the \hyperref[sec:dielectric-base]{Dielectric base} section  \\
Specular transmission  & $\color{darkblue}{f^T_\textrm{specular}}$   & specular BTDF/BSSRDF       & as in the \hyperref[sec:translucent-base]{Translucent base} section \\
Fuzz                   & $\color{darkblue}{f_\mathrm{fuzz}}$         & fuzz BRDF                  & as in the \hyperref[sec:fuzz]{Fuzz} section                        \\
Subsurface scattering  & $\color{darkblue}{f_\textrm{SSS}}$          & subsurface BSSRDF          & as in the \hyperref[sec:subsurface]{Subsurface} section            \\
Diffuse reflection     & $\color{darkblue}{f_\mathrm{diffuse}}$      & diffuse BRDF               & as in the \hyperref[sec:glossy-diffuse]{Glossy-diffuse} section    \\
\hline
\end{tabular}
\caption{Primitive BSDF/BSSRDF/EDF lobes in the OpenPBR mixture model.}
\end{table}

\clearpage

\section{Similarity theory of subsurface anisotropy}

\label{sec:similarity_theory_subsurface}

For completeness, we give here a brief derivation of the similarity relation of Equation~\ref{ss_albedo_anisotropic} for the scattering albedo $\boldsymbol{\alpha}$ of an anisotropic medium, given the isotropic medium scattering albedo $\boldsymbol{\alpha}^\star$, extinction coefficient of the anisotropic medium $\boldsymbol{\mu}_t$ and anisotropy $g$. This formulation allows the mapping between single and multi-scatter albedo to be determined for the case of an isotropic medium, then adapted to the case of any anisotropic medium.

As noted, there is the following ``similarity relation'' \cite{Zhao2014, Hyperion} between the scattering coefficients in media with different anisotropies $g$ and $g^\star$:
\begin{equation} \label{similarity_relation_albedo}
\boldsymbol{\mu}_s (1 - g) = \boldsymbol{\mu}^\star_s (1 - g^\star) \ ,
\end{equation}
where all starred quantities refer to the isotropic medium, and unstarred quantities refer to the anisotropic medium.
We can set $g^\star = 0$ (i.e., corresponding to the isotropic medium), and $g$ is the anisotropy of the medium we are interested in. The absorption coefficients of the two media can be assumed to be equal (i.e., $\boldsymbol{\mu}_a = \boldsymbol{\mu}^\star_a$).

The scattering albedo $\boldsymbol{\alpha}$ of the anisotropic medium can then be computed as follows. From the definition of the extinction coefficients in the isotropic and anisotropic cases, we have:
\begin{eqnarray}
\boldsymbol{\mu}_t       &=& \boldsymbol{\mu}_a       + \boldsymbol{\mu}_s \ , \\
\boldsymbol{\mu}^\star_t &=& \boldsymbol{\mu}^\star_a + \boldsymbol{\mu}^\star_s \ .
\end{eqnarray}
Subtracting the two equations, since the absorption coefficients are equal we have
\begin{equation}
\boldsymbol{\mu}_t - \boldsymbol{\mu}^\star_t = \boldsymbol{\mu}_s - \boldsymbol{\mu}^\star_s \ .
\end{equation}
Applying $\boldsymbol{\mu}_s = \boldsymbol{\alpha} \boldsymbol{\mu}_t$ and $\boldsymbol{\mu}^\star_s = \boldsymbol{\alpha}^\star \boldsymbol{\mu}^\star_t$:
\begin{equation} \label{similarity_relation_albedo2}
1 - \boldsymbol{\alpha} = (1 - \boldsymbol{\alpha}^\star) \frac{\boldsymbol{\mu}^\star_t}{\boldsymbol{\mu}_t} \ .
\end{equation}

Now, applying the similarity relation of Equation~\ref{similarity_relation_albedo}, we have
\begin{align*}
\boldsymbol{\mu}_t &= \boldsymbol{\mu}_a + \boldsymbol{\mu}_s \\
                   &= \boldsymbol{\mu}_a^\star + \boldsymbol{\mu}_s^\star / (1-g) \\
                   &= \boldsymbol{\mu}_t^\star \left[1 - \boldsymbol{\alpha}^\star + \frac{\boldsymbol{\alpha}^\star}{1-g}\right] \ .
\end{align*}

This allows the unknown $\boldsymbol{\mu}_t^\star$ to be eliminated from Equation~\ref{similarity_relation_albedo2},  and thus solving for $\alpha$ we obtain Equation~\ref{ss_albedo_anisotropic}, the relation between the scattering albedos of the anisotropic medium ($\boldsymbol{\alpha}$) and the isotropic medium ($\boldsymbol{\alpha}^\star$):
\[
\boldsymbol{\alpha} = \frac{\boldsymbol{\alpha}^\star}{1 - g(1 - \boldsymbol{\alpha}^\star)} \ .
\]
This allows the single-scattering albedo, which generates the required observed color $\mathbf{E}_\mathrm{multi-scatter}$, to be determined in the isotropic case (via a variety of formulas, such as Equation~\ref{hyperion_fits}), and then adapted to general anisotropy via Equation~\ref{ss_albedo_anisotropic}.

\clearpage

\section{F82-tint model fits for a variety of real metals}

\label{sec:f82_fits}

We present here fits to the F82-tint model used for the \hyperref[sec:metallic-base]{metallic} slab.
These were computed using publicly available tabulated IOR data \cite{Polyanskiy2023}.
Example programs for computing these fits can be found here:
\begin{itemize}
  \item \url{https://github.com/portsmouth/F82-tint-generator/}
  \item \url{https://github.com/peterkutz/metal-colors}
  \item \url{https://github.com/natyh/material-params}
\end{itemize}

\vspace{0.5cm}

\resizebox{\columnwidth}{!}{
\begin{tabular}{lcccccc}
\rowcolor{lightgray}
Metal & F0 (sRGB) & F0 (ACEScg) & F82-tint (sRGB) & F82-tint (ACEScg) & F0 \hspace{0.5cm} (sRGB) & F82-tint (sRGB) \\
aluminium &  0.916  0.923  0.924 &  0.918  0.922  0.923 &  0.91   0.936  0.959 &  0.921  0.934  0.955 & \cellcolor[RGB]{233,235,235} & \cellcolor[RGB]{232,238,244} \\
beryllium &  0.539  0.533  0.534 &  0.537  0.534  0.534 &  0.731  0.738  0.755 &  0.734  0.738  0.752 & \cellcolor[RGB]{137,136,136} & \cellcolor[RGB]{186,188,192} \\
brass	& 0.962 0.713 0.464	& 0.857 0.728 0.502	& 0.971 0.994 1.019	& 0.979 0.992 1.015 & \cellcolor[RGB]{245,181,118} & \cellcolor[RGB]{247,253,255} \\
caesium &  0.702  0.555  0.256 &  0.633  0.561  0.299 &  1.087  1.18   1.44  &  1.128  1.175  1.394 & \cellcolor[RGB]{178,141,65} & \cellcolor[RGB]{255,255,255} \\
chromium &  0.654  0.685  0.701 &  0.666  0.682  0.698 &  0.688  0.728  0.798 &  0.706  0.726  0.788 & \cellcolor[RGB]{166,174,178} & \cellcolor[RGB]{175,185,203} \\
cobalt &  0.699  0.704  0.671 &  0.699  0.703  0.676 &  0.727  0.772  0.823 &  0.747  0.77   0.815 & \cellcolor[RGB]{178,179,171} & \cellcolor[RGB]{185,196,209} \\
copper &  0.932  0.623  0.522 &  0.811  0.643  0.542 &  0.982  0.947  0.945 &  0.97   0.95   0.946 & \cellcolor[RGB]{237,158,133} & \cellcolor[RGB]{250,241,240} \\
germanium &  0.5    0.517  0.465 &  0.504  0.515  0.472 &  0.62   0.653  0.701 &  0.635  0.651  0.694 & \cellcolor[RGB]{127,131,118} & \cellcolor[RGB]{158,166,178} \\
gold &  1.059  0.773  0.307 &  0.929  0.788  0.374 &  0.971  1.018  0.994 &  0.987  1.013  0.997 & \cellcolor[RGB]{255,197,78} & \cellcolor[RGB]{247,255,253} \\
iridium &  0.745  0.734  0.704 &  0.739  0.734  0.709 &  0.759  0.781  0.81  &  0.768  0.78   0.806 & \cellcolor[RGB]{189,187,179} & \cellcolor[RGB]{193,199,206} \\
iron &  0.53   0.513  0.494 &  0.523  0.514  0.497 &  0.765  0.767  0.802 &  0.767  0.768  0.797 & \cellcolor[RGB]{135,130,126} & \cellcolor[RGB]{195,195,204} \\
lead &  0.626  0.64   0.693 &  0.633  0.64   0.686 &  0.758  0.773  0.799 &  0.765  0.772  0.795 & \cellcolor[RGB]{159,163,176} & \cellcolor[RGB]{193,197,203} \\
lithium &  0.916  0.89   0.807 &  0.902  0.89   0.819 &  0.985  0.998  1.027 &  0.991  0.998  1.023 & \cellcolor[RGB]{233,226,205} & \cellcolor[RGB]{251,254,255} \\
magnesium &  0.956  0.953  0.95  &  0.955  0.953  0.951 &  0.954  0.964  0.977 &  0.958  0.963  0.975 & \cellcolor[RGB]{243,243,242} & \cellcolor[RGB]{243,245,249} \\
manganese &  0.606  0.592  0.573 &  0.6    0.592  0.576 &  0.796  0.834  0.889 &  0.813  0.832  0.88  & \cellcolor[RGB]{154,150,146} & \cellcolor[RGB]{202,212,226} \\
mercury &  0.781  0.78   0.778 &  0.781  0.78   0.779 &  0.813  0.852  0.902 &  0.83   0.85   0.895 & \cellcolor[RGB]{199,198,198} & \cellcolor[RGB]{207,217,230} \\
molybdenum &  0.589  0.612  0.594 &  0.597  0.61   0.596 &  0.683  0.696  0.726 &  0.689  0.695  0.721 & \cellcolor[RGB]{150,155,151} & \cellcolor[RGB]{174,177,185} \\
nickel &  0.697  0.641  0.563 &  0.672  0.643  0.575 &  0.815  0.834  0.871 &  0.824  0.833  0.865 & \cellcolor[RGB]{177,163,143} & \cellcolor[RGB]{207,212,222} \\
palladium &  0.734  0.704  0.662 &  0.721  0.705  0.669 &  0.811  0.836  0.872 &  0.822  0.835  0.866 & \cellcolor[RGB]{187,179,168} & \cellcolor[RGB]{206,213,222} \\
platinum &  0.765  0.73   0.676 &  0.749  0.732  0.684 &  0.793  0.815  0.84  &  0.802  0.814  0.836 & \cellcolor[RGB]{194,186,172} & \cellcolor[RGB]{202,207,214} \\
potassium &  0.983  0.956  0.906 &  0.971  0.957  0.914 &  1.002  1.011  1.032 &  1.006  1.01   1.029 & \cellcolor[RGB]{250,243,231} & \cellcolor[RGB]{255,255,255} \\
rubidium &  0.919  0.859  0.747 &  0.892  0.862  0.764 &  1.016  1.042  1.105 &  1.028  1.041  1.096 & \cellcolor[RGB]{234,219,190} & \cellcolor[RGB]{255,255,255} \\
silicon &  0.345  0.369  0.426 &  0.357  0.368  0.418 &  0.72   0.701  0.663 &  0.711  0.702  0.669 & \cellcolor[RGB]{88,94,108} & \cellcolor[RGB]{183,178,169} \\
silver &  0.991  0.985  0.974 &  0.988  0.985  0.975 &  0.994  0.995  0.998 &  0.995  0.995  0.998 & \cellcolor[RGB]{252,251,248} & \cellcolor[RGB]{253,253,254} \\
sodium &  0.977  0.962  0.936 &  0.97   0.962  0.94  &  0.998  1.002  1.011 &  1.     1.002  1.01  & \cellcolor[RGB]{249,245,238} & \cellcolor[RGB]{254,255,255} \\
steel & 0.669  0.639  0.598 & 0.656  0.64   0.604 &  0.789  0.823  0.87 & 0.803  0.821  0.863 & \cellcolor[RGB]{170,162,152} & \cellcolor[RGB]{201,209,221} \\
titanium &  0.441  0.4    0.361 &  0.424  0.403  0.367 &  0.865  0.906  0.946 &  0.882  0.903  0.94  & \cellcolor[RGB]{112,102,92} & \cellcolor[RGB]{220,230,241} \\
tungsten &  0.537  0.536  0.519 &  0.536  0.535  0.522 &  0.695  0.704  0.714 &  0.699  0.703  0.713 & \cellcolor[RGB]{136,136,132} & \cellcolor[RGB]{177,179,182} \\
zinc &  0.808  0.844  0.865 &  0.822  0.842  0.861 &  0.762  0.833  0.896 &  0.793  0.829  0.887 & \cellcolor[RGB]{205,215,220} & \cellcolor[RGB]{194,212,228} \\
\end{tabular}
}

\clearpage

\section{Implementing thin‐film iridescence from first principles}

\label{sec:iridescence-from-first-principles}

In this section, we bring together the core math and physics required to implement thin-film iridescence from first principles. The exposition aims to bridge the gap between theory and practice -- combining foundational math often omitted in rendering literature with practical guidance for translating formulas into code and integrating them into a renderer.

Rather than the approach of \textcite{Belcour2017} in which the thin-film interference effects were pre-integrated over color matching functions, we favour a more direct and general ``locally spectral'' approach that computes reflectance per light path by evaluating the full Fresnel and Airy interference stack -- including complex amplitudes, polarizations, and phase shifts -- at specific wavelengths sampled per path. This can begin with fixed red, green, and blue wavelengths, but better results are achieved by stochastically sampling wavelengths from approximate camera sensitivity curves (as described in Section~\ref{sec:dispersion}). This enables convergence to neutral gray for very thick films and avoids the high-frequency color banding that fixed RGB wavelengths can produce.

The same wavelengths can also be reused to model \hyperref[sec:dispersion]{dispersion}, while all other BSDF components are free to ignore them and operate in RGB as usual. This approach uses only the Airy summation (Equation 3 from \textcite{Belcour2017}) -- a method whose origins trace back to 19th-century studies of thin-plate interference by G.B.~Airy and others -- but requires additional per-wavelength computations and assembling the necessary formulas from multiple sources rather than a single reference. We attempt to collect the necessary equations and implementation instructions here for convenience.

\begin{figure}[!tbh]
  \centering
\includegraphics[width=0.9\linewidth]{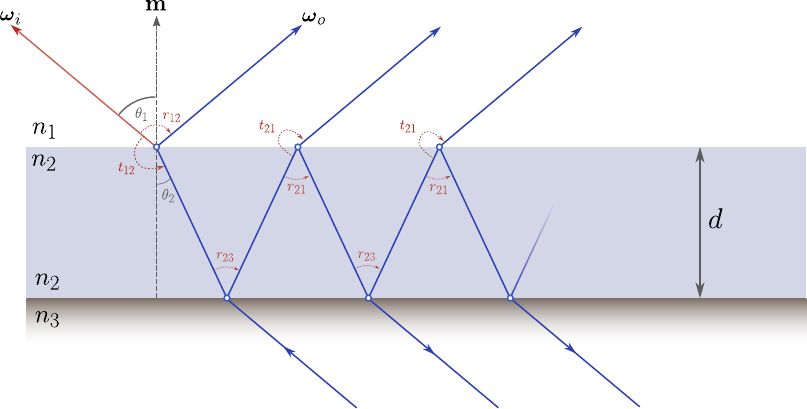}
\caption{Schematic of the thin-film model (following \cite{Belcour2017}). \label{fig:thin-film-schematic}}
\end{figure}

The geometry of the thin film is assumed to follow the model described in \textcite{Belcour2017}, as illustrated in Figure~\ref{fig:thin-film-schematic}. Table~\ref{fig:thin-film-nomenclature} summarizes the relevant mathematical quantities. The film is taken to be applied at each microfacet of the base substrate, so can be assumed to be a flat slab with normal given by the local micronormal $\mathbf{m}$, and thickness $d$. The film is assumed to have real IOR $n_2$ and thickness $d$, and to be surrounded by ambient medium above with (real) IOR $n_1$, and substrate material with IOR $n_3$ (conducting or dielectric, where $n_3$ is complex in the conducting case). The incident light from direction $\boldsymbol{\omega}_i$ is at angle $\theta_1$ to $\mathbf{m}$, while the light refracted into the film makes angle $\theta_2$ with $\mathbf{m}$. Given this setup, we need to determine the albedo of the resulting configuration for incident light with wavelength $\lambda$, accounting for interference effects.


\begin{table}[!b]
  \centering
  \begin{tabular}{ll}
    \toprule
    \textbf{Symbol} & \textbf{Description} \\
    \midrule
    $n_1$ & Refractive index of incident medium (real; e.g., air) \\
    $n_2$ & Refractive index of thin film (real) \\
    $n_3$ & Refractive index of substrate (real for dielectrics, complex for conductors) \\
    $\theta_1,\theta_2$ & Angles from surface normal in incident medium and thin film \\
    $d$ & Film physical thickness \\
    $\lambda$ & Wavelength in vacuum \\
    $r^{(s,p)}_{ij},\,t^{(s,p)}_{ij}$ & Complex Fresnel amplitude coefficients between media $i\to j$ \\
    $(s,p)$ & Polarization state index of amplitudes (s-polarized or p-polarized) \\
    $\delta$ & Round-trip phase shift in the film \\
    \bottomrule
  \end{tabular}
  \caption{Nomenclature for thin-film iridescence. \label{fig:thin-film-nomenclature}}
\end{table}


\paragraph{Prerequisites and assumptions}
\begin{itemize}
  \item A complex-number type is required, supporting addition, multiplication, complex square root, and $\exp(i\theta)$, preserving both magnitude and phase.
  \item Refractive indices (IORs) must be stored as complex values; $n_i$ may be real (for dielectrics, written as $n_i + i,0$) or complex (for conductors, $n_i + i,k_i$ where $k_i > 0$ represents the extinction coefficient).
  \item Amplitude Fresnel coefficients $r_{ij}$ and $t_{ij}$ must be stored as complex values.
  \item Input wavelengths: one or more $\lambda$ values (e.g., monochromatic, RGB triplet, or spectral samples).
  \item All formulas are evaluated per wavelength.
  \item Wavelength $\lambda$ and film thickness $d$ must use consistent units (e.g., nanometers or microns).
  \item All angles $\theta$ are in radians and measured from the surface normal.
\end{itemize}


\paragraph{1. Angles and Snell's law}
For each interface $i\to j$, given a real incidence angle $\theta_i$, compute the transmitted angle using Snell's law:
\[
  \sin\theta_j = \frac{n_i}{n_j}\,\sin\theta_i,
  \quad
  \cos\theta_j = \sqrt{1 - \sin^2\theta_j}.
\]
The square root must be evaluated using the branch such that $\operatorname{Im}(n_j \cos\theta_j) \ge 0$, ensuring that the transmitted wave behaves correctly for absorbing materials. We assume that the incident medium ($n_i$) is a real-valued dielectric and that the incidence angle $\theta_i$ is real; therefore, $\sin\theta_i$ and $\cos\theta_i$ are real-valued. However, if $n_j$ is complex or if total internal reflection occurs, then $\sin\theta_j$ and $\cos\theta_j$ may become complex (and must be computed using complex arithmetic). If $n_j$ is real and $\sin\theta_j>1$, then $\cos\theta_j$ is imaginary, indicating total internal reflection (TIR) and evanescent transmission with no net energy transport. In all cases, the results of Snell's Law can be safely input to the Fresnel equations, which will handle all regimes correctly when evaluated using complex arithmetic.


\paragraph{2. Amplitude Fresnel coefficients}
Using $\cos\theta_i$ and $\cos\theta_j$ from Step 1, compute for each polarization ($s,p$):
\begin{align}
  r_{ij}^{(s)} &= \frac{n_i\cos\theta_i - n_j\cos\theta_j}{n_i\cos\theta_i + n_j\cos\theta_j},
  &
  t_{ij}^{(s)} &= \frac{2\,n_i\cos\theta_i}{n_i\cos\theta_i + n_j\cos\theta_j}, \\
  r_{ij}^{(p)} &= \frac{n_j\cos\theta_i - n_i\cos\theta_j}{n_j\cos\theta_i + n_i\cos\theta_j},
  &
  t_{ij}^{(p)} &= \frac{2\,n_i\cos\theta_i}{n_j\cos\theta_i + n_i\cos\theta_j}.
\end{align}
Compute for $(i,j)=(1,2)$ and $(2,3)$ to get $r_{12},t_{12}$ and $r_{23},t_{23}$. In the TIR regime (for dielectrics), these formulas yield $|r_{ij}|=1$ and $t_{ij}=0$ automatically (if and only if they are performed using complex arithmetic; otherwise, TIR must be manually detected and handled with a separate code path).


\paragraph{3. Reciprocity for reverse direction}
To avoid re-running the Fresnel equations, one can use the following symmetry relations between the reflection and transmission amplitudes \cite{BornWolf}:
\begin{align}
  r_{21}^{(s,p)} &= -\,r_{12}^{(s,p)}, &
  t_{21}^{(s,p)} &= t_{12}^{(s,p)}\frac{n_2\cos\theta_2}{n_1\cos\theta_1}.
\end{align}
(Note that this works even for complex Fresnel coefficients; the minus sign accounts for the phase shift for reflection under the assumption of planar, non-magnetic media.)


\paragraph{4. Phase shift inside the film}
The round-trip phase shift is
\begin{equation}\label{eq:phase}
  \delta = \frac{4\pi\,n_2\,d\,\cos\theta_2}{\lambda}.
\end{equation}


\paragraph{5. Airy summation}
The total reflectance amplitude is computed by summing the contributions of each output reflection mode depicted in Figure~\ref{fig:thin-film-schematic}.
This is derived from the following geometric series expansion (known as \emph{Airy summation}) of the reflected wave contributions for the total reflectance amplitude $r_{\mathrm{tot}}^{(s,p)}$ for each polarization ($s,p$), where the first term is the direct reflection at the first interface, and the second term accounts for the multiple reflections within the film:
\begin{align} \label{eq:airy}
  r_{\mathrm{tot}}^{(s,p)} &= r_{12}^{(s,p)} + t_{12}^{(s,p)}\,r_{23}^{(s,p)}\,t_{21}^{(s,p)}\,e^{i\delta} + t_{12}^{(s,p)}\,r_{23}^{(s,p)}\,r_{21}^{(s,p)}\,r_{23}^{(s,p)}\,t_{21}^{(s,p)}\,e^{2i\delta} + \cdots \nonumber \\
    &= r_{12}^{(s,p)} + \sum_{k=1}^{\infty} t_{12}^{(s,p)}\,r_{23}^{(s,p)} \left(r_{21}^{(s,p)}\, r_{23}^{(s,p)}\right)^{k-1} \, t_{21}^{(s,p)} \, e^{i k\delta} \nonumber \\
  &= r_{12}^{(s,p)} + \frac{t_{12}^{(s,p)}\,r_{23}^{(s,p)}\,t_{21}^{(s,p)}\,e^{i\delta}}
          {1 - r_{21}^{(s,p)}\,r_{23}^{(s,p)}\,e^{i\delta}}.
\end{align}
The denominator ensures convergence of the series, provided that \( |r_{21}^{(s,p)}\,r_{23}^{(s,p)}\,e^{i\delta}| < 1 \).


Power reflectance per-polarization follows as
\begin{equation}\label{eq:power}
  R_{s,p} = |r_{\mathrm{tot}}^{(s,p)}|^2.
\end{equation}


\paragraph{6. Unpolarized reflectance} Assuming unpolarized incident light, the total reflectance is the average of the powers of the two polarization states:
\begin{equation}\label{eq:unpol}
  R = \tfrac12(R_s + R_p).
\end{equation}


\paragraph{Implementation tips and common pitfalls}
\begin{itemize}
  \item \emph{Complex square root branches:} When computing \(\cos\theta_j = \sqrt{1 - \sin^2\theta_j}\), choose the branch such that \(\operatorname{Im}(n_j \cos\theta_j) \ge 0\). This ensures the transmitted wave behaves correctly for absorbing materials, and is essential for energy conservation and reciprocity at interfaces.
  \item \emph{Transmission prefactor for power:} To convert amplitude transmission coefficients to power, multiply by \(\operatorname{Re}(n_j \cos\theta_j / n_i \cos\theta_i)\) for s-polarization (perpendicular), and \(\operatorname{Re}(n_j \cos^*\theta_j / n_i \cos\theta_i)\) for p-polarization (parallel). The complex conjugate in the p-term is essential for handling complex angles correctly. This prefactor ensures that \(R + T = 1\) at each interface, even when the transmitted medium is absorbing.
  \item \emph{Interface energy conservation:} Fresnel reflection and transmission at each \emph{interface} must conserve energy, even for lossy media such as metals. This means numerically verifying that \(0 \le R, T \le 1\) and \(R + T = 1\) at every boundary and for all polarizations. For conductors, the transmitted light is absorbed \emph{after} entering the medium, so interface-level energy conservation must still hold.
  \item \emph{Denominator safety:} Clamp any denominator magnitude to \(\epsilon\) before division. In particular, the denominator in Equation~\ref{eq:airy} can approach zero when the reflected waves interfere constructively (a resonance condition), even though the final reflectance remains physically bounded.
  \item \emph{Phase/magnitude extraction:} You never need to explicitly extract magnitude and phase from the complex amplitude coefficients -- compute \(r\) via complex arithmetic and convert to power with \(\lvert r\rvert^2\) at the end.
  \item \emph{Early exit for TIR:} At the exterior-film interface, if TIR occurs, reflectance is unity and transmission zero -- return \(R=1\) directly without further evaluation.
  \item \emph{Complex Fresnel unification:} Using complex \(n\) covers both dielectrics and conductors, including evanescent waves; a small TIR branch for real dielectrics (e.g., \(\sin\theta_t \ge 1\)) can shortcut evaluation (a separate function using real arithmetic may also be used for dielectrics if more performant, in which case an early exit is required).
  \item \emph{Dielectric shortcut:} Alternatively, use a dedicated real-valued Fresnel function for pure dielectrics (with its own TIR early-exit) to avoid complex arithmetic when indices are real. Snell's Law calculations can also use real numbers for dielectrics where applicable.
  \item \emph{Composite substrates:} For mixed substrates, compute \(R_{23}^{(s,p)}\) per component (steps 1--6) and either blend immediately by mixing weights, or return separate dielectric vs.\ metal reflectances for caller-side blending before Airy summation.
  \item \emph{\( t_{23}^{(s,p)} \) is never needed:} The transmitted amplitude into medium 3 produced by the Fresnel equations is not used in the Airy summation (Equation~\ref{eq:airy}) and can be discarded in all cases, regardless of whether the substrate is a dielectric or conductor.
  \item \emph{Integration in a renderer:} Use the computed \(R\) from Step 6 to replace your microfacet Fresnel term (or to modulate it via the thin-film weight in the material model).
  \item \emph{Vectorize wavelengths and polarizations:} Evaluate each formula per sample and polarization in parallel (e.g., RGB/spectral, \(s\)/\(p\)). This helps minimize branching and improves SIMD or GPU parallelism.
  \item \emph{Energy conservation across layers:} For multi-layer systems (e.g., thin film on metal), the total reflected and transmitted power must satisfy \( R + T \le 1 \). Any shortfall from 1 indicates absorption within the film or substrate. If \( R + T < 1 \) but no absorption is expected, this suggests an implementation error.
\end{itemize}


\vspace{1cm}

\paragraph{Albedo of a thin film}

\begin{figure}[!hb]
  \centering
  \begin{subfigure}{.49\textwidth}
    \includegraphics[width=\linewidth]{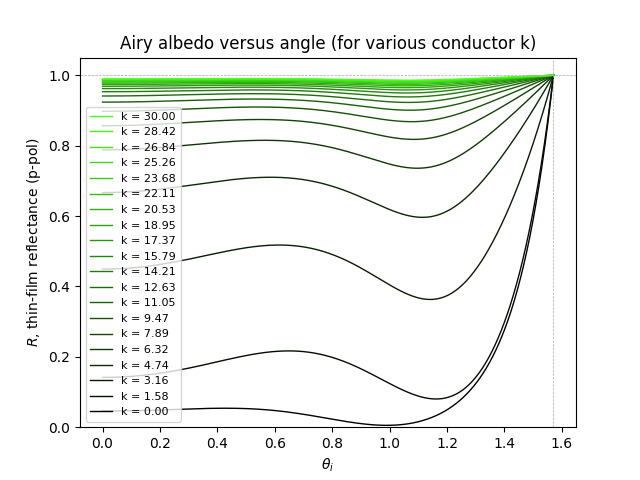}
  \end{subfigure}
   \hspace{0.1cm}
  \begin{subfigure}{.49\textwidth}
    \includegraphics[width=\linewidth]{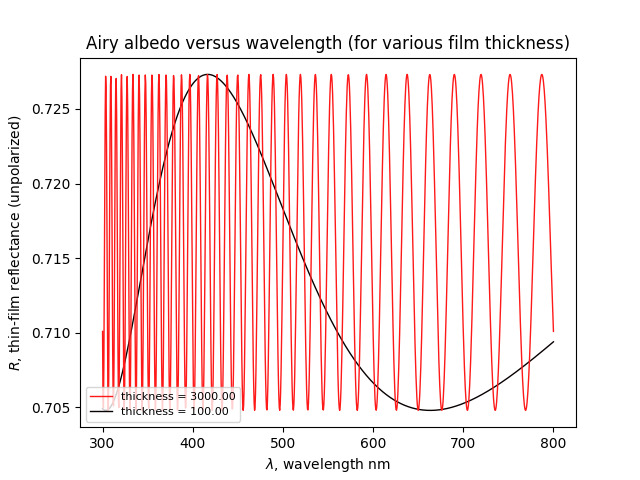}
  \end{subfigure}
  \caption{Albedo of a thin film on a conducting substrate, versus angle at 500~nm (left) and versus wavelength at normal incidence (right).}
  \label{fig:thin-film-plots}
\end{figure}

Figure~\ref{fig:thin-film-plots} plots the albedo of a thin film, computed according to the procedure described, with $n_1 = 1$, $n_2 = 1.4$, real part of $n_3 = 1.8$, and $d = 400$~nm. In the left panel, the wavelength is fixed at 500~nm, and the albedo as a function of angle of incidence shown for a variety of conductor extinction coefficients $k$.

The albedo tends to $1$ as the conductor extinction increases, which is expected physically since in this limit (also known as a \emph{Gires–Tournois etalon}) the Fresnel reflection from the metal tends to 1 leading to no energy loss. Thus a useful check of a thin-film implementation is to verify that a perfectly reflecting metal with a thin film passes a furnace test, regardless of the thin-film properties.

In the right panel, the light is incident along the normal, the conductor absorption is fixed at $5$, and the albedo is plotted as a function of wavelength for two different film thicknesses. At high thickness (relative to the wavelength), the albedo exhibits strong oscillations as a function of wavelength. This results in the color fringes disappearing at high thin-film depth (relative to optical wavelengths), as the oscillations integrated over the color matching functions integrate to gray.

\clearpage

\section{The physics of coat darkening}

\label{sec:coat_physics}

\begin{figure}[!b]
  \centering
\includegraphics[width=0.666\linewidth]{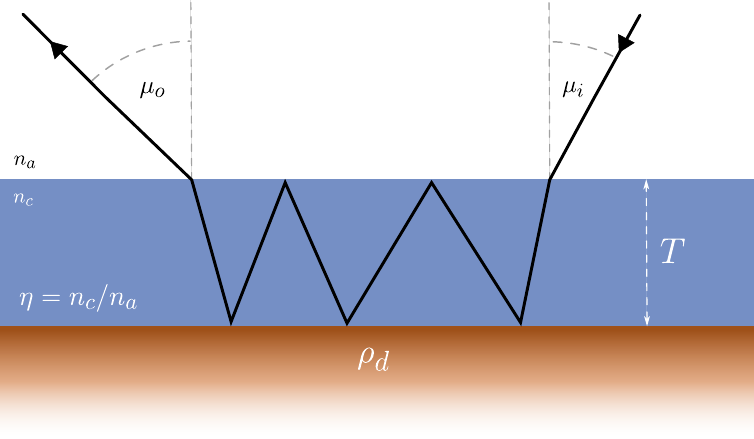}
\caption{Schematic of the light transport inside a smooth dielectric coat on top of a diffuse base substrate.}
\label{fig:coat_diagram}
\end{figure}

We present here some formulas for the physical darkening effect of light transport within a dielectric coat. In the case of a smooth dielectric coat on top of a Lambertian substrate (the so-called ``interfaced Lambertian'' model), there is an exact solution available, which we describe in detail below. This was derived using radiative transfer theory by \textcite{ELIAS20011}, and the result is also stated by \textcite{HHGTMS}. The notation used is shown in Figure~\ref{fig:coat_diagram}.

The ratio of the IOR of the coat dielectric ($n_c$) to that of the exterior medium ($n_a$) is denoted $\eta$. The Lambertian base albedo is denoted $\rho_d$. The dielectric layer absorption is parameterized by the transmittance along the normal direction, $T$.
Light is incident from the upper hemisphere $\mathcal{H}_+$ at some incident angle cosine $\mu_i$. It enters the dielectric layer, bounces within, then exits back into $\mathcal{H}_+$ with some outgoing angle cosine $\mu_o$. Due to the azimuthal symmetry, the BRDF must be a function only of these two angle cosines, i.e., $f(\mu_i, \mu_o)$. By reciprocity we must have $f(\mu_i, \mu_o) = f(\mu_o, \mu_i)$. We assume that the incident light is unpolarized, and remains unpolarized after reflection and transmission (i.e., following the usual assumption in radiative transfer\footnote{Though technically this is inconsistent, since even unpolarized incident light will become polarized after Fresnel transmission into the coat. This polarization is ignored and radiation assumed to be unpolarized, as in a standard RTE treatment of the problem.}, as employed by \textcite{ELIAS20011}). The formulas are all written assuming the spectral frequency is fixed.

We consider here only the case of a perfectly smooth dielectric coat. The extension to the rough case is briefy discussed by \textcite{HHGTMS}.

In the presence of the Lambertian base, the full BRDF has the form \cite{ELIAS20011}
\begin{equation} \label{exact_brdf_interfaced_lambertian}
f(\mu_i, \mu_o) = f_s(\mu_i, \mu_o) + f_b(\mu_i, \mu_o) \ .
\end{equation}
The first term, denoted $f_s(\mu_i, \mu_o)$, is the specular mirror reflection from the coat (technically a delta function), scaled by the \emph{Fresnel factor} (or \emph{Fresnel reflection coefficient}) $F_R(\mu, \eta)$, which gives the fraction of the energy of an incident beam from angle-cosine $\mu$ which is reflected from a smooth dielectric interface with IOR ratio $\eta$. In this case, the directional albedo $R(\mu) = \int_{\mathcal{H}_+} f(\mu, \mu_o) \,\mathrm{d} \omega^\perp_o$ reduces to $F_R(\mu, \eta)$.

The second term is given by
\begin{equation} \label{exact_brdf_interfaced_lambertian2}
f_b(\mu_i, \mu_o) = \frac{\rho_d}{\pi\eta^2} \frac{F_T(\mu_i, \eta) F_T(\mu_o, \eta) \, T^{\frac{1}{\mu^t_i} + \frac{1}{\mu^t_o}}} {1 - \rho_d K(\eta, T)} \ ,
\end{equation}
where $F_T(\mu, \eta) = 1 - F_R(\mu, \eta)$ is the \emph{Fresnel transmission coefficient}. The two $F_T$ factors account for the transmission into and out of the coat interface, for paths en route to the base. The last factor in the numerator accounts for the coat transmittance along the refracted input and output ray directions, where $\mu^t_i, \mu^t_o$ are the angle cosines of the incident and output ray refracted into the coat:
\begin{equation}
    \mu^t_i = \sqrt{1 - \frac{1-\mu_i^2}{\eta^2}} \ , \quad \mu^t_o = \sqrt{1 - \frac{1-\mu_o^2}{\eta^2}} \ .
\end{equation}
We can assume a solution exists for each, since if either the incident or outgoing directions are in a TIR configuration, then the whole second term is zero since one of the $F_T$ factors is zero.

\begin{figure}[t]
  \centering
  \begin{subfigure}{.48\textwidth}
    \centering
    \includegraphics[width=\linewidth]{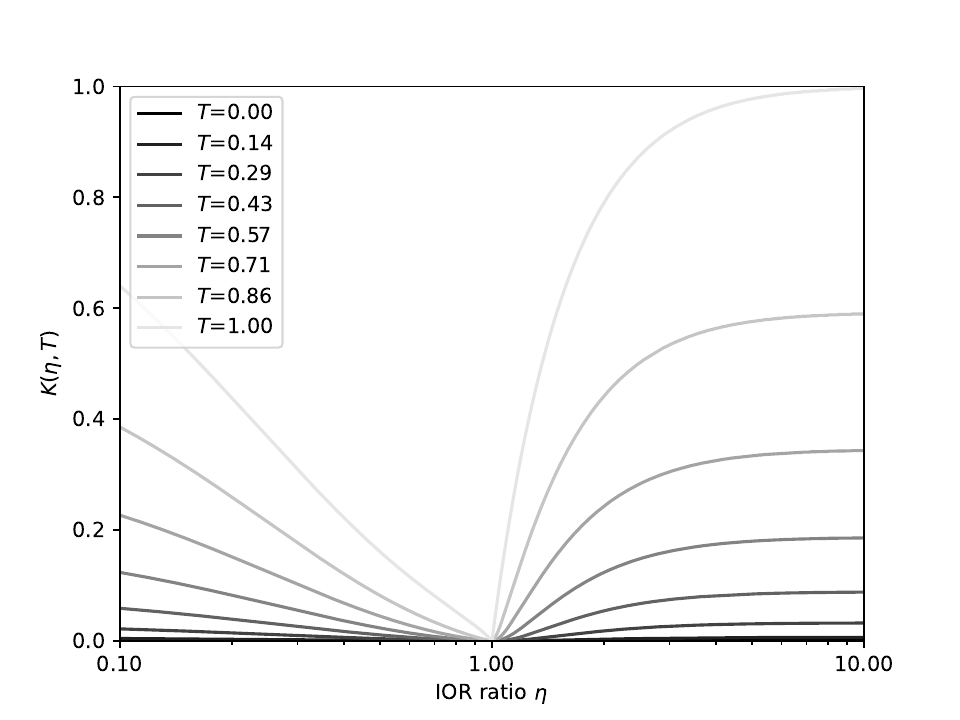}
    \label{fig:K-versus-eta}
  \end{subfigure}
  \begin{subfigure}{.48\textwidth}
    \centering
    \includegraphics[width=\linewidth]{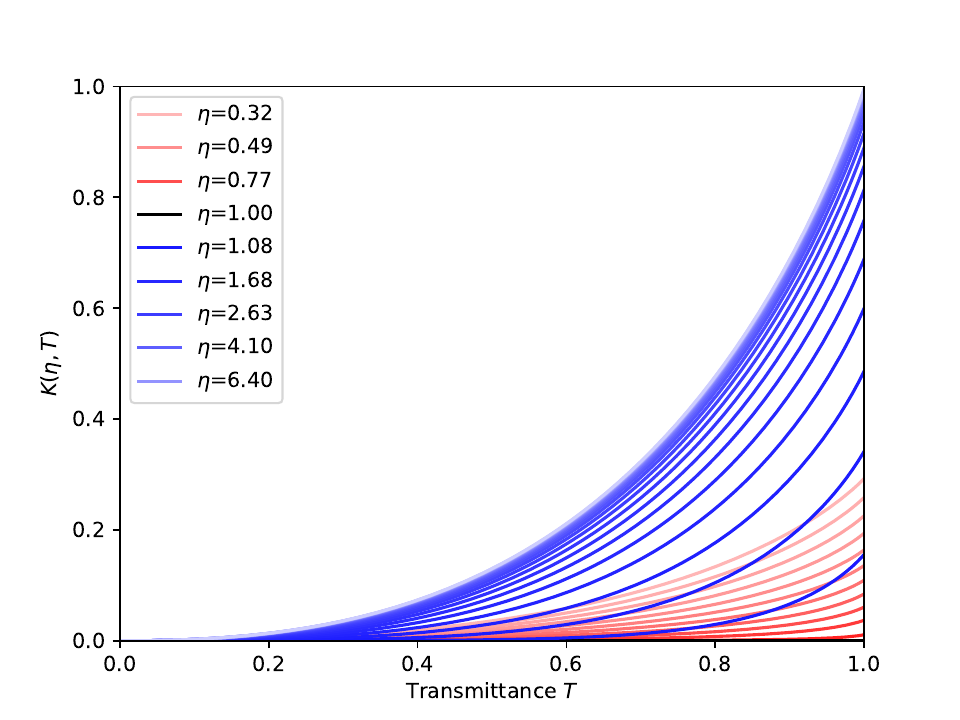}
    \label{fig:K-versus-T}
  \end{subfigure}
  \caption{$K(\eta, T)$ versus IOR ratio $\eta$ or transmittance $T$.}
  \label{fig:K-plots}
\end{figure}

The denominator accounts for the bounces due to internal reflections from the coat. The factor $K$ in the denominator is known as the \emph{internal diffuse reflection coefficient}, given by
\begin{equation}
K(\eta, T) = 2 \int_0^1 T^{2/\mu} \, F_R(\mu, 1/\eta) \, \mu \, \mathrm{d}\mu \ .
\end{equation}
It corresponds to the fraction of the energy leaving the base that returns to the base due to internal reflection from the coat. The integral for $K(\eta, T)$ does not have a simple closed-form solution. Its behavior is shown in Figure~\ref{fig:K-plots}. In the limits $\eta \rightarrow 1$ and $T \rightarrow 0$ (i.e., no coat reflection, or a completely absorbing medium), $K(\eta, T) \rightarrow 0$.

The directional albedo of the full BRDF is then given by
\begin{equation}
R(\mu_o) = F_R(\mu_o, \eta) + \frac{\rho_d}{\eta^2} F_T(\mu_o, \eta) \frac{1 - E_F(\eta)} {1 - \rho_d K(\eta, T)} \ .
\end{equation}
where $E_F(\eta) = 2 \int_0^1 F_R(\mu, \eta) \, \mu \, \mathrm{d}\mu$ is the hemispherical (or average) albedo of the Fresnel factor.
\footnote{A good approximation to $E_F(\eta)$ in the limited range $\eta \in [1,3]$ is given by \textcite{HHGTMS}, who reports it to be accurate to within 0.15\% over that IOR range:
\begin{equation} \label{hemispherical_fresnel_albedo}
  E_F(\eta) \approx \ln \biggl( \frac{10893\eta - 1438.2}{-774.4\eta^2 + 10212\eta + 1} \biggr) \ .
\end{equation}
}

\subparagraph{Clear coat}

\begin{figure}
  \centering
  \begin{subfigure}{.48\textwidth}
    \centering
    \includegraphics[width=\linewidth]{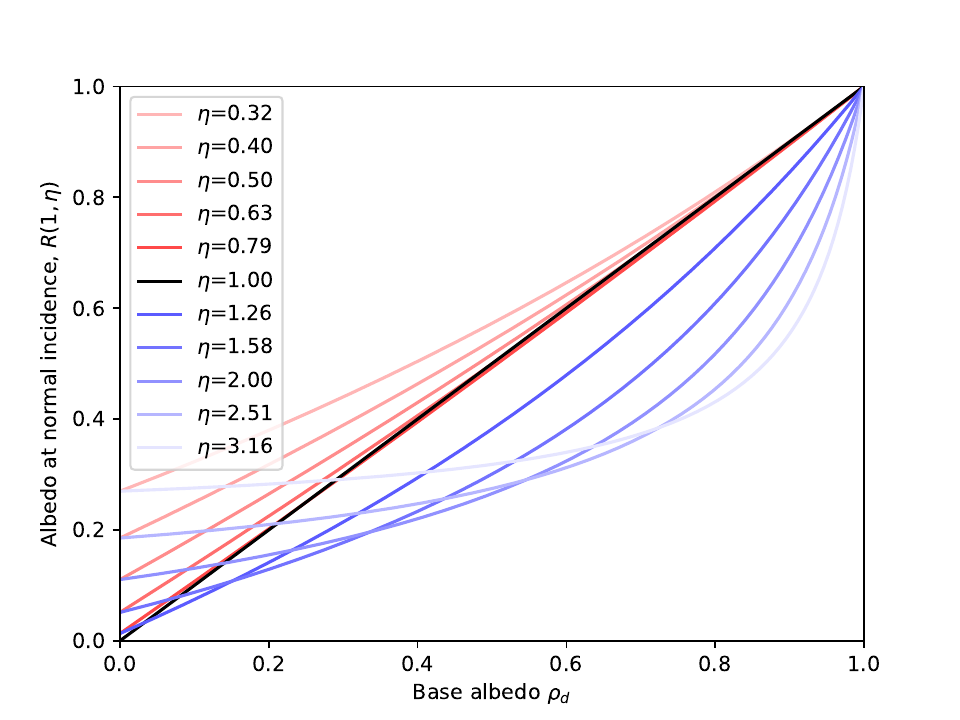}
    \caption{Exact formula, Equation~\ref{clearcoat_exact_directional_albedo}.}
    \label{fig:layer_directional_albedo_exact}
  \end{subfigure}
  \begin{subfigure}{.48\textwidth}
    \centering
    \includegraphics[width=\linewidth]{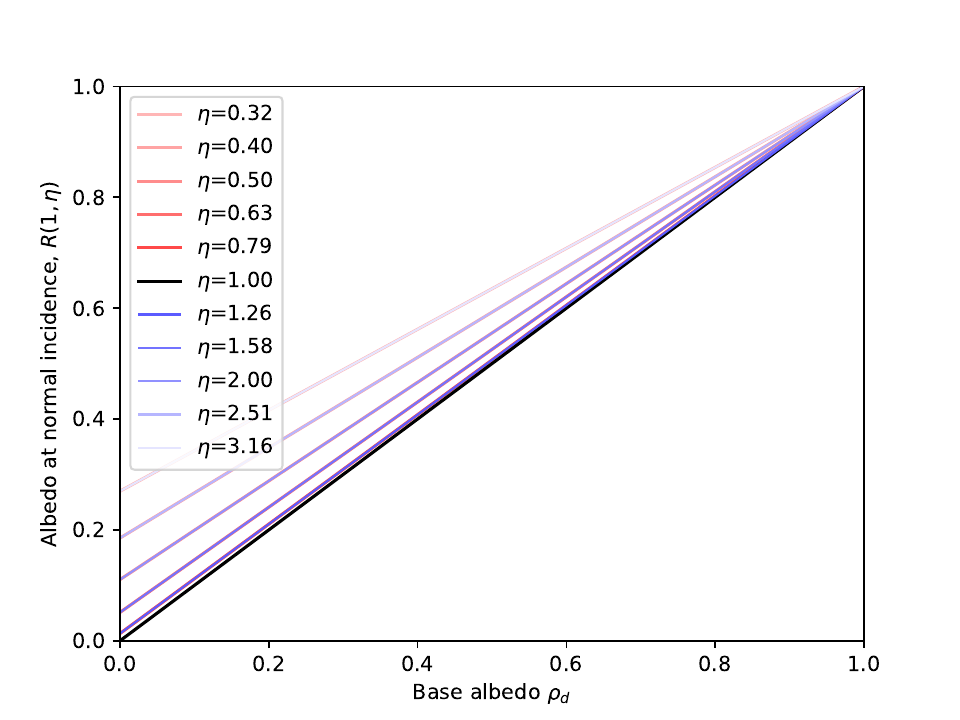}
    \caption{``Albedo-scaling'' approx., Equation~\ref{clearcoat_albedo_scaling_approx_directional_albedo}.}
    \label{fig:layer_directional_albedo_approx}
  \end{subfigure}
  \caption{Clear-coat layer directional albedo at normal incidence, $R(1)$.}
  \label{fig:layer_directional_albedo}
\end{figure}

In the limit $T\rightarrow 1$ (a transparent clear coat), the internal diffuse reflection coefficient reduces to
\begin{equation}
K_0(\eta) = 2 \int_0^1 F_R(\mu, 1/\eta) \, \mu \, \mathrm{d}\mu = 1 - 2 \int_0^1 F_T(\mu, 1/\eta) \, \mu \, \mathrm{d}\mu\ .
\end{equation}
Applying the following symmetry relation for the Fresnel hemispherical albedo:
\begin{equation} \label{fresnel_albedo_symmetry}
  1 - E_F(\eta) = \eta^2 \Bigl( 1 - E_F(1/\eta) \Bigr) \ ,
\end{equation}
this further reduces to
\begin{eqnarray}
  K_0(\eta) = 1 - (1 - E_F(\eta))/\eta^2  \ .
  \end{eqnarray}
The directional albedo thus simplifies to (as quoted by \textcite{HHGTMS}):
\begin{equation} \label{clearcoat_exact_directional_albedo}
    R(\mu_o) = F_R(\mu_o, \eta) + \frac{\rho_d}{\eta^2} \frac{\bigl(1 - F_R(\mu_o, \eta)\bigr) \bigl(1 - E_F(\eta)\bigr)} {1 - \rho_d \Bigl(1 - \bigl(1 - E_F(\eta)\bigr) / \eta^2 \Bigr)} \ .
\end{equation}
As required, $R(\mu_o) \rightarrow 1$ as base albedo $\rho_d \rightarrow 1$, since in this case no energy is absorbed. Also of course as $\eta \rightarrow 1$, $R(\mu_o) \rightarrow \rho_d$.
The corresponding hemispherical albedo is given by
\begin{equation}
    E = E_F(\eta) + \frac{\rho_d}{\eta^2} \frac{\bigl(1 - E_F(\eta)\bigr)^2} {1 - \rho_d \Bigl(1 - \bigl(1 - E_F(\eta)\bigr) / \eta^2 \Bigr)} \ .
\end{equation}

In Figure~\ref{fig:layer_directional_albedo_exact}, the directional albedo of the clear coat at normal incidence (i.e., Equation~\ref{clearcoat_exact_directional_albedo} evaluated at $\mu_0=1$) is plotted as a function of base albedo $\rho_d$. This quantity represents the observed reflected radiance at normal incidence given unit white uniform illumination. (The black line represents the reflection in the case $\eta=1$, i.e., no coat, in which case it reduces to the base albedo.)

For comparison in Figure~\ref{fig:layer_directional_albedo_approx}, we plot the so-called ``albedo-scaling'' approximation of the clear-coat layer albedo, given by
\begin{equation} \label{clearcoat_albedo_scaling_approx_directional_albedo}
    R(\mu_o) \approx F_R(\mu_o, \eta) + \rho_d (1 - F_R(\mu_o, \eta))  \ .
\end{equation}
This corresponds to assuming the clear-coat layer BRDF is approximated by
\begin{equation} \label{clearcoat_approx_brdf}
  f(\mu_i, \mu_o) \approx f_s(\mu_i, \mu_o) + \frac{\rho_d}{\pi} (1 - F_R(\mu_o, \eta)) \ .
\end{equation}
This is non-reciprocal, but at least satisfies $R(\mu_o) \rightarrow 1$ as $\rho_d \rightarrow 1$.
Note that, in this approximation, the $\eta$ and $1/\eta$ cases look identical at normal incidence. For the $\eta>1$ case, the approximation is clearly extremely poor. In both the exact formula and the approximation, for a dark base ($\rho_d=0$) the reflection is equal to the Fresnel reflection from the coat, and for a white base ($\rho_d=1$) the reflection is white (i.e., the furnace test passes).



\begin{figure}[!h]
  \captionsetup[subfigure]{labelformat=empty}
  \captionsetup[subfigure]{skip=-13pt}
  \centering
  \begin{subfigure}[t]{0.2\textwidth}
    \centering
    \includegraphics[width=\linewidth]{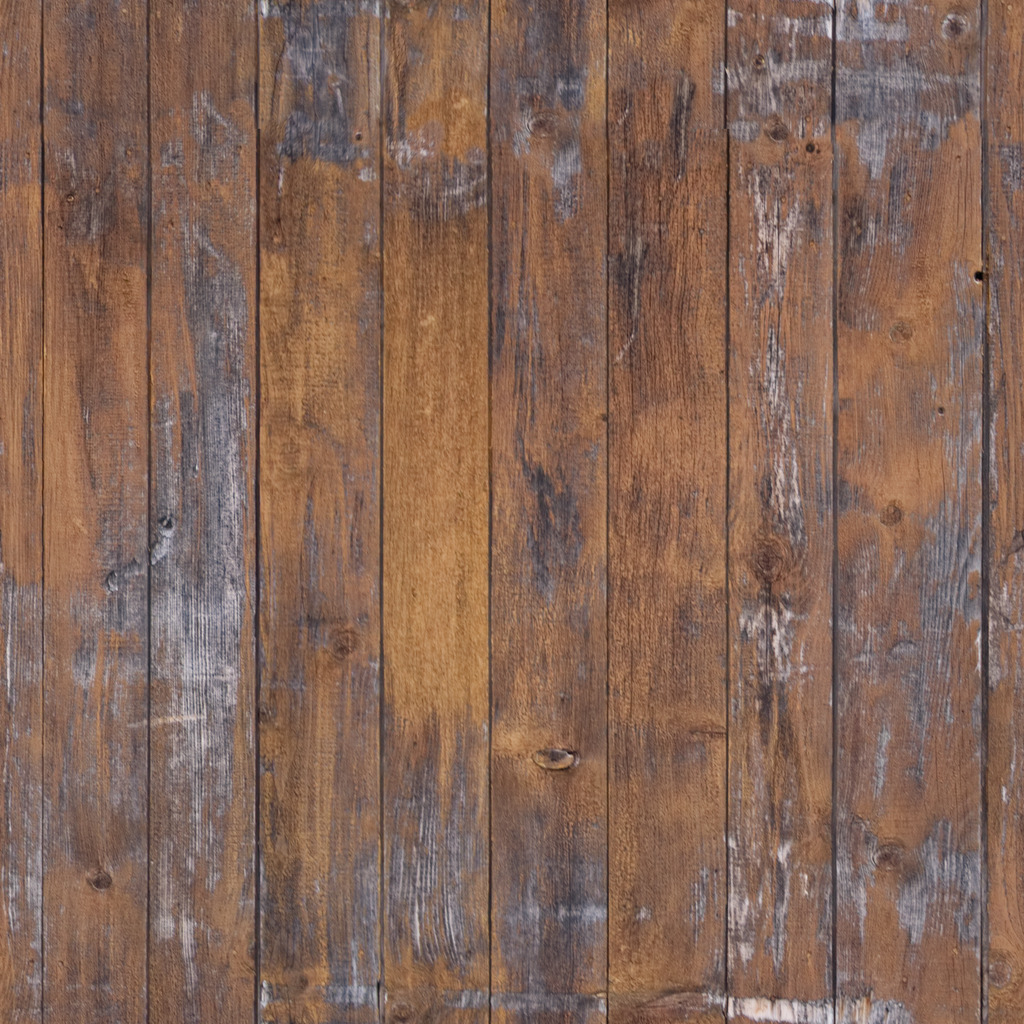}
    \caption{\color{white}$\eta=1$}
  \end{subfigure}
  \begin{subfigure}[t]{0.2\textwidth}
    \centering
    \includegraphics[width=\linewidth]{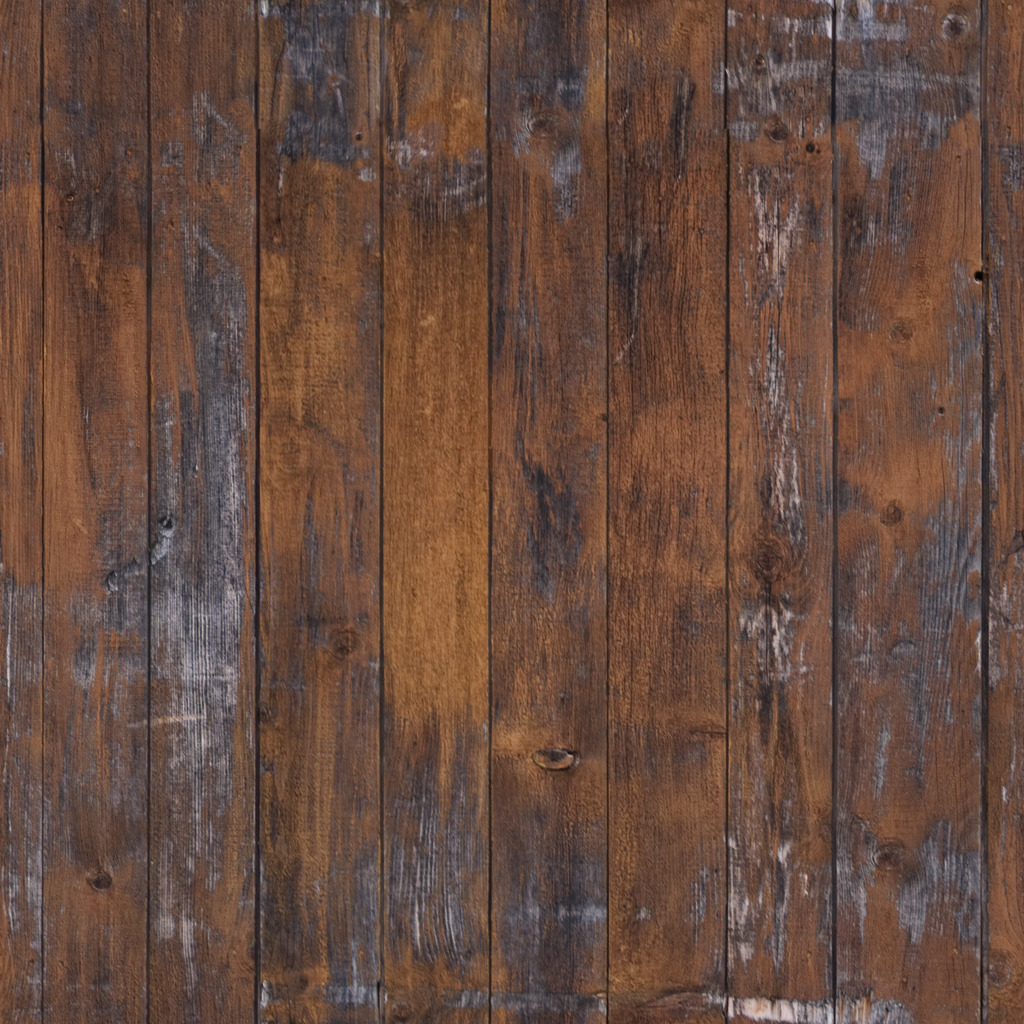}
    \caption{\color{white}$\eta=1.25$}
  \end{subfigure}
  \begin{subfigure}[t]{0.2\textwidth}
    \centering
    \includegraphics[width=\linewidth]{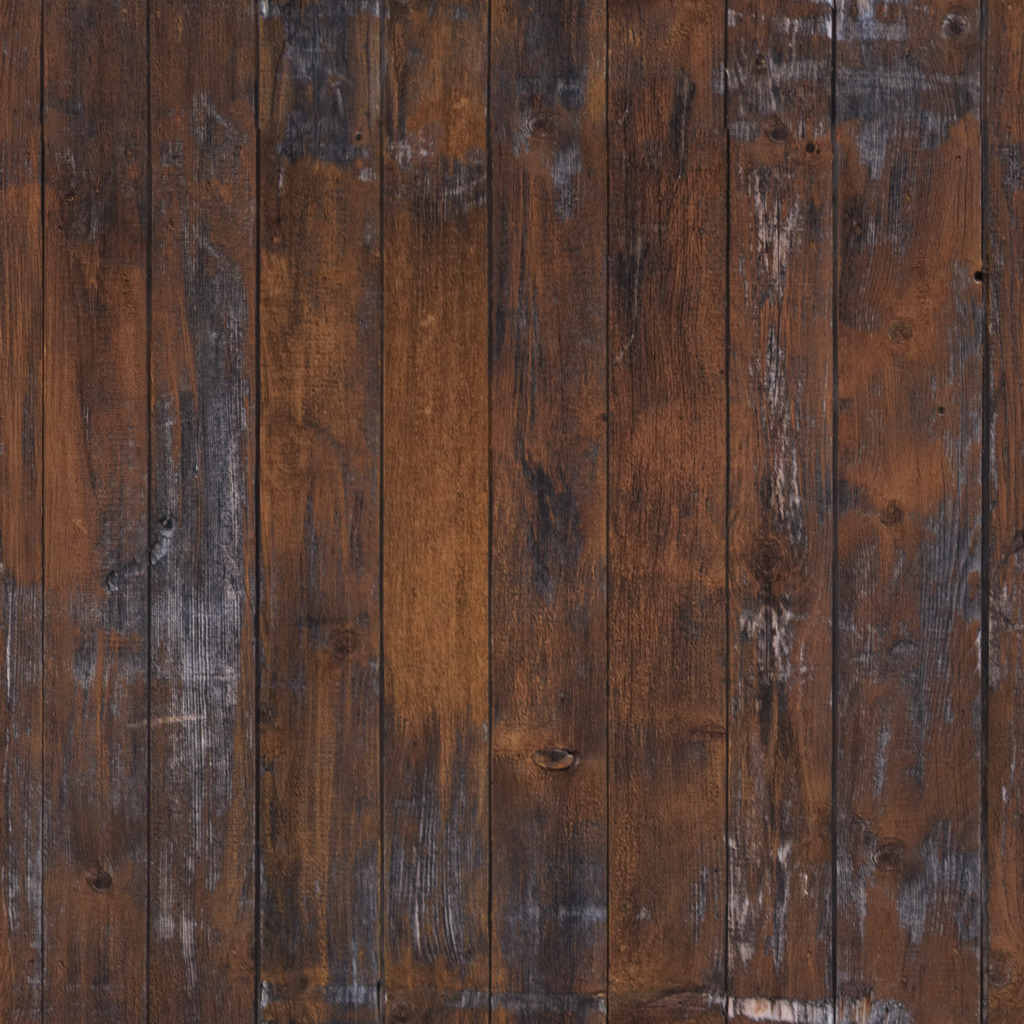}
    \caption{\color{white}$\eta=1.5$}
  \end{subfigure}
  \vfill
  \vspace{2pt}
  \begin{subfigure}[t]{0.2\textwidth}
    \centering
    \includegraphics[width=\linewidth]{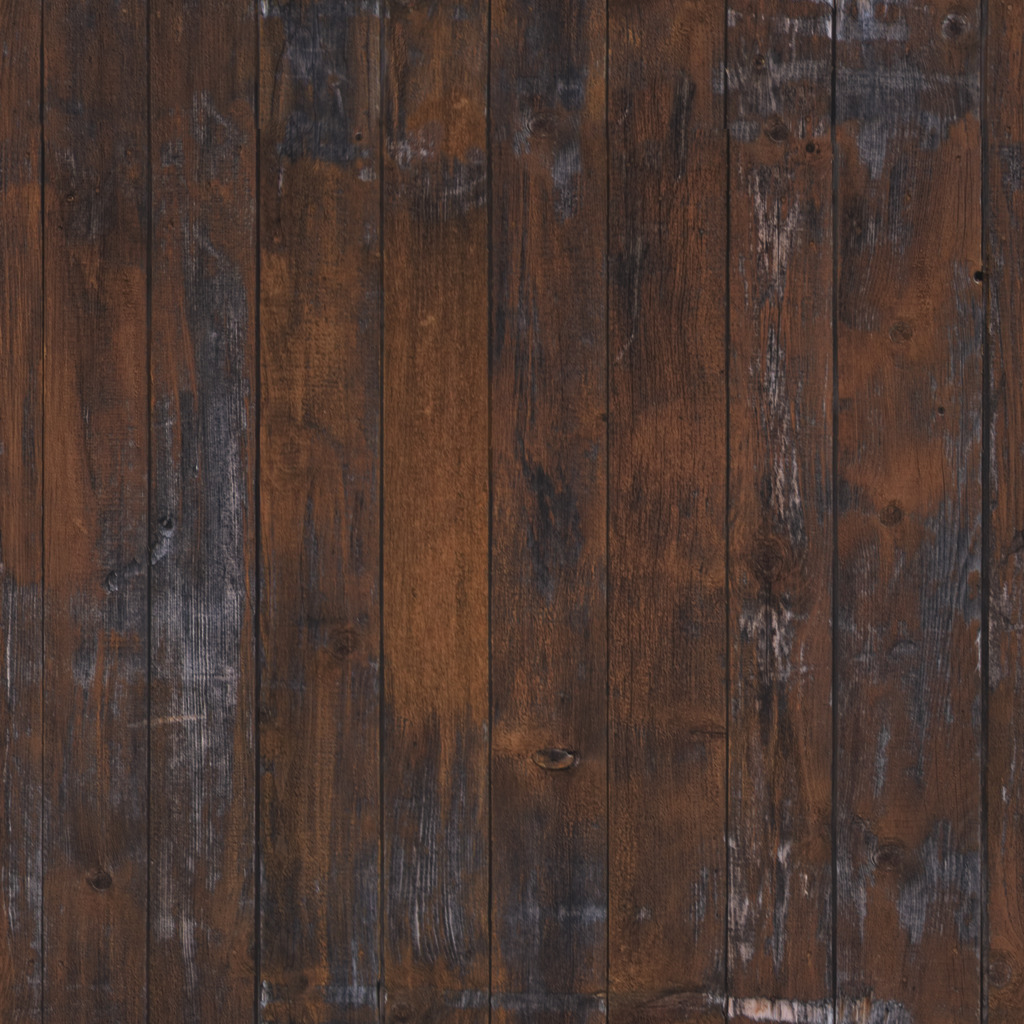}
    \caption{\color{white}$\eta=1.75$}
  \end{subfigure}
  \begin{subfigure}[t]{0.2\textwidth}
    \centering
    \includegraphics[width=\linewidth]{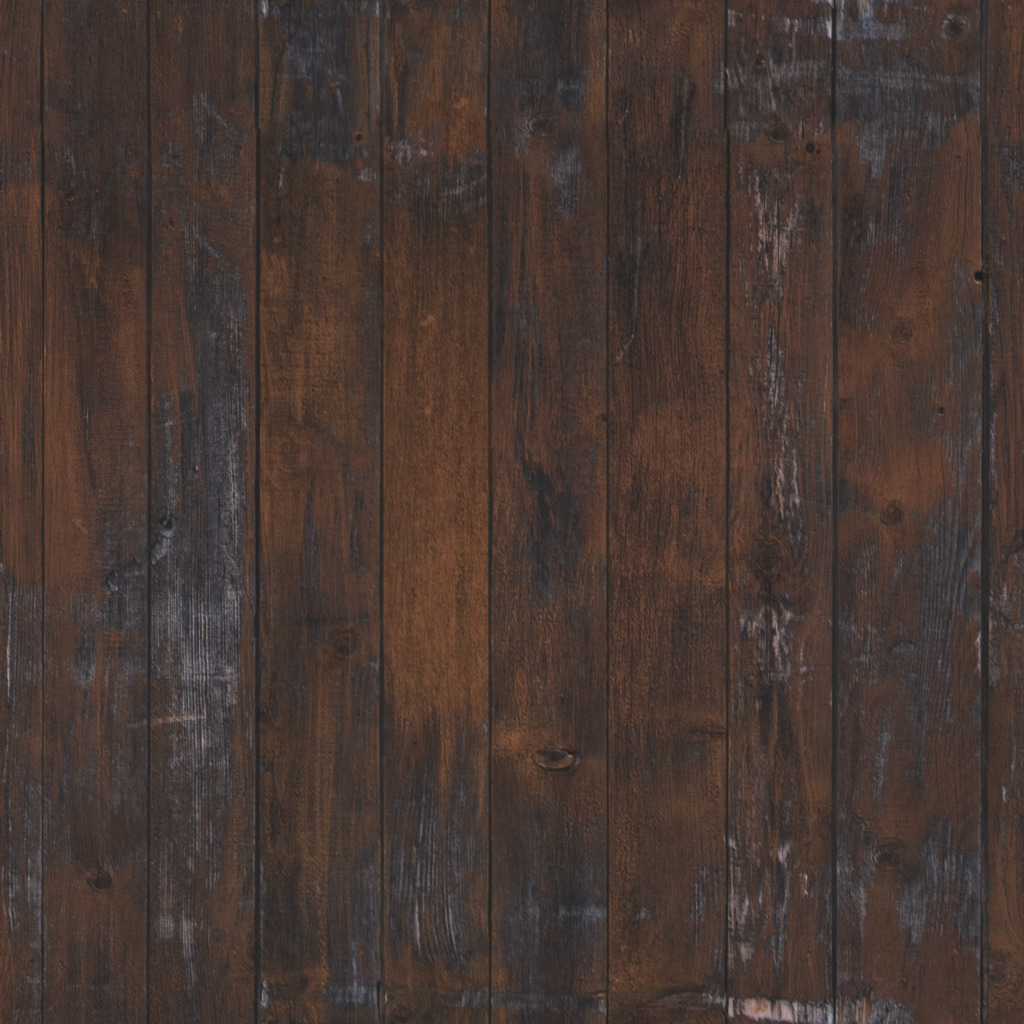}
    \caption{\color{white}$\eta=2$}
  \end{subfigure}
  \begin{subfigure}[t]{0.2\textwidth}
    \centering
    \includegraphics[width=\linewidth]{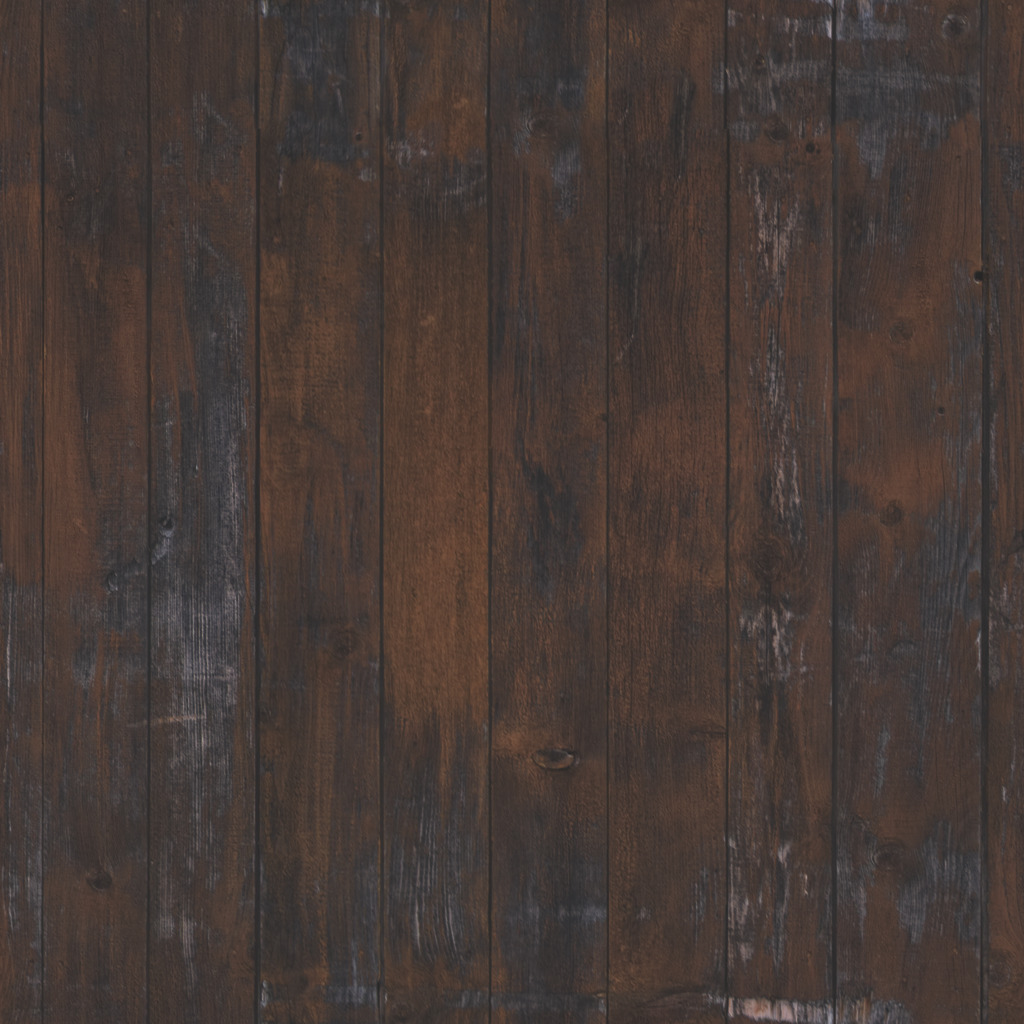}
    \caption{\color{white}$\eta=2.25$}
  \end{subfigure}
  \vfill
  \vspace{2pt}
  \begin{subfigure}[t]{0.2\textwidth}
    \centering
    \includegraphics[width=\linewidth]{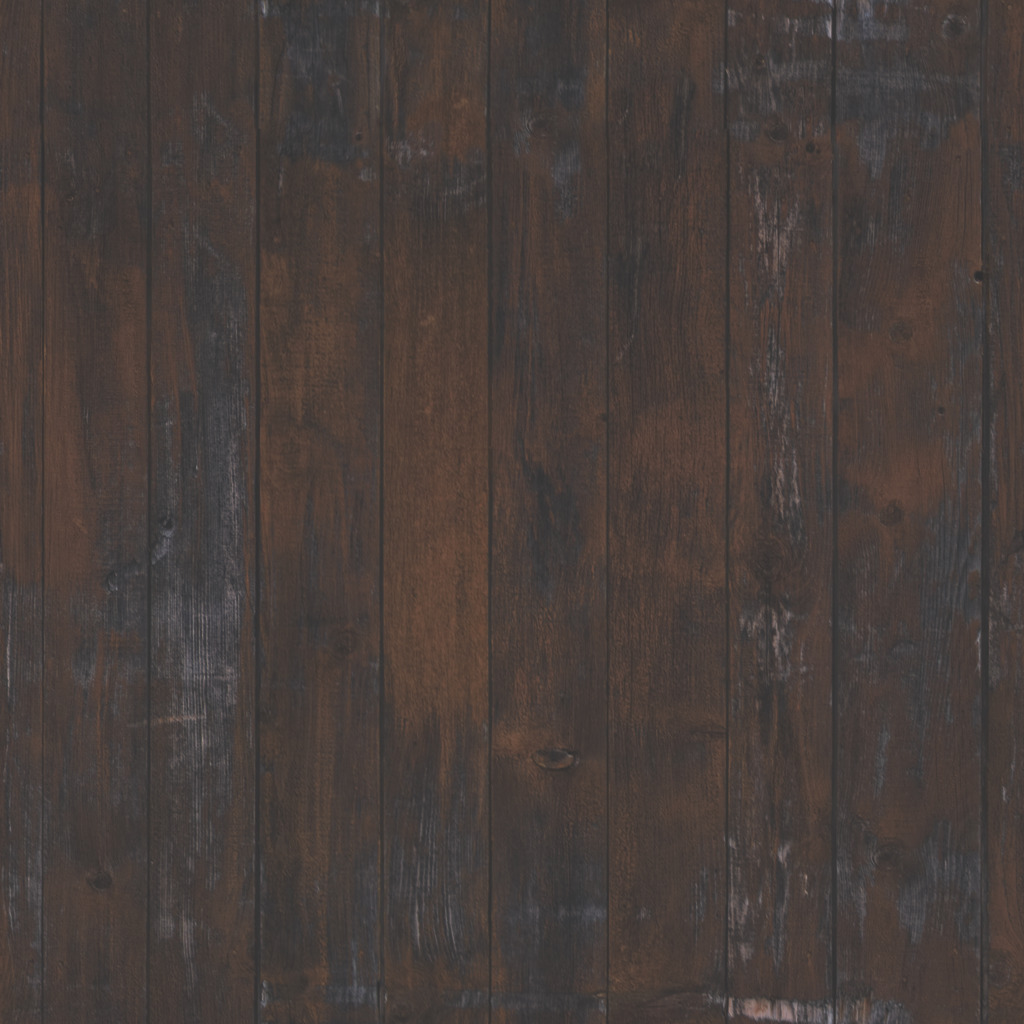}
    \caption{\color{white}$\eta=2.5$}
  \end{subfigure}
  \begin{subfigure}[t]{0.2\textwidth}
    \centering
    \includegraphics[width=\linewidth]{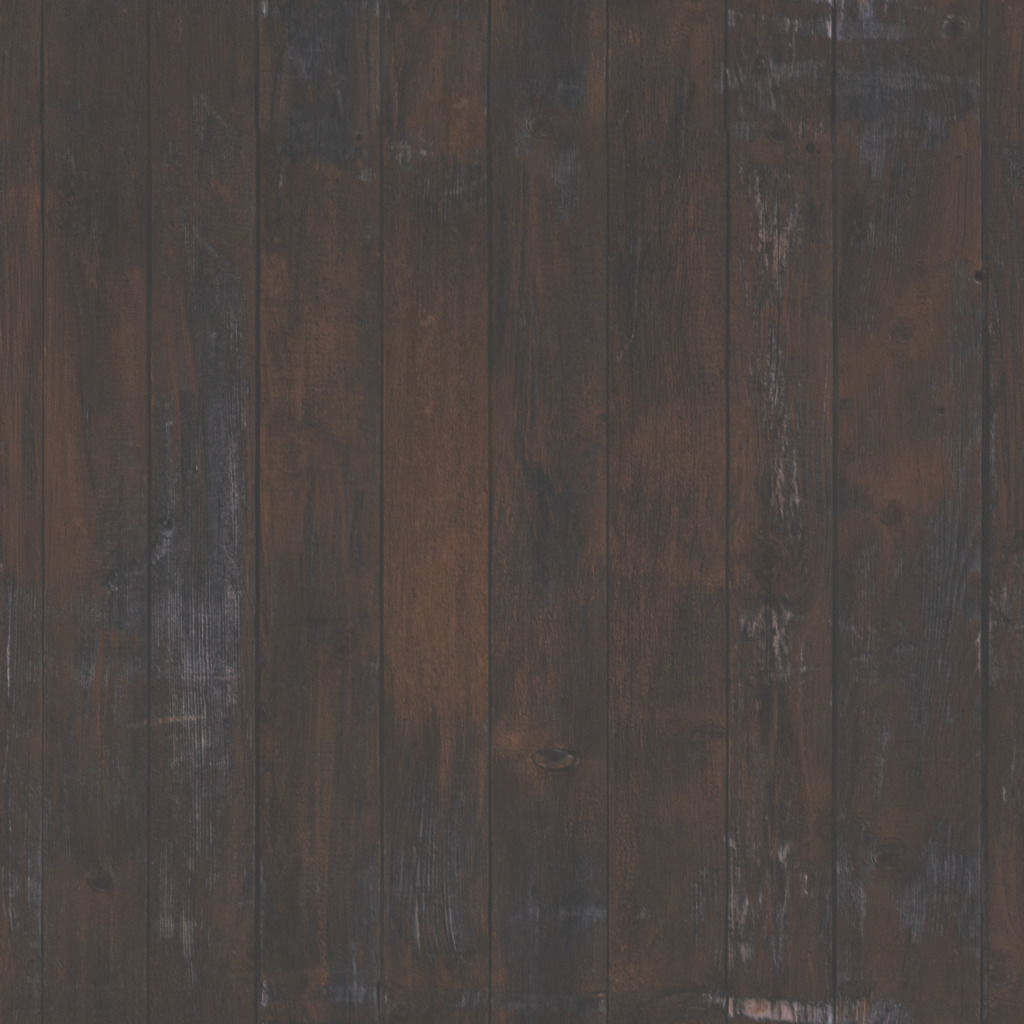}
    \caption{\color{white}$\eta=2.75$}
  \end{subfigure}
  \begin{subfigure}[t]{0.2\textwidth}
    \centering
    \includegraphics[width=\linewidth]{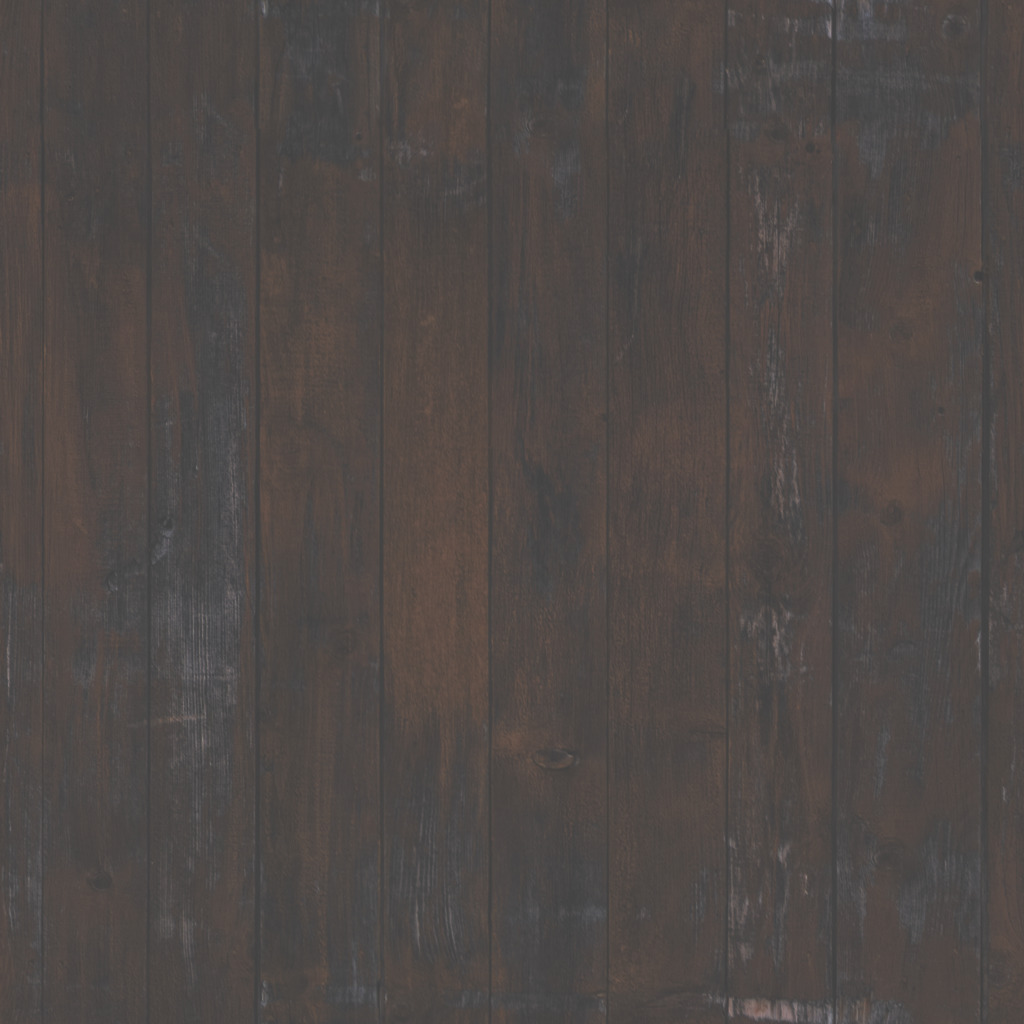}
    \caption{\color{white}$\eta=3$}
  \end{subfigure}
  \caption{Appearance of textured base with clear coats of different IOR $\eta$.}
  \label{fig:clear-coat-textured-base-images}
\end{figure}

Figure~\ref{fig:clear-coat-textured-base-images} shows the appearance at normal incidence of the exact solution for smooth clear coat (under uniform white sky illumination) if the base is textured. The effect appears as ``varnish''-like saturation and darkening for IORs below 2, while beyond this the base is increasingly masked by the Fresnel reflection from the coat.

\subparagraph{Absorbing coat}

The exact BRDF of Equation~\ref{exact_brdf_interfaced_lambertian2} can be expressed as a modification of the albedo-scaling formula for the diffuse BRDF as follows:
\begin{equation} \label{lambertian_modified_albedo_scaling_approx}
  f_b(\mu_i, \mu_o) = \frac{\rho_d}{\pi} \bigl(1 - F_R(\mu_o, \eta)\bigr) \Delta(\rho_d, \mu_i, \eta) \ ,
\end{equation}
where $\Delta$ is a darkening factor given by
\begin{equation} \label{albedo_scaling_darkening}
  \Delta(\rho_d, \mu_i, \eta) =  \frac{1}{\eta^2} \frac{F_T(\mu_i, \eta)} {1 - \rho_d K(\eta, T)}  \ .
\end{equation}

This suggests an improvement of the albedo-scaling approximation, which is to replace $\Delta(\rho_d, \mu_i, \eta)$ with its average over the hemisphere. That is, take
\begin{equation} \label{clearcoat_modified_albedo_scaling_approx}
  f(\mu_i, \mu_o) \approx f_s(\mu_i, \mu_o) + \frac{\rho_d}{\pi} \bigl(1 - F_R(\mu_o, \eta)\bigr) T^{\frac{1}{\mu^t_i} + \frac{1}{\mu^t_o}} \Delta(\rho_d, \eta) \ ,
\end{equation}
where
\begin{equation} \label{coat_darkening_factor_rough_case}
  \Delta(\rho_d, \eta) = \frac{1 - K_0(\eta)} {1 - \rho_d K(\eta, T)} \ .
\end{equation}
This BRDF, though non-reciprocal, has the same directional albedo as the exact result (i.e., Equation~\ref{clearcoat_exact_directional_albedo}).

A further, very rough approximation of $K(\eta, T)$ accounting for absorption is to ignore the angular dependence of $T$ inside the integral, producing $K(\eta, T) \approx T^2 K_0(\eta)$.

\subparagraph{Specular base}

\label{subparagraph:specular_base}

If the base has specular reflection (due to a perfectly smooth metallic or dielectric interface) then the solution described in the previous sections no longer applies.

The result can be obtained straightforwardly using a geometric series summation quite similar to the Airy summation of Equation~\ref{eq:airy}, except ignoring phase factors since we are working in the geometrical optics limit. Denoting the Fresnel factor of the reflection of the light incident to the coat as $F_R$ (and corresponding transmission coefficient $F_T = 1 - F_R$), the Fresnel factor (i.e., albedo) of the reflection of the refracted ray from the base as $\boldsymbol{\rho}_b$, and absorption factor $\mathbf{C} = T^{2/\mu^t_i}$, we can write the total directional albedo of the coated base as (using the fact that the Fresnel factors of the internal and external reflections from the coat both equal $F_R$ assuming a mirror reflection from the base):
\begin{align} \label{eq:smooth_coat_darkening}
 R  &= F_R  + F_T\,\mathbf{C}\,\boldsymbol{\rho}_b\,F_T + F_T\,\mathbf{C}\,\boldsymbol{\rho}_b\,(\mathbf{C}\,F_R\,\boldsymbol{\rho}_b)\,F_T + \cdots \nonumber \\
    &= F_R  + \sum_{k=1}^{\infty} F_T\,\mathbf{C}\,\boldsymbol{\rho}_b \left(\mathbf{C}\, F_R\, \boldsymbol{\rho}_b\right)^{k-1} \, F_T \nonumber \\
    &= F_R  + \frac{F^2_T\,\mathbf{C}\,\boldsymbol{\rho}_b}{1 - F_R\,\mathbf{C}\,\boldsymbol{\rho}_b}.
\end{align}
As required by energy conservation, $R \rightarrow 1$ as $\mathbf{C}, \boldsymbol{\rho}_b \rightarrow 1$.
Re-expressing this in the form of a darkening factor $\mathbf{\Delta}$ applied to albedo scaling, we have
\begin{equation}
  R = F_R + (1 - F_R)\,\mathbf{C}\,\boldsymbol{\rho}_b\,\mathbf{\Delta} \ ,
\end{equation}
where
\begin{equation} \label{coat_darkening_factor_smooth_case}
  \mathbf{\Delta} = \frac{1 - F_R}{1 - F_R\,\mathbf{C}\,\boldsymbol{\rho}_b}.
\end{equation}
This (exact result) has a very similar form to the approximate darkening factor in the case of a Lambertian base, Equation~\ref{coat_darkening_factor_rough_case}, except with $K_0$ replaced by the Fresnel factor of the coat, and $T^2$ replaced by the more accurate $\mathbf{C}$.

\begin{figure}
  \centering
  \begin{subfigure}{.48\textwidth}
    \centering
    \includegraphics[width=\linewidth]{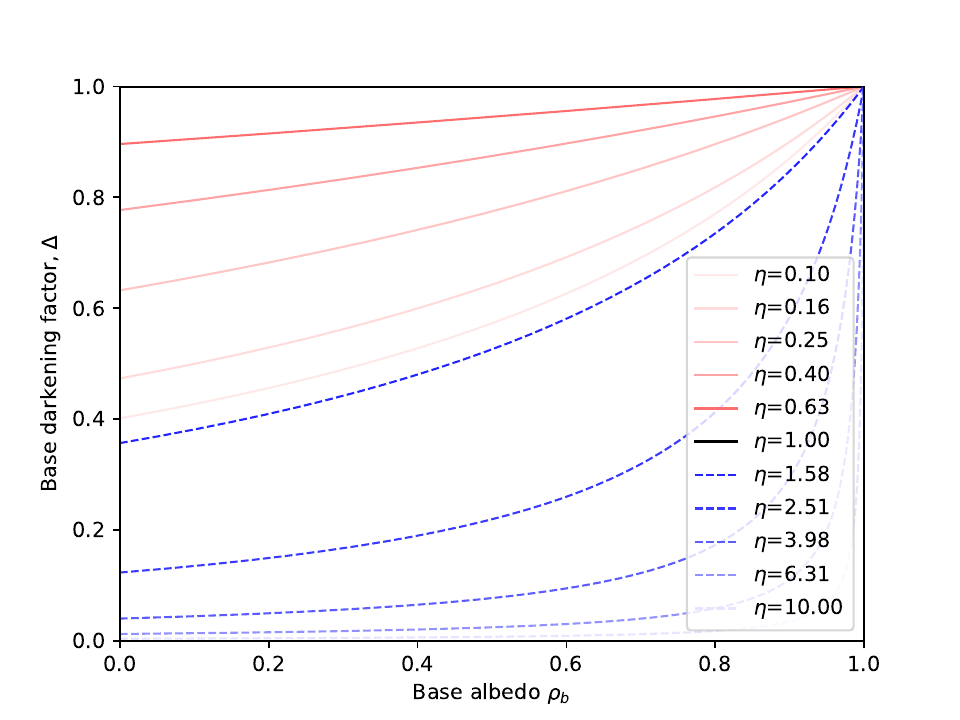}
    \caption{Rough base case, Equation~\ref{coat_darkening_factor_rough_case}}
    \label{fig:layer_darkening_factor_rough}
  \end{subfigure}
  \begin{subfigure}{.48\textwidth}
    \centering
    \includegraphics[width=\linewidth]{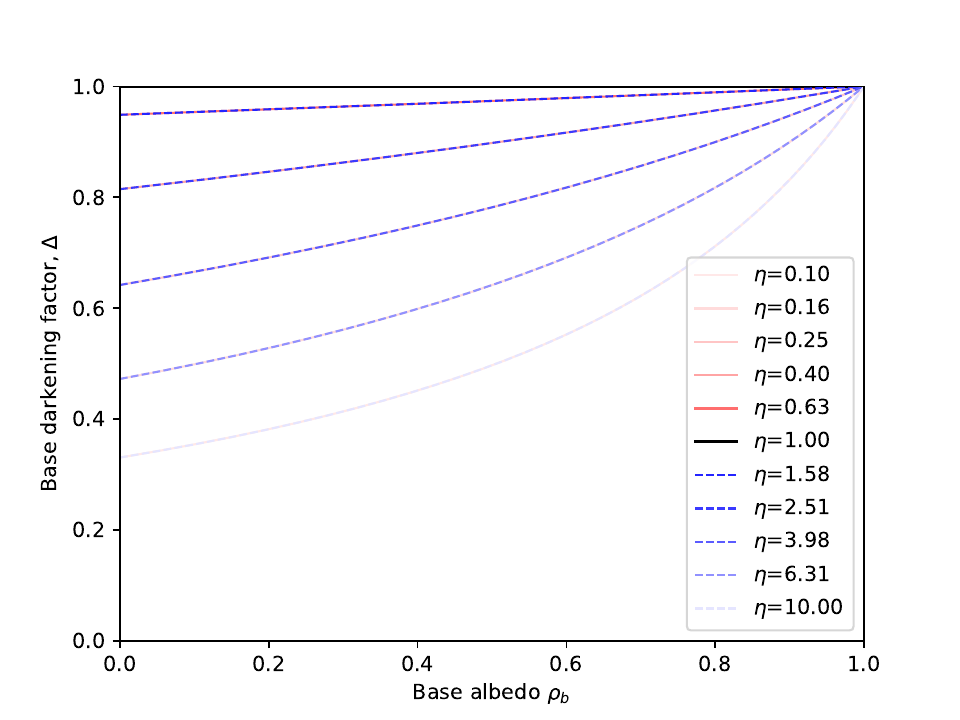}
    \caption{Smooth base case, Equation~\ref{coat_darkening_factor_smooth_case}}
    \label{fig:layer_darkening_factor_smooth}
  \end{subfigure}
  \caption{Darkening factor $\mathbf{\Delta}$ of a base substrate under a smooth clear-coat layer, for rough and smooth base.}
  \label{fig:layer_darkening_factor}
\end{figure}

Figure~\ref{fig:layer_darkening_factor} compares the coat darkening factor $\mathbf{\Delta}$ in the cases of a rough and smooth base substrate (with a clear coat), for various coat IOR ratios $\eta$. A smooth base leads to less darkening (as well as less tinting in the absorptive case) since there is no TIR at all regardless of the angle of incidence, since all rays exit the coat at the same angle they entered (which couldn't have been a TIR direction). In contrast, when diffuse scattering occurs, a very high proportion of rays end up undergoing TIR. This difference in the darkening of a coat for a smooth versus a rough base was also noted by \textcite{Weidlich2023}.

\clearpage

\section*{Revision history}

\begin{itemize}
  \item v1.0 (\textbf{Aug 10, 2025}) -- Initial version for SIGGRAPH 2025 course notes.
  \item v1.1 (\textbf{Aug 13, 2025}) -- Fixed minor issues.
  \item v1.2 (\textbf{Oct 27, 2025})
    \begin{itemize}
      \item Fixed more typos, broken links, and formatting issues.
      \item Added a \hyperref[subparagraph:specular_base]{derivation} of the coat darkening effect for a specular base.
      \item Added brass and steel to the \hyperref[sec:f82_fits]{table} of fits to measured metals.
    \end{itemize}
  \item v1.3 (\textbf{Feb 14, 2026})
      \begin{itemize}
      \item The coat darkening example code (Listing~\ref{listing:coat_darkening}) was improved to account for the view-dependent absorption as well (and some typos in it were fixed).
      \item Fixed broken alpha in some images.
    \end{itemize}
\end{itemize}

\end{document}